\documentclass[a4paper,titlepage]{scrbook}

\usepackage[utf8]{inputenc}
\usepackage[T1]{fontenc}
\usepackage{graphicx}
\usepackage{cancel}
\usepackage{slashed} 
\usepackage[ngerman]{babel}
\usepackage{enumerate}
\usepackage[ntheorem]{defoqua}
\usepackage{stmaryrd}
\usepackage{cite}
\usepackage{palatino} 
\usepackage{fancyheadings}
\input xypic
\usepackage[arrow, matrix,curve]{xy}

\newcommand{\T} {\operatorname{\mathcal{T}}}

\newcommand{\glna}[1] {\operatorname{\stackrel{#1}{=}}}

\newcommand{\Menge}[2] {\operatorname{\{#1\:\|\:#2\}}}

\newcommand{\V} {\mathbb{V}}
\newcommand{\M} {\mathbb{M}}

\newcommand{\net}[2] {\left\{#1_{\alpha}\right\}_{\alpha \in #2}}

\newcommand{\KE}[3] {\operatorname{(#1\otimes #2\otimes #3)}}
\newcommand{\wt}[1] {\widetilde{#1}}

\newcommand{\nk}[1] {\llbracket#1\rrbracket}
\newcommand{\Rm} {\operatorname{R-mod}}
\newcommand{\Rcc} {\operatorname{R-comp}}
\newcommand{\Ab} {\operatorname{Ab}}

\newcommand{\ot} {\operatorname{\otimes}}
\newcommand{\ovl}[1] {\overline{#1}}
\newcommand{\cp} {\operatorname{\circ}}
\newcommand{\pt} {\operatorname{\partial}}
\newcommand{\cpt} {\operatorname{\hat{\partial}}}

\newcommand{\ck}[1] {\operatorname{\mathit{(i_{#1}\cp\delta)(1\ot
      u_{#1}\ot 1)}}}
\newcommand{\nett}[3] {\{#1_{#2}\}_{#3}}
\newcommand{\nettt}[3] {\{#1_{#2}\}_{#2\in #3}}

\newcommand{\im} {\operatorname{\mathrm{im}}}
\newcommand{\sym} {\operatorname{\mathrm{S}^{\bullet}(\mathbb{V})}}

\newcommand{\pite} {\operatorname{\otimes_{\pi}}}

\newcommand{\C} {\operatorname{\mathrm{X}}}

\newcommand{\K} {\operatorname{\mathcal{K}}}
\newcommand{\cK} {\operatorname{\hat{\mathcal{K}}}}
\newcommand{\Ss} {\mathrm{S}}
\newcommand{\SsV} {\mathrm{S}^{\bullet}(\mathbb{V})}
\newcommand{\Tt} {\operatorname{\mathrm{T}}}
\newcommand{\Hol} {\mathrm{Hol(\V)}}

\newcommand{\p} {\operatorname{\mathfrak{p}}}
\newcommand{\q} {\operatorname{\mathfrak{q}}}
\newcommand{\Pp} {\operatorname{\mathfrak{P}}}
\newcommand{\bbot} {\operatorname{\boldsymbol{\otimes}}}
\newcommand{\Ae} {\operatorname{\mathcal{A}^{e}}}
\newcommand{\cod} {\operatorname{\mathrm{cod}}}
\newcommand{\dom} {\operatorname{\mathrm{dom}}}
\newcommand{\Ob} {\operatorname{\mathrm{Ob}}}

\newcommand{\Poly} {\operatorname{\mathrm{Pol}(\mathbb{R}^{n})}}

\newcommand{\DiffOp}[2]{\mathrm{DiffOp}_{#1}^{#2}(\mathcal{A},\mathcal{M})}
\newcommand{\DiffOpc}[2] {\mathrm{DiffOp}_{#1}^{#2,\cont}(\mathcal{A},\mathcal{M})}
\newcommand{\DiffOpS}[2] {\mathrm{DiffOp}_{#1}^{#2}(\Ss^{\bullet}(\V),\mathcal{M})}
\newcommand{\diff}{\operatorname{\mathrm{diff}}}
\newcommand{\cont}{\operatorname{\mathrm{cont}}}
\newcommand{\vv}{\operatorname{\mathrm{v}}}

\begin{document}

\pagestyle{empty}

\begin{titlepage}
    \begin{center}
    \end{center}
    \vspace{1.0cm}
    \begin{center}
        \LARGE
        Hochschild-Kohomologien\\
        von\\ Observablenalgebren\\in der\\ Klassischen  Feldtheorie
    \end{center}
    \normalsize
    \vspace{2.0cm}
    \begin{center}
        \Large Diplomarbeit\\ 
        \vspace{0.4cm}
        \large vorgelegt von\\
        \vspace{0.5cm}
        \Large Maximilian Hanusch\\
        \vspace{0.5cm}
        \normalsize September 2010
    \end{center}
    \vspace{1.0cm}
    \begin{center}
        \large Wissenschaftliche Betreuung:\\
        \vspace{0.2cm}
        \large Apl. Prof. Dr. Stefan Waldmann
    \end{center}
    \vspace{1.0cm}
    \begin{center}
        \large\sc{Physikalisches Institut\\
          Fakult\"at f\"ur Mathematik und Physik\\
          Albert-Ludwigs-Universit\"at Freiburg}
    \end{center}
    \normalsize
\end{titlepage}


\cleardoublepage

\frontmatter

\tableofcontents
\clearpage
\thispagestyle{empty}

\pagestyle{fancy}
\fancyhf{} 
\fancyhead[OR]{\leftmark} 
\fancyhead[EL]{\leftmark} 
\fancyfoot[C]{\thepage} 

\chapter{Einleitung}
Dem Wesen theoretischer Arbeiten im Bereich der mathematischen Physik
entsprechend, bereichert auch diese Diplomarbeit auf zweierlei Weisen.
Zum einen bietet sie eine Fülle an mathematischem
Abwechslungsreichtum, da hier sowohl algebraische als auch
topologischen Methoden zum Einsatz kommen, die das Erreichen
der im Rahmen dieser Arbeit erstrebten Ziele durch ihr elegantes
Zusammenspiel überhaupt erst ermöglichen. Die Kombination von 
abtrakter homologischer Algebra mit Funktional-Analytischen Konzepten
wird unter Ausnutzung explizit konstruierter Kettenabbildungen tiefere Erkenntnisse
über die gewünschten Hochschild-Kohomologien liefern, die im Rahmen des
allgemeinen algebraischen Formalismuses nicht greifbar wären und über
reine Isomorphieaussagen weit hinaus gehen.
Die physikalische Motivation ist dabei von dem innigen Wunsch getragen,
eine der grundlegendsten physikalischen Theorien, die
Quantenfeldtheorie, in ihrer Natur zu ergründen und ihr einzigartiges Zusammenwirken
mit den altbewährten klassichen Theorien besser zu verstehen. Der
wichtige Beitrag soll hierbei im Rahmen Deformationsquantisierung
geleistet werden, die als Bindeglied zwischen den klassischen und
den moderneren Quantenfeldtheorien anzusehen ist.

\section*{Motivation} 
Die Quantisierung einer klassischen Theorie ist eine 
oftmals schwer zu fassende Prozedur, die sich eher
formalen Argumenten bedient, als wirklich die grundlegenden
Zusammenhänge aufzuzeigen. Hierbei erweist sich die quantisierte Theorie
im Allgemeinen als die fundamentalere von beiden, jedoch ist nicht
ignorierbar, dass auch die klassiche Theorie die Natur im Rahmen ihres
Gültigkeitsbereiches vortrefflich beschreibt. Insofern ist es eine interessante und zudem
absolut nicht-triviale Frage, unter
welchen Bedingungen und aus welchem Grund die klassiche Theorie der allgemeineren
Quantentheorie vorzuziehen ist. Eng damit verbunden ist Frage,
inwiefern wir unsere bisherigen Ansichten über die täglich erlebte
Realität in Frage stellen müssen. Denn es ist sicher nicht klar, ob
die Quantenfeldtheorie oder spezieller die Quantenmechanik nicht auch in anderen Bereichen als dem Mikrokosmos
anwendbar ist. Imposante Beispiele sind hierbei sicher das
Konzept der Superposition von Zuständen und die Rolle des Messprozesses an sich.
Diese haben in der klassichen Mechanik keine Bedeutung, sorgen jedoch in der
Quantenmechanik dafür, dass sich die physikalische Realität
eines ganzen Quantensystems allein durch eine Messung vollkommen
verändert werden kann.
\subsection*{Deformation von Observablenalgebren}
In der klassischen Mechanik besteht der Konfigurationsraum aus Orts-
und-Impulsko\-or\-dinaten $q$ und $p$ und kann bei einem $n$-Teilchensystem als
$\mathbb{R}^{2n}$ aufgefasst werden. Geläufige assoziativen,
kommutative Observablenalgebren
sind hierbei die Polynome
$\Pol\left(\mathbb{R}^{n}\right)$ oder die glatten Funktionen
$C^{\infty}\left(\mathbb{R}^{2n}\right)$, die vermöge der total antisymmetrischen
Poisson-Klammer 
\begin{equation*}
    \{f,g\}=
    \displaystyle\sum_{k=1}^{n}\frac{\pt f}{\pt q^{k}}\frac{\pt g}{\pt
      p_{k}}-\frac{\pt g}{\pt q^{k}}\frac{\pt f}{\pt p_{k}}\qquad \forall\:f,g
    \in \Pol\left(\mathbb{R}^{2n}\right)\text{ oder } C^{\infty}\left(\mathbb{R}^{2n}\right),
\end{equation*} welche die Leibnizregel und die Jacobiidentität erfüllt,
zu einer Poisson-Algebra werden. Auf der anderen Seite besteht der
Konfigurationsraum der Quantenmechanik aus einem Hilbertraum
$\mathcal{H}$ und die Observablen aus der $^{*}$-Unteralgebra der
beschränkten Operatoren $\mathcal{B}(\mathcal{H})$, wobei die hierin
enthaltenen selbstadjungierten
Operatoren die tatsächlich physikalisch messbaren Observablen sind.

Die Deformationstheorie assoziativer Algebren legt uns nun ein effektives
Werkzeug in die Hand, Quantenobservablen aus denen der klassichen
Theorie derart zu konstruieren, dass die Nichtkommutativität der
neuen Algebramultiplikation $\star$ bereits gewährleistet ist. Hierbei
betrachtet man für
eine klassiche Observablenalgebra $(\mathcal{A},*)$
den Raum $\mathcal{A}\llbracket
\hbar\rrbracket$
der formalen Potenzreihen in dem formalen Parameter $\hbar$ mit
Koeffizienten in $\mathcal{A}$, in welchen man sich $\mathcal{A}$ im
Vektorraum-Sinne als
Monome $0$-ter Ordnung eingebettet vorstellen kann.
Eine formale
Deformation $\mu$ der Ordnung $k$ ist dann (vgl. \cite{bayen.et.al:1978a}) 
eine
$\mathbb{C}\nk{\hbar}$-bilineare Multiplikation
$\mu \colon \mathcal{A}\nk{\hbar}\times
\mathcal{A}\nk{\hbar}\longrightarrow \mathcal{A}\nk{\hbar}$
der Form            
\begin{equation*}
    \mu(a, b) = \sum_{r=0}^{k}\hbar^{r}
    \mu_{r}(a,b)\qquad\forall\:a,b\in \mathcal{A}
\end{equation*}mit $\mathbb{C}$-bilinearen Abbildungen
$\mu_{r}\colon \mathcal{A}\times \mathcal{A}\longrightarrow \mathcal{A}$,
so dass $\mu$ folgende Eigenschaften besitzt:
\begin{enumerate}
\item
    $\mu$ ist assoziativ bis zur Ordnung $k$.
\item
    $\mu_{0}(a,b)=a*b$.
\item
    $\mu_{1}(a,b)-\mu_{1}(b,a)=i\{a,b\}$.
\end{enumerate}
Hierbei ist \textit{iii.)} auch als
\emph{Korrespondezprinzip} bekannt und garantiert, dass
der total antisymmetrische Teil des Quanten-Kommutators $[a,b]:=\mu(a, b)- \mu(b, a)$ in
der Ordnung $\hbar$ mit $i\{\cdot,\cdot\}$ übereinstimmt. Ist $\hbar$
in der physikalischen Situation eine dimensions-behaftete Größe, so
sind die $\mu_{r}$ als Größen der Dimension $[(Js)^{-r}]$ zu verstehen,
womit $\mu$ eine Multiplikation
$\mathcal{A}\times \mathcal{A}\rightarrow \mathcal{A}$ definiert.
Im Falle $k=\infty$ spricht man von einem Sternprodukt und schreibt
$\star$ anstelle von $\mu$. Diese stellen die eigentlich interessanten
Objekte dar, wobei hier die physikalische  Wohldefiniertheit von $\mu$, also die
Konvergenz der Summe, ein im
Allgemeinen ungelöstes Problem darstellt. Die Existenz und Klassifikation solcher Sternprodukte sind
 für die klassiche Situation wohlverstandene Probleme und wurde für die Algebra $C^{\infty}(M)$ einer
  endlich-dimensionalen symplektischen Mannigfaltigkeit $M$ erstmals von Lecomte und de Wilde
\cite{dewilde.lecomte:1983b, dewilde.lecomte:1988a} unter
Verwendung kohomologischer Überlegungen gelöst. Einen sehr
einfachen und geometrischen Existenzbeweis liefert zudem die
Fedosov-Konstruktion, die ohne kohomologische Überlegungen auskommt
und sich lediglich "`konventioneller"' Techniken wie kovarianter
Ableitungen und dem Tensorkalkül bedient, vgl. \cite{fedosov:1996a}. Ganz allgemein für endlich-dimensionale
Poisson-Mannigfaltigkeiten wurde das Existenz- und Klassifikations-Problem von
Kontsevich (vgl. \cite{kontsevich:2003a}) gelöst.

Ein einfache Verfahren um Sternprodukte zu erhalten, ist, sich diese
Ordnung für Ordnung zu konstruieren. dabeibei sind die Assoziativität
von $\mathcal{A}$ und $\mathcal{A}\nk{\hbar}$ die entscheidenden
Faktoren. Hierfür betrachten wir eine formale Deformation $\mu=\mu_{0}+…+\mu_{k}$
der Ordnung $k$, die wir durch ein $\mathbb{C}$-bilineares $\mu_{k+1}$ zu einer formalen
Deformation $\circ=\mu + \hbar^{k+1}\mu_{k+1}$ der Ordnung $k+1$ fortsetzen wollen. Dann muss die
Bedingung $a\circ(b\circ c)=(a\circ b)\circ c$ insbesondere für alle $a,b,c \in
\mathcal{A}$ bis zur Ordnung $k+1$
erfüllt sein, in welcher wir erhalten, dass\footnote{vgl. \cite[Kapitel 2]{Weissarbeit}}
\begin{align*}
    a*\mu_{k+1}(b,c) - \mu_{k+1}(a*b,c)  + &\:\mu_{k+1}(a,b*c)-
    \mu_{k+1}(a,b)*c
\\ &=\underbrace{\sum_{r=1}^{k}\big[\mu_{r}(\mu_{k+1-r}(a,b),c)-\mu_{r}(a,\mu_{k+1-r}(b,c))\big]}_{R_{k}}
\end{align*}gilt, was mit Hilfe des Hochschild-Differentials auch in der
Form $\delta\mu_{k+1}=R_{k}$ geschrieben werden kann. Eine
längere Rechnung unter Ausnutzung der Assoziativität von $\mathcal{A}$
zeigt zudem $\delta R_{k}=0$, womit ein derartiges $\mu_{k+1}$ nur
dann gefunden werden kann, wenn $[R_{k}]$ die $0$-Klasse ist.
In diesem Sinne bilden die Elemente der dritten
Hochschild-Kohomologie $HH^{3}(\mathcal{A},\mathcal{A})$ der Algebra $\mathcal{A}$, die
Quelle von Obstruktionen für die Fortsetzbarkeit 
formaler Deformationen zu Sternprodukten und es
lässt sich zeigen, dass die zweite Hochschild-Kohomologie
die Äquivalenzklassen von infinitisimalen Deformationen,
also solchen bis zur Ordnung $1$ klassifiziert. Ist hingegen ein
Sternprodukt $\left(\mathcal{A}\nk{\hbar},\star\right)$ für eine
assoziative und
kommutative Algebra $\mathcal{A}$ vorgegeben und hat man einen $\mathcal{A}$-Modul $\mathcal{M}$, dessen Modulstruktur man
auf $\mathcal{A}\nk{\hbar}$ fortsetzen möchte, so sind
die Hoch\-schild-Ko\-ho\-mologien
$HH^{2}(\mathcal{A},\End_{\mathbb{K}}(\mathcal{M}))$ und
$HH^{1}(\mathcal{A},\End_{\mathbb{K}}(\mathcal{M}))$ von
entscheidender
Wichtigkeit\footnote{$\End_{\mathbb{K}}(\mathcal{M})$ ist
  hier als $\mathcal{A}-\mathcal{A}$-Bimodul aufzufassen} für das
Deformationsproblem. 

Die Berechnung dieser Hochschild-Kohomologien ist ein für Observablenalgebren auf endlich-dimensionalen
Vektorräume, wie es beispielsweise der Konfigurationsraum der klassichen
Mechanik ist, gut verstandenes Problem. Die Kohomologie-Gruppen 
$HH^{k}\left(\Pol\left(\mathbb{R}^{n}\right),\Pol\left(\mathbb{R}^{n}\right)\right)$
wurden erstmals von Hochschild, Kostant und Rosenberg im Rahmen des 
Hochschild-Kostant-Rosenberg-Theoremes bestimmt
(vgl. \cite{hochschild.kostant.rosenberg:1962a}), welches inbesondere
besagt, dass jede Kohomologieklasse $[\eta]\in
HH^{k}\left(\Pol\left(\mathbb{R}^{n}\right),\Pol\left(\mathbb{R}^{n}\right)\right)$
genau einem $k$-Multivektorfeld entspricht. Analoge Aussagen wurden ebenfalls
für lokalen,  stetigen und
die differentiellen Hochschild-Koho\-mo\-lo\-gien
\begin{align*}
    &HH^{k}_{\diff}(C^{\infty}(M),C^{\infty}(M)),\\
    &HH^{k}_{\mathrm{loc}}(C^{\infty}(M), C^{\infty}(M))\qquad\text{und}\\
    &HH^{k}_{\cont}(C^{\infty}(M), C^{\infty}(M))
\end{align*}
gezeigt
\cite{pflaum:1998a, connes:1994a ,cahen.gutt.dewilde:1980a}.   In
 \cite{Weissarbeit} wurden zudem die stetig-differentielle
 Hoch\-schild-Kohomologie $HH_{\mathrm{c,d}}^{k}(C^{\infty}(V),\mathcal{M})$  für
 differentielle $C^{\infty}(V)-C^{\infty}(V)$-Bimoduln $\mathcal{M}$ mit einer
 konvexen Teilmenge $V\subseteq \mathbb{R}^{n}$ berechnet.

\subsection*{Observablenalgebren in der Feldtheorie, die symmetrische Algebra}
In jeder klassichen Feldtheorie
besteht der Konfigurationsraum  aus
Feldern, die im Speziellen selbst die glatten Funktionen auf einer Mannigfaltigkeit, aber in jedem Falle unendlich-dimensionale
$\mathbb{K}$-Vektorr"aume\footnote{Hier und im Folgenden bedeutet
  $\mathbb{K}$ immer $\mathbb{R}$ oder
  $\mathbb{C}$.} sind. Ist zum Beispiel $X\subseteq \mathbb{R}^{n}$ eine
offene Teilmenge und $\mathcal{D}(X)$ der Testfunktionen-Raum
der glatten Funktionen von $X\longrightarrow\mathbb{R}$ mit Kompaktem Träger
in $X$. Dann ist man insbesondere an den Observablen der Form
\begin{equation*}
    p(\psi)=\sum_{k=0}^{n}\int_{X_{1}\times…\times X_{k}}\phi_{k}(x_{1},…,x_{k})\:\psi(x_{1})…\psi(x_{k})\:dx_{1}…dx_{k}
\end{equation*}mit $n\in \mathbb{N}$, $\psi\in \mathcal{D}(X)$, $X_{i}=X$ für alle $1\leq
i\leq k$ und $\phi_{k}\in \mathcal{E}^{\mathrm{sym}}_{\mathrm{sep}}\left(X^{k}\right)$
interessiert. Hierbei bezeichnet
{\small\begin{equation*}
      \mathcal{E}_{\mathrm{sep}}\left(X^{k}\right)=\left\{\phi\in
          \mathcal{E}\left(X^{k}\right)\:\Bigg|\:\phi(x_{1},…,x_{k})=\sum_{i=1}^{n}\phi_{1}(x_{1})…\phi_{k}(x_{k})\:\:\forall\:(x_{1},…,x_{k})\in
          X^{k}\right\}
  \end{equation*}}mit $\phi_{1},…,\phi_{k}\in \mathcal{E}(X)=C^{\infty}(X)$ für alle $1\leq
i\leq k$ den Vektorraum aller Abbildungen $X^{k}\longrightarrow \mathbb{R}$, die als endliche Summe
faktorisierender, glatter Funktionen geschrieben werden können und
$\mathcal{E}^{\mathrm{sym}}_{\mathrm{sep}}\left(X^{k}\right)$ den
Unterraum der
total symmetrische Elemente
von $\mathcal{E}_{\mathrm{sep}}$. Jedes
$\mathcal{E}_{\mathrm{sep}}\left(X^{k}\right)$ ist eine Realisierung
des $k$-fachen Tensorproduktes $\Tt^{k}(\mathcal{E}(X))$ und
$\mathcal{E}^{\mathrm{sym}}_{\mathrm{sep}}\left(X^{k}\right)$ eine
Realisierung des symmetrischen Tensorproduktes
$\Ss^{k}(\mathcal{E}(X))$. Mit Hilfe der Abbildung
\begin{equation*}
    \tau\colon (\phi,\psi)\longrightarrow \int_{X}\phi(x)\psi(x)dx
\end{equation*}und linearer Fortsetzung von
\begin{align*}
    \Delta\colon \Tt^{k}(\mathcal{E}(X))\times \mathcal{D}(X)&\longrightarrow
    \mathbb{R}\\
    (\phi_{1}\ot…\ot \phi_{k},\psi)&\longmapsto \tau(\phi_{1},\psi)…\tau(\phi_{k},\psi)
\end{align*}auf die gesamte symmetrische Algebra
$\Ss^{\bullet}(\mathcal{E}(X))$, entspricht diese gerade den
Observablen der obigen Form. Möchte man hingegen auch
unendliche Summen und symmetrische $\phi_{k}$ in ganz
$\mathcal{E}\left(X^{k}\right)$ zulassen, so ist dies möglich, indem
man $\mathcal{E}(X)$
mit der üblichen Fr\'echet-Topologie versieht und die symmetrische Algebra mit
Hilfe des Konzeptes des $\pi$-Tensorproduktes lokalkonvex
topologisiert. Durch Vervollständigung von
$\Ss^{\bullet}(\mathcal{E}(X))$ 
erhält man eine Algebra $\mathrm{Hol}(\mathcal{E}(X))$, welche die
gewünschten Observablen\footnote{Unter der Vorraussetzung, dass diese
  gewisse "`Konvergenzbedingungen"' erfüllen.} induziert. 
Sind
in diesem Rahmen Feldgleichungen auf $\mathcal{D}(X)$ durch einen
linearen Operator $\Lambda$ gegeben, so lässt sich unter gewissen
Voraussetzungen zeigen (vgl. \cite{baer.ginoux.pfaeffle:2007a}), dass
zu $\Lambda$ gehörige avancierte und retardierte Greensche
Funktionen $G_{x,y}^{+}$ und $G^{-}_{x,y}$ existieren mit denen man
durch derivative Fortsetzung von
\begin{align*}
      \{\phi,\psi\}=\Delta(\phi,\psi)\qquad \forall\:\phi,\psi\in
      \mathcal{E}(X)\quad\text{mit}\quad \Delta=G_{x,y}^{+}-G^{-}_{x,y}
\end{align*} auf ganz $\Ss^{\bullet}(\mathcal{E}(X))$, eine stetige Poisson-Klammer erhält, die  stetig bilinear auf
$\mathrm{Hol}(\mathcal{E}(X))$ fortgesetzt werden kann. Hierfür existieren bereits
eine Fülle an Beispielen für Sternprodukte (vgl,
\cite{duetsch.fredenhagen:2001a},\cite{duetsch.fredenhagen:2003a}), die
mit den Resultaten dieser Arbeit nun in einem formalen Rahmen
behandelbar sind.

Das eben behandelte Beispiel ist lediglich der Spezialfall eines
allgemeinen Konzeptes, das es erlaubt, sich Observablenalgebren mit
Hilfe der symmetrischen Algebra zu konstruieren. Hierfür benötigt man lediglich einen
Konfigurationsraum $\V$ und einen hausdorffschen, lokalkonvexen Vektorraum
$(\mathbb{U},P)$ mit einer $\mathbb{K}$-bilineare Abbildung     
$\tau\colon \V\times \mathbb{U}\colon \longrightarrow \mathbb{K}$, deren Bild sich für festes $v\in \V$ für alle $u\in
\mathbb{U}$ durch eine Halbnorm aus $P$ abschätzen lässt. Dann definiert jedes Element aus
$\mathrm{Hol}(\mathbb{U})$ eine auf ganz $\V$ konvergente
Potenzreihenfunktion. Ist beispielsweise $\V=\mathbb{C}^{n}$ und
$\mathbb{U}=\mathbb{C}^{n*}$ schwach*-topologisiert, so entspricht
$\mathrm{Hol}(\mathbb{C}^{n*})$ gerade den ganz holomorphen Funktionen auf
$\mathbb{C}^{n}$, was gleichzeitig der Grund für die
Namensgebung $\mathrm{Hol}$ ist. Im Falle $\V=\mathcal{E}(X)$ mit schwach*-topologisiertem
$\mathbb{U}=\mathcal{E}'(X)$ ist dann beispielsweise die Exponentialfunktion
\begin{equation*}
    p(\phi)=\sum_{k=0}^{\infty}\frac{1}{k!}\overbrace{\delta_{z}(\phi)…\delta_{z}(\phi)}^{k-mal}=\sum_{k=0}^{\infty}\frac{1}{k!}\phi(z)^{k}
\end{equation*}in $\mathrm{Hol}(\mathcal{E}'(X))$ enthalten und die
angeführten Beispiele bilden nur einen Bruchteil der Kombinationen, die möglich sind.

\section*{Ziele dieser Arbeit}
Die symmetrische Algebra über einem beliebigen $\mathbb{K}$-Vektorraum
$\V$ besitzt die Eigenschaft, dass sie eine Fülle an wichtigen
Observablenalgebren der klassichen Feldtheorie als Spezialfall enthält.
Das Ziel dieser Arbeit soll es daher sein, die
Elementarbausteine der Deformationsquantisierung, die
Hochschild-Kohomologien, dieser reichhaltigen Algebra zu berechnen um
hiermit die Deformationstheorie dieser Observablenalgebren
auf ein festes Fundament zu stellen. Zudem wollen wir in den wichtigen
Spezialfällen lokalkonvexer Vektorräume $\V$ die interessantere stetige
Hochschild-Kohomologien von $\SsV$ und im hausdorffschen
Fall ebenfalls die ihrer noch umfassenderen Vervollständigung $\Hol$ berechnen, welche unter anderem
solch wichtige Observablen wie die Exponentialfunktion
enthält. Hierbei ist der Nutzen der stetigen Hochschild-Kohomologien
darin zu sehen, als dass Sternprodukte, die aus stetigen bilinearen 
Komponenten $\mu_{r}$ bestehen, im Allgemeinen reguläreres Verhalten
zeigen und somit leichter zu handhaben sind.
\section*{Resultate}
Sei im Folgenden $\V$ ein beliebig-dimensionaler $\mathbb{K}$-Vektorraum mit
$\mathbb{K}=\mathbb{R}$ oder $\mathbb{C}$. Dann bezeichnet
$(\SsV,\vee)$ die symmetrische Algebra über $\V$ und $\mathcal{M}$ einen $\SsV-\SsV$-Bimodul, der
ebenfalls ein $\mathbb{K}$-Vektorraum ist. 
Befinden wir uns im lokalkonvexen Rahmen, so verstehen wir $(\V,\T_{P})$ als
lokalkonvexen Vektorraum mit erzeugendem Halbnormensystem $P$ und $\SsV$
denken wir uns dann, vermöge dem submultiplikativen Halbnormensystem
$\Pp$, bestehen aus den Elementen
\begin{equation*}
    \p(\omega)=\sum_{k=0}^{\infty}p^{k}(\omega_{k})\qquad
    p\in \tilde{P},\:\SsV\ni\omega=\sum_{k}\omega_{k}\text{ mit }\omega_{k}\in \Ss^{k}(\V), 
\end{equation*} lokalkonvex topologisiert. Hierbei bezeichnet
$\tilde{P}$ das filtrierende System aller bezüglich $\T_{P}$ stetigen
Halbnormen und wegen der Submultiplikativität ist $\vee$
stetig. Ist $(\V,\T_{P})$ hausdorffsch, so bezeichnet $(\Hol,*)$ die lokalkonvexe Algebra mit
submultiplikativem Halbnormensystem $\hat{\Pp}$, die durch
Vervollständigung von $(\SsV,\vee)$ erhalten wird. In diesem Rahmen
ist $\mathcal{M}$ zudem als lokalkonvexer Vektorraum derart zu
verstehen, dass die Modul-Multiplikationen stetig sind.\\\\ 
In jedem Fall verlangen wir, dass die Modul-Multipliaktionen
$\mathbb{K}$-bilinear sind und dass $\mathcal{M}$ verträglich ist,
dass also $1*_{L}m=m=m*_{R} 1$ für alle $m\in \mathcal{M}$ gilt.\\\\
\emph{Die Resultate dieser Arbeit sind:}
\begin{itemize}
\item
    \textbf{Satz}\\
    \begin{enumerate}
    \item
        Gegeben ein $\SsV-\SsV$-Bimodul $\mathcal{M}$, dann gilt:
        \begin{equation*}
            HH^{k}(\Ss^{\bullet}(\mathbb{V}),\mathcal{M})\cong H^{k}\left(KC(\V,\mathcal{M}),\Delta\right).
        \end{equation*}
        Ist $\mathcal{M}$ zudem symmetrisch, so ist:
        \begin{equation*}
            HH^{k}(\Ss^{\bullet}(\mathbb{V}),\mathcal{M})\cong \Hom^{a}_{\mathbb{K}}\big(\V^{k},\mathcal{M}\big).
        \end{equation*}
    \item
        Gegeben ein lokalkonvexer, $\SsV-\SsV$-Bimodul $\mathcal{M}$, dann gilt:
        \begin{equation*}
            HH^{k}_{\cont}\left(\Ss^{\bullet}(\mathbb{V}),\mathcal{M}\right)\cong
            H^{k}\big(KC^{\cont}(\V,\mathcal{M}),\Delta^{c}\big)
        \end{equation*}
        Ist $\mathcal{M}$ zudem symmetrisch, so ist:
        \begin{equation*}
            HH^{k}_{\cont}(\Ss^{\bullet}(\mathbb{V}),\mathcal{M})\cong\Hom_{\mathbb{K}}^{a,\cont}(\V^{k},\mathcal{M}).
        \end{equation*}
    \item
        Gegeben ein vollständiger, hausdorffscher, lokalkonvexer
        $\Hol-\Hol$-Bimodul $\mathcal{M}$, dann gilt:
        \begin{equation*}
            HH^{k}_{\cont}\big(\Hol,\mathcal{M}\big)\cong HH^{k}_{\cont}\big(\Ss^{\bullet}(\V),\mathcal{M}\big). 
        \end{equation*} Ist $\mathcal{M}$ zudem symmetrisch, so ist:
        \begin{equation*}        
            HH^{k}_{\cont}(\Hol,\mathcal{M})\cong \Hom_{\mathbb{K}}^{a,\cont}(\V^{k},\mathcal{M}).
        \end{equation*}
    \end{enumerate}
    Hierbei bezeichnet $\Hom_{\mathbb{K}}^{a}(\V^{k},\mathcal{M})$ die
    total antisymmetrischen, $\mathbb{K}$-multilinearen Abbildungen von
    $\V^{k}$ nach $\mathcal{M}$ und $\V^{k}$ das $k$-fache
    kartesische Produkt von $\V$. Mit
    $\left(KC(\V,\mathcal{M}),\Delta\right)$ ist der Kokettenkomplex
    mit Kettengliedern $KC^{k}(\V,\mathcal{M})=\Hom_{\mathbb{K}}^{a}(\V^{k},\mathcal{M})$
    und Kettendifferentialen 
    \begin{equation*}
        (\Delta^{k}\phi)(v_{1},\dots, v_{k+1})=
        \sum_{l=1}^{k+1}(-1)^{l-1}\:[u_{l}*_{L}-\:u_{l}\:*_{R}]\:\phi(v_{1},\dots,\blacktriangle^{l},\dots,v_{k+1})
    \end{equation*}
    gemeint. Schließlich bezeichnet $\left(KC_{cont.}(\V,\mathcal{M}),\Delta^{c}\right)$
    den stetigen Unterkomplex von
    $\left(KC(\V,\mathcal{M}),\Delta\right)$ mit
    Kettengliedern
    $\Hom_{\mathbb{K}}^{a,\cont}(\V^{k},\mathcal{M})\subseteq\Hom_{\mathbb{K}}^{a}(\V^{k},\mathcal{M})$. Die
    Isomorphien in Teil \textit{iii.)} werden hierbei durch die
    Einschränkungs-Abbildungen $\tau^{k}\colon \hat{\phi}\longrightarrow
    \hat{\phi}\big|_{\SsV^{k}}$ induziert, die einen
    Kettenisomorphismus $\tau$ zwischen dem Hochschild-Komplex von
    $\Hol$ und dem von $\SsV$ definieren. Die Isomorphie in Teil
    \textit{i.)} ist abstrakter Natur, kann aber auch durch explizite
    Kettenabbildungen $F$ und $G$ oder genauer durch die aus ihnen durch
    Anwendung des
    $\hom_{\mathcal{A}^{e}}(\cdot,\mathcal{M})$-Funktors\footnote{$\mathcal{A}=\SsV$}
    gewonnenen Kettenabbildungen $F^{*}$ und $G^{*}$ erhalten
    werden. Besagte $F$ und $G$ haben wir hierbei durch Abstraktion
    von solchen
    gewonnen\footnote{vgl. \cite{bordemann.et.al:2003a:pre},
      \cite[Sect. III.2$\alpha$]{connes:1994a}}, die
    für den endlich-dimensionalen Fall existieren. Diese 
    Kettenabbildungen induzieren nun auch die Isomorphismen in
    Teil \textit{ii.)}, wobei hier unter anderem deren
    Stetigkeit zu zeigen und explizite stetige Homotopieabbildungen $s_{k}$ zu konstruieren waren.
\item
    Eine weitergehende Analyse mit der Hilfe von $F$ und $G$
    sowie der Homotopie $s$, liefert in den symmetrischen Fällen die folgenden
    unend\-lich-dimen\-sio\-nalen Verallgemeinerungen der klassischen
    Hochschild-Kostant-Ro\-sen\-berg-The\-oreme, die nun insbesondere für
    den physikalisch relevanteren Spezialfall $\mathcal{M}=\SsV,\Hol$
    gelten.\\\\
    \textbf{Satz (Hochschild-Kostant-Rosenberg)}
    \begin{enumerate}
    \item
        Gegeben ein symmetrischer $\Ss^{\bullet}(\V)-\Ss^{\bullet}(\V)$-Bimodul $\mathcal{M}$.
        Dann besitzt jede Kohomologieklasse
        $\left[\eta\right]\in HH^{k}(\Ss^{\bullet}(\V),\mathcal{M})$ genau einen total
        antisymmetrischen Repräsentanten $\phi^{a,\eta}_{D}$. 
        Dieser ist derivativ in jedem Argument und gegeben durch
        $\phi^{a,\eta}_{D}=\mathrm{Alt}_{k}(\phi)$ f"ur beliebiges $\phi\in
        [\eta]$ mit $\phi^{a,0}_{D}=0$ für
        die $0$-Klasse $[0]$.  
        Insgesamt gilt für alle $\phi\in [\eta]$:
        \begin{equation*}
            \phi=\underbrace{\phi^{a,\eta}_{D}}_{\mathrm{Alt}_{k}(\phi)}+\underbrace{\delta^{k-1}\big(\zeta^{k-1}_{-1}s^{*}_{k-1}\zeta^{k}\phi\big)}_{\phi-\mathrm{Alt}_{k}(\phi)}.
        \end{equation*}
    \item
        Gegeben ein symmetrischer, lokalkonvexer
        $\Ss^{\bullet}(\V)-\Ss^{\bullet}(\V)$-Bimodul $\mathcal{M}$.
        Dann besitzt jedes
        $\left[\eta_{c}\right]\in HH^{k}_{\cont}(\Ss^{\bullet}(\V),\mathcal{M})$ genau einen total
        antisymmetrischen, stetigen Repräsentanten
        $\phi^{a,\eta}_{c,D}$. Dieser ist derivativ in jedem Argument und gegeben durch
        $\phi^{a,\eta}_{c,D}=\mathrm{Alt}_{k}(\phi_{c})$ f"ur beliebiges $\phi_{c}\in
        [\eta_{c}]$ mit $\phi^{a,0}_{c,D}=0$ für
        die $0$-Klasse $[0_{c}]$.
        Insgesamt gilt für alle $\phi_{c}\in [\eta_{c}]$:
        \begin{equation*}
            \phi_{c}=\underbrace{\phi^{a,\eta}_{c,D}}_{\mathrm{Alt}_{k}(\phi_{c})}+\underbrace{\delta_{c}^{k-1}\big(\zeta^{k-1}_{-1}s^{*}_{k-1}\zeta^{k}\phi_{c}\big)}_{\phi_{c}-\mathrm{Alt}_{k}(\phi_{c})}.
        \end{equation*}
    \item
        Gegeben ein vollständiger, symmetrischer, hausdorffscher, lokalkonvexer\\
        $\Hol-\Hol$-Bimodul $\mathcal{M}$. Dann besitzt jedes
        $\left[\hat{\eta}_{c}\right]\in HH^{k}_{\cont}(\Hol,\mathcal{M})$ genau einen total
        antisymmetrischen, stetigen Repräsentanten
        $\hat{\phi}^{a,\eta}_{c,D}$. Dieser ist derivativ in
        jedem Argument und gegeben durch $\hat{\phi}^{a,\eta}_{c,D}=\mathrm{Alt}_{k}\big(\hat{\phi}_{c}\big)$ f"ur beliebiges $\hat{\phi}_{c}\in
        [\hat{\eta}_{c}]$ mit $\hat{\phi}^{a,0}_{c,D}=0$ für 
        die $0$-Klasse $[\hat{0}_{c}]$.

        Insgesamt gilt für alle $\hat{\phi}_{c}\in [\hat{\eta}_{c}]$:
        \begin{equation*}
            \hat{\phi}_{c}=\underbrace{\hat{\phi}^{a,\eta}_{c,D}}_{\mathrm{Alt}_{k}(\hat{\phi}_{c})}+\underbrace{\hat{\delta}_{c}^{k-1}\widehat{\Big(\zeta^{k-1}_{-1}s^{*}_{k-1}\zeta^{k}\phi_{c}\Big)}}_{\hat{\phi}_{c}-\mathrm{Alt}_{k}(\hat{\phi}_{c})}\quad\text{ 
              mit }\quad \phi_{c}=\hat{\phi}_{c}\big|_{\SsV^{k}}.
        \end{equation*}
    \end{enumerate}
    Hierbei ist $\zeta$ ein sehr einfacher Kettenisomorphismus, zwischen dem Hochschild-Komplex und einem Hilfkomplex, in
    den auch $G^{*}$ abbildet. Mit
    $s^{*}_{k}=\hom_{\mathcal{A}^{e}}s_{k}$ ist der Pullback mit der rekursiv definierten
    Homotopieabbildung $s_{k}$ gemeint. In Teil \textit{iii.)} bezeichnet
    $\widehat{\Big(\zeta^{k-1}_{-1}s^{*}_{k-1}\zeta^{k}\phi_{c}\Big)}$
    die stetige Fortsetzung von
    $\Big(\zeta^{k-1}_{-1}s^{*}_{k-1}\zeta^{k}\phi_{c}\Big)$
    und $\phi_{c}=\hat{\phi}_{c}\big|_{\SsV}$ die Einschränkung der
    stetige Abbildung $\hat{\phi}_{c}$ auf $\SsV\subseteq \Hol$. Eine
    genauere
    Analyse liefert für die erste Hochschild-Kohomologie
    \begin{align*}
        [\eta]&=\phi_{D}^{a,\eta}\quad \text{für alle}\quad[\eta]\in HH^{1}(\SsV,\mathcal{M}),\\
        [\eta_{c}]&=\phi_{c,D}^{a,\eta}\quad\text{für
          alle}\quad[\eta]\in HH_{\cont}^{1}(\SsV,\mathcal{M}),\\
        [\hat{\eta}_{c}]&=\hat{\phi}_{c,D}^{a,\eta}\quad\text{für alle}\quad[\hat{\eta}]\in HH_{\cont}^{1}(\Hol,\mathcal{M})
    \end{align*}und für die zweite die explizite Formel
    \begin{align*}
        \left(\zeta^{1}_{-1}s^{*}_{1}\zeta^{2}\phi\right)(x)=&\:\phi(1,x)+\sum_{p=1}^{l}\Bigg[\frac{1}{p}\phi\left(x^{p},x_{p}\right)+…\nonumber
        \\ &+\binom{p}{l}^{-1}\sum_{j_{1},…j_{l}}^{p-1}\phi\left((x^{p})^{j_{1},…,j_{l}},x_{p}\right)*_{R}(x^{p})_{j_{1},…,j_{l}}+…\nonumber
        \\ &+\frac{1}{p}\phi(1,x_{p})*_{R}x^{p}\Bigg]
    \end{align*}für alle $x\in \SsV$ mit $\deg(x)=l$. Analoge, wenn
    auch weniger konkrete Aussagen für den
    nichtsymmetrischen Fall sind
    \begin{align*}
        \phi&=\overbrace{\zeta_{-1}^{k}\Omega^{*}_{k}\zeta^{k}\phi}^{=\:\tilde{\phi}\in
          [\eta]} +
        \left(\phi-\zeta_{-1}^{k}\Omega^{*}_{k}\zeta^{k}\phi\right)=\tilde{\phi}+
        \delta^{k-1}\big(\zeta_{-1}^{k-1}s^{*}_{k-1}\zeta^{k}\phi\big),\\
        \phi_{c}&=\overbrace{\zeta_{-1}^{k}\Omega^{*}_{k}\zeta^{k}\phi_{c}}^{=\:\tilde{\phi_{c}}\in
          [\eta_{c}]} +
        \left(\phi_{c}-\zeta_{-1}^{k}\Omega^{*}_{k}\zeta^{k}\phi_{c}\right)=\tilde{\phi_{c}}+
        \delta_{c}^{k-1}\big(\zeta_{-1}^{k-1}s^{*}_{k-1}\zeta^{k}\phi_{c}\big)
    \end{align*}für $\phi\in[\eta]\in HH^{k}(\SsV,\mathcal{M})$,
    $\phi_{c}\in[\eta_{c}]\in HH_{\cont}^{k}(\SsV,\mathcal{M})$ und
    $\hat{\phi}_{c}\in[\hat{\eta}_{c}]\in
    HH_{\cont}^{k}(\Hol,\mathcal{M})$ sowie
    \begin{equation*}
        \hat{\phi}_{c}=\overbrace{\widehat{\Big(\zeta_{-1}^{k}\Omega^{*}_{k}\zeta^{k}\phi_{c}\Big)}}^{\in [\hat{\eta}_{c}]} +\:
        \hat{\delta}_{c}^{k-1}\widehat{\left(\zeta_{-1}^{k-1}s^{*}_{k-1}\zeta^{k}\phi_{c}\right)}\quad\text{ 
          mit }\quad [\hat{\eta}_{c}]\ni\phi_{c}=\hat{\phi}_{c}\big|_{\SsV^{k}}
    \end{equation*}und $\Omega_{k}=F_{k}\cp G_{k}$.
    Hierfür beachte man, dass wir in obiger Formel f"ur
    $\zeta^{1}_{-1}s^{*}_{1}\zeta^{2}\phi$ explizit zwischen $*_{R}$
    und $*_{L}$ unterschieden haben, diese also auch für
    nicht\--\-sym\-me\-trischen Fall gültig ist. Besagte Formel lässt
    sich dann in der
    Situation der Deformation einer Modul-Struktur zur rekursiven
    Konstruktion nutzbringend
    einsetzen, sofern die zweite Hochschild-Kohomologie
    $HH^{2}(\mathcal{A},\End_{\mathbb{K}}(\mathcal{M}))$ verschwindet.
    Die Bezeichnungsweise
    "`Hochschild-Kostant-Rosen\-berg-Theoreme"' ist hierbei auch
    insofern gerechtfertigt als dass jedes, in allen Argumenten
    derivative $\phi \in
    HC^{k}(\SsV,\mathcal{M})$,
    die algebraische Definition eines Differentialoperators erster
    Ordnung erfüllt, also $\phi_{D}^{a,\eta}\in \DiffOpS{k}{1}$ gilt. Weiterhin ist jedes derartige Element
    $\phi\in HC^{k}_{\cont}(\SsV,\mathcal{M})$ ein stetiger
    Differentialoperator erster Ordnung und ebenso verhält es sich mit
    solchen Elementen aus
    $HC^{k}_{\cont}(\Hol,\mathcal{M})$. In diesem Sinne nehmen die
    Repräsentanten $\phi_{D}^{a,\eta}$ den Platz ein, der den Multivektorfelder
    im endlich-dimensionalen Rahmen gebührt, vgl. \cite[Prop
    6.2.8]{waldmann:2007a}.
\item
    Motiviert durch den Fakt, dass $G^{*}$ in beiden Fällen eine
    Kettenabbildung
    \begin{align*}
        \xi\colon&\left(KC(\V,\mathcal{M}),\Delta\right)\longrightarrow\left(HC(\SsV,\mathcal{M}),
            \delta\right)\qquad\qquad\quad\text{bzw.}
        \\
        \xi\colon&\left(KC_{\cont}(\V,\mathcal{M}),\Delta^{c}\right)\longrightarrow\left(HC_{\cont}(\SsV,\mathcal{M}),\delta_{c}\right)
    \end{align*}
    definiert, die Isomorphismen $\wt{\xi}^{k}$ auf Kohomologie-Niveau
    induziert und zudem differentielle Bilder hat, sind wir der
    Frage nachgegangen, ob sich mit dieser auch die
    Hochschild-Kohomologien der differen\-tiellen und
    der stetig-diffe\-rentiellen Unterkomoplexe
    $\left(HC_{\diff}(\SsV,\mathcal{M}),\delta_{\diff}\right)$ und
    $\left(HC_{\mathrm{c,d}}(\SsV,\mathcal{M}),\delta_{\mathrm{c,d}}\right)$ 
    berechnen lassen. Diese sind zunächst nur im Falle symmetrischer Bimoduln
    wohldefiniert, für den wir folgendes Korollar erhielten:\\\\ 
    \textbf{Korollar}
    \begin{enumerate}
    \item
        Sei $\mathcal{M}$ ein symmetrischer $\Ss^{\bullet}(\V)-\Ss^{\bullet}(\V)$-Bimodul. Dann induzieren $\xi$ und $\hat{\xi}$ Kettenabbildungen  $\wt{\xi^{k}}$ und $\wt{\hat{\xi}^{k}}$ zwischen $(HC_{\diff}(\SsV,\mathcal{M}),\delta_{\diff})$ und\\  $(KC(\V,\mathcal{M}),\Delta)$.
        Des Weiteren ist $\wt{\xi^{k}}$ injektiv und $\wt{\hat{\xi}^{k}}$ surjektiv.
    \item
        Sei $\mathcal{M}$ ein symmetrischer, lokalkonvexer
        $\Ss^{\bullet}(\V)-\Ss^{\bullet}(\V)$-Bimodul. Dann induzieren $\xi$
        und $\hat{\xi}$ Kettenabbildungen zwischen
        $(HC_{\mathrm{c,d}}(\SsV,\mathcal{M}),\delta_{\mathrm{c,d}})$
        und  $(KC^{\cont}(\V,\mathcal{M}),\Delta)$. Des Weiteren ist
        $\wt{\xi^{k}}$ injektiv und $\wt{\hat{\xi}^{k}}$ surjektiv.
    \item
        Gegeben ein vollständiger, symmetrischer,
        hausdorffscher, lokalkonvexer\\ $\Hol-\Hol$-Bimodul $\mathcal{M}$, so
        induzieren die Einschränkungs-Abbildung\-en einen Kettenisomorphismus:
        \begin{equation*}
            \big(HC_{\mathrm{c,d}}(\Hol,\mathcal{M}),\hat{\delta}_{\mathrm{c,d}}\big)\cong
            \big(HC_{\mathrm{c,d}}(\SsV,\mathcal{M}),\delta_{\mathrm{c,d}}\big)
        \end{equation*}und es gilt die Isomorphie:
        \begin{equation*}
            HH^{k}_{\mathrm{c,d}}(\Hol,\mathcal{M})\cong HH^{k}_{\mathrm{c,d}}(\SsV,M).
        \end{equation*}
    \end{enumerate}Hierbei bezeichnet $\hat{\xi}$ die mit Hilfe
    von $F^{*}$ definierte Kettenabbildung, welche im
    nicht-differentiellen Falle die auf Kohomologie-Niveau
    zu $\widetilde{\xi^{k}}$ inversen Isomorphismen
    $\wt{\hat{\xi}^{k}}$ induziert. 
    Eine analoge Aussage ist auch für die sogenannten differentiellen
    $\SsV-\SsV$-Bimoduln $k$-ter Ordnung herleitbar, 
    deren Rechtsmodul-Multiplikation in der Form
    \begin{equation*}
        *_{R}=*_{L}+ D_{1}+…+D_{k}
    \end{equation*}mit $\mathbb{K}$-bilinearen Abbildungen
    $D_{l}:\SsV\times \mathcal{M}\longrightarrow \mathcal{M}$ derart geschrieben
    werden kann, dass zusätzlich folgende Bedingungen erfüllt sind: 
    \begin{itemize}
    \item[\textbf{a.)}]
        Jedes $D_{l}$ ist $*_{L}$-linear im zweiten Argument.
    \item[\textbf{b.)}]
        Für $D_{l_{1},\dots,l_{p}}^{a_{1},\dots,a_{p}}=D_{l_{1}}^{a_{1}}\cp
        \dots \cp D_{l_{p}}^{a_{p}}$ mit $D_{l}^{a}(m):= D_{l}(a,m)$
        und $a\in\SsV$ ist
$D_{l_{1},\dots,l_{p}}^{a_{1},\dots,a_{p}}= 0$, falls $\sum_{i=1}^{p}l_{i}> k$.
    \item[\textbf{c.)}]
        F"ur alle $1\leq l\leq k$ gilt:
        \begin{align*}
            D_{l}(a*b,m)=&\:b*_{L}D_{l}(a,m)+ D_{1}(b,D_{l-1}(a,m))+D_{2}(b,D_{l-2}(a,m))+\dots\\ &\qquad+D_{l-2}(b,D_{2}(a,m))+ D_{l-1}(b,D_{1}(a,m))+a*_{L}D_{l}(b,m). 
        \end{align*}
    \item[\textbf{d.)}]
        F"ur jedes $m\in \mathcal{M}$ ist $D_{l}(\cdot,m)\in
        \DiffOp{1}{l}$ sowie $D_{l}(\vv,m)=0$, falls $l\geq 2$ und $\deg(\vv)=1$.
    \end{itemize}
\end{itemize}
        Für derartige Bimoduln sind die differentiellen Unterkomplexe ebenfalls
wohldefiniert und es gilt folgender Satz:\\\\
\textbf{Satz}
\begin{enumerate}
\item
    Sei $\mathcal{M}$ ein differentieller
    $\Ss^{\bullet}(\V)-\Ss^{\bullet}(\V)$-Bimodul. Dann besitzt
    jede Kohomologieklasse $[\eta]\in HH^{k}(\SsV,\mathcal{M})$
    mindestens einen differentiellen Repr"asentanten $\phi\in \DiffOpS{k}{s+1}$. Des Weiteren induzieren $\xi$ und $\hat{\xi}$ Kettenabbildungen zwischen $\left(HC_{\diff}(\SsV,\mathcal{M}),\delta_{\diff}\right)$ und  $(KC(\V,\mathcal{M}),\Delta)$.
    Hierbei ist $\wt{\xi^{k}}$ injektiv und $\wt{\hat{\xi}^{k}}$ surjektiv.
\item
    Sei $\mathcal{M}$ ein differentieller, lokalkonvexer
    $\Ss^{\bullet}(\V)-\Ss^{\bullet}(\V)$-Bimodul. Dann besitzt
    jede Kohomologieklasse $[\eta]\in
    HH_{\cont}^{k}(\SsV,\mathcal{M})$ mindestens einen
    differentiellen Repr"asentanten $\phi\in \DiffOpS{k}{s+1,\cont}$. Des Weiteren induzieren $\xi$
    und $\hat{\xi}$ wohldefinierte Kettenabbildungen zwischen
    $(HC_{\mathrm{c,d}}(\SsV,\mathcal{M}),\delta_{\mathrm{c,d}})$
    und  $(KC^{\cont}(\V,\mathcal{M}),\Delta)$. Hierbei ist
    $\wt{\xi^{k}}$ injektiv und $\wt{\hat{\xi}^{k}}$ surjektiv.
\end{enumerate}
Ein Beispiel für einen solchen Bimodul ist hierbei
der Unterraum $\mathcal{M}$ aller
        Differentialoperatoren
        $m\in\mathrm{DiffOp}_{1}^{s}(\SsV,\SsV)$, die als eine endliche
        Summe der Form
        \begin{equation*}
            m=\sum_{l=0}^{s}\sum_{|\alpha|=l}\delta_{\alpha_{1}}^{|\alpha_{1}|}…\delta_{\alpha_{k}}^{|\alpha_{k}|}
        \end{equation*}mit Derivationen
        $\delta_{\alpha_{i}}\in\mathrm{DiffOp}_{1}{1}(\SsV,\SsV)$
        geschrieben werden können. Hierbei ist in der zweiten Summe
        $\alpha \in \mathbb{N}^{k}$, wobei $k$ für jeden Summanden
        variieren darf. Mit $\delta^{|\alpha_{i}|}_{\alpha_{i}}$ ist
        die $|\alpha_{i}|$-fache Anwendung von
        $\delta_{\alpha_{i}}$ gemeint, und wegen der
        Derivationseigenschaft ist die Reihenfolge Verkettungen
        unwichtig. Der Summand für $l=0$ soll dann lediglich aus einem
        Element $m_{0}\in \SsV$ bestehen.\\\\ 
Obiger Satz und obiges Korollar stellen nun die unendlich-dimensionalen
Verallgemeinerungen der in \cite[Kapitel 5]{Weissarbeit} behandelten
Zusammenhänge dar, in welchen es sogar möglich ist, die Isomorphie
besagter Kohomologiegruppen für die Algebra $C^{\infty}(V)$ mit einer
konvexen Teilmengen $V\subseteq \mathbb{R}^{n}$ zu zeigen. Dies ist im
wesentlichen dem Fakt geschuldet, dass die Differentialoperatoren hier
besonders einfach mit Hilfe partieller Ableitungen
geschrieben werden können. Es gelten dann Kettenregeln der Form
$\pt_{y}f(tx+(1-t)y)=f'(tx +(1-t)y)(1-t)$, womit die Homotopie $s$
bzw. $s^{*}$
auch im
differentiellen Fall gewinnbringend eingesetzt werden kann,
siehe \cite[Prop 5.6.6]{Weissarbeit}. In unserem
unendlich-dimensionalen Rahmen bleibt jedoch zu hoffen, dass 
eine andere Homotopie als $s^{*}$ existiert, die letztlich
die Surjektivität von $\wt{\xi^{k}}$ und die Injektivität von
$\wt{\hat{\xi}^{k}}$ zeigt.

\section*{Aufbau}
Diese Arbeit ist wie folgt gegliedert:
\begin{itemize}
\item
    In Kapitel 1 definieren wir den zentralen Begriff der
    Hochschild-Kohomologie ganz allgemein f"ur Hochschild-Koketten mit
    Werten in Bimoduln und verwenden die im Anhang A bereitgestellten Grundlagen, um besagte Kohomologiegruppen exemplarisch 
    f"ur die Polynom-Algebra $\Pol(\mathbb{R}^{n})$ zu
    berechnen. Hierbei bedienen wir uns den Methoden aus \cite[Kapitel
    5]{Weissarbeit}. Durch Abstraktion der hier benutzten Homotopien und Kettenabbildungen sind wir schließlich in der Lage, die Hochschild-Kohomologien f"ur die symmetrische Algebra eines beliebigen $\mathbb{K}$-Vektorraumes $\V$ zu bestimmen. 
\item
    Im 2. Kapitel führen wir den Begriff des topologischen Komplexes und den der stetigen Hochschild-Kohomologie ein. Aufbauend auf Kapitel 1 und mit Hilfe der in Anhang B bereitgestellten funktional-analytischen Mittel werden die stetigen
    Hochschild-Kohomologien der geeignet topologisierten symmetrischen
    Algebra unter expliziter Verwendung der im vorherigen Kapitel
    definierten Kettenabbildungen f"ur beliebige lokalkonvexe
    Vektorr"aume $(\V,P)$ sowie lokalkonvexe Bimoduln berechnet. 

    Im letzten Abschnitt dieses Kapitels betrachten wir speziell
    lokalkonvexe Vektorräume $(\V,\T_{P})$ mit hausdorffschen
    Topologien. In diesem Fall ist die symmetrische Algebra ebenfalls
    hausdorffsch topologisiert und wir dürfen deren Vervollst"andigung $\big(\Hol,\hat{\Pp}\big)$ betrachten. 
    Wir geben hier zun"achst eine detaillierte Beschreibung dieses Raumes
    und mit Dichtheitsargumenten werden wir in der Lage sein, auch die
    Hochschild-Kohomologie dieser Algebra für hausdorffsche und zudem
    vollständige $\Hol-\Hol$-Bimoduln zu charakterisieren. Ein
    essentielles Beispiel f"ur einen solchen  Bimodul wird dann immer
    $\Hol$ selbst darstellen.
\item
    Das 3. Kapitel ist ganz der Verallgemeinerung der
    Hochschild-Kostant-Rosenberg-Theoreme auf den
    unendlich-dimensionalen Fall gewidmet, welche wir für symmetrische
    $\SsV-\SsV$-Bimoduln sowohl im rein algebraischen als auch im
    lokalkonvexen Fall und für symmetrische lokalkonvexe
    $\Hol-\Hol$-Bimoduln formulieren werden.
\item
    Im letzten Kapitel beschäftigen wir uns mit dem algebraischen
    Konzept des Multidifferentialoperators und gehen der Frage nach,
    für welche $\SsV-\SsV$- bzw. $\Hol-\Hol$-Bimoduln es möglich ist,
    den Begriff des Hochschild-Komplexes zu definieren und dessen
    Kohomologien mit Hilfe der uns zur Verfügung stehenden Mittel zu berechnen. 
\end{itemize}

\clearpage
\thispagestyle{empty}
\cleardoublepage

\pagestyle{fancy}
\fancyhf{} 
\fancyhead[OR]{\rightmark} 
\fancyhead[EL]{\leftmark} 
\fancyfoot[C]{\thepage} 

\mainmatter
\chapter{Hochschild-Kohomologien} 
\label{sec:Hochschkohoms}
In diesem Kapitel geben wir die zentrale Definition dieser Arbeit, die
der Hochschild-Kohomologie und berechnen diese exemplarisch für die
Polynomalgebra $\Pol(\mathbb{R}^{n})$, sowie als Verallgemeinerung für die symmetrische
Algebra über einem beliebigen $\mathbb{K}$-Vektorraum $\V$. Dabei wollen wir
hier und für den Rest dieser Arbeit $\mathbb{K}$ immer als $\mathbb{R}$ oder
$\mathbb{C}$ annehmen. Unter einer $\mathbb{K}$-Algebra $\mathcal{A}$ verstehen
wir im Folgenden einen $\mathbb{K}$-Vektorraum $\mathcal{A}$ mit
assoziativer, $\mathbb{K}$-bilinearer Algebramultiplikation.
Sprechen wir von einem $\mathcal{A}-\mathcal{A}$-Bimodul $\mathcal{M}$, so ist
stets ein $\mathbb{K}$-Vektorraum mit
$\mathcal{A}-\mathcal{A}$-Bimodulstruktur derart gemeint, dass sowohl
die Linksmodul-Multiplikation $*_{L}$ als auch die
Rechtsmodul-Multiplikation $*_{R}$ $\mathbb{K}$-bilineare Abbildungen sind. Ist
$\mathcal{A}$ unitär, so setzen wir immer die Verträglichkeit von
$\mathcal{M}$, also $1_{\mathcal{A}}*_{L} m=m$ und
$m*_{R}1_{\mathcal{A}}=m$ für alle $m\in \mathcal{M}$ voraus. 
Der Verständlichkeit halber werden wir jedoch die an gegebener
Stelle wichtigen Eigenschaften nochmals explizit erwähnen.
Alle im Folgenden auftretenden Tensorprodukte sind als solche
über dem jeweils verwendeten Körper $\mathbb{K}$ zu verstehen.
\section{Einführung}
\label{sec:Einf}
Gegeben eine $\mathbb{K}$-Algebra $\mathcal{A}$ und ein
$\mathcal{A}-\mathcal{A}$-Bimodul $\mathcal{M}$, so betrachten wir für $k\in \mathbb{Z}$ die $\mathbb{K}$-Vektorräume
\begin{equation*}
    HC^{k}(\mathcal{A},\mathcal{M}):=
    \begin{cases} \{0\} & k<0\\
        \mathcal{M} & k=0\\
        \Hom_{\mathbb{K}}(\underbrace{\mathcal{A}\times…\times
          \mathcal{A}}_{k-mal},\mathcal{M})& k\geq 1,
    \end{cases} 
\end{equation*} die $\mathbb{K}$-multilinearen Abbildungen von
$\mathcal{A}\times…\times \mathcal{A}$ nach $\mathcal{M}$.
Vermöge der gegebenen Links- und Rechtsmodulstruktur
definieren wir $\mathbb{K}$-lineare Abbildungen
\begin{equation*}
    \delta^{k}\colon HC^{k}(\mathcal{A},\mathcal{M})\longrightarrow
    HC^{k+1}(\mathcal{A},\mathcal{M})
\end{equation*}
durch
\begin{equation}
    \label{eq:Hochschilddelta}
    \begin{split}
        (\delta^{k}\phi)(a_{1},…,a_{k+1})=a_{1}*_{L}\phi(a_{2},…,a_{k+1})&+\sum_{i=1}^{k}(-1)^{i}\phi(a_{1},…,a_{i}*a_{i+1},…,a_{k+1})\\
        &+(-1)^{k+1}\phi(a_{1},…,a_{k})*_{R}a_{k+1}.
    \end{split}
\end{equation}
Eine elementare Rechnung zeigt $\delta^{k+1}\cp\delta^{k}=0$, und wir erhalten einen
Kokettenkomplex $(HC^{\bullet}(\mathcal{A},\mathcal{M}),\delta)$ mit
$HC^{k}(\mathcal{A},\mathcal{M})$ als $\mathbb{K}$-Moduln und 
$\delta^{k}$ als $\mathbb{K}$-Homomorphismen. Wir kommen nun zu der
für diese Arbeit zentralen Definition.\\
\begin{definition}[Hochschild-Kohomologie]
    Wir definieren die $k$-te Hochschild-Kohomologie durch:
    \begin{equation*}
        HH^{k}(\mathcal{A},\mathcal{M}):=
        \begin{cases}  
            \ker\left(\delta^{0}\right) & k=0\\
            HH^{k}(\mathcal{A},\mathcal{M})=\ker\left(\delta^{k}\right)/\im\left(\delta^{k-1}\right)&
            k\geq 1.
        \end{cases} 
    \end{equation*}
    Ungeachtet ihrer $\mathbb{K}$-Vektorraum Struktur wollen wir
    diese im Folgenden entweder als Kohomologiegruppen oder einfach als
    Kohomologien bezeichnen.
\end{definition}
Vermöge der universellen Eigenschaft des Tensorproduktes erhalten wir
einen Isomorphismus 
\begin{equation*}
    \ot_{k*}\colon
    \Hom_{\mathbb{K}}(\mathcal{A}\times…\times
    \mathcal{A},\mathcal{M})\longrightarrow \Hom_{\mathbb{K}}(\mathcal{A}\ot…\ot
    \mathcal{A},\mathcal{M}),
\end{equation*}
dessen Inverses $\ot_{k}^{*}=\ot_{k*}^{-1}$ einfach der Pullback mit $\ot_{k}$ ist.
Die Tensorvariante von \eqref{eq:Hochschilddelta} ist dann gegeben
durch lineare Fortsetzung von
\begin{equation*}
    \label{eq:THochschilddelta}
    \begin{split}
        (\delta^{k}\phi)(a_{1}\ot…\ot a_{k})=&\:a_{1}*_{L}\phi(a_{2}\ot…\ot
        a_{k+1})\\ &+\sum_{i=1}^{k}(-1)^{i}\phi(a_{1}\ot…\ot
        a_{i}a_{i+1}\ot…\ot a_{k+1})\\
        &+(-1)^{k+1}\phi(a_{1}\ot …\ot a_{k})*_{R}a_{k+1}
    \end{split}
\end{equation*}vermöge Korollar \ref{kor:WohldefTensorprodabbildungen}
und wir erhalten
\begin{equation}
    \label{eq:TensorglKetteniso}
    \delta^{k}_{\ot}\cp \ot_{k*}= \ot_{k+1*}\cp \delta^{k}_{\times},
\end{equation}
womit $\ot_{*}$ ein Kettenisomorphismus zwischen diesen
beiden Kokettenkomplexen ist. Dies bedeutet insbesondere die
Isomorphie derer Kohomologien (vgl. Lemma \ref{lemma:kettenabzu}~\textit{ii.)}), und wir dürfen uns im Folgenden
darauf beschränken, die einfacher handhabbare Tensorvariante des
Hochschild-Komplexes zu betrachten.\\\\
Wir wollen nun zunächst einsehen, dass die Kohomologiegruppen $HH^{k}(\mathcal{A},\mathcal{M})$ durch
Anwendung eines $\mathrm{Ext}$-Funktors auf die Algebra $\mathcal{A}$ erhalten
werden können, dass also
\begin{equation*}
    HH^{k}(\mathcal{A},\mathcal{M})\cong \mathrm{Ext}_{R}^{k}(\cdot,\mathcal{M})(\mathcal{A})=H^{k}(\hom_{R}(\cdot,\mathcal{M})C)=H^{k}(\Hom_{R}(C,\mathcal{M}),d^{*})
\end{equation*}gilt. Dabei bezeichnet $C$ eine projektive Auflösung
$(C,d,\epsilon)$ von $\mathcal{A}$ und $R$ einen geschickt zu
wählenden Ring. Mit Beispiel \ref{bsp:ExtBeisp} folgt dann bereits
$HH^{0}(\mathcal{A},\mathcal{M})\cong \Hom_{R}(\mathcal{A},\mathcal{M})$.
\begin{lemma}
    \label{lemma:AewirdzuunitRing}
    Gegeben eine assoziative $\mathbb{K}$-Algebra $(\mathcal{A},*)$.
    \begin{enumerate}
    \item
        Die Menge
        $\mathcal{A}^{e}=\mathcal{A}\ot \mathcal{A}$ wird vermöge
        der Multiplikation
        \begin{equation}
            \label{eq:AeRingmultdef}
            (a\ot b) *_{e} (\widetilde{a}\ot
            \widetilde{b}):=(a*\widetilde{a})\ot
            (b*^{opp}\widetilde{b})=(a*\widetilde{a})\ot
            (\widetilde{b}*b)
        \end{equation} zu einer assoziativen $\mathbb{K}$-Algebra. Ist $\mathcal{A}$
        unitär, so auch $\mathcal{A}^{e}$ vermöge $1_{\mathcal{A}^{e}}=1_{\mathcal{A}}\ot 1_{\mathcal{A}}$.
    \item 
        Jeder $\mathcal{A}-\mathcal{A}$-Bimodul $\mathcal{M}$ wird durch 
        \begin{equation*}
            a\ot b*_{e} m=a*_{L}( m*_{R} b)=(a*_{L}m)*_{R}b\qquad\quad \forall\:a\ot b\in \mathcal{A}^{e},\: m\in \mathcal{M}
        \end{equation*} zu einem $\mathcal{A}^{e}$-Linksmodul.    
    \end{enumerate}
    \begin{beweis}
        \begin{enumerate}
        \item
            Die Assoziativität folgt unmittelbar aus der Assoziativität von
            $\mathcal{A}$. Der Rest ist ebenfalls klar. Für
            die Wohldefiniertheit von $*_{e}$ definieren
            wir die Abbildung $*'_{e}$ vermöge
            Korollar \ref{kor:WohldefTensorprodabbildungen} durch lineare
            Fortsetzung von
            \begin{align*}
                *'_{e}\colon \mathcal{A}\ot \mathcal{A} \ot \mathcal{A} \ot
                \mathcal{A}&\longrightarrow \mathcal{A}\ot \mathcal{A}\\
                a\ot b \ot \tilde{a}\ot \tilde{b} & \longmapsto
                a\tilde{a}\ot b\tilde{b}
            \end{align*}und setzen
            $*_{e}= *'_{e}\cp \cong \cp\ot_{2}$ mit $\ot_{2}\colon \mathcal{A}^{e}\times
            \Ae\longrightarrow \Ae\ot \Ae$ und $\cong$ der Isomorphismus $\Ae\ot \Ae\cong\mathcal{A}\ot
            \mathcal{A}\ot \mathcal{A}\ot \mathcal{A}$.
        \item
            Für die Wohldefiniertheit beachte man, dass für festes
            $m\in \mathcal{M}$ die Abbildung $*_{m}\colon a\ot b\rightarrow
            amb$ die Bedingungen von Korollar \ref{kor:WohldefTensorprodabbildungen} erfüllt, mithin linear auf ganz
            $\mathcal{A}^{e}$ fortsetzt. Mit \textit{i.)}
            folgt
            \begin{align*}
                [(a\ot b) *_{e} (\widetilde{a}\ot\widetilde{b})]*_{e}m=a\widetilde{a}m\widetilde{b}b=(a\ot b)*_{e} (\widetilde{a}m\widetilde{b})=(a\ot b)*_{e}[(\widetilde{a}\ot\widetilde{b})*_{e}m],
            \end{align*} was die Behauptung zeigt.
        \end{enumerate}
    \end{beweis}
\end{lemma}
\begin{definition}[Bar-Komplex]
    Sei $\mathcal{A}$ eine assoziative $\mathbb{K}$-Algebra. Wir betrachten die
    $\mathbb{K}$-Vektorräume
    \begin{align*}
        \C_{k}=\mathcal{A}\ot \underbrace{\mathcal{A}\ot …
          \ot \mathcal{A}}_{k-mal} \ot \mathcal{A}
    \end{align*} $\qquad\qquad\qquad\qquad \C_{0}=\mathcal{A}\ot \mathcal{A},\quad\quad
    \C_{1}=\mathcal{A}\ot\mathcal{A}\ot \mathcal{A},\quad\quad
    \C_{2}=\mathcal{A}\ot\mathcal{A}\ot \mathcal{A}\ot \mathcal{A}$,
    \\\\ die wie in Lemma \ref{lemma:AewirdzuunitRing} durch
    \begin{equation}
        \label{eq:ModMultiplBarkom}
        a\ot b *_{e}\left(x_{0}\ot x_{1}\ot … \ot x_{k}\ot
            x_{k+1}\right)=(a x_{0})\ot x_{1}\ot … \ot x_{k}\ot\:
        (x_{k+1}b)
    \end{equation}zu $\mathcal{A}^{e}$-Linksmoduln werden.
    Des Weiteren seien $\mathcal{A}^{e}$-Homomorphismen durch lineare
    Fortsetzung von
    \begin{align*}
        d_{k}\colon\C_{k}&\longrightarrow \C_{k-1}\\
        (x_{0}\ot … \ot x_{k+1})&\longmapsto
        \sum_{j=0}^{k}(-1)^{j}x_{0}\ot…\ot
        x_{j}x_{j+1}\ot…\ot x_{k+1}
    \end{align*} für $k\geq 1$ definiert.
    Dann gilt $d_{k}\cp d_{k+1}=0$ und wir erhalten einen wohldefinierten
    Kettenkomplex $(\C,d)$, den wir im Folgenden als den zu $\mathcal{A}$
    gehörigen Bar-Komplex bezeichnen wollen.
\end{definition}
Für unitäres $\mathcal{A}$ ist $\mathcal{A}^{e}$ unitär und
Lemma \ref{lemma:unitarereRingModulnhabenfreieAufloesung} besagt, dass es dann
rein abstrakt eine projektive (sogar freie) Auflösung von
$\mathcal{A}$ als $\mathcal{A}^{e}$-Modul geben
muss. Ein essentielles Beispiel liefert folgendes Lemma.
\begin{lemma}[Bar-Auflösung für unitäre $\mathbb{K}$-Algebren]
    \label{lemma:unitAlhabeBarAufloesProj}
    Gegeben eine unitäre, assoziative $\mathbb{K}$-Algebra $\mathcal{A}$, so wird
    der Bar-Komplex vermöge der Abbildung
    \begin{align*}
        \epsilon\colon\C_{0}&\longrightarrow \mathcal{A}\\
        a\ot b&\longmapsto ab
    \end{align*} zu einer projektiven Auflösung $(\C,d,\epsilon)$ von
    $\mathcal{A}$.
    \begin{beweis}
        Zunächst ist $\epsilon$ ein wohldefinierter $\mathcal{A}^{e}$-Homomorphismus        
        \begin{equation*}
            \epsilon(a\ot b *_{e}x_{0}\ot
            x_{1})\glna{\eqref{eq:ModMultiplBarkom}}\epsilon\:(ax_{0}\ot
            x_{1}b)=ax_{0}x_{1}b\glna{\eqref{eq:ModMultiplBarkom}}a\ot b*_{e}x_{0}x_{1}=a\ot
            b*_{e}\epsilon(x_{0}\ot x_{1}).
        \end{equation*} Des Weiteren ist $\epsilon$ surjektiv, da
        $\epsilon\:(1\ot a)=a\in \mathcal{A}$, und es gilt zudem
        \begin{equation*}
            (\epsilon\circ d_{1})(x_{0}\ot x_{1}\ot
            x_{2})=\epsilon(x_{0}x_{1}\ot x_{2}-x_{0}\ot x_{1}x_{2})=0.
        \end{equation*} Für die Projektivität reicht es, die
        $\mathcal{A}^{e}$-Freiheit jedes $\C_{k}$ zu zeigen. Dafür
        beachte man, dass
        $\C_{0}\cong \mathcal{A}^{e}$, $\C_{1}\cong
        \mathcal{A}^{e}\ot \mathcal{A}$, $…$, $\C_{k}\cong
        \mathcal{A}^{e}\bigotimes^{k-2} \mathcal{A}$, womit $\C_{k}\cong \mathcal{A}^{e}\ot V_{k}$
        für $\mathbb{K}$-Vektorräume $V_{k}$. Dann liefert
        die $\mathcal{A}^{e}$-lineare Fortsetzung der Abbildung
        \begin{align*}
            \tau_{k}\colon\mathcal{A}^{e}\ot V_{k}&\longrightarrow
            (\mathcal{A}^{e})^{\dim(V_{k})}\\ 
            a^{e}\ot\vec{e}_{\alpha}&\longmapsto \oplus_{\alpha} a^{e}\qquad\qquad \forall\:a^{e}\in \mathcal{A}^{e} 
        \end{align*} mit $\{\vec{e}_{\alpha}\}_{\alpha\in I}$ eine Basis von
        $V_{k}$, einen Isomorphismus
        $\C_{k}\longrightarrow(\mathcal{A}^{e})^{\dim(V_{k})}$.\\\\ 
        Es bleibt die Exaktheit
        des Komplexes nachzuweisen. Hierfür betrachten wir die
        Kettenabbildungen
        \begin{align*}
            h_{k}\colon\C_{k}&\longrightarrow \C_{k+1}\\
            x_{0}\ot…\ot x_{k+1}&\longmapsto 1\ot
            x_{0}\ot…\ot x_{k+1}\qquad k\geq -1
        \end{align*}für welche wir erhalten, dass
        \begin{equation}
            \label{eq:Homotbar}
            \begin{split}
                \epsilon\circ h_{-1}&=\id_{A},\\
                d_{1}\cp h_{0}+h_{-1}\cp\epsilon &=\id_{\C_{0}}, \text{ sowie}\\
                d_{k+1}\cp h_{k}+h_{k-1}\cp d_{k}&=\id_{\C_{k}} \text{ für
                }k\geq 1.
            \end{split}
        \end{equation} Hiermit folgt für $\alpha\in \ker(d_{k})$ und
        $k\geq 1$:
        \begin{equation*}
            \alpha=(d_{k+1}\cp h_{k})(\alpha)+(h_{k-1}\cp
            d_{k})(\alpha)=(d_{k+1}\cp h_{k})(\alpha)\in \im(d_{k+1}),
        \end{equation*} analog für $\alpha\in \ker(\epsilon)$. Dies
        zeigt die Exaktheit.
    \end{beweis}
\end{lemma}
\begin{proposition}
    \label{prop:barauffuerunitalgebraIsomozuHochschildkohomo}
    Gegeben eine unitäre, assoziative $\mathbb{K}$-Algebra
    $\mathcal{A}$ und ein $\mathcal{A}-\mathcal{A}$-Bimodul
    $\mathcal{M}$. Bezeichne
    $(\C,d,\epsilon)$ die Bar-Auflösung über $\mathcal{A}$. Dann gilt:
    \begin{equation*}
        HH^{k}(\mathcal{A},\mathcal{M})\cong \mathrm{Ext}^{k}_{\mathcal{A}^{e}}(\cdot,
        \mathcal{M})(\mathcal{A})=H^{k}(\mathrm{hom}_{\mathcal{A}^{e}}(\cdot,\mathcal{M})\C)=H^{k}(\Hom_{\mathcal{A}^{e}}(\C,\mathcal{M}),d^{*}).
    \end{equation*}
    \begin{beweis}
        Sowohl $(HC^{\bullet}(\mathcal{A},\mathcal{M}),\delta)$, als auch
        der durch Anwendung von $\hom_{\mathcal{A}^{e}}$ auf $(\C,d)$ ge\-wonnene
        Kokettenkomplex $(\C^{*},d^{*})$ mit Kokettengliedern
        $\C^{k}=\C_{k}^{*}=\Hom_{\mathcal{A}^{e}}(\C_{k},\mathcal{M})$ und Ko-Differentialen
        $d^{k}=d_{k+1}^{*}\colon\C^{k}\longrightarrow \C^{k+1}$, sind Komplexe
        von $\mathbb{K}$-Vektorräumen. Mit den Abbildungen
        \begin{align*}
            \Xi^{k}\colon\Hom_{\mathcal{A}^{e}}(\C_{k},\mathcal{M})&\longrightarrow
            HC^{k}(\mathcal{A},\mathcal{M})\\
            \psi &\longmapsto \left(\widetilde{\psi}\colon(x_{1}\ot…\ot x_{k})\mapsto
                \psi(1\ot x_{1}\ot…\ot
                x_{k}\ot1)\right),
        \end{align*} erhalten wir aus der Verträglichkeit von
        $\mathcal{M}$ sowie der Bilinearität der Modul-Multi\-pli\-ka\-tio\-nen:
        \begin{align*}
            \Xi^{k}(\psi)(\lambda\: x_{1}\ot…\ot
            x_{k})&=\psi\:(1\ot x_{1}\ot…\ot \lambda x_{i}\ot…\ot
            x_{k}\ot1)\\ &=\psi\:(\lambda 1\ot x_{1}\ot…\ot
            x_{k}\ot1)\\ & =\lambda 1\ot 1 *_{e}\psi\:(1\ot x_{1}\ot…\ot
            x_{k}\ot1)\\ &=\lambda\: \psi\:(1\ot x_{1}\ot…\ot
            x_{k}\ot1)\\ &=\lambda\: \Xi^{k}(\psi)(x_{1}\ot…\ot
            x_{k}).
        \end{align*} Damit bilden die $\Xi^{k}$ in der Tat in die behauptete
        Menge ab, und da
        \begin{equation*}
            \Xi^{k}(\lambda \psi+\phi)=\lambda\:
            \Xi^{k}(\psi)+ \Xi(\phi),
        \end{equation*}sind diese zudem
        $\mathbb{K}$-Homomorphismen.
        Nun folgt die Injektivität obiger
        Abbildung unmittelbar aus der $\mathcal{A}^{e}$-Linearität der
        Urbilder. Für die Surjektivität betrachten wir ein
        $\widetilde{\psi}\in HC^{k}(\mathcal{A},\mathcal{M})$ und
        definieren $\Xi^{k}_{-1}\colon HC^{k}(\mathcal{A},\mathcal{M})\longrightarrow
        \Hom_{\mathcal{A}^{e}}(\C_{k},\mathcal{M})$ durch
        \begin{align*}
            \Xi^{k}_{-1}(\psi)(x_{0}\ot
            x_{1}\ot…\ot x_{k}\ot
            x_{k+1})=x_{0}\ot x_{k+1}*_{e}\psi(x_{1}\ot…\ot
            x_{k}),
        \end{align*}
        womit $\Xi^{k} \cp\Xi^{k}_{-1} =\id_{HC^{k}(\mathcal{A},\mathcal{M})}$ , also die
        $\Xi^{k}$ Isomorphismen sind.
        Nun folgt
        \begin{equation}
            \label{eq:isobarHsch}
            \Xi^{k+1}d^{*}_{k+1}=\delta^{k}\:\Xi^{k},
        \end{equation}da
        {\allowdisplaybreaks\small
          \begin{align*}
              \left(\Xi^{k+1}\cp d^{*}_{k+1}\right)&(\psi)(x_{1}\ot…\ot
              x_{k+1})=(d^{*}_{k+1}\psi)(1\ot x_{1}\ot…\ot x_{k+1}\ot 1)
              \\ =&\: \psi(x_{1}\ot…\ot x_{k+1}\ot 1)
              +\sum_{j=1}^{k}(-1)^{j}
              \psi(1\ot x_{1}\ot…\ot x_{j}x_{j+1}\ot…\ot x_{k+1}\ot 1)
              \\ &\qquad\qquad\qquad\qquad\qquad+ (-1)^{k+1}\psi(1\ot…\ot x_{k+1})
              \\=&\: x_{1}\psi(1\ot x_{2}\ot…\ot x_{k+1}\ot 1)
              +\sum_{j=1}^{k}(-1)^{j}
              \psi(1\ot x_{1}\ot…\ot x_{j}x_{j+1}\ot…\ot x_{k+1}\ot 1)
              \\ &\qquad\qquad\qquad\qquad\qquad\qquad\:\:\:+ (-1)^{k+1}\psi(1\ot x_{1}\ot…\ot x_{k}\ot 1)\:x_{k+1}
              \\=&\: x_{1}\left(\Xi^{k}\cp\psi\right)(x_{2}\ot…\ot x_{k+1})
              + \sum_{j=1}^{k}(-1)^{k}\left(\Xi^{k}\cp\psi\right)(x_{1}\ot…\ot
              x_{j}x_{j+1}\ot…\ot x_{k+1})
              \\ &\qquad\qquad\qquad\qquad\qquad\qquad\:\:\:+ (-1)^{k+1}\left(\Xi^{k}\cp \psi\right)(x_{1}\ot…\ot x_{k})\:x_{k+1}
              \\ =& \left(\delta^{k}\cp\:\Xi^{k}\right)(\psi)(x_{1}\ot…\ot x_{k+1}),
          \end{align*}}und mit der $\mathbb{K}$-Linearität der $ \Xi^{k}$
        zeigt Lemma \ref{lemma:kettenabzu}~\textit{iii)} die
        Isomorphismen-Eigenschaft der $\wt{\Xi^{k}}$.
    \end{beweis}
\end{proposition}

\section{Die Hochschild-Kohomologie der Algebra
  $\Poly$}
\label{subsec:HschKPol}
Wir wollen als erstes einfaches Beispiel die
Hochschild-Kohomologie der $\mathbb{R}$-Algebra $\mathcal{A}=\Poly$, der Polynome
auf $\mathbb{R}^{n}$, berechnen. Diese ist sicher unitär und assoziativ und wir haben gemäß
Lemma \ref{lemma:unitAlhabeBarAufloesProj} und
Proposition \ref{prop:barauffuerunitalgebraIsomozuHochschildkohomo} bereits eine
projektive Auflösung, deren Kohomologiegruppen zu den
gesuchten Hochschild-Kohomologien isomorph sind.
Als nächstes wollen wir uns eine
weitere projektive Auflösung $(C',d',\epsilon')$ von
$\mathcal{A}$ verschaffen und wissen bereits, dass dann:
\begin{equation*}
    H^{k}(\Hom_{\mathcal{A}^{e}}(\C,\mathcal{M}))\cong \mathrm{Ext}^{k}_{\mathcal{A}^{e}}(\cdot,
    \mathcal{M})(\mathcal{A})\cong
    H^{k}(\Hom_{\mathcal{A}^{e}}(C',\mathcal{M})).
\end{equation*}
\begin{definition}[Koszul-Komplex]
    Gegeben die Algebra $\mathcal{A}=\Poly$, so definieren wir den Koszul-Komplex
    $(\K,\partial)$ durch die $\mathcal{A}^{e}$-Moduln
    \begin{equation*}
        \K_{0}=\mathcal{A}^{e}\qquad\text{    sowie    }\qquad
        \K_{k}=\mathcal{A}^{e}\ot
        \Lambda^{k}(\mathbb{R}^{n*})\quad\forall\:k\geq 1
    \end{equation*}mit der offensichtlichen
    $\mathcal{A}^{e}$-Multiplikation im ersten Faktor. Dies bedeutet insbesondere 
    $\K_{k}=0$ falls $k>n$. Für $0< k\leq n$ definieren wir die
    $\mathcal{A}^{e}$-Homomorphismen:
    \begin{align*}
        \partial_{k}\colon\K_{k}&\longrightarrow \K_{k-1}\\
        \omega&\longmapsto \left[(v,w)(x_{1},…,x_{k-1})\mapsto \omega(v,w)((v-w),x_{1},…,x_{k-1})\right].
    \end{align*}Es folgt unmittelbar $\pt_{k}\cp
    \pt_{k+1}=0$, und man beachte zudem, dass die $\pt_{k}$ mit Hilfe
    der Einsetzabbildung $i_{a}(v,\omega)(x_{2},…,x_{k})=\omega(v,x_{2},…,x_{k})$:
    \begin{align*}
        &i_{a}\colon \mathbb{R}^{n}\times  \Lambda^{k}(\mathbb{R}^{n*})\longrightarrow
        \Lambda^{k-1}(\mathbb{R}^{n*})\\
        &(v, \omega^{1}\wedge…\wedge \omega^{k})\longmapsto
        \sum_{l=1}^{k}(-1)^{l-1}\omega^{l}(v)\:\omega^{1}\wedge…\blacktriangle^{l}…\wedge \omega^{k}
    \end{align*}  
    auch als
    $\pt_{k}=\displaystyle\sum_{j=1}^{n}\xi^{j}i_{a}(\vec{e}_{j},\cdot)$
    mit $\mathcal{A}^{e}\ni\xi^{j}=x^{j}\ot 1-1\ot x^{j}$ geschrieben werden können.
\end{definition}
\begin{lemma}[Koszul-Auflösung]
    Sei $\mathcal{A}=\Poly$, so wird der Koszul-Komplex
    vermöge der Abbildung
    \begin{align*}
        \epsilon\colon\K_{0}&\longrightarrow \mathcal{A}\\
        a\ot b&\longmapsto ab
    \end{align*} zu einer projektiven und sogar freien Auflösung $(\K,\partial,\epsilon)$
    von $\Poly$.
    \begin{beweis}
        Wir hatten bereits gesehen, dass $\epsilon$ ein surjektiver
        $\mathcal{A}^{e}$-Homomorphismus ist. Die
        Freiheit der $\K_{k}$ folgt ebenso wie für die
        Bar-Auflösung, da die $\Lambda^{k}(\mathbb{R}^{n*})$
        ebenfalls Vektorräume $V$ mit Basen sind.
        Für die Exaktheit definieren wir die Abbildungen
        \begin{align*}
            h_{-1}\colon\mathcal{A}&\longrightarrow
            \K_{0}\\
            p&\longmapsto [(v,w)\mapsto p(w)]
        \end{align*} und $h_{k}\colon\K_{k}\longrightarrow
        \K_{k+1}$ für $k\geq 0$, durch:
        \begin{equation}
            \label{eq:exakthAbbHvonPol}
            \begin{split}
                h_{k}(\omega)(v,w)&=\sum_{j=1}^{n}e^{j}\wedge\int_{0}^{1}dt\:
                t^{k}\frac{\partial\omega}{\partial v^{j}}(tv+(1-t)w,w)\\ &=\frac{1}{k!}
                \sum_{i_{1},…,i_{k},j=1}^{n}\int_{0}^{1}dt\:
                t^{k}\frac{\partial\omega_{i_{1},…,i_{k}}}{\partial
                  v^{j}}(tv+(1-t)w,w)\:e^{j}\wedge e^{i_{1}}\wedge
                …\wedge e^{i_{k}}\nonumber,
            \end{split}
        \end{equation}wobei
        \begin{equation*}
            \omega=\frac{1}{k!}
            \sum_{i_{1},…,i_{k}}^{n}\omega_{i_{1},…,i_{k}}\ot
            e^{i_{1}}\wedge…\wedge e^{i_{k}}\quad\text{ und }\quad
            \omega_{i_{1},…,i_{k}}\in \mathcal{A}^{e}.
        \end{equation*}Zunächst überzeugt man sich, dass besagte
        Abbildungen in der Tat nach
        \begin{equation*}
            \mathrm{Pol}(\mathbb{R}^{n})\ot\mathrm{Pol}(\mathbb{R}^{n})\ot
            \Lambda^{k}(\mathbb{R}^{n*}) =
            \mathrm{Pol}(\mathbb{R}^{n}\times\mathbb{R}^{n})\ot \Lambda^{k}(\mathbb{R}^{n*})
        \end{equation*}
        abbilden, denn es ist ja jedes
        $\frac{\partial\omega_{i_{1},…,i_{k}}}{\partial v^{j}}$ als Ableitung eines
        Polynoms nach den ersten Argumenten wieder ein
        Polynom auf $\mathbb{R}^{n}\times \mathbb{R}^{n}$. 

        Ebenso haben wir $p(t\vec{x}+(1-t)\vec{y})\in
        \mathrm{Pol}(\mathbb{R}\times \mathbb{R}^{n}\times \mathbb{R}^{n})$ für $p\in \mathrm{Pol}(\mathbb{R}^{n}\times \mathbb{R}^{n})$,
        und die Integration ist nichts weiter, als die
        $\mathcal{\mathbb{R}}$-lineare Fortsetzung der Abbildung
        \begin{equation*}
            \int_{0}^{1}dt\:t^{k}\colon t^{l}x^{n}\longmapsto
            \frac{1}{l+k+1}x^{n}.
        \end{equation*}
        Behändiges Rechnen unter Verwendung der Derivationseigenschaft
        \begin{equation*}
            i_{a}(v)(\phi\wedge \psi )=i_{a}(v)(\phi)\wedge \psi +(-1)^{deg(\phi)}\phi\wedge i_{a}(v)(\psi)
        \end{equation*}
        zeigt:
        \begin{align*}
            \epsilon\circ h_{-1}&=\id_{\mathcal{A}}, \\
            h_{-1}\circ\epsilon+\partial_{1}\circ h_{0}&=\id_{\K_{0}}\quad\quad\text{und}\\
            h_{k-1}\circ \partial_{k}+\partial_{k+1}\circ h_{k}&=\id_{\K_{k}}\quad\quad k\geq1,
        \end{align*} 
        mithin die Exaktheit von
        $(\K,\partial,\epsilon)$, vgl. \cite[Kapitel 5]{Weissarbeit}.
    \end{beweis}
\end{lemma} 
Mit obigem Lemma erhalten wir nun umgehend folgenden Satz.
\begin{satz}
    \label{satz:PolsatzHochsch}
    Sei $\mathcal{A}=\Poly$ und $\mathcal{M}$ ein
    $\mathcal{A}-\mathcal{A}$-Bimodul, dann gilt:
    \begin{equation}
        \label{eq:HochschPol11}
        HH^{k}(\mathcal{A},\mathcal{M})\cong
        H^{k}(\Hom_{\mathcal{A}^{e}}(\C,\mathcal{M}),d^{*})\cong
        H^{k}(\Hom_{\mathcal{A}^{e}}(\K,\mathcal{M}),\pt^{*}).
    \end{equation}
    Ist $\mathcal{M}$ zudem symmetrisch, so ist:
    \begin{equation}
        \label{eq:HochschPol22}
        HH^{k}(\mathcal{A},\mathcal{M})\cong \Hom_{\mathcal{A}^{e}}(\K_{k},\mathcal{M})=\mathcal{M}\ot\Lambda^{k}(\mathbb{R}^{n}).
    \end{equation}
    \begin{beweis}
        \eqref{eq:HochschPol11} ist wegen$H^{k}(\Hom_{\mathcal{A}^{e}}(\C,\mathcal{M}))\cong \mathrm{Ext}^{k}_{\mathcal{A}^{e}}(\cdot,
        \mathcal{M})(\mathcal{A})\cong
        H^{k}(\Hom_{\mathcal{A}^{e}}(\K,\mathcal{M}))$ klar und für \eqref{eq:HochschPol22} betrachten wir den Komplex
        $(\K^{*},\partial^{*})$,
        der durch Anwendung des
        $\mathrm{hom}_{\mathcal{A}^{e}}(\cdot,\mathcal{M})$-Funktors auf $(\K,\partial)$
        gewonnen wird.
        Seien weiter $\phi\in
        \K^{*}_{k}=\Hom_{\mathcal{A}^{e}}(\K_{k},\mathcal{M})$ und
        $\omega=\sum_{i_{1},…,i_{k+1}}\omega_{i_{1},…,i_{k+1}}\ot e^{i_{1}}\wedge…\wedge
        e^{i_{k+1}}\in \K_{k+1}$.\\ Dann folgt für
        $\pt^{*}_{k+1}\colon\K^{*}_{k}\longrightarrow \K^{*}_{k+1}$:
        \begin{align*}
            (\partial^{*}_{k+1}\phi)(\omega)\glna{\eqref{eq:homfktrMor}}&(\phi\circ
            \partial_{k+1})(\omega)\\=&\:\phi\left(\partial_{k+1}\left(\sum_{i_{1},…,i_{k+1}}\omega_{i_{1},…,i_{k}}\ot
                    u^{i_{1}}\wedge…\wedge
                    u^{i_{k+1}}\right)\right)\\ &=
            \sum_{i_{1},…,i_{k+1}}\phi\left(\sum_{j=1}^{n}\xi^{j}*_{e}\omega_{i_{1},…,i_{k+1}}\ot
                i_{a}\left(\vec{e}_{j},u^{i_{1}}\wedge…\wedge
                    u^{i_{k+1}}\right)\right)\\
            &=\sum_{i_{1},…,i_{k+1}}\sum_{j=1}^{n}\xi^{j}*_{e}\phi\left(\omega_{i_{1},…,i_{k+1}}\ot
                i_{a}\left(\vec{e}_{j},u^{i_{1}}\wedge…\wedge
                    u^{i_{k+1}}\right)\right) \\ &= \sum_{i_{1},…,i_{k+1}}\sum_{j=1}^{n}\:(x^{j}\ot
            1-1\ot x^{j})*_{e}\phi\left(\omega_{i_{1},…,i_{k+1}}\:i_{a}\left(\vec{e}_{j},u^{i_{1}}\wedge…\wedge
                    u^{i_{k+1}}\right)\right)\\ &=0.
        \end{align*}Dabei gilt die letzte Gleichheit wegen der 
        Symmetrie von $\mathcal{M}$.
        Dies zeigt $\ker(\partial^{*}_{k+1})= \K^{*}_{k}$ und $\im(\partial^{*}_{k})=0$, mithin
        $H^{k}(\Hom_{\mathcal{A}^{e}}(\K,\mathcal{A}))=\ker(\pt_{k+1}^{*})/\im(\pt_{k}^{*})=\Hom_{\mathcal{A}^{e}}(\K_{k},\mathcal{A})$.\\\\
        Für die letzte Gleichheit in \eqref{eq:HochschPol22} erinnern wir, dass 
        $\Lambda^{k}(\mathbb{R}^{n*})^{*}=\Lambda^{k}(\mathbb{R}^{n})$
        und erhalten für $\phi \in \Hom_{\mathcal{A}^{e}}(\K_{k},\mathcal{M})$: 
        {\allowdisplaybreaks\small
          \begin{align*}
              \phi(a^{e}\ot \omega)=\:&a^{e}*_{e}\phi\left(1^{e} \ot
                  \sum_{j_{1},…,j_{k}=1}^{n}\omega_{j_{1},…,j_{k}}e^{j_{1}}\wedge…\wedge
                  e^{j_{k}}\right)
              \\=&\: a^{e}*_{e}
              \sum^{n}_{j_{1},…,j_{k}}\omega_{j_{1},…,\omega_{k}}\ot
              1*_{e}\phi\:(1^{e}\ot e^{j_{1},…,j_{k}})
              \\=&\:
              a^{e}*_{e}\sum_{j_{1},…,j_{k}=1}^{n}\omega_{j_{1},…,j_{k}}\:\phi^{j_{1},…,j_{k}}
              \\=&\: a^{e}*_{e} \left(\sum_{j_{1},…,j_{k}=1}^{n} \phi^{j_{1},…,j_{k}}\ot e_{j_{1}}\wedge…\wedge
                  e_{j_{k}}\right)(1^{e}\ot\omega)
              \\=&\: \left(\sum_{j_{1},…,j_{k}=1}^{n} \phi^{j_{1},…,j_{k}}\ot e_{j_{1}}\wedge…\wedge
                  e_{j_{k}}\right)(a^{e}\ot\omega),
          \end{align*}} und mit der Endlichkeit der Summe in der Tat
        {\small\begin{equation*}
              \left(\sum_{j_{1},…,j_{k}=1}^{n} \phi^{j_{1},…,j_{k}}\ot e_{j_{1}}\wedge…\wedge
                  e_{j_{k}}\right)\in \mathcal{M}\ot \Lambda^{k}(\mathbb{R}^{n}).
          \end{equation*}}Dabei haben wir im zweiten Schritt wieder
        $1*_{L}m=m=m*_{R}1$ für alle $m\in \mathcal{M}$ und die Bilinearität der Modul-Multiplikationen benutzt. Die letzten beiden Schritte folgen mit
        der Konvention 
        \begin{equation}
            \label{eq:Multi} 
            m\ot \lambda\:(a^{e}\ot
            \omega):=a^{e}*_{e}m\cdot
            \omega(\lambda)\qquad\text{ mit }\qquad \lambda\in \Lambda^{k}(\mathbb{R}^{n}).
        \end{equation}
        Umgekehrt ist klar, dass jedes Element aus
        $\mathcal{M}\ot \Lambda^{k}(\mathbb{R}^{n})$ vermöge \eqref{eq:Multi} ein Element in
        $\Hom_{\mathcal{A}^{e}}(\K_{k},\mathcal{M})$ definiert.
    \end{beweis}
\end{satz}
Für $M=\Pol(\mathbb{R}^{n})$ wurde dieser Satz ursprünglich von
Hochschild, Kostant und Rosenberg bewiesen, siehe
\cite{hochschild.kostant.rosenberg:1962a}. Eine Behandlung des Falles
$\mathcal{M}=\mathcal{A}=C^{\infty}(M)$ findet man in 
\cite{cahen.gutt.dewilde:1980a}. 
\begin{bemerkung}
    Für einen expliziten Isomorphismus benötigen wir zunächst eine zu $\mu=\id_{\mathcal{A}^{e}}$ gehörige Kettenabbildung
    \begin{equation}
        \label{eq:Gpol}
        G\colon(\C,d)\rightarrow (\K,\partial)
    \end{equation} oder
    \begin{equation}
        \label{eq:FPol}
        F\colon(\K,\partial)\rightarrow (\C,d).
    \end{equation} 
    Nach dem Beweis von
    Lemma \ref{lemma:GruppenKOhomsausprojaufloesundFunktoren}~\textit{ii.)}
    erhalten wir durch Anwenden des
    $\mathrm{hom}_{\mathcal{A}^{e}}(\cdot,\mathcal{M})$-Funktors
    Abbildungen $F_{k}^{*}$ und $G_{k}^{*}$, die Isomorphismen
    \begin{align*}
        \widetilde{G_{k}^{*}}\colon&H^{k}(\Hom_{\mathcal{A}^{e}}(\K,\mathcal{M}))\longrightarrow
        H^{k}(\Hom_{\mathcal{A}^{e}}(\C,\mathcal{M}))\\
        \widetilde{F_{k}^{*}}\colon&H^{k}(\Hom_{\mathcal{A}^{e}}(\C,\mathcal{M}))\longrightarrow
        H^{k}(\Hom_{\mathcal{A}^{e}}(\K,\mathcal{M})) 
    \end{align*}auf Kohomologie-Niveau induzieren. Nach selbigem Beweis
    sind $\widetilde{F_{k}^{*}}$ und $\widetilde{G_{k}^{*}}$ sogar zueinander invers.  
    Die Verkettung $\wt{\ot_{k}^{*}} \cp\wt{\Xi^{k}}\cp\wt{G_{k}^{*}}$ 
    ist dann der gewünschten Isomorphismus nach
    $HH^{k}(\mathcal{A},\mathcal{M})$. 

    Explizite Kettenabbildungen sind beispielsweise gegeben durch \cite{bordemann.et.al:2003a:pre}:  
    \begin{align*}
        G_{k}&\colon \bigotimes^{k+2}\Poly \longrightarrow
        \mathrm{Pol}(\mathbb{R}^{n} \times \mathbb{R}^{n})\ot
        \Lambda^{k}(\mathbb{R}^{n*})\\ (G_{k}p)(v,w)&=
        \sum_{i_{1},…,i_{k}}^{n}e^{i_{1}}\wedge…\wedge
        e^{i_{k}}\int_{0}^{1}dt_{1}\int_{0}^{t_{1}}dt_{2}...\int_{0}^{t_{k-1}}dt_{k}\\
        &\frac{\partial
          p}{\partial q_{1}...\partial q_{k}}(v,t_{1}v+(1-t_{1})w,…,t_{k}v+(1-t_{k})w,w)
    \end{align*} mit $\partial_{k+1}\cp G_{k+1}=G_{k}\cp d_{k+1}$, $G_{k}=0$
    für $k>n$ und $G_{0}=\id_{\mathcal{A}^{e}}$, sowie durch die
    Abbildung aus \cite[Sect. III.2$\alpha$]{connes:1994a}:
    \begin{align*}
        F_{k}\colon  \mathrm{Pol}(\mathbb{R}^{n} \times \mathbb{R}^{n})\ot
        \Lambda^{k}(\mathbb{R}^{n*})&\longrightarrow \bigotimes^{k+2}\Poly
        \\ \omega & \longmapsto [(v,w)(x_{1},…,x_{k})\mapsto \omega(v,w)(x_{1}-v,…,x_{k}-v)],
    \end{align*} mit $d_{k+1}\cp F_{k+1}=F_{k}\cp \partial_{k+1}$,
    $F_{k}=0$ für $k>n$ und $F_{0}=\id_{\mathcal{A}^{e}}$.\\\\
    Eine einfache Rechnung zeigt dann $G_{k}\cp F_{k}= \id_{\K_{k}}$,
    also $F_{k}^{*}\cp
    G_{k}^{*}=\id_{\Hom_{\mathcal{A}^{e}}}(\K_{k},\mathcal{M})$ und
    somit $\wt{F_{k}^{*}}\cp
    \wt{G_{k}^{*}}=\id_{H^{k}(\Hom_{\mathcal{A}^{e}}(\K_{k},\mathcal{M}))}$.
    Zusammen mit der Isomorphismus-Eigenschaft von $\wt{F_{k}^{*}}$
    und  $\wt{G_{k}^{*}}$ bestätigt dies $\wt{G_{k}^{*}}^{-1}=\wt{F_{k}^{*}}$.
\end{bemerkung}
\section{Die Hochschild-Kohomologie der Algebra
  $\Ss^{\bullet}(\mathbb{V})$}
\label{subsec:HochschKohSym}
\subsection{Die symmetrische und die Graßmann Algebra}
Als abstrakte Variante des Polynom-Begriffes betrachten wir für einen
gegebenen $\mathbb{K}$-Vektorraum $\V$, die unitäre, assoziative $\mathbb{K}$-Algebra
$\left(\sym,\vee\right)$. 
Dies ist der gradierte Vektorraum
$\displaystyle\Ss^{\bullet}(\V)=\bigoplus_{k=0}^{\infty}\mathrm{S}^{k}(\mathbb{V})$
mit Untervektorräumen
$\mathrm{S}^{k}(\mathbb{V})=\im\left(\mathrm{Sym}_{k}\right)\subseteq
\Tt^{k}(\V)$ und $S^{0}(\V)=\mathbb{K}$. Hierbei bezeichnet $\mathrm{Sym}_{k}\colon\Tt^{k}(\V)\longrightarrow \Tt^{k}(\V)$ die mit
Korollar \ref{kor:WohldefTensorprodabbildungen} wohldefinierte lineare
Fortsetzung von
\begin{align*}
    v_{1}\ot\dots\ot v_{k}\longmapsto \frac{1}{k!}\sum_{\sigma\in S_{k}}v_{\sigma(1)}\ot\dots\ot v_{\sigma(k)}.
\end{align*}
Die bis auf kanonische Isomorphie kommutative, assoziative
Algebramultiplikation ist dabei definiert durch $\vee=S\cp
\ot^{\bullet}$. Hierbei bezeichnet $S\colon\Tt^{\bullet}(\V)\longrightarrow \SsV$
die Abbildung:
\begin{align*}
    \SsV\ni\sum_{l}\alpha_{l}\longmapsto \sum_{l}\mathrm{Sym}_{l}(\alpha_{l})\quad\text{ mit }\quad\mathrm{Sym}_{0/1}=\id_{\Ss^{0/1}(\V)},
\end{align*}
und
$\ot^{\bullet}\colon \Ss^{\bullet}(\V)\times
\Ss^{\bullet}(\V)\longrightarrow  \mathrm{T}^{\bullet}(\V)$ ist gegeben durch
\begin{align*}
    \ot^{\bullet}\colon\left(\sum_{l}\alpha_{l},\sum_{m}\beta_{m}\right)\longmapsto\sum_{l,m}\alpha_{l}\ot\beta_{m}\qquad\forall\:\alpha_{l}\in \Ss^{l}(\V), \beta_{m}\in \Ss^{m}(\V).
\end{align*}
Hierbei haben wir stillschweigend Lemma \ref{lemma:assTenprod}~\textit{ii.)}, also $\Tt^{l}(\V)\ot \Tt^{m}(\V)\cong \Tt^{l+m}(\V)$ benutzt. 
Weiter beachte man, dass $ \mathbb{K}\ot \mathbb{W} \cong \mathbb{W} \cong
\mathbb{W}\ot \mathbb{K}$
für beliebigen $\mathbb{K}$-Vektorraum
$\mathbb{W}$ gilt, da $\mathbb{K}$ selbst ein eindimensionaler
$\mathbb{K}$-Vektorraum ist.
Man setzt dann
\begin{equation*}
    \ot^{\bullet}(\mathbb{k},\alpha_{l})=\left(\:\cong_{\Ss^{l}(\V)}\cp\ot\right)(\mathbb{k},\alpha_{l})=\mathbb{k}\cdot
    \alpha_{l}\qquad\forall\:  \alpha_{l}\in \Ss^{l}(\V)
\end{equation*}
und erhält insbesondere $1_{\SsV}=1_{\mathbb{K}}$ als Einselement.
Obige Definition hat dabei den Vorteil, dass
\begin{equation*}
    (\alpha_{1}\vee\dots\vee\alpha_{l})\vee (\alpha_{l+1}\vee\dots\vee \alpha_{k})=\alpha_{1}\vee\dots\vee\alpha_{k}
\end{equation*}
f"ur $\alpha_{1}\vee\dots\vee \alpha_{k}:=\mathrm{Sym}_{k}(\alpha_{1}\ot…\ot
\alpha_{k})$ gilt.
Analog definieren wir die Gra"smann-Algebra $\Lambda^{\bullet}(\V)=\displaystyle\bigoplus_{k=0}^{\infty}\Lambda^{k}(\V)$ mit $\Lambda^{k}(\V)=\im\left(\mathrm{Alt}_{k}\right)$, und $\mathrm{Alt}_{k}$ die lineare Fortsetzung von $u_{1}\ot…\ot u_{k} \longmapsto \frac{1}{k!}\displaystyle\sum_{\sigma\in
  S^{k}}\sign(\sigma)\:u_{\sigma(1)}\ot…\ot u_{\sigma(k)}$. Mit
$A=\displaystyle\sum_{k=1}^{\infty}\mathrm{Alt}_{k}$ ist dann die zugeh"orige
Algebramultiplikation durch $\wedge=A\cp\ot^{\bullet}$ definiert.
Wie f"ur den symmetrischen Fall, folgt $(\alpha_{1}\wedge\dots\wedge\alpha_{l})\wedge
(\alpha_{l+1}\wedge\dots\wedge \alpha_{k})=\alpha_{1}\wedge\dots
\wedge\alpha_{k}$
mit
$\alpha_{1}\wedge\dots\wedge \alpha_{k}:=\mathrm{Alt}_{k}(\alpha_{1}\ot\dots\ot
\alpha_{k})$.
\subsection{Bestimmung der Hochschild-Kohomologie von
  $\Ss^{\bullet}(\mathbb{V})$}
Mit der Unitarität von $\mathcal{A}=\SsV$ ist die Existenz einer Bar-Auflösung $(\C,d,\epsilon)$ von
$\sym$ gesichert, und es gilt die Isomorphie $HH^{k}(\mathcal{A},\mathcal{M})\cong
H^{k}(\Hom_{\mathcal{A}^{e}}(\C,\mathcal{M}))$ für die von uns betrachteten $\mathcal{A}-\mathcal{A}$-Bimodul $\mathcal{M}$. Was wir nun noch benötigen, um die Hochschild-Kohomologie von $\SsV$ zu bestimmen, ist 
lediglich eine Koszul-Auflösung $(\K,\partial,\epsilon)$ von $\Ss^{\bullet}(\V)$, da dann wieder
\begin{equation*}
    H^{k}(\Hom_{\mathcal{A}^{e}}(\C,\mathcal{M}))\cong \mathrm{Ext}^{k}_{\mathcal{A}^{e}}(\cdot,
    \mathcal{M})(\mathcal{A})\cong
    H^{k}(\Hom_{\mathcal{A}^{e}}(C',\mathcal{M}))
\end{equation*}gilt.
Zu diesem Zwecke seien die $\K_{k}$ wie in Abschnitt
\ref{subsec:HschKPol} gegeben durch
$\mathcal{A}^{e}$-Moduln
\begin{equation*}
    \K_{0}=\mathcal{A}^{e}\qquad\text{    und    }\qquad
    \K_{k}=\mathcal{A}^{e}\ot \Lambda^{k}(\mathbb{V})\:\text{ f"ur }k\geq 1
\end{equation*} mit der bekannten $\mathcal{A}^{e}$-Multiplikation im
ersten Faktor.\\\\
F"ur  $\alpha \in \mathrm{S}^{l}(\mathbb{V})$ mit $\alpha=\alpha_{1}\vee…\vee \alpha_{l}$
bezeichne im Folgenden
$\alpha^{j}\in \Ss^{l-1}(\mathbb{V})$ das Element
$\alpha_{1}\vee…\blacktriangle^{j}…\vee \alpha_{l}$, welches durch Weglassen von
$\alpha_{j}$ aus $\alpha$ entsteht. Ebenso sei $\alpha^{j_{1},…,j_{s}}\in
\Ss^{l-s}(\mathbb{V})$ das Element, welches durch Weglassen der  $\alpha_{j_{1}},\dots,\alpha_{j_{l}}$ aus $\alpha$ hervorgeht. Ist $\deg(\alpha)=l$, so setzen wir $\alpha^{1,…,l}=1_{\SsV}$. Sinngem"a"s benutzen wir diese Konventionen f"ur die Elemente in $\Lambda^{k}(\V)$.\\
\begin{definition}
    \label{def:partialdef}
    F"ur obige $\mathcal{A}^{e}$-Moduln definieren wir Kettendifferentiale durch:
    \begin{align*}
        \partial_{k}\colon \mathcal{A}^{e}\ot
        \Lambda^{k}(\mathbb{V})&\longmapsto \mathcal{A}^{e}\ot    \Lambda^{k-1}(\mathbb{V})\\
        \omega&\longmapsto
        \big[\pt^{k}_{1}-\pt^{k}_{2}\big](\omega)
    \end{align*}
    mit        
    \begin{align*}
        \pt_{1}^{k}\colon\Ss^{\bullet}(\mathbb{V})\ot
        \Ss^{\bullet}(\mathbb{V})\ot\Lambda^{k}(\mathbb{V})&\longrightarrow \Ss^{\bullet+1}(\mathbb{V})\ot
        \Ss^{\bullet}(\mathbb{V})\ot\Lambda^{k-1}(\mathbb{V})\\
        \KE{\alpha}{\beta}{u}&\longmapsto
        \sum_{j=1}^{k}(-1)^{j-1}\:\KE{u_{j}\vee
          \alpha}{\beta}{u^{j}}\\
        \pt_{2}^{k}\colon\Ss^{\bullet}(\mathbb{V})\ot
        \Ss^{\bullet}(\mathbb{V})\ot\Lambda^{k}(\mathbb{V})&\longrightarrow \Ss^{\bullet}(\mathbb{V})\ot
        \Ss^{\bullet+1}(\mathbb{V})\ot\Lambda^{k-1}(\mathbb{V})\\
        \KE{\alpha}{\beta}{u}&\longmapsto
        \sum_{j=1}^{k}(-1)^{j-1}\:\KE{\alpha}{u_{j}\vee\beta}{u^{j}},
    \end{align*}
    wobei wir hier und im Folgenden $\mathcal{A}^{e}\ot \Lambda^{0}(\V)$ mit $\mathcal{A}^{e}$ identifizieren wollen. 
\end{definition}
\begin{bemerkung}
    Die Wohldefiniertheit obiger Abbildungen folgt wieder mit Korollar \ref{kor:WohldefTensorprodabbildungen}, da wir diese auch schreiben können, als
    \begin{align*}
        \pt^{k}_{1}&=(S \ot \id\ot \id)\cp
        \tilde{\pt}^{k}_{1}\Big|_{\K_{k}}\\
        \pt^{k}_{2}&=(\id \ot S \ot \id)\cp
        \tilde{\pt}^{k}_{2}\Big|_{\K_{k}}
    \end{align*} mit
    \begin{align*}
        \tilde{\pt}^{k}_{1}\colon\Tt^{\bullet}(\mathbb{V})\bbot
        \Tt^{\bullet}(\mathbb{V})\bbot
        \Tt^{k}(\mathbb{V})&\longrightarrow \Tt^{\bullet+1}(\mathbb{V})\bbot
        \Tt^{\bullet}(\mathbb{V})\bbot
        \Tt^{k-1}(\mathbb{V})\\
        \alpha\bbot \beta\bbot u&\longmapsto k\left(u_{1}\ot \alpha\bbot\beta\bbot u^{1}\right)\\
        \tilde{\pt}^{k}_{2}\colon\Tt^{\bullet}(\mathbb{V})\bbot
        \Tt^{\bullet}(\mathbb{V})\bbot
        \Tt^{k}(\mathbb{V})&\longrightarrow \Tt^{\bullet}(\mathbb{V})\bbot
        \Tt^{\bullet+1}(\mathbb{V})\bbot
        \Tt^{k-1}(\mathbb{V})\\
        \alpha\bbot \beta\bbot u&\longmapsto k\left(\alpha\bbot u_{1}\ot\beta\bbot u^{1}\right).
    \end{align*}
    Dabei liegt der Faktor $k$ an unserer Konvention
    \begin{equation*} 
        u_{1}\wedge\dots \wedge
        u_{k}=\frac{1}{k!}\displaystyle\sum_{\sigma\in
          S_{k}}\sign(\sigma)(u_{\sigma(1)}\ot\dots\ot u_{\sigma(k)}).
    \end{equation*} In der Tat erhalten wir f"ur $\alpha, \beta \in \Ss^{\bullet}(\V)$:
    {\allowdisplaybreaks\begin{align*}
          \wt{\pt}^{k}_{1}(\alpha\ot\beta\ot&\: u_{1}\wedge\dots\wedge u_{k})=\frac{1}{k!}\sum_{\sigma\in S_{k}}\sign(\sigma)\wt{\pt}^{k}_{1}\left(\alpha\:\boldsymbol{\ot}\:\beta\:\boldsymbol{\ot}\: u_{\sigma(1)}\ot\dots\ot u_{\sigma(k)}\right)\\
          =&\:\frac{1}{(k-1)!}\sum_{\sigma\in S_{k}}\sign(\sigma)\left(u_{\sigma(1)}\ot\alpha\:\boldsymbol{\ot}\:\beta\:\boldsymbol{\ot}\:u_{\sigma(2)}\ot\dots\ot u_{\sigma(k)}\right)
          \\=&\:\frac{1}{(k-1)!}\sum_{j=1}^{k}\sum_{\substack{\sigma\in
              S_{k}\\\sigma(1)=j}}\sign(\sigma)\left(u_{j}\ot\alpha\:\boldsymbol{\ot}\:\beta\:\boldsymbol{\ot}\:u_{\sigma(2)}\ot\dots\ot u_{\sigma(k)}\right)
          \\=&\:\frac{1}{(k-1)!}\sum_{j=1}^{k}(-1)^{j-1}\sum_{\sigma\in
            S_{k-1}}\sign(\sigma)\left(u_{j}\ot\alpha\:\boldsymbol{\ot}\:\beta\:\boldsymbol{\ot}\:\sigma^{*}\left[u_{1}\ot\dots\blacktriangle^{j}\dots\ot u_{k}\right]\right)
          \\=&\:\sum_{j=1}^{k}(-1)^{j-1}\left(u_{j}\ot\alpha\:\boldsymbol{\ot}\:\beta\:\boldsymbol{\ot}\:u_{1}\wedge\dots\blacktriangle^{j}\dots\wedge u_{k}\right),
      \end{align*}}wobei $\sigma^{*}\left[u_{1}\ot\dots\ot u_{k-1}\right]=u_{\sigma(1)}\ot\dots\ot
    u_{\sigma(k-1)}$ bedeutet.
    Symmetrisieren im ersten Argument liefert dann die gew"unschte Gleichheit. Analog folgt die Behauptung f"ur $\wt{\pt}^{k}_{2}$.
\end{bemerkung}
Folgendes Lemma zeigt, dass $(\K,\pt)$ ein Kettenkomplex ist.
\begin{lemma}
    Es gilt $\partial_{k-1}\cp\pt_{k}=0$.
    \begin{beweis}
        Seien hierf"ur abk"urzend 
        \begin{align*}
            u *_{L}\: (\alpha\ot \beta \ot \omega)&= u\vee \alpha\ot \beta \ot \omega \quad\text{ sowie }\\
            u *_{R} \:(\alpha\ot \beta \ot \omega)&= \alpha\ot  u\vee\beta \ot \omega,
        \end{align*}
        dann folgt:
        {\allowdisplaybreaks
          \begin{align*}
              \left(\partial_{k-1}\cp
                  \pt_{k}\right)(\alpha\ot&\:\beta\ot u)=\:\pt_{k-1}\left(\sum_{j=1}^{k}(-1)^{j-1}[u_{j}*_{L}-u_{j}\:*_{R}]\: \alpha\ot\beta\ot u^{j}\right)
              \\=&\: \sum_{j=1}^{k}(-1)^{j-1}\Bigg[\sum_{i=1}^{j-1}(-1)^{i-1}[u_{i}*_{L}-u_{i}\:*_{R}][u_{j}*_{L}-u_{j}\:*_{R}]\:\alpha\ot\beta\ot u^{i,j}
              \\ &\qquad\qquad\qquad +\sum_{i=j+1}^{k}(-1)^{i-2}[u_{i}*_{L}-u_{i}\:*_{R}][u_{j}*_{L}-u_{j}\:*_{R}]\:\alpha\ot\beta\ot u^{j,i}\Bigg]
              \\=&\:\sum_{j=2}^{k}\sum_{i<j}(-1)^{j+i}\:[u_{i}*_{L}-u_{i}\:*_{R}][u_{j}*_{L}-u_{j}\:*_{R}]\:\alpha\ot\beta\ot u^{i,j}
              \\ &-\sum_{j=1}^{k-1}\sum_{i>j}\overbrace{(-1)^{j+i}\:[u_{i}*_{L}-u_{i}\:*_{R}][u_{j}*_{L}-u_{j}\:*_{R}]}^{\tau_{i,j}}\alpha\ot\beta\ot u^{j,i}
              \\=&\:0,
          \end{align*}}da
        $\tau_{i,j}=\tau_{j,i}$ mit der Kommutativit"at von $\vee$, und somit
        \begin{align*}
            \sum_{j=1}^{k-1}\sum_{i>j}\tau_{i,j}\alpha\ot\beta\ot
            u^{j,i}\:\glna{\mathit{i}\leftrightarrow \mathit{j}}\:\sum_{i=1}^{k-1}\sum_{i<j}\tau_{i,j}\:\alpha\ot\beta\ot
            u^{i,j}=\:\sum_{j=2}^{k}\sum_{i<j}\tau_{i,j}\:\alpha\ot\beta\ot
            u^{i,j}.
        \end{align*}
        Dabei ist letzte Gleichheit rein kombinatorischer Natur.
    \end{beweis}
\end{lemma}
Zusammen mit der Abbildung
\begin{align*}
    \epsilon\colon\mathcal{A}\ot \mathcal{A}&\longrightarrow \mathcal{A}\\
    \alpha\ot \beta&\longmapsto \alpha\vee \beta
\end{align*} wird $(\K,\partial,\epsilon)$ zu einem
projektiven Komplex über $\SsV$, und es bleibt nun lediglich dessen Exaktheit
nachzuweisen. Hierfür definieren wir die abstrakte Variante von
\eqref{eq:exakthAbbHvonPol} durch:
\begin{definition}
    \label{def:KosSymExakthHomothidelta}
    Sei
    \begin{align*}
        h_{-1}\colon \mathcal{A}&\longrightarrow \K_{0}\\
        \alpha &\longmapsto 1\ot\alpha
    \end{align*}und
    \begin{align*}
        h_{k}\colon \Ss^{\bullet}(\mathbb{V})\ot
        \Ss^{\bullet}(\mathbb{V})\ot\Lambda^{k}(\mathbb{V})&\longrightarrow \Ss^{\bullet}(\mathbb{V})\ot
        \Ss^{\bullet}(\mathbb{V})\ot\Lambda^{k+1}(\mathbb{V})\\
        \mu
        &\longmapsto\int_{0}^{1}dt\:t^{k}(i_{t}\circ\delta)(\mu),
    \end{align*}für $k\geq 0$. Hierbei haben wir die folgenden Abbildungen benutzt:
    \begin{enumerate}
    \item
        $i_{t}:\SsV\ot\SsV\ot \Lambda^{k}(\V)\longrightarrow
        \big[\SsV\ot\SsV\ot \Lambda^{k}(\V)\big][t]$ definiert durch       
        \begin{align*}
            i_{t}\colon\KE{\alpha}{\beta}{u}\longmapsto&\:
            \:t^{l}\KE{\alpha}{\beta}{u}\:+\:t^{l-1}(1-t)\sum_{j=1}^{l}\KE{\alpha^{j}}{\alpha_{j}\vee\beta}{u} +…
            \\ &+t^{l-s}(1-t)^{s}\sum_{j_{1},…,j_{s}}^{l}\KE{\alpha^{j_{1},…,j_{s}}}{\alpha_{j_{1},…,j_{s}}\vee\beta}{u} +…
            \\ &+(1-t)^{l}\KE{1}{\alpha\vee \beta}{u}
        \end{align*}für $\deg(\alpha)=l$ und $i_{t}(1\ot \beta\ot
        u)=(1\ot \beta\ot u)$. Hierbei ist mit $\displaystyle\sum_{j_{1},…,j_{s}}$
        die Summe über alle $j_{1}\neq\dots \neq j_{s}$
        gemeint. Das Bild unter $i_{i}$ ist dann 
        als Polynom in $t$ mit Werten in $\SsV\ot\SsV\ot
        \Lambda^{k}(\V)$ zu verstehen.
    \item
        \begin{align*}
            \delta\colon\Ss^{\bullet}(\mathbb{V})\ot \Ss^{\bullet}(\mathbb{V})\ot
            \Lambda^{k}(\mathbb{V})&\longrightarrow \Ss^{\bullet-1}(\mathbb{V})\ot \Ss^{\bullet}(\mathbb{V})\ot \Lambda^{k+1}(\mathbb{V})\\
            \alpha_{l}\ot \beta \ot u&\longmapsto
            \sum_{j=1}^{l}\alpha_{l}^{j}\ot \beta\ot \:(\alpha_{l})_{j}\wedge u
        \end{align*} 
        für $\alpha_{l}\in \Ss^{l}(\V)$ und $\delta(1\ot\beta\ot u)=0$.\\
    \end{enumerate}
\end{definition}
\begin{bemerkung}
    Obige Abbildungen sind wohldefiniert, da zum einen $\delta=(\id\ot \id\ot A) \cp \tilde{\delta}\big|_{\K_{k}}$ mit
    \begin{align*}
        \tilde{\delta}\colon\Tt^{\bullet}(\mathbb{V})\bbot \Tt^{\bullet}(\mathbb{V})\bbot
        \Tt^{k}(\mathbb{V})&\longrightarrow
        \Tt^{\bullet-1}(\mathbb{V})\bbot \Tt^{\bullet}(\mathbb{V})\bbot
        \Tt^{k+1}(\mathbb{V})\\
        \alpha_{l}\bbot\beta\bbot u&\longmapsto l\left(\alpha_{l}^{1}\bbot\: \beta\bbot\: (\alpha_{l})_{1}\ot u\right),
    \end{align*}wobei der Faktor $l$ der Konvention f"ur $v_{1}\vee\dots\vee v_{l}=\frac{1}{l!}\displaystyle\sum_{\sigma\in S_{l}}v_{\sigma(1)}\ot\dots\ot v_{\sigma(l)}$\\ geschuldet ist, und da zum anderen 
    \begin{equation*}
        i_{t}=(\id\ot S\ot \id)\cp \left[\sum_{l=0}^{\infty}\sum_{s=0}^{l}\eta_{t}^{l,s}\right]\Bigg|_{\K_{k}}
    \end{equation*} mit
    \begin{align*}
        \eta^{l,s}_{t}\colon\Tt^{l}(\mathbb{V})\bbot
        \Tt^{\bullet}(\mathbb{V})\bbot
        \Tt^{k}(\mathbb{V})&\longrightarrow \Tt^{\bullet}(\mathbb{V})\bbot
        \Tt^{\bullet}(\mathbb{V})\bbot
        \Tt^{k}(\mathbb{V})\\
        \alpha_{l}\bbot \beta\bbot u&\longmapsto
        \binom{l}{s}\:t^{l-s}(1-t)^{s}\alpha_{l}^{1,…,s}\bbot\:
        (\alpha_{l})_{1,…,s}\ot \beta\bbot u,
    \end{align*}wobei $\eta_{t}^{0,0}(1\ot \beta\ot u)=(1\ot \beta\ot
    u)$.

    In der Tat erhalten wir f"ur $\alpha_{l}\in \Ss^{l}(\V)$, $\beta \in \SsV$ und $u\in \Lambda^{k}(\V)$:
    {\begin{align*}
          \eta_{t}^{l,s}(\alpha_{l}\ot \beta \ot u)
          =&\:\frac{1}{s!(l-s)!}t^{l-s}(1-t)^{s}\sum_{\sigma\in
            S_{l}}(\alpha_{l})_{\sigma(s+1)}\ot\dots
          \ot\:(\alpha_{l})_{\sigma(l)}\bbot
          \\ &\qquad\qquad\qquad\qquad\qquad\qquad\:(\alpha_{l})_{\sigma(1)}\ot\dots \ot\:(\alpha_{l})_{\sigma(s)}\ot \beta \bbot u
          \\=&\:t^{l-s}(1-t)^{s}\sum_{j_{1},\dots,j_{s}}^{l}\alpha_{l}^{j_{1},\dots,j_{l}}\bbot\: (\alpha_{l})_{j_{1},\dots,j_{l}}\ot\beta \bbot\:u.
      \end{align*}}
\end{bemerkung}
Um nun die gew"unschte Homotopieeigenschaft f"ur von $h$ nachzuweisen, ben"otigen wir zun"achst einige Rechenregeln. Sei hierf"ur  $\cdot$ die $\mathbb{K}$-bilineare Abbildung:
\begin{align*}
    \cdot\colon \mathcal{A}^{e} \ot \Lambda^{k}(\V)\times \mathcal{A}^{e} \ot \Lambda^{k'}(\V) &\longrightarrow \mathcal{A}^{e} \ot \Lambda^{k+k'}(\V)\\
    \big(\KE{\alpha}{\beta}{u},\KE{\wt{\alpha}}{\wt{\beta}}{\wt{u}}\big)&\longmapsto \KE{\alpha\vee \wt{\alpha}}{\:\beta\vee\wt{\beta}}{\:u\wedge \wt{u}}
\end{align*} und $\pt$ die Abbildung, die durch die
$\pt_{k}$ auf ganz $\Ss^{\bullet}(\V)\ot\Ss^{\bullet}(\V)\ot
\Lambda^{\bullet}(\V)$, verm"oge der Konvention $\pt|_{\mathcal{A}^{e}\ot\Lambda^{0}(\V)}=0$, induziert wird.
\begin{lemma}
    \label{lemma:EigenschHomotBaukloetze}
    \begin{enumerate}
    \item
        \label{item:deltaDerivat}
        \begin{align*}
            \delta(\KE{\alpha}{\beta}{u}\cdot
            \KE{\wt{\alpha}}{\wt{\beta}}{\wt{u}})=&\:\delta\KE{\alpha}{\beta}{u}\cdot\KE{\wt{\alpha}}{\wt{\beta}}{\wt{u}}\\
            &+
            (-1)^{\deg(u)}\KE{\alpha}{\beta}{u}\cdot\: \delta\KE{\wt{\alpha}}{\wt{\beta}}{\wt{u}};
        \end{align*}
    \item
        \label{item:partialDerivat}
        \begin{align*}
            \partial(\KE{\alpha}{\beta}{u}\cdot
            \KE{\wt{\alpha}}{\wt{\beta}}{\wt{u}})=&\:\partial\KE{\alpha}{\beta}{u}\cdot\KE{\wt{\alpha}}{\wt{\beta}}{\wt{u}}\\
            &+
            (-1)^{\deg(u)}\KE{\alpha}{\beta}{u}\cdot\: \partial\KE{\wt{\alpha}}{\wt{\beta}}{\wt{u}};
        \end{align*}
    \item
        \label{item:iFaktorisation}
        \begin{equation*}
            i_{t}(\mu\cdot\nu)=i_{t}(\mu)\cdot
            i_{t}(\nu)\quad\quad\forall\:\mu,\nu\in \Ss^{\bullet}(\mathbb{V})\ot \Ss^{\bullet}(\mathbb{V})\ot\Lambda^{\bullet}(\mathbb{V});
        \end{equation*} 
    \item
        \label{item:special_tAbleitunsRel}        
        \begin{equation*}
            \frac{d}{dt}i_{t}(\alpha\ot \beta\ot 1)=(\partial_{1}\circ i_{t}\circ
            \delta)(\alpha\ot \beta\ot 1).
        \end{equation*}
    \end{enumerate}
    \begin{beweis}  
        \begin{enumerate}
        \item            
            Wir erhalten mit $\alpha\wedge
            \beta=(-1)^{\deg(\alpha)\deg(\beta)}\:\beta\wedge \alpha$
            sowie der Assoziativität von $\wedge$:
            {\allowdisplaybreaks\begin{align*}
                  \delta&\:(\alpha\vee\wt{\alpha}\ot \beta\vee\wt{\beta}\ot u\wedge\wt{u})\\
                  &=\sum_{j=1}^{l}\KE{\alpha^{j}\vee\wt{\alpha}}{\:\beta\vee\wt{\beta}}{\:\alpha_{j}\wedge
                    u\wedge\wt{u}}
                  +\sum_{j=1}^{\wt{l}}\KE{\alpha\vee\wt{\alpha}^{j}}{\beta\vee\wt{\beta}}{\wt{\alpha}_{j}\wedge
                    u\wedge\wt{u}}
                  \\ &= \sum_{j=1}^{l}\KE{\alpha^{j}}{\beta}{\alpha_{j}\wedge
                    u}\cdot\KE{\wt{\alpha}}{\wt{\beta}}{\wt{u}}
                  +(-1)^{\deg(u)}\KE{\alpha}{\beta}{u}\cdot \sum_{j=1}^{\wt{l}}\KE{\wt{\alpha}^{j}}{\wt{\beta}}{\wt{\alpha}_{j}\wedge
                    \wt{u}}
                  \\ &= \delta\KE{\alpha}{\beta}{u}\cdot\KE{\wt{\alpha}}{\wt{\beta}}{\wt{u}}+(-1)^{\deg(u)}\KE{\alpha}{\beta}{u}\cdot\:\delta\KE{\wt{\alpha}}{\wt{\beta}}{\wt{u}}.
              \end{align*}}
        \item
            Sei $\deg(u)=k$ und $\deg(\wt{u})=\wt{k}$, dann folgt:
            {\begin{align*}
                  \partial^{1}_{k+\wt{k}}(\alpha\vee\wt{\alpha}\ot\beta\vee\wt{\beta}\ot u&\:\wedge\wt{u})\\=&\:\sum_{j=1}^{k}(-1)^{j-1}\KE{u_{j}\vee\alpha\vee\wt{\alpha}\:}{\:\beta\vee\wt{\beta}}{\:u^{j}\wedge\wt{u}}
                  \\ &\:+\sum_{j=1}^{\wt{k}}(-1)^{j+k-1}\KE{u_{j}\vee\alpha\vee\wt{\alpha}\:}{\:\beta\vee\wt{\beta}}{\:u\wedge\wt{u}^{j}}
                  \\
                  =&\:\sum_{j=1}^{k}(-1)^{j-1}\KE{u_{j}\vee\alpha}{\beta}{u^{j}}\cdot\KE{\wt{\alpha}}{\wt{\beta}}{\wt{u}}
                  \\ &\: +(-1)^{[k=\deg(u)]}\KE{\alpha}{\beta}{u}\cdot
                  \sum_{j=1}^{\wt{k}}(-1)^{j-1}\KE{\wt{u}_{j}\vee\wt{\alpha}}{\wt{\beta}}{\wt{u}^{j}}\\
                  =&\: \partial^{1}_{k}\KE{\alpha}{\beta}{u}\cdot\KE{\wt{\alpha}}{\wt{\beta}}{\wt{u}}
                  \\ &\:+\:(-1)^{\deg(u)}\KE{\alpha}{\beta}{u}\cdot\:\partial^{1}_{\wt{k}}\KE{\wt{\alpha}}{\wt{\beta}}{\wt{u}}.
              \end{align*}}Analog folgt dies für $\pt^{2}_{k+\wt{k}}$, was die Behauptung zeigt.
            
        \item
            Dies folgt unmittelbar daraus, dass jeder Summand aus $i_{t}(\mu\cdot\nu)$ eindeutig als Produkt zweier
            Summanden aus $i_{t}(\mu)$ und $i_{t}(\nu)$ geschrieben werden kann.
        \item            
            Zunächst reicht es, die Aussage für Elemente
            $\alpha\ot 1\ot 1$ zu zeigen, da man auf beiden Seiten
            der zu zeigenden Gleichung $1\ot \beta\ot 1$
            herausziehen kann. Es folgt
            {\begin{align*}
                  \frac{d}{dt}i_{t}(1\ot 1\ot1)=
                  \frac{d}{dt}(1\ot 1\ot1)=0=(\partial_{1}\circ i_{t}\circ
                  \delta)(1\ot1\ot1)  
              \end{align*} und weiter für $\deg(\vv)=1$:
              \begin{align*}
                  \frac{d}{dt}i_{t}(\vv\ot 1\ot1)=&\:\frac{d}{dt}\Big[t(\vv\ot 1\ot1)+(1-t)(1\ot\vv\ot1)\Big]
                  =(\vv\ot 1\ot1)-(1\ot\vv\ot1)
                  \\=&\: \partial_{1}(1\ot1\ot \vv)
                  =(\partial_{1}\circ i_{t}\ot1)\KE{1}{1}{v}
                  \\=&\:(\partial_{1}\circ i_{t}\circ \delta)(\vv\ot 1\ot1).
              \end{align*}}Angenommen, obige Aussage gelte für
            $\deg(\alpha)=k$, dann erhalten wir:
            {\allowdisplaybreaks
              \begin{align*}
                  \frac{d}{dt}i_{t}(\vv&\vee\alpha\ot1\ot1)=
                  \frac{d}{dt}\Big[i_{t}(\vv\ot 1\ot1)\cdot\:i_{t}(\alpha\ot1\ot1)\Big]\\
                  =&\:\frac{d}{dt}i_{t}(\vv\ot1\ot1)\cdot
                  \:i_{t}(\alpha\ot1\ot1)
                  +\:i_{t}(\vv\ot1\ot1)\cdot\:\frac{d}{dt}i_{t}(\alpha\ot1\ot1)\\
                  =&\:(\partial_{1}\circ i_{t}\circ\delta)(\vv\ot1\ot1)\cdot\:
                  i_{t}(\alpha\ot1\ot1)
                  +\:i_{t}(\vv\ot1\ot1)\cdot\:(\partial_{1}\circ
                  i_{t}\circ\delta)(\alpha\ot1\ot1)\\
                  =&\:\partial_{1}\Big[(i_{t}\circ\delta)(\vv\ot1\ot1)\cdot\:i_{t}(\alpha\ot1\ot1)
                  +\:i_{t}(\vv\ot1\ot1)\cdot\:(i_{t}\circ\delta)(\alpha\ot1\ot1)\Big]
                  \\ =&\:(\partial_{1}\circ
                  i_{t})\Big[\:\delta(\vv\ot1\ot1)\cdot(\alpha\ot1\ot1)
                  +(\vv\ot1\ot1)\cdot\:\delta(\alpha\ot1\ot1)\Big]
                  \\ =&\:(\partial_{1}\circ i_{t}\circ\delta\ot1)\Big[(\vv\ot1\ot1)\cdot(\alpha\ot1\ot1)\Big]\\=&\: (\partial_{1}\circ i_{t}\circ\delta\ot1)(\vv\vee\:\alpha\ot1\ot1).
              \end{align*}}
        \end{enumerate}
    \end{beweis}
\end{lemma}
Folgende Proposition liefert schließlich die Exaktheit von
$(\K,\partial,\epsilon)$.
\begin{proposition}
    \label{prop:ExaktheitsbewSym}
    Es gilt
    \begin{align*}
        \epsilon\circ h_{-1}&=\id_{\mathcal{A}}, \\
        h_{-1}\circ\epsilon+\partial_{1}\circ h_{0}&=\id_{\K_{0}}\quad\quad\text{und}\\
        h_{k-1}\circ \partial_{k}+\partial_{k+1}\circ
        h_{k}&=\id_{\K_{k}}\quad\quad \text{für }k\geq1.
    \end{align*}
    \begin{beweis}
        Zunächst folgt
        \begin{equation*}
            (\epsilon\cp h_{-1})(\alpha)=\epsilon(1\ot \alpha)=\alpha,
        \end{equation*} sowie mit Lemma \ref{lemma:EigenschHomotBaukloetze}~\textit{iv.)}:
        \begin{align*}
            (h_{-1}\cp \epsilon + \partial_{1}\cp h_{0})(\alpha\ot
            \beta)=&\:1\ot \alpha\vee\beta+ \int_{0}^{1}dt\:(\pt_{1}\cp i_{t}\cp\delta)(\alpha\ot
            \beta)\\\glna{\textit{iv.)}}&\: 1\ot \alpha\vee\beta + \alpha\ot \beta -1\ot
            \alpha\vee \beta\\=&\: \alpha\ot\beta.
        \end{align*}
        Das zeigt die ersten beiden Behauptungen. Für die
        dritte sei $\mu=\alpha\ot\beta\ot u\in \K_{k}$, dann folgt: 
        \begin{align*}
            (\partial_{k+1}\circ
            h_{k})(\mu)=\partial_{k+1}\left[\int_{0}^{1}dt\:t^{k}\:(i_{t}\circ\delta)(\mu)\right]=\int_{0}^{1}dt\:t^{k}\:(\partial_{k+1}\circ
            i_{t}\circ\delta)(\mu),
        \end{align*} und für den Integranden mit
        Lemma \ref{lemma:EigenschHomotBaukloetze}~\textit{ii.)}
        \begin{align*}
            (\pt_{k+1}\circ \:i_{t}\circ \delta)(\mu)=&\:(\pt_{k+1}\circ\:
            i_{t})\big[\delta\KE{\alpha}{1}{1}\cdot\KE{1}{\beta}{u}\big]\\
            =& \:\pt_{k+1}\big[(i_{t}\circ\delta)\KE{\alpha}{1}{1}\cdot\KE{1}{\beta}{u}\big]\\
            \glna{\textit{ii.)}}& \:(\pt_{k+1}\circ\:
            i_{t}\circ\delta)\KE{\alpha}{1}{1}\cdot\KE{1}{\beta}{u}\\ &+(-1)^{1}(i_{t}\circ\delta)\KE{\alpha}{1}{1}\cdot\:\pt_{k}\KE{1}{\beta}{u},
        \end{align*} womit insgesamt
        \begin{equation}
            \label{eq:TermvonGlHomotdersichschnellweghebt}
            \begin{split}
                (\partial_{k+1}\circ
                h_{k})(\mu)=&\int_{0}^{1}dt\:t^{k}\:(\partial_{k+1}\circ
                i_{t}\circ\delta)\KE{\alpha}{1}{1}\cdot\KE{1}{\beta}{u} \\ &
                -\int_{0}^{1}dt\:t^{k}\:(i_{t}\circ\delta)\KE{\alpha}{1}{1}\cdot\:\partial_{k}\KE{1}{\beta}{u}.
            \end{split}
        \end{equation}
        Für 
        \begin{align*}
            (h_{k-1}\circ\partial_{k})(\mu)=\int_{0}^{1}dt\:t^{k-1}\:(i_{t}\circ\delta\circ\partial_{k})(\mu)
        \end{align*} rechnen wir zunächst:
        {\begin{align*}
              (\delta\circ\partial_{k})(\mu)=&\:(\delta\circ\partial_{k})\big[\KE{1}{\beta}{u}\cdot\KE{\alpha}{1}{1}\big]\\
              =&\:\delta\big[\pt_{k}\KE{1}{\beta}{u}\cdot \KE{\alpha}{1}{1}\big]\\
              =& \:(\delta\circ\partial_{k})\KE{1}{\beta}{u}\cdot
              \KE{\alpha}{1}{1}+\:
              (-1)^{k-1}\partial_{k}\KE{1}{\beta}{u}\cdot\:\delta\KE{\alpha}{1}{1}\\
              = & \:(\delta\circ\partial_{k})\KE{1}{\beta}{u}\cdot
              \KE{\alpha}{1}{1}+\:
              \delta\KE{\alpha}{1}{1}\cdot\:\partial_{k}\KE{1}{\beta}{u}.
          \end{align*}} Anwenden von $i_{t}$ liefert:
        \begin{align*}
            (i_{t}\circ\delta\circ\partial_{k})(\mu)=&\:(i_{t}\circ\delta\circ \partial_{k})\KE{1}{\beta}{u}\cdot\:i_{t}\KE{\alpha}{1}{1}\\
            &+
            (i_{t}\circ\delta)\KE{\alpha}{1}{1}\cdot\:(i_{t}\circ\partial_{k})\KE{1}{\beta}{u}\\
            = &\: k \KE{1}{\beta}{u}\cdot\: i_{t}\KE{\alpha}{1}{1}\\&\: +t\:(i_{t}\circ\delta)\KE{\alpha}{1}{1}\cdot\:\partial_{k}\KE{1}{\beta}{u}. 
        \end{align*} In der Tat erhalten wir für den ersten Term in
        der letzten Gleichheit:
        \begin{align*}
            (i_{t}\circ\delta\circ\partial_{k})\KE{1}{\beta}{u}=&\:(i_{t}\circ\delta)\left(\sum_{j=1}^{k}(-1)^{j-1}\:u_{j}\ot
                \beta\ot u^{j} -\sum_{j=1}^{k}(-1)^{j-1}\:1\ot
                u_{j}\vee\beta\ot
                u^{j}\right)
            \\=&\:i_{t}\left(\sum_{j=1}^{k}(-1)^{j-1}1\ot\beta\ot u_{j}\wedge u^{j}\right)
            \\=&\:k\KE{1}{\beta}{u}
        \end{align*} und für den zweiten:
        {\allowdisplaybreaks
          \begin{align*}
              (i_{t}\circ\partial_{k})(1\ot\beta&\:\ot u)=i_{t}\left(\sum_{j}^{k}(-1)^{j-1}\left[u_{j}\ot\beta\ot u^{j}-1\ot u_{j}\vee\beta\ot u^{j}\right]\right)\\
              =&\:\sum_{j}^{k}(-1)^{j-1}\left[t\KE{u_{j}}{\beta}{u^{j}}+(1-t)\KE{1}{u_{j}\vee\beta}{u^{j}}-\KE{1}{u_{j}\vee\beta}{u^{j}}\right]
              \\ =&\: t\:\partial_{k}\KE{1}{\beta}{u}.
          \end{align*}}Das Zwischenergebnis lautet
        \begin{equation}
            \label{eq:zwischenergebnis}
            \begin{split}
                (h_{k-1}\circ\pt_{k})(\mu)=&\int_{0}^{1}dt\:k\:t^{k-1}i_{t}\KE{\alpha}{1}{1}\cdot\KE{1}{\beta}{u}\\
                &
                +\int_{0}^{1}dt\:t^{k}(i_{t}\circ\delta)\KE{\alpha}{1}{1}\cdot\:\partial_{k}\KE{1}{\beta}{u},
            \end{split}
        \end{equation} wobei der zweite Summand bereits das Negative
        vom zweiten Summanden in
        \eqref{eq:TermvonGlHomotdersichschnellweghebt} ist.\\\\ Für
        den ersten erhalten wir weiter:
        \begin{align*}
            \int_{0}^{1}dt\:k\:t^{k-1}i_{t}\KE{\alpha}{1}{1}=&\int_{0}^{1}dt\frac{d}{dt}\:\big[t^{k}i_{t}\KE{\alpha}{1}{1}\big]
            -\int_{0}^{1}dt\:t^{k}\frac{d}{dt}i_{t}\KE{\alpha}{1}{1}\\
            =&\:\KE{\alpha}{1}{1}-\int_{0}^{1}dt\:t^{k}(\partial_{k}\circ i_{t}\circ\delta)\KE{\alpha}{1}{1},
        \end{align*} dabei folgt die letzte Gleichheit mit Lemma \ref{lemma:EigenschHomotBaukloetze}~\textit{iv.)} und
        \begin{align*}
            \int_{0}^{1}dt\frac{d}{dt}\:\big[t^{k}i_{t}\KE{\alpha}{1}{1}\big]=&\:\big[t^{k}
            i_{t}\KE{\alpha}{1}{1}\big]_{0}^{1}=\:i_{t}\KE{\alpha}{1}{1}\big|_{t=1}=
            \KE{\alpha}{1}{1}.
        \end{align*} 
        Aus \eqref{eq:zwischenergebnis} wird dann
        \begin{equation} 
            \label{eq:rueckrichtwegheb}
            \begin{split}
                (h_{k-1}\circ\pt_{k})(\mu)=&\KE{\alpha}{\beta}{u}-
                \int_{0}^{1}dt\:t^{k}(\partial_{k}\circ i_{t}\circ\delta)\KE{\alpha}{1}{1}\cdot\KE{1}{\beta}{u}\\
                &
                +\int_{0}^{1}dt\:t^{k}(i_{t}\circ\delta)\KE{\alpha}{1}{1}\cdot\:\partial_{k}\KE{1}{\beta}{u},
            \end{split}
        \end{equation}
        und Addition von
        \eqref{eq:TermvonGlHomotdersichschnellweghebt} und
        \eqref{eq:rueckrichtwegheb} zeigt schließlich die Behauptung.
    \end{beweis}
\end{proposition}
Hiermit erhalten wir umgehend folgenden Satz:
\begin{satz}
    \label{satz:HochschkohmvonSym}
    Sei $\mathcal{A}=\SsV$ und $\mathcal{M}$ ein $\SsV-\SsV$-Bimodul, dann gilt: 
    \begin{equation*}
        HH^{k}(\Ss^{\bullet}(\mathbb{V}),\mathcal{M})\cong
        H^{k}\left(\Hom_{\mathcal{A}^{e}}(\C,\mathcal{M})\right)\cong
        H^{k}\left(\Hom_{\mathcal{A}^{e}}(\K,\mathcal{M})\right).
    \end{equation*}
    Ist $\mathcal{M}$ zudem symmetrisch, so ist:
    \begin{equation*}
        HH^{k}(\Ss^{\bullet}(\mathbb{V}),\mathcal{M})\cong \Hom_{\mathcal{A}^{e}}\left(\K_{k},\mathcal{M}\right).
    \end{equation*}
    \begin{beweis}
        Die erste Isomorphie hatten wir bereits eingesehen, und die
        zweite folgt mit
        Lemma \ref{lemma:GruppenKOhomsausprojaufloesundFunktoren}~\textit{ii.)} unmittelbar aus dem Fakt, dass sowohl $(\C,d,\epsilon)$, als auch $(\K,\pt,\epsilon)$ projektive Aufl"osungen von $\SsV$ sind.
        
        F"ur die zweite Behauptung sei $\phi\in
        \K^{*}_{k}=\Hom_{\mathcal{A}^{e}}(\K_{k},\mathcal{M})$ und
        $\omega=\alpha\ot \beta \ot u \in \K_{k+1}$. Dann folgt:
        {
          \begin{align*}
              (\partial^{*}_{k+1}\phi)(\omega)=&\:\phi\left(\partial_{k+1}(\alpha\ot
                  \beta \ot u)\right)\\=&\:
              \phi\left(\sum_{j=1}^{n}(-1)^{j-1}\left[u_{j}\vee \alpha
                      \ot \beta \ot u^{j}- \alpha
                      \ot u_{j}\vee\beta \ot u^{j} \right]\right)
              \\=&\sum_{j=1}^{n}(-1)^{j-1} [u_{j}\ot 1 - 1\ot u_{j}]*_{e} \phi\left(\alpha
                  \ot \beta \ot u^{j}\right)
              \\=&\:0,
          \end{align*}}womit
        $\ker(\pt_{k+1}^{*})=\Hom_{\mathcal{A}^{e}}(\K_{k},\mathcal{M})$ und
        $\im(\pt_{k}^{*})=0$, also
        \begin{equation*}
            H^{k}(\Hom_{\mathcal{A}^{e}}(\K,\mathcal{A}))=\ker(\pt_{k+1}^{*})/\im(\pt_{k}^{*})=\Hom_{\mathcal{A}^{e}}(\K_{k},\mathcal{A}).
        \end{equation*}
    \end{beweis} 
\end{satz}
Ohne zusätzliche Annahmen über $\V$ und
$\Hom_{\mathcal{A}^{e}}(\K_{k},\mathcal{M})$ erhalten wir jedoch im 
Allgemeinen kein Analogon zu \eqref{eq:HochschPol22}. Es gilt jedoch:
\begin{korollar}
    Sei $\V$ ein endlich-dimensionaler $\mathbb{K}$-Vektorraum  und
    $\mathcal{M}$ ein symmetrischer $\SsV-\SsV$-Bimodul, dann ist: 
    \begin{equation*}
        HH^{k}(\Ss^{\bullet}(\mathbb{V}),\mathcal{M})\cong
        \mathcal{M}\ot \Lambda^{k}(\V).
    \end{equation*}
    \begin{beweis}
        Dies folgt analog zum zweiten Teil von
        Satz \ref{satz:PolsatzHochsch}, da wegen der endlichen Dimension $n$
        von $\V$ mit $\V^{*}\cong\V$ ebenfalls
        $\Lambda^{k}(\V)^{*}\cong\Lambda^{k}(\V^{*})\cong\Lambda^{k}(\V)$
        gilt.
        Es folgt dann zunächst für $\phi\in
        \Hom_{\mathcal{A}^{e}}(\K_{k},\mathcal{M})$, dass
        \begin{equation*}
            \phi(a^{e}\ot
            \omega)=\left(\sum_{j_{1},…,j_{k}}^{n}\phi^{j_{1},…,j_{k}}\ot
                e^{j_{1}}\wedge…\wedge e^{j_{k}}\right)(a^{e}\ot \omega)
        \end{equation*}mit $\big\{e_{i}\big\}_{1\leq i\leq n}$ eine Basis von
        $\V$ und $\left\{e^{i}\right\}_{1\leq i\leq n}$ die hierzu duale Basis
        von $\V^{*}$. Dies zeigt
        $\Hom_{\mathcal{A}^{e}}(\K_{k},\mathcal{M})\cong
        \mathcal{M}\ot \Lambda^{k}(\V^{*})\cong \mathcal{M}\ot
        \Lambda^{k}(\V)$, wobei die zweite Isomorphie vermöge $\Lambda^{k}(\V^{*})\cong\Lambda^{k}(\V)$ am leichtesten mit einem Basis-Argument und
        Bemerkung \ref{bem:TenprodBasis} folgt.
    \end{beweis}
\end{korollar}

\subsection{Explizite Kettenabbildungen}
Wir wollen nun explizite Kettenabbildungen für die Bar- und
Koszulauflösung angeben. Seien hierf"ur abstrakte $\mathcal{A}^{e}$-lineare Varianten von \eqref{eq:Gpol} und
\eqref{eq:FPol} gegeben durch:
\begin{equation}
    \label{eq:SymF}
    \begin{split}
        F_{k}\colon\Ss^{\bullet}(\mathbb{V})\ot \Ss^{\bullet}(\mathbb{V})\ot \Lambda^{k}(\mathbb{V})&\longrightarrow \bigotimes^{k+2}\Ss^{\bullet}(\mathbb{V})\\
        \KE{\alpha}{\beta}{u}&\longmapsto\sum_{\sigma\in
          S_{k}}\mathrm{sign}(\sigma)\:  (\alpha\ot
        u_{\sigma(1)}\ot…\ot u_{\sigma(k)}\ot\beta)
    \end{split}
\end{equation}für $u=u_{1}\wedge\dots\wedge u_{k}$ sowie
\begin{equation}
    \label{eq:SymG}
    \begin{split}
        G_{k}\colon\bigotimes^{k+2}\Ss^{\bullet}(\mathbb{V})&\longrightarrow \Ss^{\bullet}(\mathbb{V})\ot
        \Ss^{\bullet}(\mathbb{V})\ot \Lambda^{k}(\mathbb{V})\\
        \omega&\longmapsto
        \int_{0}^{1}dt_{1}\int_{0}^{t_{1}}dt_{2}…\int_{0}^{t_{k-1}}dt_{k}\:(i\circ\delta)(\omega)
    \end{split}
\end{equation}mit $G_{0}=F_{0}=\id_{\mathcal{A}^{e}}$. 
Die beteiligten Komponenten sind dabei wie folgt definiert:
\begin{definition}
    \label{def:GAbb}
    \begin{enumerate}
    \item
        Seien $\mu=(\alpha\ot
        u_{1}\ot…\ot u_{m}\ot\beta\ot \omega)$ und $\nu=(\alpha'\ot
        u'_{1}\ot…\ot u'_{m}\ot\beta'\ot \omega')$, so definieren wir das
        komponentenweise Produkt
        \begin{align*}
            \cdot\colon \bigotimes^{m+2}\Ss^{\bullet}(\mathbb{V})\ot \Lambda^{k}(\V)\times
            \bigotimes^{m+2}\Ss^{\bullet}(\mathbb{V})\ot \Lambda^{k}(\V)\longrightarrow
            \bigotimes^{m+2}\Ss^{\bullet}(\mathbb{V})\ot
            \Lambda^{k}(\V)
        \end{align*} durch
        \begin{align*}
            \cdot\colon (\mu,\nu)\longmapsto
            \alpha\vee\alpha'\ot u_{1}\vee u'_{1}\ot…\ot u_{m}\vee u'_{m}\ot
            \beta\vee \beta'\ot \omega\wedge \omega'.
        \end{align*}
        Im Spezialfall $m=0$ stimmt dieses mit unserer alten Definition
        überein. Sinngemäß sei diese Abbildung auch für $k=0$ definiert.
    \item        
        \begin{align*}
            \hat{\circ}\colon\bigotimes^{l+2}\Ss^{\bullet}(\mathbb{V})\ot \Lambda^{\bullet}(\V)\times\bigotimes^{l'+2}\Ss^{\bullet}(\mathbb{V})\ot
            \Lambda^{\bullet}(\V)&\longrightarrow
            \bigotimes^{l+l'+2}\Ss^{\bullet}(\mathbb{V})\ot \Lambda^{\bullet}(\V)\\
            \Big((\alpha\ot \ovl{u}\ot \beta\ot \omega), (\alpha'\ot \ovl{u}'\ot
            \beta'\ot \omega')\Big)&\longmapsto \alpha\vee \alpha'\ot \ovl{u}\ot
            \ovl{u}'\ot \beta\vee \beta'\ot \omega\wedge \omega'.
        \end{align*}
    \item
        \begin{align*}
            i\colon\bigotimes^{k+2}\Ss^{\bullet}(\mathbb{V})\ot
            \Lambda^{\bullet}(\mathbb{V})&\longrightarrow
            \Big[\Ss^{\bullet}(\V)\ot \Ss^{\bullet}(\V)\ot
            \Lambda^{\bullet}(\V)\Big] \big[ t_{1},…,t_{k}\big]\\ 
            \alpha\ot u_{1}\ot…\ot u_{k}\ot \beta \ot\omega&\longmapsto
            (\alpha\ot\beta)*_{e} \left[\prod_{s=1}^{k}\hat{i}_{s}(1\ot u_{s}\ot1)\right]\ot\omega
        \end{align*}mit $\prod$ das Produkt $\cdot$ für den Spezialfall
        $k=0$ und $\hat{i}_{s}$ die $\mathcal{A}^{e}$-lineare
        Abbildung
        \begin{align*}
            \hat{i}_{s}\colon\bigotimes^{3}\Ss^{\bullet}(\mathbb{V}) \longrightarrow&\:  \Big[\bigotimes^{2}\Ss^{\bullet}(\mathbb{V})\Big]\big[ t_{s}\big]\\
            \alpha\ot u\ot\beta\longmapsto&\:  t_{s}^{m}u\vee\alpha\ot\beta\:+\:
            t_{s}^{m-1}(1-t_{s})\sum_{j=1}^{m}u^{j}\vee\alpha\ot
            u_{j}\vee\beta\:+…\\ 
            &+
            t_{s}^{m-l}(1-t_{s})^{l}\sum_{j_{1},…,j_{l}}^{m}u^{j_{1},…,j_{l}}\vee
            \alpha\ot u_{j_{1},…,j_{l}}\vee\beta\:+\:…\:\\ &+
            (1-t_{s})^{m}\alpha\ot u\vee\beta,
        \end{align*} mit $\deg(u)=m$ und $\hat{i}_{s}(\alpha\ot 1\ot
        \beta)=\alpha \ot \beta$.\\\\
        Für Elemente $\alpha\ot u \ot\beta\ot \omega$ schreiben wir im
        Folgenden auch $i_{s}(\alpha\ot u \ot\beta\ot \omega)$ anstelle
        $i(\alpha\ot u \ot\beta \ot\omega)$, um zu verdeutlichen, dass
        das Bild dieses Elementes nur von einer Variablen $t_{s}$
        abhängt. Ist es an gegebener Stelle angebracht, so schreiben
        wir der Deutlichkeit halber auch $i_{t_{1},…,t_{k}}$ anstatt $i$.
    \item      
        \begin{align*}
            \delta\colon\bigotimes^{\bullet+2}\Ss^{\bullet}(\mathbb{V})\longrightarrow&\:
            \bigotimes^{\bullet+2}\Ss^{\bullet}(\mathbb{V})\ot \Lambda^{\bullet}(\mathbb{V})\\
            \alpha\ot u_{1}\ot…\ot u_{k}\ot \beta\longmapsto&\:
            \sum_{j_{1}}^{n_{1}}…\sum_{j_{k}}^{n_{k}}\alpha\ot
            u_{1}^{j_{1}}\ot…\ot u_{k}^{j_{k}}\ot \beta\ot\:
            (u_{1})_{j_{1}}\wedge…\wedge (u_{k})_{j_{k}}\\ \longmapsto &\:
            (\alpha\ot\beta)*_{e}\widehat{\bigodot}_{s=1}^{k}\wt{\delta}(1\ot u_{s}\ot1)
        \end{align*} 
        mit $\deg(u_{i})=n_{i}$ und $\widehat{\bigodot}$
        das Produkt $\hat{\circ}$. Dabei bezeichnet $\wt{\delta}$ die $\mathcal{A}^{e}$-lineare Abbildung
        \begin{align*}
            \wt{\delta}\colon\Ss^{\bullet}(\mathbb{V})\ot \Ss^{\bullet}(\mathbb{V})\ot
            \Ss^{\bullet}(\mathbb{V})&\longrightarrow  \Ss^{\bullet}\ot
            \Ss^{\bullet-1}(\mathbb{V})\ot \Ss^{\bullet}(\mathbb{V})\ot \Lambda^{1}(\mathbb{V})\\
            \alpha\ot u\ot\beta&\longmapsto \sum_{j=1}^{k}\alpha\ot
            u^{j}\ot\beta\ot u_{j}
        \end{align*} mit $\deg(u)=k\:$ 
        und $\:\wt{\delta}(\alpha\ot 1\ot \beta)=0$.
    \end{enumerate}
\end{definition}
\begin{bemerkung}
    \begin{enumerate}
    \item
        Der Wohldefiniertheit wegen sei angemerkt, dass auch hier die beteiligten Komponenten als Einschr"ankungen von
        symmetrisierten und antisymmetrisierten Abbildungen auf die
        jeweiligen Unterr"aume geschrieben werden k"onnen. Besagte
        Abbildungen werden in Kapitel \ref{sec:SvonV} nachgeliefert, da wir sie dort
        auch explizit ben"otigen.
    \item
        In dem Moment, in dem wir die Kettenabbildungs-Eigenschaft von $F$ und $G$
        nachgewiesen haben zeigt der Beweis von
        Lemma \ref{lemma:GruppenKOhomsausprojaufloesundFunktoren}~\textit{ii.)},
        dass $F^{*}$ und $G^{*}$ 
        zueinander inverse Isomorphismen $\wt{F^{*}}$ und $\wt{G^{*}}$ auf
        Kohomologie-Niveau induzieren.
    \end{enumerate}
\end{bemerkung}
Der zweite Teil des folgenden Lemmas liefert uns die Kettenabbildungs-Eigenschaft von
$F$. Die Bedeutung des ersten Teils wird am Ende
dieses Kapitels klar werden.
\begin{lemma}
    \label{lemma:Fkettenabb}
    Es gilt:
    \begin{enumerate}
    \item        
        $G_{k}\cp F_{k}= \id_{\K_{k}}$
    \item
        $d_{k}\circ F_{k}=F_{k-1}\circ \partial_{k}$.
    \end{enumerate}
    \begin{beweis} 
        \begin{enumerate}
        \item
            Mit $\int_{0}^{1}dt_{1}…\int_{0}^{t_{k-1}}dt_{k}=\frac{1}{k!}$
            und $\alpha\ot\beta\ot u \in \K_{k}$
            folgt:
            {\allowdisplaybreaks\small\begin{align*}
                  (G_{k}\circ F_{k})(\alpha\ot\beta\ot
                  u)
                  =&\:\int_{0}^{1}dt…\int_{0}^{t_{k-1}}dt_{k}\:(i\cp\delta)(F_{k}(\alpha\ot
                  \beta\ot u))
                  \\=&\int_{0}^{1}dt…\int_{0}^{t_{k-1}}dt_{k}\:(i\cp\delta)\left(\sum_{\sigma\in
                        S_{k}}\mathrm{sign}(\sigma)(\alpha\ot u_{\sigma(1)}\ot…\ot
                      u_{\sigma(k)}\ot\beta)\right)
                  \\=&\int_{0}^{1}dt…\int_{0}^{t_{k-1}}dt_{k}\:i\left(\sum_{\sigma\in
                        S_{k}}\mathrm{sign}(\sigma)(\alpha\ot\beta\ot u_{\sigma(1)}\wedge…\wedge
                      u_{\sigma(k)})\right)
                  \\=& \int_{0}^{1}dt…\int_{0}^{t_{k-1}}dt_{k}\:i\left(\sum_{\sigma\in
                        S_{k}}\mathrm{sign}(\sigma)\:\mathrm{sign}(\sigma)(\alpha\ot\beta\ot
                      u)\right)
                  \\=&\: k! \int_{0}^{1}dt…\int_{0}^{t_{k-1}}dt_{k}\:(\alpha\ot\beta\ot
                  u)
                  \\=&\: (\alpha\ot\beta\ot u).
              \end{align*}}
        \item 
            Sei zunächst $\mu=\alpha\ot \beta\ot \mathrm{v}$ mit
            $\deg(\mathrm{v})=1$, dann folgt mit $F_{0}=\id_{\mathcal{A}^{e}}$
            \begin{align*}
                (d_{1}\cp F_{1})(\alpha\ot\beta\ot \mathrm{v})=\alpha\vee \mathrm{v}\ot\beta
                -\alpha\ot \mathrm{v}\vee \beta=(F_{0}\cp \pt_{1})(\alpha\ot\beta\ot \mathrm{v}), 
            \end{align*}und für $k>1$
            erhalten wir 
            {\small\allowdisplaybreaks    
              \begin{align*}
                  (d_{k}\circ F_{k})(\mu)=&\:(-1)^{0}\sum_{\sigma\in
                    S_{k}}\mathrm{sign}(\sigma)\:
                  (\alpha\vee u_{\sigma(1)}\ot u_{\sigma(2)}\ot…\ot u_{\sigma(k)}\ot\beta)\\ &+ (-1)^{k}\sum_{\sigma\in
                    S_{k}}\mathrm{sign}(\sigma)\:
                  (\alpha\ot u_{\sigma(1)}\ot…\ot
                  u_{\sigma(k-1)} \ot
                  u_{\sigma(k)}\vee\beta)\\ &+\:
                  \underbrace{\sum_{j=1}^{k-1}(-1)^{j}\sum_{\sigma\in
                      S_{k}}\mathrm{sign}(\sigma)\:(\alpha\ot
                    u_{\sigma(1)}\ot…\ot u_{\sigma(j)}\vee
                    u_{\sigma(j+1)}\ot…\ot
                    u_{\sigma(k)}\ot\beta}_{0})
                  \\= &\:\sum_{j=1}^{k}\sum_{\substack{\sigma\in
                      S_{k}\\\sigma(1)=j}}\mathrm{sign}(\sigma)\:
                  (\alpha\vee u_{j}\ot u_{\sigma(2)}\ot…\ot
                  u_{\sigma(k)}\ot\beta)\\ &\quad+ (-1)^{k}\sum_{j=1}^{k}\sum_{\substack{\sigma\in
                      S_{k}\\\sigma(k)=j}}\mathrm{sign}(\sigma)\:
                  (\alpha\ot u_{\sigma(1)}\ot…\ot u_{\sigma(k-1)}\ot
                  u_{j}\vee\beta) \\ = &\: 
                  \sum_{j=1}^{k}\:\mathrm{sign}(\pi_{1\shortleftarrow j})\sum_{\sigma\in
                    S_{k-1}}\mathrm{sign}(\sigma)\:
                  (\alpha\vee u_{j}\ot\sigma^{*}u^{j}\ot\beta)\\ &\quad+ (-1)^{k}\sum_{j=1}^{k}\mathrm{sign}(\pi_{j\shortrightarrow k})\sum_{\sigma\in
                    S_{k-1}}\mathrm{sign}(\sigma)\:
                  (\alpha\ot \sigma^{*}u^{j}\ot
                  u_{j}\vee\beta).
              \end{align*}}Dabei bedeutet $\sigma^{*}u^{j}$ lediglich die Permutation
            $\sigma$, angewandt auf das Element
            \begin{equation*}
                \Tt^{k-1}(\V)\ni u^{j}=u_{1}\ot…\blacktriangle^{j}…\ot u_{k}.
            \end{equation*}
            $\pi_{1\shortleftarrow j}$ bezeichnet die Permutation, die
            $u_{j}$ sukzessive durch Transpositionen an die erste Stelle schiebt, sinngemäß für $\pi_{j\shortrightarrow
              k}$. Es folgt
            {\allowdisplaybreaks\small
              \begin{align*}
                  (d_{k}\circ F_{k})(\mu)= 
                  &\:\sum_{j=1}^{k}(-1)^{j-1}\sum_{\sigma\in
                    S_{k-1}}\mathrm{sign}(\sigma)\:
                  (\alpha\vee u_{j}\ot\sigma^{*}u^{j}\ot\beta)\\ &+ \sum_{j=1}^{k}(-1)^{k}(-1)^{k-j}\sum_{\sigma\in
                    S_{k-1}}\mathrm{sign}(\sigma)\:
                  (\alpha\ot \sigma^{*}u^{j}\ot
                  u_{j}\vee\beta)\\ =
                  &\:\sum_{j=1}^{k}(-1)^{j-1}\sum_{\sigma\in
                    S_{k-1}}\mathrm{sign}(\sigma)\:
                  (\alpha\vee u_{j}\ot\sigma^{*}u^{j}\ot\beta)\\ &\quad-\sum_{j=1}^{k}(-1)^{j-1}\sum_{\sigma\in
                    S_{k-1}}\mathrm{sign}(\sigma)\:
                  (\alpha\ot \sigma^{*}u^{j}\ot
                  u_{j}\vee\beta)\\ =&\:
                  F_{k-1}\left(\sum_{j=1}^{k}(-1)^{j-1}\Big[u_{j}\vee\alpha\ot\beta\ot
                      u^{j}-\alpha\ot u_{j}\vee\beta\ot
                      u^{j}\Big]\right)\\ = &\: (F_{k-1}\circ\partial_{k})(\mu).
              \end{align*}}
        \end{enumerate}
    \end{beweis}
\end{lemma}
Für die Kettenabbildungs-Eigenschaft von $G$ benötigen wir zunächst
einige Rechen-regeln. 
\begin{lemma}
    \begin{enumerate}
    \item
        \begin{align}
            \label{eq:faktIs}
            \hat{i}_{s}(v\cdot w)&=\hat{i}_{s}(v)\cdot \hat{i}_{s}(w) \\
            \label{eq:InormalesProdFaktor}
            i(\mu\cdot\nu)&=i(\mu)\cdot i(\nu),
        \end{align} für $v,w \in \bigotimes^{3}\Ss^{\bullet}(\mathbb{V})$
        sowie $\mu,\nu\in \displaystyle\bigotimes^{m+2}\Ss^{\bullet}(\V)\ot \Lambda^{\bullet}$.
    \item
        \begin{equation}
            \label{eq:DeltaDerivaufuElem}
            \delta(v\cdot w)=\delta(v)\cdot w\ot1 +
            v\ot 1\cdot\delta(w), 
        \end{equation} für $v,w \in \bigotimes^{3}\Ss^{\bullet}(\mathbb{V})$.
    \item
        \begin{align}
            \label{eq:iFaktor}
            i(\mu\:\hat{\circ}\:\nu)&=i(\mu)\cdot i(\nu)\\
            \label{eq:deltaFakt}
            \delta(\mu\:\hat{\circ}\:\nu)&=\delta(\mu)\:\hat{\circ}\:\delta(\nu)
        \end{align}
    \item        
        \begin{align}
            \label{eq:tAbli}
            \frac{d}{ds} \hat{i}_{s}(\alpha\ot u\ot\beta)&=(\partial_{1}
            \cp i_{s}\circ \delta)(\alpha\ot u\ot \beta),\\
            \label{eq:partialmultisuperderiv}
            \pt_{k}\left[\prod_{i=1}^{k}(1\ot1\ot u_{i})\right]&=\sum_{j=1}^{k}(-1)^{j-1}\pt_{1}(1\ot
            1\ot u_{j})\cdot
            \prod_{i\neq j}(1\ot 1\ot u_{i})
        \end{align} für $u_{i}\in \Lambda^{1}(\mathbb{V})$.
    \end{enumerate}
    \begin{beweis}  
        \begin{enumerate}
        \item
            \eqref{eq:faktIs} folgt wie
            Lemma \ref{lemma:EigenschHomotBaukloetze}~\textit{iv)} mit der
            Kommutativität von $\vee$.

            Für \eqref{eq:InormalesProdFaktor} seien 
            $\mu=(1\ot u_{1}\ot…\ot u_{m}\ot1\wedge \omega)$ und $\nu=(1\ot
            u'_{1}\ot…\ot u'_{m}\ot 1\wedge \omega')$, dann erhalten
            wir
            \begin{align*}
                i(\mu\cdot\nu)=&
                \:\left[\prod_{s=1}^{m}\hat{i}_{s}(1\ot u_{s}\vee u'_{s}\ot
                    1)\right]\ot
                \omega\wedge\omega'\\
                =&  \:\left[\prod_{s=1}^{m}\hat{i}_{s}(1\ot u_{s}\ot 1)\cdot \hat{i}_{s}(1\ot u'_{s}\ot
                    1)\right]\ot
                \omega\wedge\omega'\\
                =&  \left(\left[\prod_{s=1}^{m}\hat{i}_{s}(1\ot u_{s}\ot 1)\right]\ot
                    \omega \right)\cdot \left(\left[\prod_{s=1}^{m}\hat{i}_{s}(1\ot u'_{s}\ot 1)\right]\ot
                    \omega' \right)\\
                = &\: i(\mu)\cdot i(\nu).
            \end{align*} 
        \item            
            Wir rechnen
            {\small\allowdisplaybreaks\begin{align*}
                  \delta\big[(1\ot u\ot1)&\:\cdot(1\ot u'\ot1)\big]
                  \\=&\:\sum_{j=1}^{m+m'}\left(1\ot\: [u\vee u']^{j}\ot1\ot\: [u\vee u']_{j}\right)
                  \\=&\:\sum_{j=1}^{m}(1\ot u^{j}\ot1\ot u_{j})\cdot(1\ot u'\ot1\ot 1)+(1\ot
                  u\ot1\ot 1)\cdot\sum_{j=1}^{m'}(1\ot u'^{j}\ot1\ot u'_{j})
                  \\=&\:\delta(1\ot u\ot1)\cdot (1\ot u'\ot1\ot 1)+(1\ot u\ot1\ot 1) \cdot
                  \delta(1\ot u'\ot1).
              \end{align*}}Zusammen mit der
            $\mathcal{A}^{e}$-Linearität von $\delta$ zeigt dies \eqref{eq:DeltaDerivaufuElem}.
        \item
            Für \eqref{eq:iFaktor} seien 
            $\mu=(\alpha\ot\ovl{u}\ot \beta\ot \omega)$ und
            $\nu=(\alpha'\ot\ovl{u}'\ot \beta'\ot\omega')$,
            dann folgt:
            {\small\allowdisplaybreaks
              \begin{align*}
                  i(\mu\:
                  \hat{\cp}\:\nu)=&\:(\alpha\vee\alpha'\ot\beta\vee\beta')*_{e}\left[\prod_{s=1}^{m+m'}\hat{i}_{s}(1\ot\:(\ovl{u}\ot\ovl{u}')_{s}\ot
                      1)\right]\ot\:
                  \omega\wedge\omega'\\ 
                  =&\:(\alpha\ot\beta)*_{e} (\alpha'\ot\beta')*_{e}\left[\prod_{s=1}^{m}\hat{i}_{s}(1\ot\:\ovl{u}_{s}\ot
                      1)\cdot \prod_{s=1}^{m'}\hat{i}_{s}(1\ot\:\ovl{u}'_{s}\ot
                      1)\right]\ot\:
                  \omega\wedge\omega'\\ 
                  =& \left((\alpha\ot
                      \beta)*_{e}\left[\prod_{s=1}^{m}\hat{i}_{s}(1\ot\ovl{u}\ot1)\right]\ot \omega\right)
                  \cdot \left((\alpha'\ot
                      \beta')*_{e}\left[\prod_{s=1}^{m'}\hat{i}_{s}(1\ot\ovl{u}'\ot1)\right]\ot
                      \omega'\right)\\ =&\: i(\mu)\cdot i(\nu).
              \end{align*}}
            \eqref{eq:deltaFakt} folgt mit $\mu=(\alpha\ot\ovl{u}\ot \beta)$ und
            $\nu=(\alpha'\ot\ovl{u}'\ot \beta')$ analog zu \eqref{eq:iFaktor} f"ur $\widehat{\bigodot}$ anstelle von $\prod$ , $i$ anstelle $\delta$ und $\wt{\delta}$ anstelle $i_{s}$:
            {\small\allowdisplaybreaks
              \begin{align*}
                  \delta(\mu\:
                  \hat{\cp}\:\nu)=&\:(\alpha\vee\alpha'\ot\beta\vee\beta')*_{e}\left[\widehat{\bigodot}_{l=1}^{m+m'}\wt{\delta}(1\ot\:(\ovl{u}\ot\ovl{u}')_{l}\ot
                      1)\right]\\ 
                  =&\:(\alpha\ot\beta)*_{e} (\alpha'\ot\beta')*_{e}\left[\widehat{\bigodot}_{l=1}^{m}\wt{\delta}(1\ot\:\ovl{u}_{l}\ot
                      1)\:\hat{\circ}\widehat{\bigodot}_{l=1}^{m'}\wt{\delta}(1\ot\:\ovl{u}'_{l}\ot
                      1)\right]\\ 
                  =& \left((\alpha\ot
                      \beta)*_{e}\left[\widehat{\bigodot}_{l=1}^{m}\wt{\delta}(1\ot\ovl{u}\ot1)\right]\right)\hat{\circ} \left((\alpha'\ot
                      \beta')*_{e}\left[\widehat{\bigodot}_{l=1}^{m'}\wt{\delta}(1\ot\ovl{u}'\ot1)\right]\right)\\ =&\: \delta(\mu)\:\hat{\circ}\: \delta(\nu).
              \end{align*}}
        \item
            F"ur \eqref{eq:tAbli} erhalten wir  analog zu
            Lemma \ref{lemma:EigenschHomotBaukloetze}~\textit{iv)}: 
            {\begin{equation*}
                  \frac{d}{ds}\hat{i}_{s}(\alpha\ot 1\ot \beta)=0=(\pt_{1}\cp
                  i_{s}\cp \delta)(\alpha\ot1\ot \beta)
              \end{equation*}}sowie
            {\begin{align*}
                  \frac{d}{ds}\hat{i}_{s}(\alpha\ot \mathrm{v}
                  \ot\beta)=&\:\frac{d}{ds}\big[s(\mathrm{v}\vee\alpha\ot1)+(1-s)(1\ot\mathrm{v}\vee\beta)\big]
                  \\=&\: (\mathrm{v}\vee\alpha\ot1-1\ot\mathrm{v}\vee\beta)
                  \\=&\:\pt_{1}\:(\alpha\ot\beta\ot\mathrm{v} )
                  \\=&\:(\pt_{1}\cp i_{s})(\alpha\ot1\ot\beta\ot\mathrm{v})
                  \\=&\:(\pt_{1}\cp i_{s}\cp\delta)(\alpha\ot\mathrm{v}\ot\beta),
              \end{align*}}und mit der
            $\mathcal{A}^{e}$-Linearität beider Seiten induktiv:
            {\allowdisplaybreaks
              \begin{align*} 
                  \frac{d}{ds}\hat{i}_{s}&(1\ot \mathrm{v}\vee u\ot
                  1)=\:\frac{d}{ds}\big[\hat{i}_{s}(1\ot \mathrm{v}\ot 1)\cdot
                  \hat{i}_{s}(1\ot u\ot 1)\big]\\ 
                  =&\:\frac{d}{ds}\hat{i}_{s}(1\ot \mathrm{v}\ot 1)\cdot
                  \hat{i}_{s}(1\ot u\ot 1)+\hat{i}_{s}(1\ot \mathrm{v}\ot 1)\cdot
                  \frac{d}{ds}\hat{i}_{s}(1\ot u\ot 1)
                  \\ =&\:(\pt_{1}\cp
                  i_{s}\cp \delta)(1\ot \mathrm{v}\ot 1)\cdot \hat{i}_{s}(1\ot u\ot 1)+\hat{i}_{s}(1\ot \mathrm{v}\ot 1)\cdot(\pt_{1}\cp
                  i_{s}\cp \delta)(1\ot u\ot 1)\\ 
                  =& \:\pt_{1}\Big[(i_{s}\cp \delta)(1\ot \mathrm{v}\ot
                  1)\cdot i_{s}(1\ot u\ot 1\ot 1) +i_{s}(1\ot \mathrm{v}\ot 1\ot 1)\cdot( i_{s}\cp
                  \delta)(1\ot u\ot 1)\Big]\\
                  =&\:(\pt_{1}\cp i_{s})\Big[\delta(1\ot
                  \mathrm{v}\ot 1)\cdot(1\ot u\ot 1\ot1)+(1\ot \mathrm{v}\ot 1\ot1)\cdot\delta\:(1\ot u\ot 1)\Big]
                  \\=&\: (\pt_{1}\cp i_{s}\cp\delta)(1\ot\mathrm{v}\vee u\ot1).
              \end{align*}}
            \eqref{eq:partialmultisuperderiv} erhält man mit
            {\allowdisplaybreaks\small\begin{align*}
                  \pt_{k}\left[\prod_{i=1}^{k}(1\ot 1\ot u_{i}
                      )\right]=& \sum_{j=1}^{k}(-1)^{j-1}\left[\prod_{i=1}^{j-1}(1\ot 1\ot u_{i})\cdot \pt_{1}(1\ot 1\ot u_{i})\cdot \prod_{i=j+1}^{k}(1\ot 1\ot u_{i})\right]
                  \\=&\sum_{j=1}^{k}(-1)^{j-1}\pt_{1}(1\ot 1\ot u_{i})\cdot
                  \prod_{i\neq j}(1\ot 1\ot u_{i}).
              \end{align*}}
        \end{enumerate}
    \end{beweis}
\end{lemma}
Folgende Proposition zeigt schließlich die gewünschte Eigenschaft von
$G$.
\begin{proposition}
    \label{prop:GcirFistidundGKettenabb}
    Es gilt:
    \begin{equation*}
        \pt_{k}\cp G_{k}=G_{k-1}\cp d_{k}.
    \end{equation*}
    \begin{beweis}  
        Wir beginnen mit  {\allowdisplaybreaks{\footnotesize
            \begin{align*}
                (\pt_{k}\cp&\: G_{k})(1\ot\ovl{u}\ot 1)
                \\\glna{\substack{\eqref{eq:iFaktor}\\\eqref{eq:deltaFakt}}}&\:
                \pt_{k}\left[\int_{0}^{1}dt_{1}…\int_{0}^{t_{k-1}}dt_{k}\:(i_{1}\cp\delta)(1\ot
                    u_{1}\ot 1)\cdot…\cdot \ck{k}\right]
                \\\glna{\eqref{eq:partialmultisuperderiv}}&\:\sum_{j=1}^{k}(-1)^{j-1}\int_{0}^{1}dt_{1}…\int_{0}^{t_{k-1}}dt_{k}\:
                (\pt_{1}\cp i_{j}\cp\delta)(1\ot u_{j}\ot
                1)\cdot\prod_{i\neq j}\ck{i}
                \\\glna{\eqref{eq:tAbli}}&\:\sum_{j=1}^{k}\:(-1)^{j-1}\int_{0}^{1}dt_{1}…\int_{0}^{t_{k-1}}dt_{k}\:
                \frac{d}{dt_{j}}i_{j}(1\ot u_{j}\ot 1\ot1)\cdot\prod_{i\neq
                  j}\ck{i}
                \\=&\:\sum_{j=1}^{k-1}\:\int_{0}^{1}dt_{1}…\int_{0}^{t_{k-1}}dt_{k}\:
                \frac{d}{dt_{j}}\left[i_{j}(1\ot u_{j}\ot
                    1\ot1)\cdot\ck{j+1}\right]\cdot\prod_{\substack{i\neq
                    j\\ i\neq j+1}}\ck{i}
                \\ &+ (-1)^{k-1} \int_{0}^{1}dt_{1}…\int_{0}^{t_{k-1}}dt_{k}\:
                \frac{d}{dt_{k}}i_{k}(1\ot u_{k}\ot 1\ot1)\cdot\prod_{i=1}^{k-1}\ck{i}.
            \end{align*}}} Nun folgt durch Anwendung von {\footnotesize$\displaystyle\int_{0}^{t_{j-1}}dt_{j}$} auf {\footnotesize
          \begin{align*}
              \int_{0}^{t_{j}}dt_{j+1}\frac{d}{dt_{j}}f(t_{j},t_{j+1})=\frac{d}{dt_{j}}\left[\int_{0}^{t_{j}}dt_{j+1}f(t_{j},t_{j+1})\right]-f(t_{j},t_{j}),
          \end{align*}} dass{\footnotesize
          \begin{equation*}
              \int_{0}^{t_{j-1}}dt_{j}\int_{0}^{t_{j}}dt_{j+1}\frac{d}{dt_{j}}f(t_{j},t_{j+1})=\int_{0}^{t_{j-1}}dt_{j+1}f(t_{j-1},t_{j+1})-\int_{0}^{t_{j-1}}dt_{j}f(t_{j},t_{j}),
          \end{equation*}} mithin für $2\leq j\leq k-1$: {\footnotesize
          \begin{align*}
              \int_{0}^{t_{j-1}}dt_{j}\int_{0}^{t_{j}}dt_{j+1}&\:\frac{d}{dt_{j}}\left[i_{j}(1\ot
                  u_{j}\ot 1\ot 1)\cdot \ck{j+1}\right]\int_{0}^{t_{j+1}}dt_{j+2}…
              \\=&\int_{0}^{t_{j-1}}dt_{j+1}\left[i_{j-1}(1\ot
                  u_{j}\ot 1\ot 1)\cdot \ck{j+1}\right]\int_{0}^{t_{j+1}}dt_{j+2}…\\
              &-\int_{0}^{t_{j-1}}dt_{j}\left[i_{j}(1\ot
                  u_{j}\ot 1\ot 1)\cdot (i_{j}\cp\delta)(1\ot
                  u_{j+1}\ot 1)\right]\int_{0}^{t_{j}}dt_{j+2}…\:.
          \end{align*}}Für $j=1$ gilt nun obige Formel ebenfalls
        mit $t_{0}\simeq t_{j-1}=1$, und wir erhalten 
        {\allowdisplaybreaks  {\footnotesize
            \begin{align*}
                (\pt_{k}\cp G_{k})&(1\ot\ovl{u}\ot 1)
                \\=&\: \overbrace{i_{0}(1\ot u_{1}\ot 1\ot
                  1)\big|_{t_{0}=1}}^{u_{1}\ot1\ot1}\cdot\int_{0}^{1}dt_{2}\int_{0}^{t_{2}}dt_{3}…\int_{0}^{t_{k-1}}dt_{k}\prod_{2\leq
                  i\leq k}\ck{i}
                \\ &- \int_{0}^{1}dt_{1}\:i_{1}(1\ot u_{1}\ot1\ot1)\cdot
                (i_{1}\cp \delta)(1\ot u_{2}\ot
                1)\cdot\int_{0}^{t_{1}}dt_{3}…\int_{0}^{t_{k-1}}dt_{k}\prod_{3\leq i\leq k}\ck{i}
                \\ &+\sum_{j=2}^{k-1}\:(-1)^{j-1}\int_{0}^{1}dt_{1}…\int_{0}^{t_{j-2}}dt_{j-1}\bold{\int_{0}^{t_{j-1}}dt_{j+1}}\int_{0}^{t_{j+1}}dt_{j+2}…\int_{0}^{t_{k-1}}dt_{k}\\
                &\quad\quad \left[i_{j-1}(1\ot u_{j}\ot1\ot1)\cdot\prod_{i\neq
                      j}\ck{i}\right]
                \\ &-\sum_{j=2}^{k-1}\:\int_{0}^{1}dt_{1}…\int_{0}^{t_{j-2}}dt_{j-1}\int_{0}^{t_{j-1}}dt_{j}\bold{\int_{0}^{t_{j}}}dt_{j+2}\int_{0}^{t_{j+2}}dt_{j+3}…\int_{0}^{t_{k-1}}dt_{k}
                \\ &\quad\quad \left[i_{j}(1\ot u_{j}\ot1\ot1)\cdot
                    (i_{j}\cp\delta)(1\ot u_{j+1}\ot1)\cdot\prod_{\substack{i\neq
                        j\\ i\neq j+1}}\ck{i}\right]
                \\ &+ (-1)^{k-1}
                \int_{0}^{1}dt_{1}…\int_{0}^{t_{k-2}}dt_{k-1}\prod_{i=1}^{k-1}\ck{i}\:\:\cdot
                \underbrace{\Big[i_{k}(1\ot u_{k}\ot
                  1\ot1)\Big]_{0}^{t_{k-1}}}_{i_{k-1}(1\ot
                  u_{k}\ot1\ot1)-1\ot u_{k}\ot1}.
            \end{align*}}Durch Umbenennung der
          $t_{j}$-Variablen in jedem Summanden zu $t_{1},…,t_{k-1}$,
          folgt: {\allowdisplaybreaks\footnotesize
            \begin{align*}
                (\pt_{k}&\:\cp G_{k})(1\ot\ovl{u}\ot 1)
                \\=& \int_{0}^{1}dt_{1}…\int_{0}^{t_{k-2}}dt_{k-1}\:(i\cp\delta)(u_{1}\ot
                u_{2}\ot…\ot u_{k}\ot 1)
                \\ &- \int_{0}^{1}dt_{1}…\int_{0}^{t_{k-2}}dt_{k-1}\:i_{1}(1\ot
                u_{1}\ot 1\ot 1)\:\cdot \prod_{i=1}^{k-1}(i_{i}\cp\delta)(1\ot u_{i+1}\ot 1)
                \\
                &+\sum_{j=2}^{k-1}\:(-1)^{j-1}\int_{0}^{1}dt_{1}…\int_{0}^{t_{k-2}}dt_{k-1}\:\Bigg[i_{j-1}(1\ot
                    u_{j}\ot1\ot1)\:\cdot\prod_{i=1}^{j-1}\ck{i}\:\cdot
  \\ & \qquad\qquad\qquad\qquad\qquad\qquad\qquad\qquad\qquad\qquad\qquad\qquad\qquad\qquad\qquad\qquad\qquad\prod_{i=j}^{k-1}(i_{i}\cp\delta)(1\ot
                    u_{i+1}\ot1)\Bigg]\displaybreak
                \\ & - \sum_{j=2}^{k-1}\:\int_{0}^{1}dt_{1}…\int_{0}^{t_{k-2}}dt_{k-1}\:\Bigg[i_{j}(1\ot u_{j}\ot1\ot1)\cdot
                (i_{j}\cp\delta)(1\ot
                u_{j+1}\ot1)\cdot\prod_{i=1}^{j-1}\ck{i}\cdot
                \\ & \qquad\qquad\qquad\qquad\qquad\qquad\qquad\qquad\qquad\qquad\qquad\qquad\qquad\qquad\qquad\qquad\qquad\prod_{i=j+1}^{k-1}(i_{i}\cp\delta)(1\ot
                u_{i+1}\ot1)\Bigg]
                \\ &+(-1)^{k-1}\int_{0}^{1}dt_{1}…\int_{0}^{t_{k-2}}dt_{k-1}\:\prod_{i=1}^{k-1}\ck{i}\cdot\:
                i_{k-1}(1\ot u_{k}\ot 1\ot1)
                \\ &+ (-1)^{k}\int_{0}^{1}dt_{1}…\int_{0}^{t_{k-2}}dt_{k-1}\:(i\cp\delta)(1\ot
                u_{1}\ot…\ot u_{k})
                \\=&   \int_{0}^{1}dt_{1}…\int_{0}^{t_{k-2}}dt_{k-1}\:(i\cp\delta)(u_{1}\ot
                u_{2}\ot…\ot u_{k}\ot 1)
                \\ &+ (-1)^{1} \int_{0}^{1}dt_{1}…\int_{0}^{t_{k-2}}dt_{k-1}\:i_{1}(1\ot
                u_{1}\ot 1\ot1)\cdot (i_{t_{1},…,t_{k}}\cp\delta)(1\ot u_{2}\ot…\ot u_{k}\ot1)
                \\
                &-\sum_{j=2}^{k-1}(-1)^{j}\int_{0}^{1}dt_{1}…\int_{0}^{t_{k-2}}dt_{k-1}\:\left[\prod_{i=1}^{j-1}\ck{i}\cdot\:
                    i_{j-1}(1\ot
                    u_{j}\ot1\ot1)\cdot\prod_{i=j}^{k-1}(i_{i}\cp\delta)(1\ot
                    u_{i+1}\ot1)\right]
                \\ & +
                \sum_{j=2}^{k-1}(-1)^{j}\int_{0}^{1}dt_{1}…\int_{0}^{t_{k-2}}dt_{k-1}\:\Bigg[\prod_{i=1}^{j-1}\ck{i}\cdot\:
                i_{j}(1\ot u_{j}\ot1\ot1)\cdot \prod_{i=j}^{k-1}(i_{i}\cp\delta)(1\ot
                u_{i+1}\ot1)\Bigg]
                \\&-(-1)^{k}\int_{0}^{1}dt_{1}…\int_{0}^{t_{k-2}}dt_{k-1}\:\prod_{i=1}^{k-1}\ck{i}\cdot\:
                i_{k-1}(1\ot u_{k}\ot 1\ot1)
                \\ &+ (-1)^{k}\int_{0}^{1}dt_{1}…\int_{0}^{t_{k-2}}dt_{k-1}\:(i\cp\delta)(1\ot
                u_{1}\ot…\ot u_{k})
                \\=&\int_{0}^{1}dt_{1}…\int_{0}^{t_{k-2}}dt_{k-1}\:(i\cp\delta)(u_{1}\ot
                u_{2}\ot…\ot u_{k}\ot 1)
                \\ & + \sum_{j=1}^{k-1}(-1)^{j}\int_{0}^{1}dt_{1}…\int_{0}^{t_{k-2}}dt_{k-1}\:\Bigg[\prod_{i=1}^{j-1}\ck{i}\cdot\:
                i_{j}(1\ot u_{j}\ot1\ot1)\cdot \prod_{i=j}^{k-1}(i_{i}\cp\delta)(1\ot
                u_{i+1}\ot1)\Bigg]
                \\ &-\sum_{j=2}^{k}(-1)^{j}\int_{0}^{1}dt_{1}…\int_{0}^{t_{k-2}}dt_{k-1}\:\left[\prod_{i=1}^{j-1}\ck{i}\cdot\:
                    i_{j-1}(1\ot
                    u_{j}\ot1\ot1)\cdot\prod_{i=j}^{k-1}(i_{i}\cp\delta)(1\ot
                    u_{i+1}\ot1)\right]
                \\ & +(-1)^{k}\int_{0}^{1}dt_{1}…\int_{0}^{t_{k-2}}dt_{k-1}\:(i\cp\delta)(1\ot
                u_{1}\ot…\ot u_{k})
                \\=&\int_{0}^{1}dt_{1}…\int_{0}^{t_{k-2}}dt_{k-1}\:(i\cp\delta)(u_{1}\ot
                u_{2}\ot…\ot u_{k}\ot 1)
                \\ & + \sum_{j=1}^{k-1}(-1)^{j}\int_{0}^{1}dt_{1}…\int_{0}^{t_{k-2}}dt_{k-1}\:\Bigg[\prod_{i=1}^{j-1}\ck{i}\cdot\:
                i_{j}(1\ot u_{j}\ot1\ot1)\cdot \prod_{i=j}^{k-1}(i_{i}\cp\delta)(1\ot
                u_{i+1}\ot1)\Bigg]
                \\ &+\sum_{j=1}^{k-1}(-1)^{j}\int_{0}^{1}dt_{1}…\int_{0}^{t_{k-2}}dt_{k-1}\:\left[\prod_{i=1}^{j}\ck{i}\cdot\:
                    i_{j}(1\ot
                    u_{j+1}\ot1\ot1)\cdot\prod_{i=j+1}^{k-1}(i_{i}\cp\delta)(1\ot
                    u_{i+1}\ot1)\right]
                \\ & +(-1)^{k}\int_{0}^{1}dt_{1}…\int_{0}^{t_{k-2}}dt_{k-1}\:(i\cp\delta)(1\ot
                u_{1}\ot…\ot u_{k})
                \\=&\int_{0}^{1}dt_{1}…\int_{0}^{t_{k-2}}dt_{k-1}\:(i\cp\delta)(u_{1}\ot
                u_{2}\ot…\ot u_{k}\ot 1)\displaybreak
                \\ & +
                \sum_{j=1}^{k-1}(-1)^{j}\int_{0}^{1}dt_{1}…\int_{0}^{t_{k-2}}dt_{k-1}\:\Bigg[\prod_{i=1}^{j-1}\ck{i}\cdot
                \\ &\quad\quad\quad\quad
                \left[i_{j}(1\ot u_{j}\ot1\ot1)\cdot(i_{j}\cp\delta)(1\ot
                    u_{j+1}\ot1)+ \ck{j} \cdot\: i_{j}(1\ot
                    u_{j+1}\ot1\ot1)\right]\cdot 
                \\ &\qquad\qquad\qquad\qquad\qquad\qquad\qquad\qquad\qquad\qquad\qquad\qquad\qquad\qquad\qquad\qquad
                \prod_{i=j+1}^{k-1}(i_{i}\cp\delta)(1\ot
                u_{i+1}\ot1)\Bigg]
                \\ & +(-1)^{k}\int_{0}^{1}dt_{1}…\int_{0}^{t_{k-2}}dt_{k-1}\:(i\cp\delta)(1\ot
                u_{1}\ot…\ot u_{k}) 
                \\=&  \int_{0}^{1}dt_{1}…\int_{0}^{t_{k-2}}dt_{k-1}\:(i\cp\delta)(u_{1}\ot
                u_{2}\ot…\ot u_{k}\ot 1)
                \\ & +
                \sum_{j=1}^{k-1}(-1)^{j}\int_{0}^{1}dt_{1}…\int_{0}^{t_{k-2}}dt_{k-1}\:\Bigg[\prod_{i=1}^{j-1}\ck{j}\cdot\:(i_{j}\cp
                \delta)(1\ot
                u_{j}\vee
                u_{j+1}\ot 1)\cdot
                \\ &\qquad\qquad \qquad \qquad \qquad \qquad \qquad \qquad \qquad  \qquad\qquad \qquad \qquad \qquad \qquad\qquad
                \prod_{i=j+1}^{k-1}(i_{i}\cp\delta_{k})(1\ot
                u_{i+1}\ot1)\Bigg]
                \\ &+ (-1)^{k}\int_{0}^{1}dt_{1}…\int_{0}^{t_{k-2}}dt_{k-1}\:(i\cp\delta)(1\ot
                u_{1}\ot…\ot u_{k})
                \\=& \int_{0}^{1}dt_{1}…\int_{0}^{t_{k-2}}dt_{k-1}(i\cp\delta\cp d)(1\ot
                \ovl{u}\ot 1)
                \\=&\: (G_{k-1}\cp d_{k})(1\ot \ovl{u}\ot 1).
            \end{align*}}}Ebenso folgt für den Fall
        $\alpha\ot u\ot \beta\in \C_{1}$, dass
        \begin{align*}
            (\pt_{1}\cp G_{1})(\alpha\ot u\ot \beta)=&\:\int_{0}^{1}dt\: (\pt_{1}\cp i_{t}\cp\delta)(\alpha\ot u\ot \beta)=\:\hat{i}_{t}(\alpha\ot u\ot \beta)\big|_{0}^{1}
            \\=&\:d_{1}(\alpha\ot u\ot
            \beta)=(G_{0}\cp d_{1})(\alpha\ot u\ot \beta).
        \end{align*}
    \end{beweis}
\end{proposition}
\begin{bemerkung}[Die Kettenabbildungen $F$ und $G$]
    \label{bem:HoimotAbstrIsosUnterk}
    Es gibt auch durchaus abstraktere Möglichkeiten, die Hochschild-Kohmologie
    der symmetrischen Algebra zu berechnen, vgl.
    \cite{cartan.eilenberg:1999a}. 
    Der von uns beschrittene Weg bietet
    jedoch den großen Vorteil, dass uns nun explizite Kettenabbildungen
    $F$ und $G$ zwischen Bar- und Koszul-Komplex zur Verfügung
    stehen, die zueinander inverse Isomorphismen    
    \begin{align*}
        \wt{F_{k}^{*}}\colon H^{k}(\Hom_{\mathcal{A}^{e}}(\C,\mathcal{M}))&\longrightarrow H^{k}(\Hom_{\mathcal{A}^{e}}(\K,\mathcal{M}))\qquad\text{und}\\
        \wt{G_{k}^{*}}\colon
        H^{k}(\Hom_{\mathcal{A}^{e}}(\K,\mathcal{M}))&\longrightarrow
        H^{k}(\Hom_{\mathcal{A}^{e}}(\C,\mathcal{M}))      
    \end{align*}
    induzieren. Diese können auf verschiedene Arten nutzbringend
    eingesetzt werden:
    \begin{enumerate}
    \item
        Mit obigen
        Kettenabbildungen ist es möglich, die Isomorphien 
        der Ko\-ho\-mo\-lo\-gie-Grup\-pen von Unterkomplexen $(\mathcal{X},d^{*})$ von
        $(\C^{*},d^{*})$ und $(\mathit{K},\pt^{*})$ von $(\K^{*},\pt^{*})$
        zu zeigen. Dabei ist ein Unterkomplex
        $(\mathcal{X},d^{*})$ von $(\C^{*},d^{*})$ ein
        Kokettenkomplex derart, dass $\mathcal{X}^{k}\subseteq \C^{*}_{k}$
        und $d_{k+1}^{*}: \mathcal{X}^{k}\longrightarrow \mathcal{X}^{k+1}$
        gilt. Analog für $(\mathit{K},\pt^{*})$ und $(\K^{*},\pt^{*})$.
        Hierfür ist zunächst nachzuweisen, dass
        \begin{align*}
            F_{k}^{*}\colon \mathcal{X}^{k}&\longrightarrow \mathit{K}^{k}\qquad\text{und}\\
            G_{k}^{*}\colon \mathit{K}^{k}&\longrightarrow \mathcal{X}^{k}, 
        \end{align*} also
        $\mathrm{F}^{k}=F^{*}_{k}\big|_{\mathcal{X}^{k}}$ gilt und somit $\mathrm{G}^{k}=G_{k}^{*}\big|_{\mathit{K}^{k}}$ wohldefinierte
        Kettenabbildungen zwischen besagten Unterkomplexen
        sind. Lemma \ref{lemma:Fkettenabb}~\textit{i.)} zeigt dann, dass 
        \begin{equation*}
            F_{k}^{*}\cp
            G_{k}^{*}=\hom_{\mathcal{A}^{e}}(\cdot,\mathcal{M})(G_{k}\cp
            F_{k})=\id_{\K^{*}_{k}},
        \end{equation*}also $\mathrm{F}^{k}\cp
        \mathrm{G}^{k}=\id_{\mathrm{K}^{k}}$ und somit
        $\wt{\mathrm{F}^{k}}\cp\wt{\mathrm{G}^{k}}=\wt{\mathrm{F}^{k}\cp
          \mathrm{G}^{k}}=\id_{H^{k}(\mathrm{K},d^{*})}$ gilt. Hiermit
        folgt die
        Injektivität von $\wt{\mathrm{G}^{k}}$ und die Surjektivität von
        $\wt{\mathrm{F}^{k}}$.
        Für die umgekehrte Aussage beachten man, dass
        $\mathrm{G}\cp \mathrm{F}\colon 
        (\mathcal{X},d^{*})\longrightarrow (\mathcal{X},d^{*})$ 
        eine Kettenabbildung ist. Können wir dann $\mathrm{G}\cp
        \mathrm{F}\sim\id_{\mathcal{X}^{k}}$ vermöge Homotopieabbildungen $\mathrm{s}^{k}\colon
        \mathcal{X}^{k}\longrightarrow\mathcal{X}^{k-1}$ nachweisen, so
        zeigt Lemma \ref{lemma:tildeabbeind}, dass
        $\wt{\mathrm{G}^{k}}\cp\wt{\mathrm{F}^{k}}=\wt{\mathrm{G}^{k}\cp\mathrm{F}^{k}}=\id_{H^{k}(\mathcal{X},d^{*})}$
        gilt. Dies liefert die Surjektivität von $\wt{\mathrm{G}^{k}}$ und die Injektivität von
        $\wt{\mathrm{F}^{k}}$, also $H^{k}(\mathcal{X},d^{*})\cong
        H^{k}(\mathrm{K},\pt^{*})$ vermöge den zueinander inversen
        Isomorphismen $\wt{\mathrm{G}^{k}}$ und
        $\wt{\mathrm{F}^{k}}$. Hierbei beachte man, dass die Existenz einer derartigen Homotopie
        nicht offensichtlich ist. In der Tat besagt zwar
        Satz \ref{satz:AufluProjKompKettab}, dass $F\cp G\sim
        \id_{\C_{k}}$ vermöge $\mathcal{A}^{e}$-linearen Homotopieabbildungen
        $s_{k}\colon \C_{k}\longrightarrow \C_{k+1}$ und somit $G^{*}\cp F^{*}\sim
        \id_{\K^{*}_{k}}$ vermöge $s_{k}^{*}$. Jedoch ist in keiner Weise
        gewährleistet, dass sich die $s_{k}^{*}$ auf Abbildungen zwischen
        den Unterkomplexen einschränken lassen und die gewünschte Homotopie
        $\mathrm{s}$ liefern.

        Ist man schließlich an den Kohomologie-Gruppen eines speziellen Unterkom-plexes
        $HC_{\circ}(\SsV,\mathcal{M})$ von
        $HC(\SsV,\mathcal{M})$ interessiert, so muss jetzt nur noch
        sichergestellt werden, dass sich der Kettenisomorphismus $\Xi$ auf eine Isomorphismus
        zwischen $(\mathcal{X},d^{*})$ und $HC_{\circ}(\SsV,\mathcal{M})$
        einschränken lässt. Ein essentielles Beispiel ist hierbei der
        stetige Hochschild-Komplex $HC_{\cont}(\SsV,\mathcal{M})$,
        den wir im nächsten Kapitel kennen lernen werden.
    \item
        Obige Kettenabbildungen erlauben es, tiefere Erkenntnisse
        über die Natur von $HH^{k}(\SsV,\mathcal{M})$ und
        $HH^{k}_{\cont}(\SsV,\mathcal{M})$  zu gewinnen. Im Falle 
        symmetrischer Bimoduln beispielsweise erhalten wir  Analoga zu dem bekannten
        Hochschild-Kostant-Rosenberg-Theorem\footnote{vgl. \cite[Prop
          6.2.48]{waldmann:2007a}, \cite{cahen.gutt.dewilde:1980a}} für
        $HH^{k}(\SsV,\mathcal{M})$ und
        $HH^{k}_{\cont}(\SsV,\mathcal{M})$, siehe Kapitel
        \ref{sec:HKRTheos}. Auch für die Klasse der differentiellen
        Bimoduln, welche die symmetrischen als Spezialfall enthalten, lassen sich
        mit Hilfe der Abbildung $G$ ähnliche Aussagen ableiten, siehe
        Kapitel \ref{cha:DiffHochK}. Selbst für den Fall, dass in der
        Situation von \textit{i.)} nicht klar ist, dass eine Homotopie $\mathrm{s}$
        existiert, sind $\wt{\mathrm{G}^{k}}$ injektiv und
        $\wt{\mathrm{F}^{k}}$ surjektiv und liefern nützliche
        Informationen über $HH^{k}_{\circ}(\SsV,\mathcal{M})$. Dies
        wird beispielsweise für den differentiellen
        $HH^{k}_{\diff}(\SsV,\mathcal{M})$ und den
        stetig-differentiellen Unterkomplex
        $HH^{k}_{\mathrm{c,d}}(\SsV,\mathcal{M})$ der Fall sein, welchen
        wir in Kapitel \ref{cha:DiffHochK} begegnen werden.
    \item
        Die Kettenabbildungen $F$ und $G$ werden es uns erlauben, die
        Hochschild-Koho-mologie des stetigen Unterkomplexes
        $HC_{\cont}(\SsV,\mathcal{M})$, den wir im nächsten Kapitel
        für beliebige lokalkonvexe Algebren definieren werden, zu berechnen. Hieraus erhalten
        wir die Hochschild-Kohomologien
        $HH_{\cont}^{k}(\Hol,\mathcal{M})$ des stetigen Unterkomplexes
        $HC_{\cont}(\Hol,\mathcal{M})$ für vollständige
        lokalkonvexe\footnote{vgl. Kapitel \ref{cha:TopKompl}}
        Bimoduln $\mathcal{M}$. Hierbei ist $(\Hol,*)$, die durch
        Vervollständigung von $(\SsV,\vee)$ erhaltene Algebra (vgl.
        Kapitel \ref{sec:StetHKHol}), für welche wir ebenfalls ein
        Hochschild-Kostant-Rosenberg-Theorem erhalten werden. Hierfür beachte man, dass die Berechnung der Hochschild-Kohomologie des Kokettenkomplexes
        $HH^{k}(\Hol,\mathcal{M})$ ein im Allgemeinen schwieriges
        Problem darstellt, die in Rahmen der Deformationsquantisierung
        weitaus interessantere, stetige Hochschild-Kohomologie mit dem
        hier gewählten Zugang aber ausgesprochen einfach zu erhalten ist.
    \end{enumerate}
\end{bemerkung}

%
%
%
%
\chapter{Topologische Komplexe und stetige Hochschild-Kohomologien}
\label{cha:TopKompl}
Wie bereits in Bemerkung \ref{bem:HoimotAbstrIsosUnterk} erwähnt, wollen wir
in diesem Kapitel die stetigen Hochschild-Kohomologien
$HC_{\cont}(\SsV,\mathcal{M})$ und $HC_{\cont}(\Hol,\mathcal{M})$ der
lokalkonvex topologisierten Algebren $\SsV$ und $\Hol$ für
lokalkonvexe Bimoduln berechnen.
Hierbei heißt ein $\mathbb{K}$-Vektorraum lokalkonvex, wenn er ein topologischer
Vektorraum ist (Vektorraumoperationen sind stetig) und
seine Topologie durch ein Halbnormensystem $P$ erzeugt wird. Dabei
soll $P$ im Folgenden immer als filtrierend\footnote{vgl. Definition \ref{def:Halbnormensysteme}~\textit{ii.)}} voraussetzen werden. Eine
lokalkonvexe $\mathbb{K}$-Algebra ist dann ein lokalkonvexer
$\mathbb{K}$-Vektorraum mit stetiger Algebramultiplikation. Dies ist
gleichbedeutend damit (vgl. Satz \ref{satz:stetmultabb}), dass für jede
Halbnorm $q\in P$ eine Konstante
$c_{*}>0$ und Halbnormen $p_{*1},p_{*2}\in P$ derart existieren, dass für
alle $a,b\in \mathcal{A}$ die Abschätzung $p(a*b)\leq c_{*}\:
p_{*1}(v)\:p_{*2}(w)$ erfüllt ist. Sei $\mathcal{A}$ lokalkonvex, so
verstehen wir unter einem lokalkonvexen
$\mathcal{A}-\mathcal{A}$-Bimodul $(\mathcal{M},*_{L},*_{R})$ einen
lokalkonvexen Vektorraum $(\mathcal{M},Q)$ mit stetigen, $\mathbb{K}$-bilinearen
Modul-Multiplikationen. Dies bedeutet, dass für
jedes $q\in Q$ 
Konstanten $c_{L},c_{R}>0$ sowie Halbnormen $p_{L},p_{R}\in P$ und
$q_{L},q_{R}\in Q$ existieren, so dass für alle $a\in \mathcal{A}$ und
alle $m\in \mathcal{M}$ die Abschätzungen $q(a*_{L}m)\leq c_{L}\:
p_{L}(a)\:q_{L}(m)$ und $q(m*_{R}a)\leq c_{R}\: p_{R}(a)\:q_{R}(m)$
gelten. Ist $\mathcal{A}$ unitär, so wollen wir wieder
$1_{\mathcal{A}}*_{L}m=m=m*_{R}1_{\mathcal{A}}$ für alle $m\in
\mathcal{M}$ voraussetzen.

Nach Bemerkung \ref{bem:HoimotAbstrIsosUnterk}~\textit{i.)} besteht die
Aufgabe nun zunächst darin, einen
Unterkomplex $(\mathcal{X},d^{*})$ von $(\C^{*},d^{*})$ für
$\mathcal{A}=\SsV$ derart zu
finden, dass $\Xi$ einen Kettenisomorphismus
$HC_{\cont}(\mathcal{A},\mathcal{M})\longrightarrow (\mathcal{X},d^{*})$
induziert. Dies wird mit Hilfe des folgenden Abschnittes sogar für beliebige
lokalkonvexe Algebren $\mathcal{A}$ erreichbar sein.   
\section{Vorbereitung}
\label{subsec:Vorber}
Gegeben eine lokalkonvexe Algebra $(\mathcal{A},*)$ und ein lokal
konvexer $\mathcal{A}-\mathcal{A}$-Bimodul $\mathcal{M}$, so betrachten wir die $\mathbb{K}$-Vektorräume
\begin{equation}
    \label{eq:stetHSCHK}
    HC_{\cont}^{k}(\mathcal{A},\mathcal{M}):=
    \begin{cases} \{0\} & k<0\\
        \mathcal{M} & k=0\\
        \Hom^{\cont}_{\mathbb{K}}(\underbrace{\mathcal{A}\times…\times
          \mathcal{A}}_{k-mal},\mathcal{M})& k\geq 1,
    \end{cases} 
\end{equation} die stetigen $\mathbb{K}$-multilinearen Abbildungen von
$\mathcal{A}^{k}$ nach $\mathcal{M}$. Mit
Satz \ref{satz:stetmultabb} sind dies wieder gerade die Elemente $\phi\in
\Hom_{\mathbb{K}}(\mathcal{A}\times…\times
\mathcal{A},\mathcal{M})$ für welche $q\in Q$
vorgegeben, eine Konstante $c>0$ und Halbnormen
$p_{1},…,p_{k}\in P$ derart existieren, dass\\
\begin{equation}
    \label{eq:phiAbsch}
    q(\phi(a_{1},…a_{k}))\leq c\:
    p_{1}(a_{1})…p_{k}(a_{k})\qquad\qquad\forall\:a_{1},…,a_{k}\in \mathcal{A}
\end{equation}gilt.
Vermöge \eqref{eq:Hochschilddelta} seien $\mathbb{K}$-lineare Abbildungen
\begin{equation*}
    \delta^{k}_{c}\colon HC_{\cont}^{k}(\mathcal{A},\mathcal{M})\longrightarrow HC_{\cont}^{k+1}(\mathcal{A},\mathcal{M})
\end{equation*} definiert, und es ist zunächst zu zeigen, dass besagtes
Bild unter $\delta^{k}_{c}$ stetig ist.\\ Wir haben:
\begin{align*}
    (\delta_{c}^{k}\phi)(a_{1},…,a_{k+1})=a_{1}*_{L}\phi(a_{2},…,a_{k+1})&+\sum_{i=1}^{k}(-1)^{i}\phi(a_{1},…,a_{i}*a_{i+1},…,a_{k+1})\\
    &+(-1)^{k+1}\phi(a_{1},…,a_{k})*_{R}a_{k+1}.
\end{align*}
Für die Stetigkeit des ersten Summanden rechnen wir mit den Abschätzungen für $*_{L}$
und $\phi$:
\begin{align*}
    q\left(a_{1}*_{L}\phi(a_{2},…,a_{k+1})\right)\leq&
    \:c_{L}\:p_{L}(a_{1})\:q_{L}(\phi(a_{2},…,a_{k+1}))
    \\\leq&\: \hat{c}\:p_{L}(a_{1})\:p_{2}(a_{2})…p_{k+1}(a_{k+1}).
\end{align*}mit $\hat{c}=c_{L}c$ und $c$, $p_{2},…,p_{k+1}$ die zu
$q_{L}$ gehörigen Halbnormen aus \eqref{eq:phiAbsch}.
Satz \ref{satz:stetmultabb} zeigt dann die Stetigkeit, und die des letzten
Summanden folgt analog. Ebenso erhalten wir für den mittleren
Summanden, dass
\begin{align*}
    q(\phi(a_{1},…,a_{i}a_{i+1}&,…,a_{k+1}))
    \\\leq&\:c\:
    p_{1}(a_{1})…p_{i-1}(a_{i-1})\:p_{i}(a_{i}*a_{i+1})\:p_{i+1}(a_{i+2})…p_{k}(a_{k+1})
    \\\leq &
    \:\hat{c}\:p_{1}(a_{1})…p_{i-1}(a_{i-1})\:p_{*1}(a_{i})\:p_{*2}(a_{i+1})\:p_{i+1}(a_{i+2})…p_{k}(a_{k+1})
\end{align*}mit $\hat{c}=c_{*}c$ und $p_{i}(a_{i}*a_{i+1})\leq
c_{*}p_{*1}(a_{i})\:p_{*2}(a_{i+1})$. Die Stetigkeit von $\delta_{c}^{k}(\phi)$ folgt
nun unmittelbar mit der Stetigkeit der Vektorraumaddition in
$\mathcal{M}$, da kartesische Produkte stetiger Funktionen bezüglich
den zugehörigen Produkttopologien ebenfalls stetig sind.

Mit $\delta_{c}^{k+1}\cp\delta^{k}_{c}=0$ liefert uns dies einen 
Koketten-Unterkomplex $(HC_{\cont}^{\bullet}(\mathcal{A},\mathcal{M}),\delta_{c})$ von
$(HC^{\bullet}(\mathcal{A},\mathcal{M}),\delta)$ und definieren die
$k$-te stetige Hochschild-Kohomologien durch
\begin{equation*}
    HH_{\cont}^{k}(\mathcal{A},\mathcal{M}):=
    \begin{cases}  
        \ker\left(\delta_{c}^{0}\right) & k=0\\
        HH_{\cont}^{k}(\mathcal{A},\mathcal{M})=\ker\left(\delta_{c}^{k}\right)/\im\left(\delta_{c}^{k-1}\right)&
        k\geq 1.
    \end{cases} 
\end{equation*}
Für die Tensorvariante des Hochschild-Komplexes erhalten wir analoge
Aussagen. Dabei folgt die Stetigkeit des Bildes unter
$\delta_{c\ot}^{k}$ zum einen durch elementare Rechnung, oder aber auch
durch Rechtskomposition \eqref{eq:TensorglKetteniso} mit $\ot_{k}^{*}$, da die $\ot_{k*}$ vermöge
der Definition der $\pi$-Topologie, also insbesondere der Stetigkeit
der Abbildungen $\ot_{k}$, bijektiv stetige auf stetige Elemente abbilden.  
Insbesondere bedeutet dies, dass $\ot_{*}$ ein Ketten-isomorphismus 
zwischen diesen beiden Unterkomplexen ist, was die Isomorphie derer
Kohomologiegruppen impliziert. 
Wir dürfen uns also wieder auf die Tensorvariante des besagten
stetigen Hochschild-Komplexes beschränken, und es soll nun unter anderem darum gehen, die Isomorphie 
\begin{equation*}
    HH_{\cont}^{k}(\mathcal{A},\mathcal{M})\cong H^{k}\left(\Hom^{\cont}_{\mathcal{A}^{e}}(\C_{c},\mathcal{M}),d_{c}^{*}\right)
\end{equation*}einzusehen. Dabei bezeichnet $(\C_{c},d_{c})$ den topologische
Bar-Komplex, welchen wir bald kennen lernen werden.\\
\begin{bemerkung}
    Gegeben lokalkonvexe Vektorräume $(\V_{1},P_{1}),…,(\V_{k},P_{k}),(\mathbb{W},Q)$
    und eine stetige $\mathbb{K}$-multilineare
    Abbildung $\phi\colon \V_{1}\times…\times\V_{k}\longrightarrow \mathbb{W}$, so sind im Folgenden bei Stetigkeitsabschätzungen
    $q(\phi\:(v_{1},…,v_{k}))\leq c\:p(v_{1})…p(v_{k})$ mit $c$ und
    $p_{1},…,p_{k}$ immer die nach Satz \ref{satz:stetmultabb} zu $q\in
    Q$ gehörige Konstante und die zu $q$ gehörigen Halbnormen
    gemeint. Hierfür mache man sich noch einmal explizit klar, dass
    wir Halbnormensysteme am Anfang dieses Kapitels immer als
    filtrierend vorausgesetzt haben 
\end{bemerkung}
Wir wollen an dieser Stelle an die
$\pi_{k}$-Topologie erinnern:
\begin{definition}[$\pi$-Topologie]
    Gegeben lokalkonvexe Vektorräume
    $(\V_{1},P_{1}),…,(\V_{k},P_{k})$, so ist die $\pi_{k}$-Topologie
    die durch das System
    $\Pi_{P_{1}\times…\times
      P_{k}}$, bestehend aus Halbnormen
    \begin{equation*}  
        p_{1}\ot…\ot p_{k} (z):=\inf\left\{\sum_{i=1}^{n}p_{1}(x^{i}_{1})…p_{k}(x^{i}_{k})\right\},
    \end{equation*}auf $\V_{1}\ot…\ot\V_{k}$ induzierte
    lokalkonvexe Topologie. 
    Hierbei ist das Infimum über alle Zerlegungen
    $z=\displaystyle\sum_{i=1}^{n}x^{i}_{1}\ot…\ot
    x^{i}_{k}$ von $z$ zu nehmen. 
    Der so gewonnenen lokalkonvexen
    Vektorraum sei im Folgenden mit
    $\V_{1}\pite…\pite\V_{k}$ bezeichnet. Ist es an
    gegebener Stelle der Lesbarkeit zuträglich, so
    benutzen wir das Symbol $\pi_{p_{1},…,p_{k}}$ anstelle von
    $p_{1}\ot…\ot p_{k}$.
\end{definition}
\begin{bemerkung}
    \label{bem:PiTopArBem}
    Die $\pi_{k}$-Topologie besitzt die folgenden wichtigen
    Eigenschaften, siehe Kapitel \ref{subsec:TenprodLkvVr}:
    \begin{enumerate}
    \item
        Es gilt $p_{1}\ot…\ot
        p_{k}\:(x_{1}\ot…\ot x_{k})=p_{1}(x_{1})…p_{k}(x_{k})$ für alle
        separablen Elemente
        $x_{1}\ot…\ot x_{k}\in \V_{1}\pite…\pite \V_{k}$.
    \item
        $\pi_{k}$ ist genau dann hausdorffsch, wenn alle $(\V_{i},P_{i})$
        mit $1\leq i\leq k$ hausdorffsch sind.
    \item
        Sind $P_{1},\dots, P_{k}$ filtrierend, so auch $\prod_{P_{1},…,P_{k}}$.
    \item
        Eine lineare Abbildung $\phi\colon \mathbb{V}_{1} \pite…\pite
        \mathbb{V}_{k}\longrightarrow \mathbb{M}$ ist genau dann stetig, wenn die Abbildung $\phi\cp \ot_{k}$ bezüglich der Produkttopologie auf $\mathbb{V}_{1}\times…\times
        \mathbb{V}_{k}$ stetig ist, also für jedes $q\in Q$ ein $c>0$ und
        $p_{i}\in P_{i}$ mit $1\leq i\leq k$ existieren, so dass
        \begin{equation*}
            q(\phi(v_{1}\ot…\ot v_{k}))\leq\:c\:p(v_{1})…p(v_{k})\qquad\forall\:v_{i}\in \V_{i},1\leq i\leq k
        \end{equation*}gilt. 
    \end{enumerate}
\end{bemerkung}
\begin{lemma}
    \label{lemma:AezuunittopRing}
    Gegeben eine assoziative, lokalkonvexe $\mathbb{K}$-Algebra $(\mathcal{A},*)$.
    \begin{enumerate}
    \item
        Dann wird die Menge
        $\mathcal{A}^{e}=\mathcal{A}\pite \mathcal{A}$, versehen mit der
        distributiven Fortsetzung der Multiplikation
        \begin{equation*}
            (a\ot b) *_{e} (\tilde{a}\ot
            \tilde{b})=(a*\tilde{a})\ot
            (b*^{opp}\tilde{b})=(a*\tilde{a})\ot
            (\tilde{b}*b)
        \end{equation*} auf ganz
        $\mathcal{A}\pite \mathcal{A}$, zu einer lokalkonvexen
        Algebra. Ist $\mathcal{A}$ unitär, so auch $\mathcal{A}^{e}$.
    \item 
        Jeder lokalkonvexe $\mathcal{A}-\mathcal{A}$-Bimodul $\mathcal{M}$ wird vermöge 
        \begin{equation*}
            a\ot b*_{e} m=a*_{L}(m *_{R}b)=(a*_{L} m)*_{R}b\quad\quad a\ot b\in \mathcal{A}^{e},\: m\in \mathcal{M}
        \end{equation*} zu einem lokalkonvexen $\mathcal{A}^{e}$-Linksmodul.  
    \end{enumerate}  
    \begin{beweis}
        Die Algebra- und Moduleigenschaften folgen wie in Lemma \ref{lemma:AewirdzuunitRing}, die Stetigkeit
        der Ringaddition ist die Stetigkeit der Vektorraumaddition in
        $(\mathcal{A}^{e},\pi_{2})$ als lkVR und
        die Stetigkeit der Algebra-Multiplikation erhalten wir mit der
        Stetigkeit von $*$, da
        \begin{align*}
            p_{1}\ot p_{2}\:(z*_{e}\tilde{z})=\:& p_{1}\ot p_{2}\:\bigg(\sum_{i}a_{i}\ot b_{i}*_{e}\sum_{j}\tilde{a}_{j}\ot \tilde{b}_{j}\bigg)
            \leq\: \sum_{i,j}p_{1}\ot p_{2}\:\left(a_{i}*\tilde{a}_{j}\ot
                \tilde{b}_{j}*b_{i}\right)
            \\=&\: \sum_{i,j}p_{1}\big(a_{i}*\tilde{a}_{j}\big)\:p_{2}\big(b_{i}*\tilde{b}_{j}\big)
            \leq\:c\sum_{i,j}p'_{1}\big(a_{i}\big)\:p''_{1}\big(\tilde{a}_{j}\big)\:p'_{2}\big(b_{i}\big)\:p''_{2}\big(\tilde{b}_{j}\big)
            \\=&\:c\:\bigg(\sum_{i}p'_{1}(a_{i})p'_{2}(b_{i})\bigg)\bigg(\sum_{j}p''_{1}(\tilde{a}_{j})p''_{2}(\tilde{b}_{j})\bigg)
        \end{align*} für alle Zerlegungen von $z,\tilde{z}\in \mathcal{A}^{e}$ und somit
        \begin{align*}
            \pi_{p,q}\:(z*_{e}\tilde{z})\leq&\: c\: \inf\bigg(\sum_{i}p'_{1}(a_{i})\:p'_{2}(b_{i})\bigg)\:\inf\bigg(\sum_{j}p''_{1}(\wt{a}_{j})\:p''_{2}(\wt{b}_{j})\bigg)
            =\:c\: \pi_{p'_{1},p'_{2}}\:(z)\:p''_{1}\ot p''_{2}\:(\tilde{z}).
        \end{align*} 
        Dies zeigt die Stetigkeit von $*$ und folglich \textit{i.)}.\\\\
        Für \textit{ii.)} sei $m\in \mathcal{M}$ und $\mathcal{A}^{e}\ni z=
        \sum_{i}a_{i}\ot b_{i}$. Dann folgt für alle
        Zerlegungen $\sum_{i}a_{i}\ot b_{i}$ von $z$, dass
        \begin{align*}
            q(z*_{e}m) =q\bigg(\sum_{i}a_{i}(m b_{i})\bigg)
            \leq&\:\sum_{i}q(a_{i}(m b_{i}))\leq c\sum_{i}p_{1}(a_{i})\:q'(m
            b_{i})
            \\\leq&\:\hat{c}\sum_{i}p_{1}(a_{i})\:q''(m)\:p_{2}(b_{i})
            =\: \hat{c}\: q''(m)\sum_{i}p_{1}(a_{i})\:p_{2}(b_{i})
        \end{align*} und somit $p(zm)\leq \hat{c}\:\pi_{p_{1},p_{2}}(z)\: q''(m)$.
    \end{beweis}
\end{lemma}
\begin{definition}[Topologischer Bar-Komplex]
    \label{def:topBarkompl}
    Gegeben eine lokalkonvexe, assoziative Algebra $\mathcal{A}$, so definieren wir
    den topologischen Bar-Komplex $(\C_{c},d_{c})$ durch die
    $\mathcal{A}^{e}$-Moduln
    \begin{align*}
        \C^{c}_{k}=\mathcal{A}\pite \underbrace{\mathcal{A}\pite …
          \pite \mathcal{A}}_{k-mal} \pite \mathcal{A}
    \end{align*} $\qquad\qquad\qquad\quad \C_{0}=\mathcal{A}\pite \mathcal{A},\quad\quad
    \C_{1}=\mathcal{A}\pite\mathcal{A}\pite \mathcal{A},\quad\quad \C_{2}=\mathcal{A}\pite\mathcal{A}\pite \mathcal{A}\pite \mathcal{A}$ \\\\ mit $\mathcal{A}^{e}$-Multiplikation
    \begin{equation*}
        (a\ot b)(x_{0}\ot x_{1}\ot … \ot x_{k}\ot
        x_{k+1}):=(ax_{0})\ot x_{1}\ot … \ot x_{k}\ot
        (x_{k+1}b)
    \end{equation*} und $\mathcal{A}^{e}$-Homomorphismen
    \begin{align*}
        d^{c}_{k}\colon\C^{c}_{k}&\longrightarrow \C^{c}_{k-1}\\
        (x_{0}\ot … \ot x_{k+1})&\longmapsto
        \sum_{j=0}^{k}(-1)^{k}x_{0}\ot…\ot
        x_{j}x_{j+1}\ot…\ot x_{k+1}
    \end{align*} für $k\geq1$ mit $d^{c}_{k}\cp d^{c}_{k+1}=0$.
\end{definition}
Die relevanten Eigenschaften klärt folgende Proposition:
\begin{proposition}
    \label{prop:topBarKomplexprop}
    \begin{enumerate}
    \item
        Die $X^{c}_{k}$ sind lokalkonvexe $\mathcal{A}^{e}$-Moduln.
    \item
        Die $d^{c}_{k}$ sind stetig, ebenso die
        exaktheitsliefernde Homotopieabbildungen:
        \begin{equation*}
            h^{c}_{k}\colon x_{0}\ot…\ot x_{k+1}\longmapsto 1\ot
            x_{0}\ot…\ot x_{k+1}.
        \end{equation*}
    \item 
        Ist $\mathcal{A}$ unitär und $\mathcal{M}$ ein lokalkonvexer
        $\mathcal{A}-\mathcal{A}$-Bimodul, so gilt:
        \begin{equation*}
            HH_{\cont}^{k}(\mathcal{A},\mathcal{M})\cong H^{k}\left(\Hom^{\cont}_{\mathcal{A}^{e}}(\C_{c},\mathcal{M}),d^{*}_{c}\right).
        \end{equation*}
    \end{enumerate}
    \begin{beweis}  
        \begin{enumerate}
        \item
            Sei $z\in \mathcal{A}^{e}$ und $x\in X^{c}_{k}$, dann folgt:
            \begin{align*}
                \pi_{p_{0},…,p_{k+1}}&\:(zx)=\pi_{p_{0},…,p_{k+1}}\bigg(\sum_{i}a_{i}\ot
                b_{i}\cdot\sum_{j}x_{0}^{j}\ot…\ot x^{j}_{k+1}\bigg)
                \\ \leq&\:
                \sum_{i,j}p_{0}\big(a_{i}x_{0}^{j}\big)\:p_{k+1}\big(x^{j}_{k+1}b_{i}\big)\:\pi_{p_{1},…,p_{k}}\big(x_{1}^{j}\ot…\ot
                x_{k}^{j}\big)
                \\\leq&\:
                c\sum_{i,j}p'_{0}(a_{i})\:p''_{0}(x_{0}^{j})\:\:p'_{k+1}(x_{k+1}^{j})\:p''_{k+1}(b_{i})\:\:\pi_{p_{1},…,p_{k}}(x_{1}^{j}\ot…\ot
                x_{k}^{j})
                \\=&\:c\:\sum_{i}p'_{0}(a_{i})\:p''_{k+1}(b_{i})\:\: \sum_{j}
                p''_{0}(x_{0}^{j})\:\pi_{p_{1},…,p_{k}}\big(x_{1}^{j}\ot…\ot
                x_{k}^{j}\big)\:p'_{k+1}(x_{k+1}^{j}). 
            \end{align*} Dies zeigt 
            \begin{equation*}
                \pi_{p_{0},…,p_{k+1}}(zx)\leq c\: \pi_{p'_{0},p''_{k+1}}(z)\:\pi_{p''_{0},…,p'_{k+1}}(x)
            \end{equation*} und somit die Stetigkeit der
            Modul-Multiplikation. Die Stetigkeit der Addition in den
            $\C^{c}_{k}$ ist klar, da diese vermöge $\pi_{k+2}$
            topologische Vektorräume sind. 
        \item
            Mit der Stetigkeit von $+$ sind wieder Summen stetiger Funktionen stetig und es reicht
            daher, die Stetigkeit der Abbildungen:
            \begin{equation*}
                x_{0}\ot…\ot
                x_{k+1}\longmapsto x_{0}\ot…\ot x_{i}x_{i+1}\ot…\ot
                x_{k+1}
            \end{equation*}
            nachzuweisen. Wir erhalten diese mit dem üblichen
            Infimums-Argument aus
            \begin{align*}
                p_{0}\ot…\ot p_{k}\:(x_{0}\ot…\ot x_{i}x_{i+1}&\ot…\ot
                x_{k+1})\\ &\leq c\: p_{0}(x_{0})…p'_{i}(x_{i})\:p''_{i}(x_{i+1})…p_{k}(x_{k+1}),
            \end{align*}oder unmittelbar mit
            Bemerkung \ref{bem:PiTopArBem}. Die Stetigkeit der $h^{c}_{k}$
            folgt auf die gleiche 
            Weise vermöge:
            \begin{align*}
                p\ot p_{0}\ot…\ot p_{k}\:(1\ot x_{0}\ot x_{k+1}) =
                p(1)\: p_{0}(x_{0})…p_{k+1}(x_{k+1}).
            \end{align*}
        \item 
            Mit der Stetigkeit der $d_{k}^{c}$ ist sofort
            einsichtig, dass in der Tat
            \begin{equation*}
                d_{k+1}^{c*}\colon
                \Hom^{cont}_{\mathcal{A}^{e}}(\C^{c}_{k},\mathcal{M})\longrightarrow \Hom^{\cont}_{\mathcal{A}^{e}}(\C^{c}_{k+1},\mathcal{M})
            \end{equation*} und somit $(\C^{*}_{c},d^{*}_{c})$ ein
            wohldefinierter Kokettenkomplex ist. 
            Es bleibt dann lediglich nachzuweisen, dass die Isomorphismen
            $\Xi^{k}$ aus
            Proposition \ref{prop:barauffuerunitalgebraIsomozuHochschildkohomo},
            in beide Richtungen stetige auf stetige Homomorphismen
            abbilden. In der Tat sind dann deren Einschränkungen
            $\Xi^{k}_{c}=\Xi^{k}|_{HC^{k}_{\cont}(\mathcal{A},\mathcal{M})}$
            ebenfalls Isomorphismen und mit
            \begin{equation*}
                \Xi^{k+1}_{c}d^{c*}_{k+1}=\delta^{k}_{c}\:\Xi^{k}_{c}
            \end{equation*}zudem  Kettenabbildungen. Dies zeigt, dass
            die $\wt{\Xi^{k}_{c}}$ Isomorphismen sind.

            Für $\Xi^{k}$ folgt die gewünschte Eigenschaft mit
            stetigem $\psi$ aus:
            \begin{align*}
                q\left(\Big(\Xi^{k}\psi\Big)(\omega_{k})\right)=&\:
                q\big(\psi(1\ot \omega_{k}\ot 1)\big)\leq c\:
                \pi_{p_{0},…,p_{k+1}}(1\ot\omega_{k}\ot 1)
                \\=&\:\underbrace{c\:
                  p_{0}(1)\:p_{k+1}(1)}_{\hat{c}}\:\pi_{p_{1},…,p_{k}}(\omega_{k}).
            \end{align*} 
            Umgekehrt erhalten wir für $\phi\in
            HC^{k}_{\cont}(\mathcal{A},\mathcal{M})$ mit Lemma \ref{lemma:AezuunittopRing}~\textit{ii.)}:
            \begin{align*}
                q(x_{0}\ot x_{k+1}*_{e}\phi(x_{1}\ot…\ot x_{k}))\leq&\: c\:
                \pi_{p_{0},p_{k+1}}(x_{0}\ot x_{k+1})\:q'(\phi(x_{1}\ot…\ot x_{k}))
                \\\leq& \:\hat{c}\:\prod_{i=0}^{k+1}p_{i}(x_{i}).
            \end{align*}
        \end{enumerate}
    \end{beweis}
\end{proposition}
%
\section{Die stetige Hochschild-Kohomologie der Algebra
  $\Ss^{\bullet}(\mathbb{V})$}
\label{sec:SvonV}
In diesem Abschnitt wollen wir die stetige Hochschild-Kohomologie der Algebra $\SsV$ berechnen. Hierf"ur ist es zun"achst notwendig, diese mit einer lokalkonvexen Topologie derart auszustatten, dass die Algebramultiplikation $\vee$ stetig ist.
Sei hierfür $(\mathbb{V},P)$ ein lokalkonvexer Vektorraum und bezeichne
$\tilde{P}$ das filtrierende System aller bezüglich $\T_{P}$
stetigen Halbnormen\footnote{vgl. Korollar \ref{kor:HNTop}~\textit{ii.)}}. F"ur jedes $p\in \tilde{P}$ und jede positive Konstante
$|c|>0$ ist dann insbesondere die Halbnorm $|c|\:p$ in $\tilde{P}$ enthalten.\\ 
Jedes $\Ss^{l}(\mathbb{V})$ sei nun $\pi_{l}$-topologisiert bez"uglich des Halbnormensystemes $\tilde{P}$. Dann ist es insbesondere ausreichend, das Teilsystem $\{p^{l}\}_{p\in
  \tilde{P}}=\{\pi_{p,…,p}\}_{p\in \hat{P}}\subseteq \tilde{P}$
zu betrachten. Denn mit der Filtrationseigenschaft existiert zu jedem Satz von Halbnormen $p_{1},\dots,p_{l}\in \tilde{P}$ ein $p\in \tilde{P}$ derart, dass $p \geq
p_{i}\:\forall\:1\leq i\leq k$ und folglich
\begin{equation*}
    p_{1}\ot…\ot p_{l}\leq \overbrace{p\ot…\ot p}^{l-mal}=p^{l}
\end{equation*}
gilt. Ebenso zeigt man die umgekehrte Absch"atzbarkeit, und da besagtes Teilsystem ebenfalls filtrierend ist, zeigt Korollar \ref{kor:HNTop}~\textit{iv.)}, dass beide
Halbnormensysteme die selbe Topologie auf $\Ss^{l}(\mathbb{V})$
definieren. Diese ist gerade die durch $(\Tt^{l}(\V),\pi_{l})$ auf
$\Ss^{l}(\V)$ induzierte Teilraumtopologie.\\\\
Um nun die direkte Summe, also $\SsV$ lokalkonvex zu topologisieren,
betrachten wir das auf $\Tt^{\bullet}(\V)$ und somit auch auf $\SsV$
definierte System $\Pp$, bestehend aus den Halbnormen
\begin{align*}
    \p\colon\sum_{l}\omega_{l}\longmapsto \sum_{l=0}^{\infty}p^{l}(\omega_{l})
\end{align*}mit $p^{0}=||_{\mathbb{K}}$. Dieses ist ebenfalls
filtrierend und insbesondere ist klar, dass dann die Teilraumtopologien auf den $\Ss^{l}(\mathbb{V})$
gerade mit den $\pi_{l}$-Topologien übereinstimmen.
Des Weiteren sei darauf hingewiesen, dass die Halbnormen
$\tilde{\p}=\displaystyle\sum_{l=0}^{\infty}p_{i}^{l}$
mit paarweise verschiedenen $p_{i}\in \tilde{P}$ in der Tat
eine andere Topologie definieren, da die Summe nicht endlich ist und
wir 
im Allgemeinen kein $p\in \tilde{P}$ derart finden, dass $p\geq
p_{i}\:\forall\:i\in \mathbb{N}$. Den Hauptgrund für unsere Wahl liefert das n"achste Lemma.
Essentiell an besagtem Halbnormensystem ist zudem, dass mit $\p \in \Pp$, per
Konstruktion, ebenfalls die Halbnorm
\begin{align*}
    \p_{c}(v)=\sum_{l=0}^{\infty}|c|^{l}p^{l}
\end{align*} in $\Pp$ enthalten ist. Dies wird für spätere Stetigkeitsabschätzungen von hohem Nutzen sein. Das folgende Lemma macht $(\SsV,\vee)$ schlie"slich zu einer lokalkonvexen Algebra:
\begin{lemma}
    Vermöge $\vee$ wird $(\Ss^{\bullet}(\mathbb{V}),\Pp)$ zu einer
    assoziativen, unitären, lokalkonvexen Algebra mit
    submultiplikativem Halbnormensystem.
    \begin{beweis}
        Zun"achst erinnern wir, dass eine lokalkonvexe Algebra $(\mathcal{A},*,P)$ submultiplikativ gennant wird, falls $p(a*b)\leq p(a)\:p(b)$ f"ur alle $a,b\in \mathcal{A}$ und alle $p\in P$.
        Da dies insbesondere die Stetigkeit von $*$ impliziert, reicht es, diese Relation f"ur $(\SsV,\vee, \Pp)$ nachzuweisen.

        Zun"achst erhalten wir 
        \begin{equation}
            \label{eq:fopliu}
            \begin{split}
                \p\big(\ot^{\bullet}(\alpha,\beta)\big)=&\p\left(\sum_{l,m}\alpha_{l}\ot\beta_{m}\right)=\sum_{k}p^{k}\left(\sum_{l+m=k}\alpha_{l}\ot\beta_{m}\right)
                \\ \leq&\: \sum_{l,m}p^{l+m}\left(\alpha_{l}\ot\beta_{m}\right)
                \leq \sum_{l,m,i_{l},j_{m}}p^{l+m}\left(\alpha^{i_{l}}_{l}\ot\beta^{j_{m}}_{m}\right)
                \\=&\: \sum_{l,m,i_{l},j_{m}}p^{l}(\alpha^{i_{l}}_{l})\:p^{m}(\beta^{j_{m}}_{m})
                =
                \left(\sum_{l,i_{l}}p^{l}\big(\alpha_{l}^{i_{l}}\big)\right)\left(\sum_{m,j_{m}}p^{m}\left(\beta_{m}^{j_{m}}\right)\right)
            \end{split}
        \end{equation} für alle Zerlegungen
        $\Tt^{l}(\V)\ni\alpha_{l}=\displaystyle\sum_{i_{l}}\alpha^{i_{l}}_{l}$
        in separable $\alpha_{l}^{i_{l}}$ und 
        $\Tt^{m}(\V)\ni\beta_{m}=\displaystyle\sum_{j_{m}}\beta^{j_{m}}_{m}$
        in separable $\beta_{m}^{j_{m}}$. Dies zeigt
        \begin{align*}
            \p\left(\ot^{\bullet}(\alpha,\beta)\right)\leq&
            \:\Bigg[\sum_{l}\inf\bigg(\sum_{i}p^{l}\big(\alpha^{i}_{l}\big)\bigg)\Bigg]\:\Bigg[\sum_{m}\inf\bigg(\sum_{j}p^{m}\big(\beta^{j}_{m}\big)\bigg)\Bigg]
            \\=&\: \p\Bigg(\sum_{l}\alpha_{l}\Bigg)\:\p\Bigg(\sum_{m}\beta_{m}\Bigg),
        \end{align*}wobei wieder das Infimum über alle Zerlegungen von
        $\alpha_{l}$ und $\beta_{m}$ gemeint ist.\\\\
        F"ur $S\colon\Tt^{\bullet}(\mathbb{V})\longrightarrow
        \Ss^{\bullet}(\mathbb{V})\subseteq \Tt^{\bullet}(\mathbb{V})$ folgt
        \begin{equation*}
            p^{l}(\mathrm{Sym}_{l}(\alpha_{l}))=p^{l}\Bigg(\frac{1}{l!}\sum_{\sigma\in S_{l}}\sigma^{*}\alpha_{l}\Bigg)\leq
            \frac{1}{l!}\sum_{\sigma\in S_{l}}\:p^{l}(\alpha_{l})=p^{l}(\alpha_{l}),
        \end{equation*}
        womit 
        \begin{align*}
            \p\left(S(\alpha)\right)=\p\left(\sum_{l}\mathrm{Sym}_{l}(\alpha_{l})\right)=\sum_{l}p^{l}(\mathrm{Sym}_{l}(\alpha_{l}))\leq \sum_{l}p^{l}(\alpha_{l})=\p(\alpha).
        \end{align*}
        Mit $\vee=S\cp \ot^{\bullet}$ (vgl. Abschnitt
        \ref{subsec:HochschKohSym}) zeigt dies
        \begin{align*}
            \p(\alpha\vee
            \beta)=\p((S\cp\ot^{\bullet})(\alpha,\beta))\leq
            \p(\ot^{\bullet}(\alpha,\beta))\leq \p(\alpha)\p(\beta)
        \end{align*}f"ur alle $\alpha,\beta \in \Tt^{\bullet}(\V)$ und somit die Behauptung, da obige Ungleichung dann insbesondere f"ur alle $\alpha,\beta \in\SsV\subseteq \Tt^{\bullet}(\V)$ korrekt ist.
    \end{beweis}
\end{lemma}
Als Resultat dieses Lemmas erhalten wir mit  Proposition \ref{prop:topBarKomplexprop}~\textit{iii.}), dass
\begin{equation*}
    HH^{k}_{\cont}(\Ss^{\bullet}(\mathbb{V}),\mathcal{M})\cong H^{k}\left(\Hom_{\mathcal{A}^{e}}^{\cont}(\C_{c},\mathcal{M}),d^{*}_{c}\right)
\end{equation*} für jeden lokalkonvexen $\SsV-\SsV$-
Bimodul $\mathcal{M}$. Hierbei bezeichnet $\C_{c}$ den zu $\Ss^{\bullet}(\mathbb{V})$
gehörigen, topologischen Bar-Komplex mit
$\mathcal{A}^{e}=\Ss^{\bullet}(\mathbb{V})\pite \Ss^{\bullet}(\mathbb{V})$.\\\\
Wir wollen nun den in Abschnitt \ref{subsec:HochschKohSym} betrachteten
Koszul-Komplex in geeigneter Weise derart topologisieren, dass die
Kettendifferentiale $\partial_{k}$ stetige Abbildungen sind und wir somit in
wohlbegründeter Weise vom topologischen Koszul-Komplex
$(\K_{c},\pt_{c})$ und folglich auch vom stetigen Kokettenkomplex
$(\Hom_{\mathcal{A}^{e}}^{\cont}(\K_{c},\mathcal{M}),\pt_{c}^{*})$
sprechen dürfen. Des Weiteren werden wir nachweisen, dass dann unsere
Kettenabbildungen $F$ und $G$ in den gegebenen Topologien ebenfalls
stetig sind und somit Kettenabbildungen zwischen
$(\C_{c}^{*},d_{c}^{*})$ und $(\K_{c}^{*},\pt_{c}^{*})$ induzieren. 
\begin{definition}[Topologischer Koszul-Komplex]
    Den Koszul-Komplex aus \ref{subsec:HochschKohSym} vor Augen,
    definieren wir die topologischen Räume
    \begin{equation*}
        \K_{k}^{c}=\Ss^{\bullet}(\mathbb{V})\pite
        \Ss^{\bullet}(\mathbb{V})\pite \Lambda^{k}(\mathbb{V})
    \end{equation*} und erhalten ein erzeugendes 
    System $\Pp_{k}$, vermöge den Halbnormen: 
    \begin{equation*}
        \p_{k}=\p\ot\p\ot p^{k}=\p^{2}\ot p^{k}.
    \end{equation*} 
    Mit $\pt^{c}_{k}$ bezeichnen wir die durch
    Definition \ref{def:partialdef} auf den $\K_{k}^{c}$ induzierten
    Homomorphismen, von denen wir im Folgenden nachweisen werden, dass
    sie stetig sind.
    $\Tt^{c}_{k}\supseteq \K_{k}^{c}$ sei der mit selbigem
    Halbnormensystem ausgestatteten Raum
    \begin{equation*}
        \Tt^{\bullet}(\mathbb{V})\pite
        \Tt^{\bullet}(\mathbb{V})\pite \Tt^{k}(\mathbb{V}),
    \end{equation*} 
    der obige Topologie als
    Teilraumtopologie auf $\K_{k}^{c}$ induziert. 
\end{definition}

\begin{bemerkung}
    \label{bem:Teilraumargument}
    Will man die Stetigkeit einer Abbildung $\phi\colon\K^{c}_{k}\longrightarrow
    \K^{c}_{k'}$ nachweisen, so reicht es, diese für eine Abbildung
    $\tilde{\phi}\colon\Tt^{c}_{k}\longrightarrow \Tt^{c}_{k'}$ zu zeigen, 
    die $\phi$ auf $\K^{c}_{k}$ induziert, f"ur die also
    $\tilde{\phi}\big|_{\K^{c}_{k}}=\phi$ gilt.
    Dies sieht man sofort
    daran, dass die Stetigkeitsabschätzungen für $\tilde{\phi}$
    insbesondere f"ur die besagten Unterräume g"ultig sind. 
\end{bemerkung}
\begin{proposition}
    \label{prop:wichpropKoszStetSym}
    \begin{enumerate}
    \item
        Eine lineare Abbildung $\phi\colon\Tt^{c}_{k}\rightarrow
        \Tt^{c}_{k'}$ ist genau dann stetig, wenn für jedes
        $\q_{k'}\in \Pp_{k'}$ ein $\p_{k}\in \Pp_{k}$ derart
        existiert, dass
        \begin{equation}
            \label{eq:stetrelSummenKos}
            \q_{k'}(\phi(\alpha_{l}\ot \beta_{m}\ot u))\leq c\: \p_{k}(\alpha_{l}\ot \beta_{m}\ot u)
        \end{equation}für alle separablen $\alpha_{l}\in \Tt^{l}(\mathbb{V})$, $\beta_{m}\in
        \Tt^{m}(\mathbb{V})$ und $u\in \Tt^{k}(\mathbb{V})$ gilt.
    \item
        \label{item:partkStetig}
        Es sind alle $\pt^{c}_{k}$ stetig.
    \item
        Es sind alle $h_{k}$ stetig.
    \item
        Es sind alle $F_{k}$ stetig.
    \item
        Es sind alle $G_{k}$ stetig.
    \end{enumerate}
    \begin{beweis}  
        \begin{enumerate}
        \item
            Ist $\phi$ stetig, so gilt besagte Relation sogar f"ur alle Elemente in $\Tt_{c}^{k}$.\\ 
            Für die umgekehrte Richtung sei $\alpha\ot \beta\ot u \in
            \Tt^{c}_{k}$. Dann folgt
            \begin{align*}
                \q_{k'}(\phi(\alpha\ot\beta\ot
                u))=&\:\q_{k'}\left(\sum_{l,m,i_{l},j_{m},s}\phi\left(\alpha^{i_{l}}_{l}\ot\beta^{j_{m}}_{m}\ot
                        u^{s}\right)\right)
                \\\leq&\sum_{l,m,i_{l},j_{m},s}\q_{k'}\left(\phi\left(\alpha_{l}^{i_{l}}\ot\beta_{m}^{j_{m}}\ot
                        u^{s}\right)\right)
                \\\leq&\sum_{l,m,i_{l},j_{m},s}c\:\p_{k}\left(\alpha_{l}^{i_{l}}\ot\beta_{m}^{j_{m}}\ot
                    u^{s}\right)
                \\=&\:c\sum_{l,m,i_{l},j_{m},s}p^{l}\big(\alpha_{l}^{i_{l}}\big)\:p^{m}\left(\beta_{m}^{j_{m}}\right)\:p^{k}(u^{s})
            \end{align*} für alle Zerlegungen von
            $\alpha_{l}$, $\beta_{m}$ und $u$ in separable Summanden
            $\alpha_{l}^{i_{l}}$, $\beta_{m}^{j_{m}}$, $u^{s}$ Dies
            zeigt:
            {\begin{align*}
                  \label{eq:stetmuhmuh}
                  \q_{k'}(\phi(\alpha\ot&\: \beta\ot u))
                  \\\leq&\:c\sum_{l}\inf\left(\sum_{i}
                      p^{l}\big(\alpha_{l}^{i}\big)\right)\sum_{m}\inf\left(\sum_{i}p^{m}\left(\beta_{m}^{i}\right)\right)\inf\left(\sum_{s}\left(p^{k}(u^{s})\right)\right)\\=&\:c\:
                  \p(\alpha)\p(\beta)\:p^{k}(u),
              \end{align*}}also die Stetigkeit von $\phi\cp
            \ot_{3}$ und somit die von $\phi$ in $\pi_{3}$.
        \item 
            In Abschnitt \ref{subsec:HochschKohSym} hatten wir eingesehen, dass
            \begin{align*}
                (S\ot S\ot A)\cp
                \tilde{\pt}^{k}_{1}\Big|_{\K^{c}_{k}}&=\pt^{k}_{1}\\
                (S\ot S\ot A)\cp
                \tilde{\pt}^{k}_{2}\Big|_{\K^{c}_{k}}&=\pt^{k}_{2}.
            \end{align*} Hierbei haben wir die zus"atzlichen Symmetrisierungen und
            Antisymmetrisierungen aus reiner Bequemlichkeit
            eingef"ugt, was wegen $S|_{\SsV}=\id_{\SsV}$ und
            $A|_{\Lambda^{\bullet}(V)}=\id_{\Lambda^{\bullet}(V)}$
            ohne weiteres möglich ist.
            
            Nun ist $S\ot S\ot A$ stetig in $\Tt^{c}_{k}$
            nach \textit{i.)}, denn für
            $\alpha_{l}\ot\beta_{m}\ot u\in \Tt_{k}^{c}$
            mit
            separablen Faktoren folgt:
            \begin{align*}
                \p_{k}((S\ot S\ot A)(\alpha_{l}\ot \beta_{m} \ot
                u))=&\:\p_{k}\left(S(\alpha_{l})\ot
                    S(\beta_{m})\ot A(u)\right)\\
                =&\:p^{l}(S(\alpha_{l}))\:p^{m}(S(\beta_{m}))\:p^{k}(A(u))
                \\\leq&\:p^{l}(\alpha_{l})\:p^{m}(\beta_{m})\:p^{k}(u)
                \\=&\: \p_{k}(\alpha_{l}\ot \beta_{m} \ot u).
            \end{align*} 
            Es bleiben die Stetigkeiten von
            $\tilde{\pt}^{k}_{1}$ und $\tilde{\pt}^{k}_{2}$ zu zeigen. Diese
            folgen mit \textit{i.)} und
            \begin{align*}
                \p_{k-1}\left(\tilde{\pt}^{k}_{1}\left(\alpha_{l}\bbot\beta_{m}\bbot
                        u\right)\right)
                =&\:k\p_{k-1}\left(u_{1}\ot\alpha_{l}\bbot\beta_{m}\bbot
                    u^{1}\right)
                \\=&\:k\:p^{l+1}(u_{1}\ot
                \alpha_{l})\:p^{m}(\beta_{m})\:p^{k-1}(u^{1})
                \\=&\:k\:p^{l}(\alpha_{l})\:p^{m}(\beta_{m})\:p^{k}(u)
                \\=&\:k\p_{k}(\alpha_{l}\bbot\beta_{m}\bbot u)
            \end{align*}
            sowie einer analogen Rechnung für $\tilde{\pt}^{k}_{2}$.
            Bemerkung \ref{bem:Teilraumargument} zeigt dann die Stetigkeit von
            $\pt^{k}_{1}$ und $\pt^{k}_{2}$ und folglich die von $\pt^{c}_{k}$.
        \item
            Wir erinnern, dass $h_{k}=\int_{0}^{1}dt\:t^{k}\:
            i_{t}\cp\delta$, und zeigen zunächst die Stetigkeit von
            $\delta$. Nun war $(S\ot S\ot A) \cp
            \tilde{\delta}\big|_{\K^{c}_{k}}=\delta$ mit
            \begin{align*}
                \tilde{\delta}\colon\Tt^{\bullet}(\mathbb{V})\ot \Tt^{\bullet}(\mathbb{V})\ot
                \Tt^{k}(\mathbb{V})&\longrightarrow
                \Tt^{\bullet-1}(\mathbb{V})\ot \Tt^{\bullet}(\mathbb{V})\ot
                \Tt^{k+1}(\mathbb{V})\\
                \alpha_{l}\bbot\beta\bbot u&\longmapsto l\left(\alpha_{l}^{1}\ot \beta\bbot\: (\alpha_{l})_{1}\bbot u\right),
            \end{align*}und für separable $\alpha_{l}\in \Tt^{l}(\mathbb{V})$, $\beta_{m}\in
            \Tt^{m}(\mathbb{V})$ sowie $u\in \Tt^{k}(\mathbb{V})$ folgt:
            \begin{align*}
                \p_{k+1}\left(\:\tilde{\delta}(\alpha_{l}\ot\beta_{m}\ot
                    u)\right)=&\p^{2}\left(l\:\alpha^{1}_{l}\ot
                    \beta_{m}\right)p^{k+1}((\alpha_{l})_{1}\ot u)
                \\=&\:l\:p^{l}(\alpha_{l})\:p^{m}(\beta_{m})\:p^{k}(u)
                \\\leq&\:
                2^{l}2^{m}2^{k}\:p^{l}(\alpha_{l})\:p^{m}(\beta_{m})\:p^{k}(u)
                \\=&\:\hat{p}^{l}(\alpha_{l})\:\hat{p}^{m}(\beta_{m})\:\hat{p}^{k}(u)
                \\=&\:\hat{\p}_{k}(\alpha_{l}\ot\beta_{m}\ot u)
            \end{align*} 
            mit $\hat{p}=2p$.
            Bemerkung \ref{bem:Teilraumargument} sowie \textit{i.)} zeigen die Stetigkeit von
            $\delta$.\\\\
            Um die Stetigkeit von $\int_{0}^{1}dt\:t^{k}i_{t}$
            nachzuweisen, erinnern wir daran, dass
            \begin{equation*}
                \int_{0}^{1}dt\:t^{k}\:i_{t}=\left.\left[(S\ot S\ot A)\cp \int_{0}^{1}dt\:t^{k}\underbrace{\left[\sum_{l=0}^{\infty}\sum_{s=0}^{l}\eta_{t}^{l,s}\right]}_{\hat{i}_{t}}\right]\right|_{\K^{c}_{k}}
            \end{equation*}
            mit
            \begin{align*}
                \eta^{l,s}_{t}\colon\Tt^{\bullet}(\mathbb{V})\ot
                \Tt^{\bullet}(\mathbb{V})\ot
                \Tt^{k}(\mathbb{V})&\longrightarrow \Tt^{\bullet}(\mathbb{V})\ot
                \Tt^{\bullet}(\mathbb{V})\ot
                \Tt^{k}(\mathbb{V})\\
                \alpha_{l}\ot \beta\ot u&\longmapsto
                \binom{l}{s}\:t^{l-s}(1-t)^{s}\alpha_{l}^{1,…,s}\ot\:
                (\alpha_{l})_{1,…,s}\ot \beta\ot u.
            \end{align*}
            Wir rechnen für separable $\alpha_{l},\:\beta_{m}$ und
            $u$:
            {\allowdisplaybreaks
              \begin{align*}
                  \p_{k}\Bigg(\int_{0}^{1}dt\:t^{k}\:&\hat{i}_{t}\left(\alpha_{l}\ot\beta_{m} \ot
                      u\right)\Bigg)=\:\p_{k}\left(\int_{0}^{1}dt\:t^{k}\sum_{l'=0}^{\infty}\sum_{s=0}^{l'}\eta_{t}^{l',s}(\alpha_{l}\ot
                      \beta_{m}\ot
                      u)\right)
                  \\=&\:\p_{k}\left(\int_{0}^{1}dt\:t^{k}\sum_{s=0}^{l}\eta_{t}^{l,s}(\alpha_{l}\ot\beta_{m}\ot
                      u)\right)
                  \\=&\:\p_{k}\left(\sum_{s=0}^{l}\underbrace{\int_{0}^{1}dt\:t^{k}\:t^{l-s}(1-t)^{s}\binom{l}{s}}_{\tau^{l}_{s}}
                      \left(\alpha_{l}^{1,…,s}\ot\: (\alpha_{l})_{1,…,s}\ot\beta_{m}\ot
                          u\right)\right)
                  \\=&\:\p^{2}\left(\sum_{s=0}^{l}\tau^{l}_{s}\:\alpha_{l}^{1,…,s}\ot\:
                      (\alpha_{l})_{1,…,s}\ot\beta_{m}\right)p^{k}(u)
                  \\\leq&\:\sum_{s=0}^{l}\tau^{l}_{s}\:\p^{2}\left(\alpha_{l}^{1,…,s}\ot\:
                      (\alpha_{l})_{1,…,s}\ot\beta_{m}\right)p^{k}(u)
                  \\=&\:\sum_{s=0}^{l}\tau^{l}_{s}\:\:p^{l-s}\left(\alpha_{l}^{1,…,s}\right)\:p^{s}\Big((\alpha_{l})_{1,…,s}\ot\beta_{m}\Big)\:p^{k}(u)
                  \\=&\:\left[\sum_{s=0}^{l}\tau^{l}_{s}\right]\:p^{l}(\alpha_{l})\:p^{m}(\beta_{m})\:p^{k}(u)
                  \\=&\:\frac{1}{k+1}\p_{k}(\alpha_{l}\ot \beta_{m}\ot u).
              \end{align*}}In der Tat erhalten wir die letzte Gleichheit mit
            {\allowdisplaybreaks
              \begin{align*}
                  \sum_{s=0}^{l}\tau^{l}_{s}=&
                  \int_{0}^{1}dt\:t^{k}\sum_{s=0}^{l}\binom{l}{s}t^{l-s}(1-t)^{s}=\int_{0}^{1}dt\:t^{k}[t+(1-t)]^{l}=\int_{0}^{1}dt\:t^{k}=\frac{1}{k+1}.
              \end{align*}}Die Stetigkeit von $\int_{0}^{1}dt\:t^{k}\:
            i_{t}$ folgt abermals mit \textit{i.)} und
            Bemerkung \ref{bem:Teilraumargument} liefert schließlich die Behauptung.
        \item
            Es ist $F_{k}=\tilde{F}_{k}\big|_{\K^{c}_{k}}$ f"ur
            \begin{align*}
                \tilde{F}_{k}\colon\Tt^{c}_{k}&\longrightarrow \bigotimes^{k+2}\Tt^{\bullet}(\mathbb{V})\\
                \alpha\bbot \beta \bbot u_{1}\ot…\ot u_{k}&\longmapsto k!\left(\alpha\ot
                    u_{1}\ot…\ot u_{k} \ot \beta\right)
            \end{align*} und
            \allowdisplaybreaks{
              \begin{align*}
                  \p^{k+2}\left(\tilde{F}_{k}(\alpha \ot \beta\ot
                      u)\right)=&\:k!\:\p^{k+2}\left(\sum_{l,m,i_{l},j_{m},i}\alpha^{i_{l}}_{l}\ot u^{i}_{1}\ot…\ot
                      u^{i}_{k}\ot\beta^{j_{m}}_{m}\right)
                  \\\leq&\:k! \sum_{l,m,i_{l},j_{m},i}\p^{k+2}\left(\alpha^{i_{l}}_{l}\ot u^{i}_{1}\ot…\ot
                      u^{i}_{k}\ot\beta^{j_{m}}_{m}\right)
                  \\=&\:k!\sum_{l,m,i_{l},j_{m},i} p^{l}\left(\alpha^{i_{l}}_{l}\right)\:p\Big(u^{i}_{1}\Big)…p\Big(u^{i}_{k}\Big)\:p^{m}\bigg(\beta^{j_{m}}_{m}\Big)
                  \\=&\:k! \left(\sum_{l,i_{l}}p^{l}\left(\alpha^{i_{l}}_{l}\right)\right)\left(\sum_{m,j_{m}}p^{m}\Big(\beta^{j_{m}}_{m}\Big)\right)\:\left(\sum_{i}p^{k}\Big(u^{i}\Big)\right)
              \end{align*}}f"ur alle Zerlegungen der $\alpha_{l}$, $\beta_{m}$ und von $u$. Hiermit folgt
            \begin{align*}
                \p^{k+2}\Big(\tilde{F}_{k}(\alpha \ot \beta&\:\ot
                u)\Big)
                \\\leq&
                \:k!\:\sum_{l}\inf\left(\sum_{i}p^{l}\left(\alpha_{l}^{i}\right)\right)\:\sum_{m}\inf\left(\sum_{i}p^{l}\left(\beta_{m}^{i}\right)\right)\:\inf\left(\sum_{i}p^{k}\left(u^{i}\right)\right)
                \\=&\:k!\:\p(\alpha)\p(\beta)\:p^{k}(u),
            \end{align*}
            also die Stetigkeit von $\tilde{F}_{k}$ in $\pi_{3}$. Die von
            $F_{k}$ folgt analog zu Bemerkung \ref{bem:Teilraumargument}, da obige Ungleichung insbesondere f"ur alle Elemente aus $\K_{k}^{c}$ korrekt ist. 
        \item
            Wir zeigen dies wieder schrittweise. Sei hierfür
            $\bigotimes^{k+2}\Tt^{\bullet}(\mathbb{V})\ot
            \Tt^{k}(\mathbb{V})$ topologisiert vermöge den Halbnormen
            $\p^{k+2}\ot p^{k}$ und
            $\bigotimes^{k+2}\Tt^{\bullet}(\mathbb{V})$ vermöge
            $\p^{k+2}$. Wir definieren sinngemäß zu Definition \ref{def:GAbb}~\textit{iv)}:
            \begin{align*}
                \tilde{\delta}_{k}\colon\bigotimes^{k+2}\Tt^{\bullet}(\V)&\longrightarrow
                \bigotimes^{k+2}\Tt^{\bullet}(\V)\bbot\Tt^{k}(\mathbb{V})\\
                \alpha\ot u_{1}\ot…\ot u_{k}\ot\beta &\longmapsto \left[\prod_{i=1}^{k}n_{i}\right]\:
                \alpha\ot u_{1}^{1}\ot…\ot u_{k}^{1}\ot\beta\bbot \:(u_{1})_{1}\ot…\ot (u_{k})_{1}
            \end{align*}für $\deg(u_{i})=n_{i}$ und $\tilde{\delta}_{k}(\alpha \ot u_{1}\ot\dots \ot u_{j-1}\ot
            1 \ot u_{j+1} \ot …u_{k}\ot \beta)=0$, womit
            \begin{equation*}
                \delta=\left(\bigotimes^{k+2} \id\ot A\cp
                    \tilde{\delta}_{k}\right)\Bigg|_{\C^{c}_{k}}.
            \end{equation*}
            Für die Stetigkeit von
            $\tilde{\delta}_{k}$ sei $\alpha\ot u_{1}\ot…\ot
            u_{k}\ot\beta\in
            \Tt^{\bullet}(\mathbb{V})\ot\displaystyle\bigotimes^{k}_{n=1}\Tt^{n_{k}}(\mathbb{V})\ot
            \Tt^{\bullet}(\mathbb{V})$ mit
            $u_{i}$ für $1\leq i \leq k$ separabel. Dann folgt
            {\allowdisplaybreaks
              \begin{align*}
                  \big(\p^{k+2}\ot p^{k}\big)&\left(\tilde{\delta}(\alpha\ot u_{1}\ot…\ot
                      u_{k}\ot \beta)\right)
                  \\ &=\left(\p^{k+2}\ot
                      p^{k}\right)\left(\left[\prod_{i=1}^{k}n_{i}\right]\alpha\ot\:u_{1}^{1}\ot…\ot\:
                      u_{k}^{1}\ot \beta\:\bbot\:
                      (u_{1})_{1}\ot…\ot\:(u_{k})_{1}\right)
                  \\ &= \left[\prod_{i=1}^{k}n_{i}\right]\p(\alpha)\p\left(u_{1}^{1}\right)…\p\left(u_{k}^{1}\right)\p(\beta)\:p^{k}((u_{1})_{1}\ot…\ot\:(u_{k})_{1})
                  \\ &=\left[\prod_{i=1}^{k}n_{i}\right]\p(\alpha)\:p^{n_{1}-1}\left(u_{1}^{1}\right)…p^{n_{k}-1}\left(u_{k}^{1}\right)\p(\beta)\:p^{k}((u_{1})_{1}\ot…\ot\:(u_{k})_{1})
                  \\ &=\left[\prod_{i=1}^{k}n_{i}\right]\p(\alpha)\:p^{n_{1}}\left(u_{1}\right)…p^{n_{k}}\left(u_{k}\right)\p(\beta)
                  \\ &\leq\tilde{\p}(\alpha)\:\tilde{p}^{n_{1}}\left(u_{1}\right)…\tilde{p}^{n_{k}}\left(u_{k}\right)\tilde{\p}(\beta)
                  \\ &= \tilde{\p}^{k+2}(\alpha\ot u_{1}\ot…\ot
                  u_{k}\ot \beta)
              \end{align*}}mit $\tilde{p}=2p$.
            Den allgemeinen Fall
            erhalten wir vermöge obiger Ungleichung mit
            \begin{align*}
                \big(\p^{k+2}\ot p^{k}\big)\Big(\tilde{\delta}(\alpha\ot&\: u_{1}\ot…\ot
                u_{k}\ot \beta)\Big)\\=&\:\big(\p^{k+2}\ot
                p^{k}\big)\left(\sum_{\substack{n_{1},…,n_{k}\\i_{n_{1}},…,i_{n_{k}}}}\tilde{\delta}\left(\alpha\ot\:
                        (u_{1})^{i_{n_{1}}}_{n_{1}}\ot…\ot\:(u_{k})^{i_{n_{k}}}_{n_{k}}\ot
                        \beta\right)\right)
                \\\leq&\sum_{\substack{n_{1},…,n_{k}\\i_{n_{1}},…,i_{n_{k}}}}\big(\p^{k+2}\ot p^{k}\big)\left(\tilde{\delta}\left(\alpha\ot\:
                        (u_{1})^{i_{n_{1}}}_{n_{1}}\ot…\ot\:(u_{k})^{i_{n_{k}}}_{n_{k}}\ot
                        \beta\right)\right)
                \\\leq&\sum_{\substack{n_{1},…,n_{k}\\i_{n_{1}},…,i_{n_{k}}}}\tilde{\p}^{k+2}\left(\alpha\ot\:
                    (u_{1})^{i_{n_{1}}}_{n_{1}}\ot…\ot\:(u_{k})^{i_{n_{k}}}_{n_{k}}\ot
                    \beta\right)
                \\=&\:
                \tilde{\p}(\alpha)\sum_{\substack{n_{1},…,n_{k}\\i_{n_{1}},…,i_{n_{k}}}}\tilde{p}^{n_{1}}\left((u_{1})_{n_{1}}^{i_{n_{1}}}\right)…\tilde{p}^{n_{k}}\left((u_{k})_{n_{k}}^{i_{n_{k}}}\right)\tilde{\p}(\beta)
                \\=&\:\tilde{\p}(\alpha)\left(\sum_{n_{1}}\sum_{i_{n_{1}}}\tilde{p}^{n_{1}}\left((u_{1})^{i_{n_{1}}}_{n_{1}}\right)\right)…\left(\sum_{n_{k}}\sum_{i_{n_{k}}}\tilde{p}^{n_{k}}\left((u_{1})^{i_{n_{k}}}_{n_{k}}\right)\right)\tilde{\p}(\beta)
            \end{align*}
            für alle Zerlegungen der $(u_{i})_{n_{i}}\in \Tt^{n_{i}}(\V)$, und es
            folgt
            \begin{equation*}
                \big(\p^{k+2}\ot p^{k}\big)\left(\tilde{\delta}(\alpha\ot u_{1}\ot…\ot
                    u_{k}\ot \beta)\right)\leq \tilde{\p}(\alpha)\:\tilde{\p}(u_{1})…\tilde{\p}(u_{k})\:\tilde{\p}(\beta).
            \end{equation*}Dies zeigt die Stetigkeit von
            $\tilde{\delta}$ in $\pi_{k+2}$, und mit dem üblichen
            Teilraumargument ebenso die von $\delta$.\\\\
            Um die Stetigkeit von
            $\int_{0}^{1}dt_{1}…\int_{0}^{t_{k-1}}dt_{k}\:i_{t_{1},…,t_{k}}$
            nachzuweisen, definieren wir
            \begin{align*}
                \eta^{m,l}\colon\Tt^{\bullet}(\mathbb{V})\ot \Tt^{m}(\mathbb{V})\ot \Tt^{\bullet}(\mathbb{V})&\longrightarrow
                \bigotimes^{2}\Tt^{\bullet}(\mathbb{V})\\
                \alpha\ot u \ot \beta&\longmapsto u^{1,…,l}\ot \alpha \ot u_{1,…,l}\ot\beta
            \end{align*} mit $l\leq m$ sowie $\eta^{0,0}(\alpha\ot
            1\ot \beta)=(\alpha\ot \beta)$. Dann folgt für
            \begin{align*}
                \hat{i}'_{t}=\sum_{m=0}^{\infty}\sum_{l=0}^{m}\binom{m}{l}t^{m-l}(1-t)^{l}\eta^{m,l}
            \end{align*}
            sowie
            \begin{align*}
                \tilde{i}_{t_{1},…,i_{t_{k}}}(\alpha\ot u_{1}\ot…\ot
                u_{k}\ot\beta\ot \omega)=\alpha\ot\beta \:*_{e}\:\left[
                    \prod_{s=1}^{k}\hat{i}'_{t_{s}}(1\ot u_{s}\ot 1)\right]\ot \omega
            \end{align*} mit $\prod$ das Produkt
            $(\alpha\ot u\ot \beta)\cdot (\alpha'\ot u'\ot \beta')=
            \alpha\ot \alpha'\bbot u\ot u'\bbot \beta\ot \beta'$
            und

            $\alpha \ot \beta *_{e} \hat{\alpha}\bbot \hat{\beta}\bbot\omega=\alpha\ot\hat{\alpha}\bbot \hat{\beta}\ot\beta\bbot\omega$, dass
            \begin{align*}
                i_{t_{1},…,t_{k}}=\Big[(S\ot S\ot \id)\cp
                \tilde{i}_{t_{1},…,t_{k}}\Big]\Big|_{\bigotimes^{k+2}\Ss^{\bullet}(\mathbb{V})\ot
                  \Lambda^{k}(\mathbb{V})}.
            \end{align*}
            Sei nun $\alpha_{p}\ot u_{1}\ot…\ot u_{k}\ot
            \beta_{q}\ot \omega\in
            \displaystyle\bigotimes^{k+2}\Tt^{\bullet}(\mathbb{V})\ot
            \Tt^{k}(\mathbb{V})$ mit $\deg(u_{i})=p_{i}$,\\
            $\deg(\alpha_{p})=p$ und $\deg(\beta_{q})=q$.
            Für
            $\alpha_{p},\:\beta_{q},\:u_{i}$ separabel und $\omega\in
            \Tt^{k}(\V)$ beliebig erhalten wir:
            {\footnotesize
              \begin{align*}
                  \p_{k}\bigg(&\int_{0}^{1}dt_{1}…\int_{0}^{t_{k-1}}dt_{k}\:\tilde{i}_{t_{1},…,t_{k}}(\alpha_{p}\ot u_{1}\ot…\ot u_{k}\ot
                  \beta_{q}\ot \omega)\bigg)
                  \\=&\p^{2}\left(\int_{0}^{1}dt_{1}…\int_{0}^{t_{k-1}}dt_{k}\:\alpha_{p}\ot\beta_{q}*_{e}\prod_{i=1}^{k}\hat{i}'_{t_{i}}(1\ot
                      u_{i}\ot1)\right)p^{k}(\omega)
                  \\=&\:p^{k}(\omega)\p^{2}\left(\alpha_{p}\ot\beta_{q}*_{e}\left[\sum_{\substack{m_{i}=0\\1\leq
                              i\leq
                              k}}^{\infty}\:\sum_{\substack{l_{i}=0\\1\leq i\leq
                              k}}^{m_{i}}\underbrace{\int_{0}^{1}dt_{1}…\int_{0}^{t_{k-1}}dt_{k}\prod_{i=1}^{k}\binom{m_{i}}{l_{i}}t_{i}^{m_{i}-l_{i}}(1-t_{i})^{l_{i}}
                          }_{\tau^{m_{1},…,m_{k}}_{l_{1},…,l_{k}}}\cdot\right.\right.
                  \\ &\qquad\qquad\qquad\qquad\qquad\qquad\qquad\qquad\qquad\qquad\qquad\quad \left.\left. \phantom{\underbrace{\int_{0}^{1}dt_{1}…\int_{0}^{t_{k-1}}dt_{k}}_{\tau^{m_{1},…,m_{k}}_{l_{1},…,l_{k}}}}
                          \eta^{m_{i},l_{i}}(1\ot u_{i}\ot 1)\right]\right)
                  \\=&\:p^{k}(\omega)\p^{2}\left(\alpha_{p}\ot\beta_{q}*_{e}\left[\sum_{\substack{l_{i}=0\\1\leq i\leq
                              k}}^{p_{i}} \tau^{p_{1},…,p_{k}}_{l_{1},…,l_{k}}\prod_{i=1}^{k}u_{i}^{1,…,l_{i}}\ot
                          \:(u_{i})_{1,\dots,l_{i}}\right]\right)
                  \\=&\:p^{k}(\omega)\p^{2}\left(\sum_{\substack{l_{i}=0\\1\leq i\leq
                          k}}^{p_{i}}
                      \tau^{p_{1},…,p_{k}}_{l_{1},…,l_{k}}\alpha_{p}\ot u_{1}^{1,…,l_{1}}\ot…\ot
                      u_{k}^{1,…,l_{k}} \bbot 
                      \beta_{q}\ot
                      \:(u_{1})_{1,…,l_{1}}\ot…\ot
                      \:(u_{k})_{1,…,l_{k}}\right)
                  \\\leq&\:p^{k}(\omega)\sum_{\substack{l_{i}=0\\1\leq i\leq
                      k}}^{p_{i}}
                  \tau^{p_{1},…,p_{k}}_{l_{1},…,l_{k}}\p^{2}\left(\alpha_{p}\ot u_{1}^{1,…,l_{1}}\ot…\ot
                      u_{k}^{1,…,l_{k}} \bbot 
                      \beta_{q}\ot
                      \:(u_{1})_{1,…,l_{1}}\ot…\ot
                      \:(u_{k})_{1,…,l_{k}}\right)
                  \\=&\:p^{k}(\omega)\sum_{\substack{l_{i}=0\\1\leq i\leq
                      k}}^{p_{i}}
                  \tau^{p_{1},…,p_{k}}_{l_{1},…,l_{k}}p^{p}\Big(\alpha_{p}\Big)\:p^{p_{1}-l_{1}}\left(u_{1}^{1,…,l_{1}}\right)…p^{p_{k}-l_{k}}\left(u_{k}^{1,…,l_{k}}\right)\cdot \\
                  &\qquad\qquad\qquad\qquad\qquad\qquad\qquad\qquad\qquad\qquad\qquad\:
                  p^{q}\Big(\beta_{q}\Big)\:p^{l_{1}}\Big((u_{1})_{1,…,l_{1}}\Big)…p^{l_{k}}\Big((u_{k})_{1,…,l_{k}}\Big)\displaybreak
                  \\=&\:p^{k}(\omega)\left[\sum_{\substack{l_{i}=0\\1\leq i\leq
                          k}}^{p_{i}}
                      \tau^{p_{1},…,p_{k}}_{l_{1},…,l_{k}}\right]p^{p}(\alpha_{p})\:p^{p_{1}}(u_{1})…p^{p_{k}}(u_{k})\:p^{q}(\beta_{q})
                  \\ =&\:\frac{1}{k!}\big(\p^{k+2}\ot p^{k}\big)(\alpha_{p}\ot u_{1}\ot…\ot
                  u_{k}\ot\beta_{q}\ot \omega).
              \end{align*}} In der Tat erhalten wir die letzte Gleichheit mit:
            \begin{align*}
                \sum_{\substack{l_{i}=0\\1\leq i\leq k}}^{p_{i}}\tau^{l_{1},…,l_{k}}_{p_{1},…,p_{k}}=&\int_{0}^{1}dt_{1}…\int_{0}^{t_{k-1}}dt_{k}\prod_{i=1}^{k}\sum_{l_{i}=0}^{p_{i}}\binom{p_{i}}{l_{i}}t^{p_{i}-l_{i}}(1-t_{i})^{l_{i}}
                \\=&\:\int_{0}^{1}dt_{1}…\int_{0}^{t_{k-1}}dt_{k}=\frac{1}{k!}.
            \end{align*}
            Sei $\phi=\int_{0}^{1}dt_{1}…\int_{0}^{t_{k-1}}dt_{k}\:\tilde{i}_{t_{1},…,t_{k}}$, so folgt f"ur beliebige $\alpha=\sum_{p}\alpha_{p}$, $\beta=\sum_{q}\beta_{q}$

            und 
            $u_{i}=\sum_{i}(u_{i})_{p_{i}}$ mit $\alpha_{p}\in \Tt^{p}(\V)$, $\beta_{q}\in \Tt^{p}(\V)$ sowie$(u_{i})_{p_{i}}\in \Tt^{p_{i}}(\V)$:
            {\small\begin{align*}
                  \p_{k}(&\phi(\alpha\ot u_{1}\ot…\ot
                  u_{k}\ot\beta\ot \omega))\\=&\:\p_{k}\left(\sum_{\substack{p,q,p_{j}\\i_{0},\dots,i_{k+1}}}\phi\left(\alpha^{i_{0}}_{p}\ot\:
                          (u_{1})^{i_{1}}_{p_{1}}\ot…\ot\:(u_{k})^{i_{k}}_{p_{k}}\ot
                          \beta^{i_{k+1}}_{q}\ot \omega\right)\right)
                  \\\leq&\sum_{\substack{p,q,p_{j}\\i_{0},\dots,i_{k+1}}}\p_{k}\left(\phi\left(\alpha^{i_{0}}_{p}\ot\:
                          (u_{1})^{i_{1}}_{p_{1}}\ot…\ot\:(u_{k})^{i_{k}}_{p_{k}}\ot
                          \beta^{i_{k+1}}_{q}\ot \omega\right)\right)
                  \\\leq&\sum_{\substack{p,q,p_{j}\\i_{0},\dots,i_{k+1}}}\left(\p^{k+2}\ot p^{k}\right)\left(\alpha^{i_{0}}_{p}\ot\:
                      (u_{1})^{i_{1}}_{p_{1}}\ot…\ot\:(u_{k})^{i_{k}}_{p_{k}}\ot
                      \beta^{i_{k+1}}_{q}\ot \omega\right)
                  \\=&\:\Bigg(\sum_{p,i_{0}}p^{p}\Big(\alpha^{i_{0}}_{p}\Big)\Bigg)\Bigg(\sum_{p_{1},i_{1}} p^{p_{1}}\Big((u_{1})^{i_{1}}_{p_{1}}\Big)\Bigg)…\Bigg(\sum_{p_{k},i_{k}}p^{p_{k}}\Big((u_{k})^{i_{k}}_{p_{k}}\Big)\Bigg)\Bigg(\sum_{q,i_{k+1}}p^{q}\left(\beta^{i_{k+1}}_{q}\right)\Bigg)\:p^{k}(\omega),
              \end{align*}}und somit
            \begin{equation*}
                \p_{k}(\phi(\alpha\ot u_{1}\ot…\ot
                u_{k}\ot\beta\ot \omega))\leq \p(\alpha)\p(u_{1})…\p(u_{k})\p(\beta)\:p^{k}(\omega).
            \end{equation*}
        \end{enumerate}
    \end{beweis}
\end{proposition}
Wir befinden uns nun in folgender Situation:
\begin{bemerkung}
    \label{bem:StetIso}
    Mit Proposition \ref{prop:wichpropKoszStetSym}~\textit{ii.)} ist
    der Kokettenkomplex
    $(\Hom_{\mathcal{A}^{e}}^{\cont}(\K_{c},\mathcal{M}),\pt_{c}^{*})$ ein Unterkomplex von
    $(\Hom_{\mathcal{A}^{e}}(\K,\mathcal{M}),\pt^{*})$, und ebenso war
    $(\Hom_{\mathcal{A}^{e}}^{\cont}(\C_{c},\mathcal{A}),d_{c}^{*})$
    ein Unterkomplex von $(\Hom_{\mathcal{A}^{e}}(\C,\mathcal{A}),d^{*})$ . Mit
    Proposition \ref{prop:wichpropKoszStetSym}
    sind $F$ und $G$ stetige Kettenabbildungen. Folglich bilden
    $F^{*}$ und $G^{*}$ stetige auf stetige Homomorphismen ab und
    definieren somit Kettenabbildungen
    $\mathrm{F}^{k}=F_{k}^{*}\big|_{\C^{*}_{k}}$ und $\mathrm{G}^{k}=G_{k}^{*}\big|_{\K^{*}_{k}}$  zwischen obigen stetigen
    Unterkomplexen. Nach
    Bemerkung \ref{bem:HoimotAbstrIsosUnterk}~\textit{i.)} bedeutet dies die
    Injektivität von $\wt{\mathrm{F}^{k}}$ und die
    Surjektivität von $\mathrm{G}^{k}$.\\
    Für die
    umgekehrte Aussage nehmen wir an, die Kettenabbildung $\Omega =F\cp
    G$ mit  $\Omega_{k} =F_{k}\cp
    G_{k}\colon\C_{k}\longmapsto \C_{k}$ und 
    \begin{equation}
        \label{eq:thetarel}
        \Omega_{k}\cp d_{k+1}=d_{k+1}\cp \Omega_{k+1}
    \end{equation}
    wäre homotop zu der
    Identität auf $\C$, vermöge einer stetigen, $\mathcal{A}^{e}$-linearen
    Homotopie $s$.
    Dies war eine
    Familie von stetigen $\mathcal{A}^{e}$-Homomorphismen $\{s_{k}\}_{k\in \mathbb{N}}$ mit $s_{k}\colon \C_{k}^{c}\longrightarrow \C_{k+1}^{c}$ derart, dass:
    \begin{align}
        \label{eq:Homotopa}
        F_{k}\cp G_{k}- \id_{\C_{k}}=d_{k+1}s_{k}+s_{k-1}d_{k}\qquad\qquad \forall\: k\in \mathbb{N}
    \end{align}gilt.
    Mit der $\mathcal{A}^{e}$-Linearit"at liefert
    Anwenden des $\hom_{\mathcal{A}^{e}}(\cdot,\mathcal{M})$-Funktors, dass
    \begin{align*}
        G^{*}_{k}\cp
        F^{*}_{k}-\id_{\Hom_{\mathcal{A}^{e}}\left(\C_{k},\mathcal{M}\right)}=d_{k}^{*}s^{*}_{k-1}+s^{*}_{k}d^{*}_{k+1}
    \end{align*}und die Stetigkeit zeigt:
    \begin{align*}
        \mathrm{G}^{k}\cp
        \mathrm{F}^{k}-\id_{\Hom^{\cont}_{\mathcal{A}^{e}}\left(\C^{c}_{k},\mathcal{M}\right)}=d_{k}^{*}s^{*}_{k-1}+s^{*}_{k}d^{*}_{k+1}.
    \end{align*}
    Mit den Definitionen $d^{k}= d_{k+1}^{*}$ und $s^{k}=s^{*}_{k-1}$
    bedeutet dies
    \begin{equation}
        \label{eq:homotalg}
        \mathrm{G}^{k}\cp
        \mathrm{F}^{k}-\id_{\Hom^{\cont}_{\mathcal{A}^{e}}(\C^{c}_{k},\mathcal{M})}=d^{k-1}s^{k}+s^{k+1}d^{k},
    \end{equation}
    also $\mathrm{G}^{k}\cp \mathrm{F}^{k}\sim
    \id_{\Hom^{\cont}_{\mathcal{A}^{e}}(\C^{c}_{k},\mathcal{M})}$ und folglich $\wt{\mathrm{G}^{k}}\cp
    \wt{\mathrm{F}^{k}}=
    \id_{H^{k}\left(\left(\Hom^{cont}_{\mathcal{A}^{e}}(\C_{c},\mathcal{M}),d_{c}^{*}\right)\right)}$.\\\\    
    Dies bedeutet $\wt{\mathrm{G}^{k}}\cp \wt{\mathrm{F}^{k}}=
    \id_{H^{k}\left(\left(\Hom_{\mathcal{A}^{e}}^{\cont}(\C_{c},\mathcal{M},d_{c}^{*}\right))\right)}$, also die Surjektivität von $\wt{\mathrm{G}^{k}}$ und die
    Injektivität von $\wt{\mathrm{F}^{k}}$.
\end{bemerkung}
Um besagte Homotopie $s$ zu konstruieren gehen wir den in
\cite[Kapitel 5]{Weissarbeit} beschrittenen Weg. Hierf"ur ben"otigen wir das Konzept der $\mathcal{A}^{e}$-Linearisierung von
$\mathbb{K}$-linearen Abbildungen $\phi\colon\C_{k}\longrightarrow \C_{k'}$
zwischen Bar-Moduln.\\
\begin{definition}
    Gegeben eine $\mathbb{K}$-lineare Abbildung $\phi\colon\C_{s}\rightarrow \C_{r}$, so
    ist die $\mathcal{A}^{e}$-Linearisierung von $\phi$ definiert durch:
    \begin{equation}
        \begin{split}
            \label{eq:AelinausClin}
            \ovl{\phi}\colon \C_{s}&\longrightarrow \C_{r}\\
            v\ot \alpha_{s}\ot w&\longmapsto v\ot w *_{e} \phi(1\ot
            \alpha_{s}\ot 1).
        \end{split}
    \end{equation}Diese ist $\mathbb{K}$-linear, also mit
    Korollar \ref{kor:WohldefTensorprodabbildungen} durch \eqref{eq:AelinausClin}
    wohldefiniert. Sei $\C'_{s}=\mathcal{A}\ot \C_{s}\ot \mathcal{A}$,
    dann definieren wir die $\mathcal{A}^{e}$- sowie $\mathbb{K}$-linearen Abbildungen:
    \begin{align*}
        \phi'\colon\C'_{s}&\longrightarrow \C'_{r}\\
        v\ot v'\ot \alpha_{s}\ot w'\ot w&\longmapsto v\ot \phi(v'\ot
        \alpha_{s}\ot w')\ot w,
    \end{align*}
    \begin{align*}
        p_{s}\colon\C_{s}&\longrightarrow \C'_{s}\\
        v\ot \alpha_{s}\ot  w&\longmapsto v\ot 1\ot
        \alpha_{s}\ot 1\ot w,
    \end{align*}
    \begin{align*}
        i_{r}\colon\C'_{r}&\longrightarrow \C_{r}\\
        v\ot v'\ot \alpha_{s}\ot w'\ot w&\longmapsto v\ot w *_{e} v'\ot
        \alpha_{s}\ot w'.
    \end{align*}Hiermit l"asst sich $\ovl{\phi}$ auch schreiben als $\ovl{\phi}=i_{r}\cp \phi'\cp p_{s}$. Weiterhin folgt unmittelbar
    $i_{t}\cp p_{t}= \id_{\C_{t}}$.
\end{definition} 
Folgende Proposition liefert uns einige wichtige Eigenschaften.
\begin{proposition}
    \label{prop:gedoens}
    Gegeben seien $\mathbb{K}$-lineare Abbildungen
    $\psi\colon\C_{s}\longrightarrow \C_{t}$ und
    $\phi\colon\C_{t}\longrightarrow \C_{r}$. Dann gilt:
    \begin{enumerate}
    \item
        Es gilt $\ovl{\id}_{\C_{s}}=\id_{\C_{s}}$, zudem ist die
        $\mathcal{A}^{e}$-Linearisierung aufgefasst als Abbildung 
        \begin{equation*}
            \ovl{\phantom{u}}\colon
            \Hom_{\mathbb{K}}(\C_{s},\C_{t})\longrightarrow
            \Hom_{\mathcal{A}^{e}}(\C_{s},\C_{t})
        \end{equation*}$\mathbb{K}$-linear.
    \item
        Für $\mathcal{A}^{e}$-lineares $\phi$ ist
        $\ovl{\phi}=\phi$, also insbesondere 
        $\ovl{d}_{k}=d_{k}$ und
        $\ovl{\Omega}_{k}=\Omega_{k}$.
    \item
        Ist $i_{r}\cp \phi'=\phi\cp i_{t}$, so gilt $\ovl{\phi}=\phi$ und  $\phi\cp
        \ovl{\psi}=\ovl{\phi\cp \psi}$. Im Speziellen ist dies für
        alle $d_{k}$ der Fall.
    \item
        Im lokalkonvexen Fall gilt: Ist $\phi$ stetig, so auch $\ovl{\phi}$.
    \end{enumerate}
    \begin{beweis}
        \textit{i.)} und \textit{ii.)} sind unmittelbar klar und
        \textit{iii.)} folgt mit
        \begin{equation*}
            \ovl{\phi}=i_{r}\cp \phi' \cp p_{t}=\phi \cp i_{t}\cp p_{t}=\phi\cp \id_{\C_{t}}=\phi.
        \end{equation*}Um die zweite Aussage zu zeigen, rechnen wir
        \begin{align*}
            (\phi\cp\psi)'(v\ot v'\ot \alpha_{s}\ot w'\ot
            w)=&\:v\ot \big[(\phi\cp\psi)(v'\ot \alpha_{s}\ot w')\big]\ot
            w
            \\=&\:\phi'(v\ot \psi(v'\ot \alpha_{s}\ot w')\ot
            w)
            \\=&\:(\phi'\cp \psi')\:(v\ot v'\ot \alpha_{s}\ot w'\ot
            w),
        \end{align*} und erhalten
        \begin{equation*}
            \ovl{\phi\cp \psi}= i_{r}\cp
            \phi'\cp \psi'\cp p_{s}=\phi\cp i_{t}\cp \psi'\cp
            p_{s}=\phi\cp \ovl{\psi}.
        \end{equation*}
        Die letzte Behauptung folgt mit
        \begin{align*}
            (i_{k-1}\cp d'_{k})(v\ot x_{0}\ot…\ot x_{k+1}\ot w)=&\: i_{k-1}\left(v\ot d_{k}(x_{0}\ot…\ot
                x_{k+1})\ot w\right)
            \\=&\:v\ot w *_{e}\sum_{j=0}^{k}(-1)^{k}\: x_{0}\ot…\ot
            x_{j}x_{j+1}\ot…\ot x_{k+1}
            \\=&\:(d_{k}\cp i_{k})(v\ot x_{0}\ot…\ot x_{k+1}\ot
            w).
        \end{align*}
        Für \textit{iv.)} sei $\phi$ stetig, dann folgt mit der Stetigkeit von $*_{e}$:
        \begin{align*}
            q^{k+2}\left(\ovl{\phi}\:(v\ot \alpha_{1}\ot…\ot \alpha_{k}\ot
                w)\right)=&\:q^{k+2}(v\ot w *_{e} \phi(1\ot\alpha_{1}\ot…\ot \alpha_{k}\ot1))
            \\\leq&\:c\: p_{1}^{2}(v\ot w)\:
            p_{2}^{k+2}(1\ot\alpha_{1}\ot…\ot \alpha_{k}\ot 1)
            \\=&\:cp_{2}(1)^{2}\:p_{1}(v)\:p_{1}(w)\:p_{2}(\alpha_{1})…p_{2}(\alpha_{k}).
        \end{align*}Dies zeigt die Stetigkeit in $\pi_{k+2}$ und beendet den Beweis.
    \end{beweis}
\end{proposition} 
Folgendes Lemma liefert uns schließlich die erwünschte Homotopie
$s$.
\begin{lemma}
    \label{lemma:Homotopiejdfgjkf}
    Es ist $\id_{\C_{k}}-\:\Omega_{k}=d_{k+1} s_{k}+s_{k-1} d_{k}$, also $\Omega\sim\id_{\C}$ vermöge der $\mathcal{A}^{e}$-linearen Homotopie
    $s_{k}\colon \C_{k}\longrightarrow \C_{k+1}$, die für $k\geq 0$ rekursiv definiert ist durch:
    \begin{equation*}
        s_{k}=\ovl{h_{k}
          (\id_{\C_{k}}-\:\Omega_{k}\:-\:s_{k-1}
          d_{k})}\qquad\text{mit}\qquad s_{0}=0.
    \end{equation*} 
    Hierbei bezeichnet $h$ die exaktheitsliefernde Homotopie aus
    Proposition \ref{prop:topBarKomplexprop}~\textit{ii.)}. Zudem ist 
    $s_{k}:\C^{c}_{k}\longrightarrow \C^{c}_{k+1}$, aufgefasst als Abbildung zwischen den
    lokalkonvexen Vektorräumen $\C^{c}_{k}$ und $\C^{c}_{k+1}$ stetig.
    \begin{beweis}
        Die Stetigkeit der $s_{k}$ folgt unmittelbar aus der
        Stetigkeit der definierenden Abbildungen\footnote{vgl. Proposition \ref{prop:wichpropKoszStetSym}} und
        Proposition \ref{prop:gedoens}~\textit{iv.)}. Die $\mathcal{A}^{e}$-Linearität
        ist ebenfalls klar.\\ Für den Induktionsanfang rechnen wir mit
        Proposition \ref{prop:gedoens}:
        {\large\begin{align*}
              d_{2}s_{1}-\cancel{s_{0} d_{1}}=&\:d_{2}
              \ovl{h_{1}\left(\id_{\C_{1}}-\:\Omega_{1}\:-\:\cancel{s_{0}
                      d_{1}}\right)}\glna{\textit{iii.)}}
              \ovl{(d_{2}h_{1})
                \left(\id_{\C_{1}}-\:\Omega_{1}\right)}
              \\\glna{\eqref{eq:Homotbar}}&\:\ovl{(\id_{\C_{1}}-\:h_{0}d_{1})
                \left(\id_{\C_{1}}-\:\Omega_{1}\right)}\glna{\textit{i.),ii.)}}\id_{\C_{1}}-\:\Omega_{1}-\ovl{h_{0}d_{1}}+\ovl{h_{0}d_{1}\Omega_{1}}
              \\\glna{\eqref{eq:thetarel}}&\id_{\C_{1}}-\:\Omega_{1},
          \end{align*}}wobei wir im letzten Schritt zudem
        $\Omega_{0}=\id_{\mathcal{A}^{e}}$ benutzt haben. Für die
        höheren Grade folgt:
        {\large\begin{align*}
              d_{k+1} s_{k}=&\:\ovl{(d_{k+1}
                h_{k})
                (\id_{\C_{k}}-\:\Omega_{k}\:-\:s_{k-1}
                d_{k})}
              \\=&\:\ovl{(\id_{\C_{k}}-h_{k-1} d_{k})
                (\id_{\C_{k}}-\:\Omega_{k}\:-\:s_{k-1}
                d_{k})}
              \\=&\:\id_{\C_{k}}- \:\Omega_{k} - \ovl{s_{k-1}
                d_{k}} -
              \ovl{h_{k-1}(d_{k}\:-\:d_{k}\Omega_{k}\:-\:(d_{k}
                s_{k-1}) d_{k})}
              \\=&\:\id_{\C_{k}}- \:\Omega_{k} - s_{k-1}
              d_{k} -
              \ovl{h_{k-1}(d_{k}\:-\:\Omega_{k-1} d_{k}\:-\:(\id_{\C_{k}}-\:\Omega_{k-1}\:-\:s_{k-2}
                d_{k-1}) d_{k})}
              \\=&\:\id_{\C_{k}}- \:\Omega_{k} - s_{k-1}
              d_{k}.
          \end{align*}}
    \end{beweis}
\end{lemma}
Dies zeigt schließlich folgenden Satz:
\begin{satz}
    \label{satz:stetigHochschSym}
    Gegeben ein lokalkonvexer $\SsV-\SsV$-Bimodul $\mathcal{M}$, dann gilt: 
    \begin{equation*}
        HH^{k}_{\cont}\Big(\Ss^{\bullet}(\mathbb{V}),\mathcal{M}\Big)\cong
        H^{k}\Big(\Hom_{\mathcal{A}^{e}}^{\cont}(\C_{c},\mathcal{M}),d_{c}^{*}\Big)\cong
        H^{k}\Big(\Hom_{\mathcal{A}^{e}}^{\cont}(\K_{c},\mathcal{M}),\pt_{c}^{*}\Big).
    \end{equation*} Ist $\mathcal{M}$ zudem symmetrisch, so ist:
    \begin{equation*}
        HH^{k}_{\cont}(\Ss^{\bullet}(\mathbb{V}),\mathcal{M})\cong \Hom_{\mathcal{A}^{e}}^{\cont}\big(\K^{c}_{k},\mathcal{M}\big).
    \end{equation*}
    \begin{beweis}
        Die erste Isomorphie hatten wir bereits eingesehen. Die zweite
        folgt nun unmittelbar mit Bemerkung \ref{bem:StetIso}, da wir mit
        Lemma \ref{lemma:Homotopiejdfgjkf} die benötigte stetige Homotopie 
        gefunden haben. Die letzte Aussage folgt wie für
        Satz \ref{satz:HochschkohmvonSym}, da auch hier
        $\ker(\partial^{*}_{k+1})= \K^{*}_{k}$ und
        $\im(\partial^{*}_{k})=0$ erfüllt ist.
    \end{beweis}
\end{satz}
%
\section{Die stetige Hochschild-Kohomologie der Algebra $\Hol$}
\label{sec:StetHKHol}
Sei im Folgenden $\V$ ein hausdorffscher, lokalkonvexer
Vektorraum. Wir beginnen mit der folgenden, klärenden Proposition:
\begin{proposition}
    \label{prop:HollkAlgebra}
    \begin{enumerate}
    \item 
        Gegeben ein hlkVR $\V$, so existiert eine bis auf lineare
        Homöomorphie eindeutig bestimmte vollständige, hausdorffsche, submultiplikative, lokalkonvexe
        Algebra $(\Hol,\Pp_{H},*)$, die $(\Ss^{\bullet}(\V),\Pp,\vee)$ im
        isometrischen Sinne als dichte Unteralgebra enthält. Diese ist zudem
        assoziativ und unitär.
    \item
        Jeder hausdorffsche, lokalkonvexe $\Ss^{\bullet}(\V)-\Ss^{\bullet}(\V)$-Bimodul $(\mathcal{M},*_{L},*_{R})$ vervollständigt
        zu einem hausdorffschen, lokalkonvexen $\Hol-\Hol$-Bimodul $(\hat{\mathcal{M}},\hat{*}_{L},\hat{*}_{R})$.
    \end{enumerate}
    \begin{beweis}
        \begin{enumerate}
        \item 
            Zunächst ist nach Satz \ref{satz:PiTopsatz}~\textit{v.)} mit $\V$
            ebenfalls jedes $\left(\Ss^{k}(\V),\pi_{k}\right)$ hausdorffsch und es ist 
            offensichtlich, dass dies dann ebenfalls für
            $(\Ss^{\bullet}(V),\Pp)$ der Fall ist.
            Es folgt mit Satz \ref{satz:vervollsthlkVR}, dass die bis auf lineare Homöomorphie eindeutig
            bestimmte Vervollständigung
            $\Big(\Hol,\Pp_{H}\Big)=\left(\widehat{\Ss^{\bullet}(\V)},\hat{\Pp}\right)$
            existiert und ebenfalls hausdorffsch ist. 
            Weiter folgt, dass die bilineare Abbildung
            \begin{align*}
                \tilde{\vee}\colon i(\Ss^{\bullet}(\V))\times i(\Ss^{\bullet}(\V))&\longrightarrow
                i(\Ss^{\bullet}(\V))\subseteq\Hol\\
                (x,y)&\longmapsto i(i^{-1}(x) \vee i^{-1}(y)) 
            \end{align*} als Verkettung stetiger Funktionen stetig
            ist. Hierbei haben wir benutzt, dass die stetige Isometrie
            $i$ aus Satz \ref{satz:vervollsthlkVR} ein Homöomorphismus
            zwischen $(\SsV,\Pp)$ und $(i(\SsV),\hat{\Pp})$ ist. 
            Mit
            $\ovl{i(\Ss^{\bullet}(\V))}=\Hol$ liefert uns 
            Satz \ref{satz:stetfortsBillphi} eine eindeutig bestimmte stetige, bilineare Fortsetzung 
            \begin{align*}
                *\colon \Hol\times \Hol\longrightarrow \Hol,
            \end{align*}
            und wegen 
            \begin{equation*}
                \hat{\p}(x * y)=\displaystyle\lim_{\alpha\times\beta}\p\left(x_{\alpha}\tilde{\vee}
                    y_{\beta}\right)\leq
                \displaystyle\lim_{\alpha\times\beta}\p(x_{\alpha})\p(y_{\beta})=\hat{\p}(x)\:\hat{\p}(y)
            \end{equation*}ist $(\Hol,\hat{\Pp},*)$ zudem
            submultiplikativ. Hierbei ist $x,y\in \Hol$ mit Netzen
            $i(\SsV)\supseteq\net{x}{I}\rightarrow x$,
            $i(\SsV)\supseteq\nettt{y}{\beta}{J}\rightarrow y$.\\\\
            Für die Unitarität betrachten wir das Element $\Hol\ni
            \tilde{1}:= i(1_{\Ss^{\bullet}(\V)})$ und erhalten für ein
            Netz $i(\Ss^{\bullet}(\V))\supseteq\net{x}{I}\rightarrow x\in \Hol$ sowie
            $\hat{1}=\left\{\tilde{1}\right\}$ die konstante Folge $\tilde{1}$, dass: 
            \begin{align*} 
                \hat{1}* x = & \lim_{n\times \alpha}
                \{i(1_{\Ss^{\bullet}(\V)})\}\:\tilde{\vee}\: x_{\alpha}=\lim_{\alpha}\:
                x_{\alpha}=x.
            \end{align*}
            Spätestens hier ist nun auch klar, dass wir
            vermöge $i$ die Räume $(\Ss^{\bullet}(\V),\vee)$ und
            $i(\Ss^{\bullet}(\V),\tilde{\vee})$ identifizieren dürfen.
            Für die
            Assoziativität rechnen wir daher in Kurzschreibweise mit
            $x,y,z\in \Hol$:
            \begin{align*}
                x*(y*z)=\lim_{\alpha\times (\beta\times
                  \gamma)}x_{\alpha}\vee (y_{\beta}\vee z_{\gamma})=\lim_{(\alpha\times \beta)\times
                  \gamma}(x_{\alpha}\vee y_{\beta})\vee z_{\gamma}=(x*y)*z,
            \end{align*}da definitionsgemäß $\{y_{\beta}\vee
            z_{\gamma}\}_{\beta\times \gamma\in J\times L}\rightarrow y*z$
            und $\{x_{\alpha}\vee
            y_{\beta}\}_{\alpha\times \beta\in I\times J}\rightarrow x*y$.
        \item
            Zunächst ist wieder klar, dass für jeden solchen Bimodul
            $\mathcal{M}$ eine Vervollständigung $\hat{\mathcal{M}}$
            existiert. Des Weiteren induzieren $*_{L}$ und $*_{R}$ stetige, bilineare Abbildungen
            auf den dichten Teilräumen $i(\Ss^{\bullet}(\V))\subseteq
            \Hol$ und $i'(M)\subseteq \hat{M}$, womit stetige
            bilineare Fortsetzungen $\hat{*}_{L}$ und $\hat{*}_{R}$ existieren. Die
            $\Hol$-Verträglichkeit, also $\hat{1}\hat{*}_{L}
            \hat{m}=\hat{m}=\hat{m}\hat{*}_{R}\hat{1}$ für alle
            $\hat{m}\in \hat{\mathcal{M}}$ folgt dann wie
            die Unitarität in \textit{i.)}, und die Vererbung der
            Bimoduleigenschaft wie die Assoziativität in \textit{i.)}.
        \end{enumerate}
    \end{beweis}
\end{proposition} 
Punkt \textit{ii.)} ist unter anderem als Motivation dafür gedacht, dass
überhaupt derartige $\Hol-\Hol$-Bimoduln existieren. Als wichtiges
Resultat aus \textit{i.)} erhalten wir umgehend:
\begin{korollar}
    \label{kor:HolBarBimodul}
    Gegeben ein hlkVR $\V$ und ein lokalkonvexer
    $\mathrm{Hol}(\V)-\mathrm{Hol}(\V)$-Bimodul $\mathcal{M}$, so gilt:
    \begin{align*}
        HH_{\cont}^{k}(\Hol,\mathcal{M})\cong
        H^{k}\left(\Hom^{\cont}_{\mathcal{A}^{e}}(\C_{c},\mathcal{M}),d^{*}_{c}\right).
    \end{align*}
    Hierbei bezeichnet $(\C_{c},d_{c})$ den zu $\mathcal{A}=\Hol$ gehörigen,
    topologischen Bar-Komplex.
    \begin{beweis}
        Dies folgt schon wie im letzten Abschnitt aus
        Proposition \ref{prop:topBarKomplexprop}~\textit{iii.)}, da\\ $(\Hol,\Pp_{H},*)$ mit
        Proposition \ref{prop:HollkAlgebra} eine
        lokalkonvexe, unitäre und assoziative $\mathbb{K}$-Algebra ist.
    \end{beweis}
\end{korollar}
Wir wollen den Raum $\Hol$ zunächst mit ein wenig Anschauung füllen. Hierfür definieren wir:
\begin{definition}
    \label{def:Potenzr}
    Gegeben ein hlkVR $(\V,P)$ sowie die topologischen Räume $\left(\widehat{\Ss^{k}(\V)}, \hat{p^{k}}\right)$.\\
    Wir bezeichnen mit $\left(\prod_{\widehat{\Ss^{\bullet}(\V)}}, \Pp_{\times}\right)$ den hlkVR
    \begin{equation*}
        \textstyle\prod_{\widehat{\Ss^{\bullet}(\V)}}=\left\{\displaystyle\prod_{k=0}^{\infty}\widehat{\Ss^{k}(\V)}\ni\hat{\omega}= (\hat{\omega}_{0}\:,\hat{\omega}_{1}\:,\hat{\omega}_{2}\:,\dots)\:\Bigg| \p_{\times}(\hat{\omega})=\sum_{k=0}^{\infty}\hat{p^{k}}(\hat{\omega}_{k})<\infty, \:\forall\: p\in \tilde{P}\right\}.
    \end{equation*}
\end{definition}
Zusammen mit Satz \ref{satz:vervollsthlkVR} zeigt der Folgende, dass wir $\left(\Hol,\Pp_{H}\right)$ mit dem Potenzreihenraum $\left(\prod_{\widehat{\Ss^{\bullet}(\V)}}, \Pp_{\times}\right)$ identifizieren dürfen.
\begin{satz}[Potenzreihen]
    \label{satz:Potenz}
    Gegeben ein hlkVR $(\V,P)$, dann gilt:
    \begin{enumerate}
    \item
        $\left(\prod_{\widehat{\Ss^{\bullet}(\V)}}, \Pp_{\times}\right)$ ist vollständig.
    \item
        $\displaystyle\bigoplus_{k=0}^{\infty}\widehat{\Ss^{k}(\V)}$, topologisiert vermöge $\Pp_{\times}$, ist folgendicht in $\left(\prod_{\widehat{\Ss^{\bullet}(\V)}}, \Pp_{\times}\right)$.
    \item
        $\left(\Ss^{\bullet}(\V),\Pp\right)$ ist dicht in
        $\left(\prod_{\widehat{\Ss^{\bullet}(\V)}},\Pp_{\times}\right)$
        vermöge isometrischer Einbettung:
        \begin{align*}
            i\colon\sum_{k}\omega_{k}&\longmapsto \prod_{k=0}^{\infty}i_{k}(\omega_{k})\\
            \sum_{k=0}^{\infty}p^{k}&\longmapsto \sum_{k=0}^{\infty}\hat{p^{k}}.
        \end{align*}
        Dabei bezeichnen die $i_{k}$ die Isometrien
        $i_{k}\colon\Ss^{k}(\V)\hookrightarrow
        \widehat{\Ss^{k}(\V)}$.
    \end{enumerate}
    \begin{beweis}
        \begin{enumerate}
        \item
            Sei $\{\hat{\omega}_{\alpha}\}_{\alpha\in I}\subseteq
            \prod_{\widehat{\Ss^{\bullet}(\V)}}$ ein Cauchynetz, dann
            existiert für jedes $\epsilon\geq 0$ ein $\alpha_{\epsilon}\in I$, so dass:
            \begin{equation*}
                \sum_{k=0}^{\infty}\hat{p^{k}}\left(\hat{\omega}_{\alpha}^{k}-\hat{\omega}_{\beta}^{k}\right)< \epsilon\qquad\quad\forall\:\alpha,\beta \geq \alpha_{\epsilon}\in I.
            \end{equation*} 
            Damit ist insbesondere
            $\left\{\hat{\omega}_{\alpha}^{k}\right\}_{\alpha\in I}\subseteq
            \widehat{\Ss^{k}(\V)}$ für jedes $k\in \mathbb{N}$ ein Cauchynetz, womit

            $\left\{\hat{\omega}_{\alpha}^{k}\right\}_{\alpha\in
              I}\rightarrow \hat{\omega}^{k}$ mit eindeutigem
            $\hat{\omega}^{k}\in \widehat{\Ss^{k}(\V)}$ gilt. 

            Wir definieren
            $\hat{\omega}=\displaystyle\prod_{k=0}^{\infty}\hat{\omega}^{k}$
            und behaupten, dass dann
            $\p_{\times}(\hat{\omega})<\infty$ für alle
            $p\in\tilde{P}$ sowie
            $\left\{\hat{\omega}_{\alpha}\right\}_{\alpha\in
              I}\rightarrow \hat{\omega}$ erfüllt ist.

            Nun gilt 
            \begin{equation*}
                \p_{\times}(\hat{\omega}_{\alpha})\leq \p_{\times}(\hat{\omega}_{\alpha}-\hat{\omega}_{\alpha_{\epsilon}})+\p_{\times}(\hat{\omega}_{\alpha_{\epsilon}})<\epsilon + \p_{\times}(\hat{\omega}_{\alpha_{\epsilon}})= \wt{c}\qquad\forall\:\alpha\geq \alpha_{\epsilon}\in I
            \end{equation*}
            und folglich: 
            \begin{equation*}
                \sum_{k=0}^{n}\hat{p^{k}}\left(\hat{\omega}_{\alpha}^{k}\right)
                \leq \p_{\times}(\hat{\omega}_{\alpha})
                < \wt{c}\qquad\quad\forall\: \alpha\geq \alpha_{\epsilon}\in I,\:\forall \: n\in \mathbb{N}.
            \end{equation*}
            Hiermit erhalten wir
            \begin{equation*}
                \tau_{n}=\sum_{k=0}^{n}\hat{p^{k}}\left(\hat{\omega}^{k}\right)= \sum_{k=0}^{n}\lim_{\alpha}\:\hat{p^{k}}\left(\hat{\omega}_{\alpha}^{k}\right)= \lim_{\alpha}\:\sum_{k=0}^{n}\hat{p^{k}}\left(\hat{\omega}_{\alpha}^{k}\right)\leq \hat{c}\qquad\quad\forall\:n\in \mathbb{N},
            \end{equation*} womit
            $\left\{\tau_{n}\right\}_{n\in \mathbb{N}}$ eine
            Cauchyfolge ist, da alle Summanden positiv sind. 
            Die zeigt die Existenz des Limes $n\rightarrow \infty$ und es folgt:
            \begin{equation*}
                \p_{\times}(\hat{\omega})=\lim_{n}\:\sum_{k=0}^{n}\hat{p^{k}}\left(\hat{\omega}^{k}\right)=\lim_{n}\tau_{n}\leq \hat{c}<\infty.
            \end{equation*} 
            Für die Konvergenzaussage beachten wir, dass
            \begin{equation*}
                \mu_{n}=\sum_{k=0}^{n}\hat{p^{k}}\left(\hat{\omega}_{\alpha}^{k}-\hat{\omega}_{\beta}^{k}\right)
                \leq\p_{\times}\Big(\hat{\omega}_{\alpha}-\hat{\omega}_{\beta}\Big)< \epsilon\qquad\forall\:\alpha,\:\beta\geq \alpha_{\epsilon},
            \end{equation*}
            womit $\mu_{n}$ eine Cauchyfolge ist und der Limes
            existiert. Es folgt:
            \begin{equation*}
                \sum_{k=0}^{n}\hat{p^{k}}\left(\hat{\omega}^{k}-\hat{\omega}_{\beta}^{k}\right)
                =\lim_{\alpha}\:\sum_{k=0}^{n}\hat{p^{k}}\left(\hat{\omega}_{\alpha}^{k}-\hat{\omega}_{\beta}^{k}\right)\leq \epsilon\qquad\forall\:\beta\geq \alpha_{\epsilon},\:\forall\:n\in \mathbb{N}.
            \end{equation*}
            Dies zeigt
            \begin{equation*}
                \p_{\times}\Big(\hat{\omega}-\hat{\omega}_{\beta}\Big)=\lim_{n}\sum_{k=0}^{n}\hat{p^{k}}\left(\hat{\omega}^{k}-\hat{\omega}_{\beta}^{k}\right)
                \leq \epsilon\qquad\forall\:\beta\geq \alpha_{\epsilon},
            \end{equation*}
            also $\{\hat{\omega}_{\alpha}\}_{\alpha\in I}\rightarrow \hat{\omega}\in \prod_{\widehat{\Ss^{\bullet}(\V)}}$.
        \item
            Sei $\prod_{\widehat{\Ss^{\bullet}(\V)}}\ni\hat{\omega}=\left(\hat{\omega}^{0},\:\hat{\omega}^{1},\:\hat{\omega}^{2},\dots\right)$ und
            \begin{equation*}
                \displaystyle\bigoplus_{k=0}^{\infty}\widehat{\Ss^{k}(\V)}\ni\hat{\omega}_{n}=\left(\hat{\omega}^{0},\:\hat{\omega}^{1},\dots,\:\hat{\omega}^{n},0,0,0,\dots\right).
            \end{equation*}
            Dann ist
            $\displaystyle\lim_{n}\p_{\times}(\hat{\omega}_{n})=\displaystyle\lim_{n}\:\sum_{k=0}^{n}\hat{p^{k}}(\hat{\omega}_{k})=\p_{\times}(\hat{\omega})=c<\infty$,
            also $\left\{\p_{\times}(\hat{\omega}_{n})\right\}_{n\in
              \mathbb{N}}$ eine Cauchyfolge, womit
            $\p_{\times}(\hat{\omega}_{m}-\hat{\omega}_{n})=\displaystyle\sum_{k=n+1}^{m}\hat{p^{k}}(\hat{\omega}^{k})
            =|\p_{\times}(\hat{\omega}_{m})-\p_{\times}(\hat{\omega}_{n})|<\epsilon\:$ f"ur alle $m\geq n\geq N_{\epsilon}$.
            Es folgt
            \begin{equation*}
                \p_{\times}(\hat{\omega}-\hat{\omega}_{n})=\lim_{m}\p_{\times}(\hat{\omega}_{m}-\hat{\omega}_{n})=\lim_{m}|\p_{\times}(\hat{\omega}_{m})-\p_{\times}(\hat{\omega}_{n})|\leq\epsilon\qquad\quad\forall\:n\geq N_{\epsilon}
            \end{equation*}
            und somit $\left\{\hat{\omega}_{n}\right\}_{n\in \mathbb{N}}\rightarrow \hat{\omega}$.
        \item
            Seien $\hat{\omega}$ und $\hat{\omega}_{n}$ wie in
            \textit{ii.)}. Wir fassen $P\times \mathbb{N}$ als
            gerichtete Menge auf\footnote{vgl. Definition \ref{def:kanNetzIsoetc}~\textit{ii.)}} und wählen für jedes Element $(p,n)$ ein $k_{p,n}\in \mathbb{N}$ derart, dass
            \begin{equation*}
                \p_{\times}(\hat{\omega}-\hat{\omega}_{k})< \frac{1}{2n}\qquad\quad\forall\:k\geq k_{p,n}
            \end{equation*}
            gilt, was nach \textit{ii.)} ohne Einschränkung möglich ist.

            Ferner denken wir uns $\Ss^{\bullet}(\V)\subseteq \displaystyle\bigoplus_{k=0}^{\infty}\widehat{\Ss^{k}(\V)}$ isometrisch eingebettet und finden für jedes $k\in \mathbb{N}$ ein Netz $\left\{\omega^{k}_{\alpha_{k}}\right\}_{\alpha_{k} \in I_{k}}\subseteq \Ss^{k}(\V)$ mit $\left\{\omega^{k}_{\alpha_{k}}\right\}_{\alpha_{k} \in I_{k}}\rightarrow \hat{\omega}^{k}\in \widehat{\Ss^{k}(\V)}$.\\\\
            Für besagtes $(p,n)$ und $0\leq k \leq k_{p,n}$ bedeutet dies die Existez von Indizes $\alpha_{k}\in I_{k}$ derart, dass 
            \begin{equation*}
                \hat{p^{k}}\left(\hat{\omega}^{k}-\omega_{\alpha_{k}}^{k}\right)<\frac{1}{2n(k_{p,n}+1)}.
            \end{equation*}
            Wir definieren $\omega_{p,n}=\left(\omega^{0}_{\alpha_{1}},\:\omega^{1}_{\alpha_{2}},\dots,\: \omega^{k_{p,n}}_{\alpha_{k_{p,n}}},0,0,0,\dots\right)$, womit
            \begin{equation*}
                \p_{\times}\left(\hat{\omega}_{k_{p,n}}-\omega_{p,n}\right)=\:\sum_{k=0}^{k_{p,n}}\:\hat{p}^{k}\left(\hat{\omega}^{k}-\omega^{k}_{\alpha_{k}}\right)< \frac{1}{2n}
            \end{equation*}
            und folglich für alle $(p',n')\geq (p,n)$:
            \begin{align*}
                \p_{\times}(\hat{\omega}-\omega_{p',n'})\leq&\: \p_{\times}\left(\hat{\omega}-\hat{\omega}_{k_{p',n'}}\right)+\p_{\times}\left(\hat{\omega}_{k_{p',n'}}-\omega_{p',n'}\right)
                \\\leq&\: \p'_{\times}\left(\hat{\omega}-\hat{\omega}_{k_{p',n'}}\right)+\p'_{\times}\left(\hat{\omega}_{k_{p',n'}}-\omega_{p',n'}\right)
                \\<& \:\frac{1}{2n'}+\frac{1}{2n'}\:\leq\: \frac{1}{n}.
            \end{align*}Dies zeigt $\{\omega_{p,n}\}_{P\times
              \mathbb{N}}\rightarrow \hat{\omega}$ und mit Bemerkung \ref{bem:Netzbem}~\textit{iii.)} die Behauptung.
        \end{enumerate}
    \end{beweis}
\end{satz}
\begin{bemerkung}
    Obiger Satz besagt also insbesondere, dass die Vervollständigung
    von $(\SsV,\Pp)$ bereits durch die Vervollständigungen der
    $\left(\Ss^{k},\pi_{k}\right)$ festgelegt ist. Des Weiteren ist es sogar
    möglich, jedes $\hat{\omega}\in\widehat{\mathrm{S}^{\bullet}(\mathbb{V})}$ durch eine Folge in
    $\{\omega_{n}\}_{n\in
      \mathbb{N}}\subseteq\displaystyle\bigoplus_{k=0}^{\infty}\widehat{\Ss^{k}(\V)}$
    zu approximieren. Die Schwierigkeit liegt hierbei also in der
    Vervollständigung der $\left(\Ss^{k},\pi_{k}\right)$, die für
    unendlich-dimensionales $\V$ im Allgemeinen alles andere als trivial
    ist. Für endlich-dimensionales $\V$ hingegen ist
    $\left(\Ss^{k},\pi_{k}\right)$ bereits vollständig, vgl. Beispiel \ref{bsp:holomorpheFunkts}~\textit{i.)}.
\end{bemerkung}
\begin{beispiel}[Holomorphe Funktionen]
    \label{bsp:holomorpheFunkts}
    \begin{enumerate}
    \item 
        Wir versehen den Vektorraum $\mathbb{C}^{n*}$
        mit der üblichen euklidischen Normtopologie. Mit der
        Äquvivalenz aller Normen auf $\mathbb{C}^{n*}$ können wir uns
        wahlweise auf das System, bestehend aus allen bezüglich der
        Maximumsnorm
        \begin{equation*}
            p_{\mathrm{max}}(x)=\sum_{i=1}^{n}|x_{i}|\quad \text{mit}\quad
            x=\sum_{i=1}^{n}x_{i}\: e^{i}
        \end{equation*} 
        stetigen Halbnormen festlegen, und verschaffen uns so
        ein filtrierendes System $\tilde{P}$ auf
        $\mathbb{C}^{n*}$. Dieses enthält dann insbesondere wieder alle
        Normen der Form $|c|\:p_{\mathrm{max}}$ für positive Konstanten $|c|$. Die
        symmetrische Algebra sei wie in Abschnitt \ref{subsec:Vorber}
        topologisiert vermöge $\Pp$.

        Wir behaupten zunächst, dass
        \begin{equation}
            \label{eq:pitopendlCn}
            p^{k}_{\mathrm{max}}(z)=\sum_{i_{1},…,i_{k}}^{n}|a_{i_{1}},…,a_{i_{k}}| 
        \end{equation}
        für alle $\Tt^{\bullet}(\mathbb{C}^{n*})\ni
        z=\displaystyle\sum_{i_{1},…,i_{k}}^{n}
        a_{i_{1},…,i_{k}}e^{i_{1}}\ot…\ot e^{i_{k}}$ gilt. Hierbei beachte
        man, dass dann \eqref{eq:pitopendlCn} mit unserer
        Konvention $e^{i_{1}}\vee…\vee e^{i_{k}}=\frac{1}{k!}\sum_{\sigma\in
              S_{k}}e^{\sigma(i_{1})}\ot…\ot e^{\sigma(i_{k})}$ ebenso für alle $\Ss^{\bullet}(\mathbb{C}^{n*})\ni
            z=\displaystyle\sum_{i_{1},…,i_{k}}^{n}
            a_{i_{1},…,i_{k}}e^{i_{1}}\vee…\vee e^{i_{k}}$
               richtig ist.
        Bezeichne hierfür $p_{\ot}^{k}$ die durch
        \eqref{eq:pitopendlCn} charakterisierte Norm, dann gilt 
        \begin{equation*}
            p_{\mathrm{max}}^{k}(z)\leq
            \sum_{i_{1},…,i_{k}}^{n}|a_{i_{1},…,a_{i_{k}}}e^{i_{1}}|\cdot |e^{i_{2}}|…|e^{i_{k}}|=p_{\ot}^{k}(z)
        \end{equation*}per Definition von $p_{\mathrm{max}}^{k}$. Mit der Normeigenschaft von $p^{\ot k}$ folgt weiter, dass
        \begin{equation*}
            p_{\ot}^{k}(z)\leq \sum_{i}p_{\ot}^{k}(z^{i})=\sum_{i}p_{\mathrm{max}}^{k}(z^{i})
        \end{equation*} 
        für alle Zerlegungen $z=\displaystyle\sum_{i}
        z^{i}$ in separable $z^{i}=x_{1}^{i}\ot…\ot x_{k}^{i}$ richtig
        ist. Dabei
        folgt die zweite Gleichheit mit $x_{j}=\displaystyle\sum_{i_{j}}(x_{j})_{i_{j}}e^{i_{j}}$ aus:
        \begin{align*}
            p_{\ot}^{k}(x_{1}\ot…\ot x_{k})=&\sum_{i_{1},…,i_{k}}^{n}|(x_{1})_{i_{1}}\cdot…\cdot(x_{k})_{i_{k}}|=\sum_{i_{1},…,i_{k}}^{n}|(x_{1})_{i_{1}}|\cdot…\cdot|(x_{k})_{i_{k}}|
            \\=&\:p_{\mathrm{max}}(x_{1})\cdot…\cdot p_{\mathrm{max}}(x_{k})=\:p_{\mathrm{max}}^{k}(x_{1}\ot…\ot x_{k}).
        \end{align*}
        \\ 
        Um die Vervollständigung
        $(\mathrm{Hol}(\mathbb{C}^{n*}),\hat{\Pp})$ von
        $(\Ss^{\bullet}(\mathbb{C}^{n*}),\Pp)$ zu charakterisieren beachten wir, dass
        für festes $k\in \mathbb{N}$ die Topologie auf
        $\Ss^{k}(\mathbb{C}^{n*})$ bereits durch die Norm $p^{k}_{\max}$
        erzeugt wird. Dies folgt unmittelbar aus
        Korollar \ref{kor:HNTop}~\textit{iii.)}, da mit
        Satz \ref{satz:wichtigerSatzueberHalbnormentopologien}~\textit{vi.)}
        $p\leq |c|\: p_{\max}$ für alle $p\in \wt{P}$ und somit
        ebenfalls $p^{k}\leq |c|^{k} p_{\max}^{k}$ für alle $p\in
        \wt{P}$ gilt. Bemerkung \ref{bem:Netzbem}~\textit{i.)} zeigt dann,
        dass wir lediglich Folgenvollständigkeit nachweisen müssen,
        wenn wir
        $\widehat{\Ss^{k}(\mathbb{C}^{n*})}=\Ss^{k}(\mathbb{C}^{n*})$
        zeigen wollen. Sei hierfür $\{z_{n}\}_{n\in \mathbb{N}}\subseteq
        \Ss^{k}(\mathbb{C}^{n*})$ eine Cauchyfolge, dann ist
        \begin{align*}
            \sum_{i_{1},…,i_{k}}^{n}|a^{m}_{i_{1},…,i_{k}}-a^{n}_{i_{1},…,i_{k}}|=p^{k}_{\max}(z_{m}-z_{n})< \epsilon\qquad\forall\:m,n\geq N_{\epsilon},
        \end{align*}und mit der Vollständigkeit von $\mathbb{C}$ zeigt
        dies, dass $a^{n}_{i_{1},…,i_{k}}\longrightarrow a_{i_{1},…,i_{k}}\in
        \mathbb{C}$. Eine analoge Abschätzung liefert $z_{n}\longrightarrow
        \displaystyle\sum_{i_{1},…,i_{k}}^{n}a_{i_{1},…,i_{k}}e^{i_{1}}\vee…\vee
        e^{i_{k}}\in \Ss^{k}(\mathbb{C}^{n*})$, was die
        Folgenvollständigkeit beweist. Hierbei ist wesentlich
        eingegangen, dass $\Ss^{k}(\mathbb{C}^{n*})$ eine endliche
        Basis besitzt.

        Nach Satz \ref{satz:Potenz} ist dann
        \begin{equation*}
            \mathrm{Hol}(\mathbb{C}^{n*})=\left\{\displaystyle\prod_{k=0}^{\infty}\Ss^{k}(\mathbb{C}^{n*})\ni\hat{\omega}= (\omega_{0}\:,\omega_{1}\:,\omega_{2}\:,\dots)\:\Bigg| \sum_{k=0}^{\infty}{p}^{k}(\omega_{k})<\infty, \:\forall\: p\in \tilde{P}\right\},
        \end{equation*}und für $p_{z}=|z|p_{\max}$ bedeutet dies:
        \begin{equation}
            \label{eq:Potreiendl}
            \begin{split}
                \p_{z}(\omega)=&\:\p_{z}\left(\prod_{k=0}^{\infty}\sum_{i_{1},…,i_{k}}^{n}a_{i_{1},…,i_{k}}e^{i_{1}}\vee…\vee e^{i_{k}}\right)
                \\=&\:\sum_{k=0}^{\infty}p_{z}^{k}\left(\sum_{i_{1},…,i_{k}}^{n}a_{i_{1},…,i_{k}}e^{i_{1}}\vee…\vee e^{i_{k}}\right)
                \\=&\:\sum_{k=0}^{\infty}\sum_{i_{1},…,i_{k}}^{n}|a_{i_{1},…,i_{k}}||z|^{k}
                \\<&\:\infty.
            \end{split}
        \end{equation}
        Vermöge der bijektiven Zuordnung
        \begin{equation}
            \label{eq:isomPolSym}
            \begin{split}
                \Pol^{k}(\mathbb{C}^{n})&\longleftrightarrow \Ss^{k}(\mathbb{C}^{n*})\\
                a_{i_{1},…,i_{k}}z^{i_{1}}…z^{i_{k}}&\longleftrightarrow
                a_{i_{1},…,i_{k}} e^{i_{1}}\vee…\vee e^{i_{k}},
            \end{split}
        \end{equation}ist jedes $\Ss^{k}(\mathbb{C}^{n*})$
        isomorph zu $\Pol^{k}(\mathbb{C}^{n})$, dem Raum der Polynome
        $k$-ten Grades auf $\mathbb{C}^{n}$. Für $\left(\begin{array}{c} z_{1} \\ \vdots\\
                z_{n} \end{array}\right)\in \mathbb{C}^{n}$ sei
        $|z|=\displaystyle\max_{0\leq i\leq n}|z_{i}|$. Dann zeigen
        \eqref{eq:Potreiendl} und \eqref{eq:isomPolSym}, dass jedes
        Element aus $\mathrm{Hol(\mathbb{C}^{n*})}$ einer absolut
        konvergenten Potenzreihe auf $\mathbb{C}^{n}$
        entspricht. Umgekehrt ist für jede derartige Potenzreihe
        \begin{equation*}^{}
            \sum_{k=0}^{\infty}\sum_{i_{1},…,i_{k}}^{n}|a_{i_{1},…,i_{k}}||z|^{k}<
            \infty\qquad\forall\:|z|\geq 0,
        \end{equation*}und da $p\leq |z|\:p_{max}$ für alle $p\in
        \wt{P}$, folgt:
        \begin{equation*}
            \p\left(\sum_{k=0}^{\infty}\sum_{i_{1},…,i_{k}}^{n}a_{i_{1},…,i_{k}}e^{i_{1}}\vee…\vee
                e^{i_{k}}\right)\leq \sum_{k=0}^{\infty}\sum_{i_{1},…,i_{k}}^{n}|a_{i_{1},…,i_{k}}||z|^{k}
            < \infty.
        \end{equation*}Insgesamt zeigt dies, dass wir
        $\big(\mathrm{Hol}(\mathbb{C}^{n*}),\hat{\Pp}\big)$ mit den
        auf $\mathbb{C}^{n}$ absolut konvergenten Potenzreihen, also
        mit dem Raum der ganz holomorphen Funktionen $\mathit{Hol}(\mathbb{C}^{n})$ 
        identifizieren dürfen und liefert die Begründung für die
        Wahl der Bezeichnung $\mathrm{Hol}$. 

        Die stetige Fortsetzung
        der Halbnorm $\p_{z}$ ist dann in
        Multiindexschreibweise auch darstellbar in der Form: 
        \begin{equation*}
            \hat{\p}_{z}(\phi)=\sum_{k=0}^{\infty}\frac{|z|^{k}}{\alpha!}\left|\frac{\pt^{k}\phi}{\pt
                  x^{\alpha_{1}}…\pt x^{\alpha_{n}}}(0)\right|\quad\quad
            \forall\:\phi\in \mathit{Hol}(\mathbb{C}^{n}). 
        \end{equation*}
    \item
        Sei $\mathbb{V}=\mathbb{K}^{|\mathbb{N}|}$ oder ein anderer
        unendlichdimensionaler $\mathbb{K}$-Vektorraum und
        %
        $\V^{*}$ schwach* topologisiert,
        vermöge den Halbnormen $P^{*}=\left\{p_{v}\right\}_{v\in \V}$ mit
        \begin{equation*}
            p_{v}(\phi)=|\phi(v)|\quad\quad \forall\:\phi\in
            \V^{*}.
        \end{equation*}
        Insbesondere ist dann bereits $|c|\:p_{v}=p_{cv}$ in $P^{*}$
        enthalten und 
        $p_{\max}=\displaystyle\sum_{i=1}^{n}p_{e_{i}}$ zeigt, dass wir es hier in
        der Tat mit einer Verallgemeinerung von \textit{i.)} zu tun
        haben. Bezeichne wieder $\tilde{P}^{*}$ das filtrierende und
        separierende System aller bezüglich
        dieser Topologie stetigen Halbnormen.
        Wegen Satz \ref{satz:Potenz} dürfen wir uns
        $\mathrm{Hol}(\V^{*})$ als unendliche Potenzreihen mit
        Summanden in den $\widehat{\Ss^{k}(\V^{*})}$ vorstellen, und
        wir wollen im Folgenden zeigen, dass jedes $h\in
        \mathrm{Hol}(\V^{*})$ sogar eine komplexwertige Potenzreihe im
        Funktionensinne auf $\V$ definiert.

        Sei hierfür $\left\{h_{\alpha}\right\}_{\alpha\in I}\subseteq\Ss^{\bullet}(\V^{*})$ mit
        $\left\{h_{\alpha}\right\}_{\alpha\in I}\rightarrow h$. Dann
        ist $\left\{h_{\alpha}\right\}_{\alpha\in I}$ insbesondere ein
        Cauchynetz und somit $\p_{v}\left(h_{\alpha}-h_{\beta}\right)<
        \epsilon$ für alle $\alpha,\beta\geq \alpha_{\epsilon}$.

        Für $u_{1}\ot…\ot u_{k}\in \Tt^{k}(\V^{*})$ und $v\in \V$ ist $\tau_{v}\colon
        u_{1}\ot…\ot u_{k} \longmapsto u_{1}(v)…u_{k}(v)$, vermöge linearer
        Fortsetzung durch Korollar \ref{kor:WohldefTensorprodabbildungen}, auf
        ganz $\Tt^{k}(\V^{*})$ wohldefiniert. Sei weiter
        $\left(\Delta_{k}^{*}u^{k}\right)(v):=\tau_{v}\left(u^{k}\right)$,
        so folgt $\left(\Delta_{k}^{*}u_{1}\vee…\vee u_{k}\right)(v)=u_{1}(v)…u_{k}(v)$.
        
        Nun ist $h_{\alpha}=\displaystyle\sum_{k}u_{\alpha}^{k}$ mit
        $u_{\alpha}^{k}\in \Ss^{k}(\V^{*})$ und endlicher Summe, und
        wir definieren:
        \begin{align*}
            h(v)=\lim_{\alpha}\:\sum_{k}\left(\Delta^{*}_{k}u_{\alpha}^{k}\right)(v)\qquad \forall\:v\in \V
        \end{align*}
        Um
        die Wohldefiniertheit dieser Abbildung zu zeigen, sei $u^{k}\in \Ss^{k}(\V^{*})$, dann folgt 
        \begin{equation}
            \label{eq:abskonv}
            \begin{split}
                \left(\Delta_{k}^{*}u^{k}\right)(v)=&\sum_{i=1}^{n}u^{i}_{1}(v)…u^{i}_{k}(v)\leq \left|\sum_{i=1}^{n}u^{i}_{1}(v)…u^{i}_{k}(v)\right| 
                \\\leq&
                \sum_{i=1}^{n}\left|u^{i}_{1}(v)…u^{i}_{k}(v)\right|=
                \sum_{i=1}^{n}p_{v}(u^{i}_{1})…p_{v}(u^{i}_{k})
            \end{split}
        \end{equation}
        für alle Zerlegungen
        $\displaystyle\sum_{i=1}^{n}u^{i}_{1}\ot…\ot u^{i}_{k}$ von
        $u^{k}$ und somit $\left(\Delta_{k}^{*}u^{k}\right)(v)\leq
        p_{v}^{k}(u)$.\\
        Dies bedeutet
        \begin{align*}
            |h_{\alpha}(v)-h_{\beta}(v)|=&\:\left|\:\sum_{k}\left(\Delta^{*}_{k}\left[u_{\alpha}^{k}-u_{\beta}^{k}\right]\right)(v)\:\right|
            \leq \:\sum_{k}p_{v}^{k}\left(u_{\alpha}^{k}-u_{\beta}^{k}\right)
            \\=&\:\p_{v}(h_{\alpha}-h_{\beta})<\:\epsilon
        \end{align*}
        für alle $\alpha,\beta\geq \alpha_{\epsilon}$.  Damit ist
        $\left\{h_{\alpha}(v)\right\}_{\alpha\in I}$ ein Cauchynetz in $\mathbb{C}$
        und $h(v)=\displaystyle\lim_{\alpha} h_{\alpha}(v)$
        existiert. Für die Unabhängigkeit obiger
        Definition von der Wahl des Netzes sei
        $\Ss^{\bullet}(\V^{*})\supseteq\big\{h'_{\beta}\big\}_{\beta\in J}$ ein weiteres Netz
        mit $\big\{h'_{\beta}\big\}_{\beta\in J}\rightarrow
        h$. Dann gilt
        \begin{align*}
            |h(v)-h'_{\beta}(v)|\leq&\:
            |h(v)-h_{\alpha}(v)|+|h_{\alpha}(v)-h'_{\beta}(v)|
            \\=&\:|h(v)-h_{\alpha}(v)|+\p_{v}(h_{\alpha}-h'_{\beta}).
        \end{align*}Wegen $\left\{h_{\alpha}(v)\right\}_{\alpha\in
          I}\rightarrow h(v)$ existiert ein $\alpha_{\epsilon}\in I$, so
        dass $|h(v)-h_{\alpha}(v)|\leq \frac{\epsilon}{2}$ für alle
        $\alpha\geq \alpha_{\epsilon}$ und wegen
        $\left\{h_{\alpha}(v)\right\}_{\alpha\in
          I}\sim\big\{h'_{\beta}\big\}_{\beta\in J}$
        (vgl. Definition \ref{def:kanNetzIsoetc}~\textit{i.)}) ein $(\alpha',\beta')\in
        I\times J$, so dass $\p_{v}(h_{\alpha}-h'_{\beta})\leq
        \frac{\epsilon}{2}$ für alle $(\alpha,\beta)\geq
        (\alpha',\beta')$. Insgesamt zeigt dies
        $|h(v)-h'_{\beta}(v)|\leq \epsilon$ für alle $\alpha\geq
        \wt{\alpha}$ mit $\wt{\alpha}\geq \alpha_{\epsilon},\alpha'$, also $\big\{h'_{\beta}\big\}_{\beta\in J}\rightarrow
        h$.

        Als Beispiel sei $\hat{u}\in \widehat{\V^{*}}$, dann ist
        $\overbrace{\hat{u}\vee\dots\vee \hat{u}}^{k}\in
        \widehat{\Ss^{k}(\V^{*})}$ wegen $\widehat{\Ss^{k}\left(\V^{*}\right)}\cong
        \widehat{\Ss^{k}\big(\widehat{\V^{*}}\big)}$ nach Satz \ref{satz:PiTopsatz}~\textit{vi.)}, und wir definieren
        $\exp(\hat{u})=\displaystyle\prod_{k=0}^{\infty}\frac{1}{k!}\:\overbrace{\hat{u}\vee\dots\vee
          \hat{u}}^{k}$.\\ Dann folgt für $q=p_{v}$\::
        \begin{align*} 
            \q_{\times}(\exp(\hat{u}))=&\:\sum_{k=0}^{\infty}\frac{1}{k!}\:\hat{p_{v}^{k}}\left(\hat{u}\vee\dots\vee\hat{u}\right)=\:\sum_{k=0}^{\infty}\frac{1}{k!}\:\hat{p}_{v}^{k}\left(\hat{u}*\dots*\hat{u}\right)
            \\\leq& \:\sum_{k=0}^{\infty}\frac{1}{k!}\:\left(\hat{p}_{v}\left(\hat{u}\right)\right)^{k}<\infty,
        \end{align*}
        also $\exp(\hat{u})\in \mathrm{Hol}(\V^{*})$. Dabei folgt die letzte Ungleichung mit der
        Submultiplikativität von $(\mathrm{Hol},\hat{\Pp},*)$ nach
        Proposition \ref{prop:HollkAlgebra}~\textit{i.)}.

        Allgemein funktioniert diese Konstruktion für alle absolut konvergenten Potenz-reihen
        $\displaystyle\sum_{k=0}^{\infty}a_{k}z^{k}$ auf
        $\mathbb{C}$. 

        Ganz allgemein können wir also die Elemente aus $\Hol$ immer als so
        etwas, wie ganz holomorphen Funktionen auf dem
        Prädualraum $\V_{*}$\footnote{$(\V_{*})^{*}=\V$} auffassen, sofern er
        existiert.
    \item
        Das Resultat aus \textit{ii.)} lässt sich auch allgemeiner
        formulieren. Seien hierfür $\V$ ein $\mathbb{K}$-Vektorraum
        und $(\mathbb{U},P)$ ein hausdorffscher, lokalkonvexer $\mathbb{K}$-Vektorraum
        derart, dass eine $\mathbb{K}$-bilineare Abbildung    
        $\tau\colon \V\times \mathbb{U}\colon \longrightarrow \mathbb{K}$
        existiert, deren Bild sich für festes $v\in \V$ 
        in der Form 
        \begin{equation*}
            \tau(v,u) \leq p_{v}(u)\qquad p_{v}\in P,\:\forall\:u\in \mathbb{U}
        \end{equation*}abschätzen lässt. Sei
        $\Ss^{\bullet}(\mathbb{U})$, in gewohnter Weise, durch das System $\tilde{P}$ aller bezüglich $P$ stetigen Halbnormen
        topologisiert und $\left(\Delta_{k}^{*}u^{k}\right)\colon
        v\longrightarrow \mathbb{K}$ durch lineare Fortsetzung von
        \begin{equation*}
            \left(\Delta_{k}^{*}u_{1}\ot…\ot u_{k}\right)(v)=\tau(v,u_{1})…\tau(v,u_{k})\qquad\forall\: u_{1}\ot…\ot u_{k} \in \Tt^{k}(\mathbb{U})
        \end{equation*}auf ganz $\Tt^{k}(\mathbb{U})$ definiert.
        Dann folgt wie in \textit{ii.)}, dass
        $\left(\Delta_{k}^{*}u^{k}\right)(v)\leq p_{v}^{k}(u^{k})$ für
        alle $u^{k}\in \Ss^{k}(\mathbb{U})$ gilt und wir erhalten durch
        \begin{align*}
            h(v)=\lim_{\alpha}\:\sum_{k}\left(\Delta^{*}_{k}u_{\alpha}^{k}\right)(v)\qquad \forall\:v\in \V
        \end{align*}mit $h\in \mathrm{Hol}(\mathbb{U})$ und
        $\Ss^{\bullet}(\mathbb{U})\supseteq \{h_{\alpha}\}_{\alpha\in
          I}\rightarrow h$ wieder eine wohldefinierte $\mathbb{K}$-wertige
        Potenzreihenfunktion auf $\V$. Die restlichen Aussagen aus
        \textit{ii.)} gelten dann analog.

        Physikalisch relevant sind beispielsweise die Kombinationen
        
        \begin{table*}[h]
            \centering
            \begin{tabular}{|c|c|}
                $\V$ & $\mathbb{U}$\\\hline
                $\mathcal{D}(X)$ &
                $\big(\mathcal{E}(X),\T_{\mathcal{E}}\big)$, $\big(\mathcal{D}(X),\T_{\mathcal{E}}\big)$, $\big(\mathcal{D}(X),\T_{\mathcal{D}}\big)$\\
                $\mathcal{D}'(X)$ &
                $\big(\mathcal{D}(X),\T_{\mathcal{D}}\big)$, $\big(\mathcal{D}_{K}(X),\T_{\mathcal{D}_{K}}\big)$\\
                $\mathcal{E}'(X)$ &
                $\big(\mathcal{E}(X),\T_{\mathcal{E}}\big)$,
                $\big(\mathcal{D}(X),\T_{\mathcal{D}}\big)$, $\big(\mathcal{D}(X),\T_{\mathcal{E}}\big)$\\          
            \end{tabular}
        \end{table*}
        mit einer offenen Teilmenge $X\subseteq
        \mathbb{R}^{n}$.
        Hierbei bezeichnet $\mathcal{E}(X)$ die
        glatten Funktionen $X\longrightarrow \mathbb{R}$,
        $\mathcal{D}(X)\subseteq \mathcal{E}(X)$ die glatten 
        Funktionen $X\longrightarrow \mathbb{R}$ mit kompakten Träger
        $K\subseteq X$ und $\mathcal{D}_{K}(X)\subseteq \mathcal{D}(X)$ die glatten Funktionen
        $X\longrightarrow \mathbb{R}$ mit kompaktem Träger
        $K'\subseteq K\subseteq X$, wobei hier $K$ ein fest gewähltes
        Kompaktum ist.

        $\T_{\mathcal{E}}$ ist die durch das filtrierende,
        abzählbare System $P_{\mathcal{E}}$, bestehend aus den Halbnormen
        \begin{equation*}
            p_{K,l}(\phi)=\sup_{\substack{|\alpha|\leq l\\ x\in
                K}}\left|\pt^{\alpha}\phi(x)\right|\qquad
            \forall\:\phi\in \mathcal{E}(X)
        \end{equation*}mit $\alpha\in \mathbb{N}^{n}$ ein Multiindex, $l\in \mathbb{N}$, $K\subseteq X$ kompakt sowie $\displaystyle\pt^{\alpha}=\frac{\pt^{|\alpha|}}{\pt^{\alpha_{1}}
          x_{1}…\pt^{\alpha_{n}}x_{n}}$, erzeugte lokalkonvexe
        Topologie.

        $\T_{\mathcal{D}}$ ist die lokalkonvexe
        $\mathcal{D}(X)$-Raum Topologie (vgl. \cite[Def
        6.3]{rudin:1991a}), deren erzeugendes Halbnormensystem
        eher formaler Natur
        ist\footnote{vgl. Minkowski-Funktional:
          Bemerkung \ref{bem:Minkowski}}. 
        $\T_{\mathcal{D}_{K}}$ ist die durch dass 
        Halbnormensystem $P_{K,l}=\{p_{K,l}\}_{l\in
          \mathbb{N}}$ erzeugte, lokalkonvexe Topologie, wobei
        hier $K$ wieder fest vorgegeben ist. 

        $\mathcal{D}'(X)$ ist der
        zu $(\mathcal{D}(X),\T_{\mathcal{D}})$
        topologische Dualraum, weshalb nach Satz \ref{satz:stetmultabb}              
        \begin{equation*}
            \tau(v,u):=v(u)\leq|v(u)|\leq
            \overbrace{|c|\:\tilde{p}(u)}^{\tilde{p}'}\qquad
            v\in \mathcal{D}'(X),\:
            \forall\:u\in \mathcal{D}(X)
        \end{equation*}für eine bezüglich $\T_{\mathcal{D}}$
        stetigen Halbnorm $\tilde{p}\in
        \tilde{P_{\mathcal{D}}}$ und ein $|c|> 0$ gilt. Man beachte, dass dann $\tilde{p}'$ ebenfalls
        wieder stetig ist. Nun lässt sich zeigen (vgl. \cite[Thm
        6.6]{rudin:1991a}), dass eine lineare Abbildung
        $\phi\colon \mathcal{D}(X)\longrightarrow \mathbb{M}$
        in einen weiteren lokalkonvexen Vektorraum
        $(\mathbb{M},Q)$ genau dann stetig bezüglich
        $\T_{\mathcal{D}}$ ist, wenn für jedes Kompaktum
        $K\subseteq X$ die Einschränkung
        $\phi\big|_{\mathcal{D}_{K}}$ stetig bezüglich
        $\T_{\mathcal{D}_{K}}$ ist. Dies zeigt         
        \begin{equation*}
            \tau(v,u):=v(u)\leq|v(u)|\leq p_{K,l}(u)\qquad\forall\: u\in
            \mathcal{D}_{K}(X) 
        \end{equation*}und somit die zweite Zeile obiger
        Tabelle. 

        $\mathcal{E}'(X)$ ist der topologische
        Dualraum von $\mathcal{E}(X)$, womit wir
        Definitionsgemäß die Abschätzbarkeit
        \begin{equation*}
            \tau(v,u):=v(u)\leq |v(u)|\leq |c|\:p_{K,l}(u)\qquad\forall\: u\in \mathcal{E}(X)
        \end{equation*}mit $v\in \mathcal{E}'(X)$ und
        $|c|\:p_{K,l}\in \tilde{P}_{\epsilon}$ erhalten. Der
        Rest der dritten Zeile folgt dann unmittelbar aus dem
        bereits gezeigten sowie
        $\mathcal{E}'(X)\subseteq
        \mathcal{D}'(X)$\footnote{Dies sind gerade die
          Elemente aus $\mathcal{D}'(X)$ mit kompakten Träger,
          wie bsp. $\delta_{z}\colon \phi\longmapsto \phi(z)$.} und
        $\mathcal{D}(X)\subseteq \mathcal{E}(X)$. Dabei
        beachte man, dass
        $\big(\mathcal{D}(X),\T_{\mathcal{E}}\big)$ im
        Gegensatz zu
        $\big(\mathcal{E}(X),\T_{\mathcal{E}}\big)$ und\\
        $\big(\mathcal{D}(X),\T_{\mathcal{D}}\big)$ weder
        vollständig noch Folgen vollständig ist.

        Für die erste Zeile erhalten wir mit 
        $\mathcal{D}(X)\ni v\neq0$ und $\supp(v)=K\subseteq X$, dass
        \begin{equation}
            \label{eq:stetab}
            \begin{split}
                \tau_{\alpha}(v,u):=&\:\int_{X}
                v(x)\pt^{\alpha}u(x)dx
                \\=&\:\int_{K}
                v(x)\pt^{\alpha}u(x)\leq \overbrace{\sup_{x\in
                    K}|v(x)|\mathrm{Vol}(K)}^{|c|>0}\:p_{K,|\alpha|}(u)
            \end{split}
        \end{equation}für alle $u\in \mathcal{E}(X)$ und somit
        auch für alle $u\in \mathcal{D}(X)$ gilt. Wegen
        $|c|\:p_{K,l}\in \tilde{P}_{\mathcal{E}}$ begründet dies
        die ersten beiden Kombinationen. Für die letzte
        überlegt man sich, dass die Halbnormen $p_{l}(u)=
        \sup_{\substack{x\in X\\ |\alpha|\leq
            l}}|\pt^{\alpha}u(x)|$ mit $l\in \mathbb{N}$ und
        somit auch $c\: p_{l}$  in $\tilde{P}_{\mathcal{D}}$
        enthalten ist. 
        Nun gilt
        $\mathcal{D}'_{K}(X)\subseteq
        \mathcal{D}^{*}_{K}(X)$, $\mathcal{D}'(X)\subseteq
        \mathcal{D}^{*}(X)$,
        $\mathcal{E}'(X)\subseteq \mathcal{E}^{*}(X)$ und mit
        \eqref{eq:stetab} erhalten wir die Stetigkeitsabschätzung: 
        \begin{equation*}
            \phi\colon \psi\longmapsto \tau(\phi,\psi)\leq |c|\:
            p_{K,l}(\psi)\qquad\forall\:\psi\in \mathcal{E}(X).
        \end{equation*} Hiermit ist jedes $\phi\in \mathcal{D}(X)$ auch als Element
        in $\mathcal{E}'(X)$ auffassbar. Umgekehrt
        überlegt man sich, dass auch jedes $\psi\in
        \mathcal{E}(X)$ als Element in $\mathcal{D}_{K}'(X)$ und mit dem Stetigkeitskriterium der
        $\mathcal{D}(X)$-Raum Topologie dann ebenfalls als Element in $\mathcal{D}'(X)$ aufgefasst
        werden kann. Zusammen mit \textit{ii.)} liefert dies Kombinationen der Form: 

        \begin{table}[h]  
            \centering
            \begin{tabular}{|c|c|}
                $\V$ & $\mathbb{U}$\\\hline
                $\mathcal{D}_{K}(X)$ &
                $\big(\mathcal{D}'_{K}(X),\T^{*}\big)$, $\big(\mathcal{E}(X),\T^{*}\big)$ \\
                $\mathcal{D}(X)$ &
                $\big(\mathcal{D}'(X),\T^{*}\big)$, $\big(\mathcal{E}(X),\T^{*}\big)$\\
                $\mathcal{D}(X)$, $\mathcal{E}(X)$ &
                $\big(\mathcal{E}'(X),\T^{*}\big)$, $\big(\mathcal{D}(X),\T^{*}\big)$ \\   
            \end{tabular}
        \end{table}
        Hierbei bezeichnen $\T^{*}$ die
        jeweiligen schwach*-Topologien. In der letzten Zeile haben wir
        $\mathcal{E}'(X)\subseteq\mathcal{D}'(X)$
        benutzt und insgesamt sind hier natürlich noch sehr viel mehr
        Kombinationsmöglichkeiten erlaubt. 

        Ein einfaches Beispiel für $\V=\mathcal{D}(X)$ und
        $\mathbb{U}=\big(\mathcal{E}(X), \T_{\mathcal{E}}\big)$ ist
        dann nach \textit{ii.)}
        \begin{align*}
            \exp(\phi)(\psi)=&\:\sum_{k=0}^{\infty}\frac{1}{k!}\big(\Delta_{k}^{*}\overbrace{\phi\vee…\vee
              \phi}^{k-mal}\big)(\psi)
            \\=&\:\sum_{k=0}^{\infty}\frac{1}{k!}\overbrace{u(\psi,\phi)…u(\psi,\phi)}^{k-mal} 
            =\sum_{k=0}^{\infty}\frac{1}{k!}\left(\int_{X}\phi(x)\psi(x)dx\right)^{k}
        \end{align*}mit $\phi\in \mathcal{E}(X)$ und $\psi\in
        \mathcal{D}(X)$ oder für $\V=\mathcal{D}(X),\mathcal{E}(X)$ und $\mathbb{U}=\big(\mathcal{E}'(X), \T^{*}\big)$:
        \begin{equation*}
            \exp(\delta_{z})(\psi)=\sum_{k=0}^{\infty}\frac{1}{k!}\left(\delta_{z}(\psi)\right)^{k}=\sum_{k=0}^{\infty}\frac{1}{k!}\psi(z)^{k}
        \end{equation*}mit $\psi\in \mathcal{E}(X),\mathcal{D}(X)$ und
        $\delta_{z}\in \mathcal{E}'(X)\subseteq \mathcal{D}'(X)$.
        Im Allgemeinen sind natürlich auch sehr viel komplexere
        Summanden erlaubt. Wie diese im konkreten aussehen dürfen,
        hängt dann natürlich auch stark davon ab, ob $\mathbb{U}$
        vollständig ist oder nicht. In den vollständigen Fällen
        $\big(\mathcal{E}(X),\T_{\mathcal{E}}\big)$,
        $\big(\mathcal{D}(X),\T_{\mathcal{D}}\big)$,
        $\big(\mathcal{D}_{K}(X),\T_{\mathcal{D}_{K}}\big)$
        beispielsweise, braucht man $\Ss^{k}(\mathcal{E}(X))$,
        $\Ss^{k}(\mathcal{D}(X))$ und $\Ss^{k}(\mathcal{D}_{K}(X))$ nach
        Satz \ref{satz:PiTopsatz}~\textit{vii.)} nur in den $\pi_{k}$-Topologien
        vervollständigen und man kann sich überlegen, dass diese als
        Teilräum von $\mathcal{E}\left(X^{k}\right)$, $\mathcal{D}\left(X^{k}\right)$
        bzw. $\mathcal{D}_{K}\left(X^{k}\right)$ auffassbar sind. Als
        Realisierung des Tensorproduktes nimmt man dann beispielsweise
        den Teilraum        
        {\small\begin{equation*}
              \mathcal{E}_{\mathrm{sep}}\left(X^{k}\right)=\left\{\phi\in
                  \mathcal{E}\left(X^{k}\right)\:\Bigg|\:\phi(x_{1},…,x_{k})=\sum_{i=1}^{n}\phi_{1}(x_{1})…\phi_{k}(x_{k})\:\:\forall\:(x_{1},…,x_{k})\in
                  X^{k}\right\}
          \end{equation*}}und rechnet für separable $\phi_{1}\dots \phi_{k}\in
        \mathcal{E}(X)$ nach, dass
        \begin{equation*}
            p_{K_{1}\times…\times
              K_{k},l}(\phi_{1}…\phi_{k})\leq
            p_{K_{1},l}(\phi_{1})…p_{K_{k},l}(\phi_{k})
        \end{equation*}
        und somit ebenfalls $p_{K_{1}\times…\times
          K_{k},l}(\phi_{\mathrm{sep}})\leq
        p^{k}(\phi_{\mathrm{sep}})$ für alle $\phi_{\mathrm{sep}}\in\mathcal{E}_{\mathrm{sep}}(X)$ gilt. Dies bedeutet, dass die
        $\pi_{k}$-Topologie auf
        $\mathcal{E}_{\mathrm{sep}}\left(X^{k}\right)$ feiner ist, 
        als die durch $\left(\mathcal{E}\left(X^{k}\right)
            \T_{\mathcal{E}}\right)$ auf
        $\mathcal{E}_{\mathrm{sep}}\left(X^{k}\right)$ induzierte
        Teilraumtopologie. Da $\left(\mathcal{E}\left(X^{k}\right),
            \T_{\mathcal{E}}\right)$ vollständig ist, bedeutet dies,
        dass die Vervollständigung von
        $\mathcal{E}_{\mathrm{sep}}\left(X^{k}\right)$ bezüglich
        $\pi_{k}$ in $\mathcal{E}\left(X^{k}\right)$ enthalten sein muss. 
        In der Tat lässt sich sogar zeigen, dass diese wegen der
        Nuklearität von $\mathcal{E}(X)$
        übereinstimmen, siehe \cite[Thm 51.6]{treves:1967a}. Die
        Vervollständigung von $\Ss^{k}(\mathcal{E}(X))$
        besteht dann gerade aus allen total symmetrischen $\phi\in
        \mathcal{E}\left(X^{k}\right)$. Hiermit lässt sich zeigen,
        dass dann ebenfalls alle
        Potenzreihen der Form
        \begin{equation*}
            p(\psi)=\sum_{k=0}^{\infty}\int_{X_{1}\times…\times X_{k}}\phi_{k}(x_{1},…,x_{k})\psi(x_{1})…\psi(x_{k})d_{x_{1}}…d_{x_{k}}
        \end{equation*}
        mit total symmetrischen Elementen $\phi_{k}\in
        \mathcal{E}\left(X^{k}\right)$ und
        $\sum_{k=0}^{\infty}\hat{p^{k}}(\phi_{k})<\infty$
        durch Elemente aus
        $\Hol(\mathcal{E}(X))$ induziert werden können.
        Die
        gleiche Aussage erhalten wir ebenfalls für $(\mathcal{D}_{K}(X),\T_{\mathcal{D}_{K}})$.
        Im Falle $(\mathcal{D}(X),\T_{\mathcal{E}})$
        ist allerdings auch die Vervollständigung $\hat{\mathcal{D}}(X)$ von
        $\mathcal{D}(X)$ bezüglich $\T_{\mathcal{E}}$
        zu berücksichtigen und dies gilt natürlich auch für die
        obigen schwach*-topologisierten Varianten.



    \end{enumerate}
\end{beispiel}
Folgender Satz klärt die Gestalt der
Hochschild-Kohomologien von $\Hol$ für vollständige, hausdorffsche,
lokalkonvexe $\Hol-\Hol$-Bimoduln $\mathcal{M}$. Dabei stellt
die Vollständigkeit für uns in der Tat eine unverzichtbare
Grundvoraussetzung dar. Man beachte, dass dann der wichtige
symmetrische Spezialfall $\mathcal{M}=\Hol$ in diesem Rahmen komplett
behandelbar sein wird.\\
\begin{satz}
    \label{satz:HochschildHol}
    Sei $\mathcal{M}$ ein vollständiger, hausdorffscher, lokalkonvexer
    $\Hol-\Hol$-Bimodul und
    \begin{align*}
        \tau^{k}\colon HC^{k}_{\cont}\big(\Hol,\mathcal{M}\big)&\longrightarrow HC^{k}_{\cont}\big(\Ss^{\bullet}(\V),\mathcal{M}\big)\\
        \hat{\phi}&\longmapsto \hat{\phi}\big|_{HC^{k}_{\cont}\big(\Ss^{\bullet}(\V),\mathcal{M}\big)}
    \end{align*}die durch (\ref{eq:Hochschilddelta}) für
    $\mathcal{A}=\Hol$ bzw. $\mathcal{A}=\SsV$ definierten Kettendifferentiale.
    Dann induzieren die Abbildungen
    \begin{align*}
        \tau^{k}\colon HC^{k}_{\cont}\big(\Hol,\mathcal{M}\big)&\longrightarrow HC^{k}_{\cont}\big(\Ss^{\bullet}(\V),\mathcal{M}\big)\\
        \hat{\phi}&\longmapsto \hat{\phi}\big|_{HC^{k}_{\cont}\big(\Ss^{\bullet}(\V),\mathcal{M}\big)}
    \end{align*} einen Kettenisomorphismus
    $\tau:\big(HC_{\cont}\big(\Hol,\mathcal{M}\big),\hat{\delta}_{c}^{k}\big)\longrightarrow
    \big(HC_{\cont}\big(\Ss^{\bullet}(\V),\mathcal{M}\big),\delta_{c}^{k}\big)$
    und es gilt:
    \begin{equation*}
        HH^{k}_{\cont}\big(\Hol,\mathcal{M}\big)\cong HH^{k}_{\cont}\big(\Ss^{\bullet}(\V),\mathcal{M}\big). 
    \end{equation*} Ist $\mathcal{M}$ zudem symmetrisch, so ist
    \begin{equation*}        
        HH^{k}_{\cont}(\Hol,\mathcal{M})\cong \Hom_{\mathcal{A'}^{e}}^{\cont}\big(\K^{c}_{k},\mathcal{M}\big)
    \end{equation*} mit $(\K_{c},\pt_{c})$ der Koszul-Komplex über $A'=\Ss^{\bullet}(\V)$.
    \begin{beweis}
        Da jeder $\Hol-\Hol$-Bimodul insbesondere ein
        $\Ss^{\bullet}(\V)-\Ss^{\bullet}(\V)$-Bimodul ist und sich alle
        angeführten Eigenschaften auf die Unteralgebra übertragen, folgt
        die zweite Isomorphie mit Satz \ref{satz:stetigHochschSym} aus
        der ersten, und für diese reicht es nach Lemma \ref{lemma:kettenabzu}~\textit{ii.)}, die
        Kettenisomorphismus-Eigenschaft von $\tau$ nachzuweisen.\\\\
        Zunächst folgt mit Satz \ref{satz:stetfortsBillphi} und Bemerkung \ref{bem:stetfortmult}~\textit{ii.)}, dass die $\tau^{k}$ Isomorphismen mit
        stetiger Fortsetzung $\tau^{k}_{-1}$ als Umkehrabbildung sind.
        Dabei folgt $\widehat{\phi+\psi}=\hat{\phi}+\hat{\psi}$, also die
        Linearität von $\tau_{-1}^{k}$, sofort mit der Stetigkeit der
        Addition.

        Die Kettenabbildungs-Eigenschaft erhalten wir unmittelbar aus der Definition der
        $\Hol$-Algebramultiplikation $*$, da hiermit 
        \begin{equation}
            \label{eq:muhkuhmilch}
            \hat{\delta}^{k}_{c}\Big(\hat{\phi}\Big)\Big|_{\SsV^{k+1}}=\delta^{k}_{c}\left(\hat{\phi}\big|_{\SsV^{k}}\right), 
        \end{equation}
        also $\tau^{k+1}\cp \hat{\delta}^{k}_{c}= \delta^{k}_{c}\cp
        \tau^{k}$ gilt.
        Dies zeigt die Behauptung.
    \end{beweis}
\end{satz}
Wir wollen die Isomorphien in Satz \ref{satz:HochschildHol} noch ein
wenig n"aher betrachten.\\
\begin{definition}[Koszul-Komplex und Vervollständigter Koszul-Komplex]
    \label{def:vervollstBarKoszulkompl}
    \begin{enumerate}
    \item
        Bezeichne $\left(\K'_{c},\pt'_{c}\right)$ den topologischen Koszul-Komplex der Algebra
        $\mathcal{A}'=\Ss^{\bullet}(\V)$ und $*_{S}$ die zugeh"orige $\Ss^{\bullet}(\V)\pite
        \Ss^{\bullet}(\V)$-Modul-Multiplikation.

        Mit Hilfe von Satz \ref{satz:PiTopsatz}~\textit{vi.)} definieren wir:
        {\allowdisplaybreaks
          \begin{align*}
              \cK_{k}^{c}=\widehat{\K'^{c}_{k}}=&\widehat{\Bigg(\Ss^{\bullet}(\V)\pite
                \Ss^{\bullet}(\V) \pite
                \Lambda^{k}(\V)\Bigg)}=\:\widehat{\Bigg(\widehat{\Ss^{\bullet}(\V)}\pite
                \widehat{\Ss^{\bullet}(\V)}\pite
                \widehat{\Lambda^{k}(\V)}\Bigg)}
              \\=& \:\widehat{\Bigg(\Hol\pite \Hol \pite
                \widehat{\Lambda^{k}(\V)}\Bigg)}
              =\:\widehat{\Bigg(\Hol\pite \Hol \pite
                \Lambda^{k}(\V)\Bigg)}
          \end{align*}}sowie
        \begin{equation*}
            \cK_{0}^{c}=\widehat{\Big(\Ss^{\bullet}(\V)\pite
              \Ss^{\bullet}(\V)\Big)}=\widehat{\Big(\Hol\pite \Hol\Big)}= \hat{\mathcal{A}^{e}}.
        \end{equation*}

        Wie in Proposition \ref{prop:HollkAlgebra}~\textit{ii.)} werden diese,
        vermöge stetiger Fortsetzung $\hat{*}_{S}$ von $*_{S}$, zu
        hausdorffschen, lokalkonvexen
        $\hat{\mathcal{A}^{e}}$-Linksmoduln. Zudem erhalten wir stetige Fortsetzungen
        $\hat{\pt}^{c}_{k}$ der Kettendifferentiale $\pt'^{c}_{k}$ , die
        wegen
        \begin{align*}
            \hat{\pt}^{c}_{k}\:\left(\hat{a}^{e}\:\hat{*}_{S}\:
                \hat{\kappa}^{k}\right)=&
            \:\hat{\pt}^{c}_{k}\left(\lim_{\alpha\times\beta}\left[a^{e}_{\alpha}\:*_{S}\:
                    \kappa^{k}_{\beta}\right]\right)=\lim_{\alpha\times\beta}\pt'^{c}_{k}\left(a^{e}_{\alpha}\:*_{S}\:
                \kappa^{k}_{\beta}\right)
            \\=&\lim_{\alpha\times\beta}\left[a^{e}_{\alpha}\:*_{S}\:
                \pt'^{c}_{k}\big( \kappa^{k}_{\beta}\big)\right]=\hat{a}^{e}\:\hat{*}_{S}\: \hat{\pt}^{c}_{k}\big(\hat{\kappa}^{k}\big)
        \end{align*}$\hat{\mathcal{A}}^{e}$-linear sind.
        Den so gewonnenen topologischen Kettenkomplex
        $\big(\cK_{c},\cpt_{c}\big)$ bezeichnen wir als vervollständigten
        Koszul-Komplex über $\mathcal{A}=\Hol$.
    \item 
        Mit $\K^{c}_{k}$ benennen wir die
        hausdorffschen, lokalkonvexen $\mathcal{A}^{e}$-Linksmoduln
        \begin{equation*}
            \K^{c}_{0}=\Hol\pite\Hol \quad\text{ sowie }\quad \K^{c}_{k}=\Hol\pite\Hol\pite \Lambda^{k}(\V)
        \end{equation*}
        mit der offensichtlichen
        $\mathcal{A}^{e}$-Multiplikation $*_{Hol}$ in den ersten beiden
        Faktoren. Diese induziert dann ebenfalls eine stetige
        $\hat{\mathcal{A}}^{e}$-Multiplikation $\hat{*}_{Hol}$ auf
        $\cK_{k}^{c}$, und da
        \begin{equation*}
            *_{Hol}\big|_{\Ss^{\bullet}(\V)\pite \Ss^{\bullet}(\V)\times
              \Lambda^{k}(\V)}=*_{S}=\hat{*}_{S}\big|_{\Ss^{\bullet}(\V)\pite \Ss^{\bullet}(\V)\times
              \Lambda^{k}(\V)},
        \end{equation*} stimmen beide auf einer dichten Teilmenge
        von $\cK_{c}^{k}$ überein. Mit der Eindeutigkeit der
        stetigen Fortsetzung von $*_{S}$ ist 
        \begin{equation}
            \label{eq:Modulmultsallegleich}
            \hat{*}_{S}=\hat{*}_{Hol}\qquad\text{und somit}\qquad\hat{*}_{S}\big|_{\K^{c}_{k}}=*_{Hol}.
        \end{equation}
        $(\K^{c}_{k},\pt_{c})$ bezeichne dann den
        Kettenkomplex mit Kettendifferentialen
        $\pt^{c}_{k}=\hat{\pt}^{c}_{k}\big|_{\K^{c}_{k}}$, für dessen
        Wohldefiniertheit wir zeigen müssen, dass die $\pt^{c}_{k}$
        ausschließlich nach $\K^{c}_{k-1}\subseteq \hat{\K}^{c}_{k-1}$
        abbilden, und nicht in $\hat{\K}^{c}_{k-1}\backslash \K^{c}_{k-1}$
        landen. Hierbei bedeutet $\backslash$ die mengentheoretische Differenz. 

        Sei hierfür  
        $\Ss^{\bullet}(\V)\supseteq\net{x}{I}\rightarrow x\in \Hol$ und
        $\Ss^{\bullet}(\V)\supseteq\{y_{\beta}\}_{\beta\in
          J}\rightarrow y\in \Hol$. Dann folgt $\{x_{\alpha}\pite
        y_{\beta}\pite u\}_{\alpha\times \beta \in I\times J}\rightarrow
        x\pite y\pite u \in \K^{c}_{k}$, vermöge zweimaliger Anwendung
        der Dreiecksungleichung und der Definition der $\pi_{3}$-Halbnormen
        (siehe auch Beweis zu
        Satz \ref{satz:PiTopsatz}~\textit{vi.)}). Wir erhalten
        \begin{align*}
            \pt^{c}_{k}(x&\pite y \pite u)=\lim_{\alpha\times
              \beta}\pt'^{c}_{k}\:(x_{\alpha}\pite y_{\beta}\pite u)\\=&\:\lim_{\alpha\times
              \beta}\:\sum_{j=1}^{k}(-1)^{j-1}(u_{j}\vee x_{\alpha}\pite
            y_{\beta}\pite u^{j}) - \lim_{\alpha\times
              \beta}\:\sum_{j=1}^{k}(-1)^{j-1}(x_{\alpha}\pite
            u_{j}\vee y_{\beta}\pite u^{j})
            \\=&\:    \sum_{j=1}^{k}(-1)^{j-1}(u_{j}* x \pite
            y \pite u^{j}) - \sum_{j=1}^{k}(-1)^{j-1}(x\pite
            u_{j}* y\pite u^{j})\in \K^{c}_{k-1}
        \end{align*}
        mit gleicher Argumentation wie oben, da definitionsgemäß $\{u_{j}\vee
        x_{\alpha}\}_{\alpha\in I}\rightarrow u_{j}* x$ und $\{u_{j}\vee
        y_{\beta}\}_{\beta\in J}\rightarrow u_{j}* y$.
    \end{enumerate}
\end{definition} 
\begin{lemma}
    \label{lemma:fgh}
    Gegeben ein vollständiger, hausdorffscher, lokalkonvexer,
    $\Hol-\Hol$-Bimodul $\mathcal{M}$, so ist
    \begin{equation*}
        \Big(\Hom^{\cont}_{\mathcal{A}^{e}}(\K_{c},\mathcal{M}),\pt^{*}_{c}\Big)\cong \Big(\Hom^{\cont}_{\hat{\mathcal{A}}^{e}}(\cK_{c},\mathcal{M}),\hat{\pt}^{*}_{c}\Big)\cong \Big(\Hom^{\cont}_{\mathcal{A}'^{e}}(\K'_{c},\mathcal{M}),\pt'^{*}_{c}\Big),
    \end{equation*}wobei $\cong$ Kettenisomorphie vermöge
    Einschränkung und stetiger Fortsetzung bedeutet. Des Weiteren gilt:
    \begin{equation*}
        H^{k}\Big(\Hom^{\cont}_{\mathcal{A}^{e}}(\K_{c},\mathcal{M}),\pt^{*}_{c}\Big)\cong H^{k}\Big(\Hom^{\cont}_{\hat{\mathcal{A}}^{e}}(\cK_{c},\mathcal{M}),\hat{\pt}^{*}_{c}\Big)\cong H^{k}\Big(\Hom^{\cont}_{\mathcal{A}'^{e}}(\K'_{c},\mathcal{M}),\pt'^{*}_{c}\Big).
    \end{equation*} 
    \begin{beweis}      
        Für jedes $\phi \in
        \Hom^{\cont}_{\mathcal{A}^{e}}(\K^{c}_{k},\mathcal{M})$ existiert eine
        eindeutige lineare Fortsetzung $\hat{\phi}\in
        \Hom^{\cont}_{\hat{\mathcal{A}}^{e}}(\hat{\mathcal{K}}_{k}^{c},\mathcal{M})$
        mit $\hat{\phi}\big|_{\K^{c}_{k}}=\phi$. 
        Umgekehrt erhalten wir nach \eqref{eq:Modulmultsallegleich} aus jedem $\hat{\phi}\in
        \Hom^{\cont}_{\hat{\mathcal{A}}^{e}}(\hat{\mathcal{K}}_{k}^{c},\mathcal{M})$,
        vermöge Einschränkung, ein $\phi\in
        \Hom^{\cont}_{\mathcal{A}^{e}}(\K^{c}_{k},\mathcal{M})$,
        dessen stetige lineare Fortsetzung $\hat{\phi}$ ist. Dies zeigt
        $\Hom^{\cont}_{\mathcal{A}^{e}}(\K_{c},\mathcal{M})\cong
        \Hom^{\cont}_{\hat{\mathcal{A}}^{e}}(\hat{\mathcal{K}}_{c},\mathcal{M})$, und wegen      
        \begin{equation*}
            \hat{\pt}^{c*}_{k+1}\Big(\hat{\phi}\Big)\Big|_{\mathcal{K}^{c}_{k+1}}=
            \left(\hat{\phi}\:\cp\:
                \hat{\pt}^{c}_{k+1}\right)\Big|_{\mathcal{K}^{c}_{k+1}}=
            \hat{\phi}\big|_{\mathcal{K}^{c}_{k}}\cp\: \hat{\pt}^{c}_{k+1}\big|_{\mathcal{K}^{c}_{k+1}}= 
            \hat{\phi}\big|_{\mathcal{K}^{c}_{k}}\cp\: \pt^{c}_{k+1} =\pt^{c*}_{k+1}\left(\hat{\phi}\big|_{\mathcal{K}^{c}_{k}}\right),
        \end{equation*}haben wir es hierbei wieder mit einem
        Kettenisomorphismus zu tun. Dies zeigt die jeweils erste Isomorphie,
        und die zweite folgt ganz analog.
    \end{beweis}
\end{lemma}
%
\begin{korollar}
    \label{kor:hgf}
    Mit den Voraussetzungen aus Satz \ref{satz:HochschildHol} gilt:
    \begin{equation*}
        HH^{k}_{\cont}\Big(\Hol,\mathcal{M}\Big)\cong
        H^{k}\Big(\Hom_{\mathcal{A}'^{e}}^{\cont}(\K'_{c},\mathcal{M}),\pt'^{*}_{c}\Big)\cong
        H^{k}\Big(\Hom_{\mathcal{A}^{e}}^{\cont}(\K_{c},\mathcal{M}),\pt^{*}_{c}\Big).
    \end{equation*} Ist $\mathcal{M}$ symmetrisch, so ist:
    \begin{equation*}        
        HH^{k}_{\cont}\big(\Hol,\mathcal{M}\big)\cong  \Hom_{\mathcal{A}'^{e}}^{\cont}\big(\K'^{c}_{k},\mathcal{M}\big)\cong \Hom_{\mathcal{A}^{e}}^{\cont}\big(\K^{c}_{k},\mathcal{M}\big).
    \end{equation*}
\end{korollar}
Lemma \ref{lemma:fgh} und Korollar \ref{kor:hgf} kann man auch so auffassen, dass
es für vollständige, hausdorffsche $\Hol-\Hol$-Bimoduln es egal ist, ob wir $\hom^{\cont}_{\mathcal{A}'^{e}}(\cdot,\mathcal{M})$
auf $(\K'_{c},\pt'^{*}_{c})$, $\hom^{\cont}_{\mathcal{A}^{e}}(\cdot,\mathcal{M})$
auf $(\K_{c},\pt^{*}_{c})$ oder $\hom^{\cont}_{\hat{\mathcal{A}^{e}}}(\cdot,\mathcal{M})$
auf $(\hat{\K}_{c},\hat{\pt}^{*}_{c})$ anwenden. Wir erhalten in jedem
Fall den "`gleichen"' Kokettenkomplex mit 
den "`gleichen"' Kohomologien-Gruppen.
\begin{Bemerkung}[Nicht vollständige Bimoduln]
    Abschließend wollen wir noch erklären, warum die Berechnung der
    Hochschild-Koho-mologien für nicht vollständige Bimoduln $\mathcal{M}$ mit Hilfe
    der uns in diesem Rahmen zur Verfügung stehenden Mittel
    fehlschlägt.

    Zunächst ist der zu $\mathcal{A}=\Hol$
    gehörige Bar-Komplex sowohl projektiv, als auch exakt. Für
    $(\mathcal{K}_{c},\pt_{c})$ ist jedoch nur die Projektivität
    unmittelbar einsichtig, da die Einschränkungen
    $\hat{h}^{c}_{k}\big|_{\K^{c}_{k}}$ der stetigen Fortsetzungen der exaktheitsliefernden
    Homotopieabbildungen $h^{c}_{k}$ des topologischen Koszul-Komplexes
    $(\K_{c},\pt_{c})$, im Gegensatz zu den Einschränkungen
    der Kettendifferentiale
    $\hat{\pt}^{c}_{k}\big|_{\K^{c}_{k}}$ im Allgemeinen nicht
    ausschließlich in die Unterräume $\K^{c}_{k}\subseteq \hat{\mathcal{K}}^{c}_{k}$
    abbilden, sondern in der Tat in den Vervollständigungen
    $\hat{\mathcal{K}}^{c}_{k}$ landen. 
    Algebraisch gesehen haben wir damit nichts in der Hand, um besagte Isomorphie zu begründen. Nun
    könnte man versuchen mit Hilfe der Einschränkungen der stetigen
    Fortsetzungen $\hat{F}$ und $\hat{G}$
    zu argumentieren. Jedoch bildet auch $\hat{G}$ im Allgemeinen nicht in die
    Unterräume, sondern in die jeweilige Vervollständigung
    ab. Mit Hilfe von Lemma \ref{lemma:Homotopiejdfgjkf}, der stetigen Fortsetzung der Homotopie $s$ aus
    Lemma \ref{lemma:Fkettenabb}~\textit{i.)} sowie $\widehat{\big(G_{k}\cp F_{k}\big)}=\hat{G}_{k}\cp \hat{F}_{k}$ und
    $\widehat{\big(d_{k+1}s_{k}+s_{k-1}d_{k}\big)}=\hat{d}_{k+1}\hat{s}_{k}+\hat{s}_{k-1}\hat{d}_{k}$
    folgt dann zwar unmittelbar: 
    \begin{equation*}
        H^{k}\left(\Hom_{\hat{\mathcal{A}^{e}}}\big(\hat{\C}_{c},\mathcal{M}\big),
            \hat{d}_{c}^{*}\right)\cong H^{k}\left(\Hom_{\hat{\mathcal{A}^{e}}}\big(\hat{\K}_{c},\mathcal{M}\big),\hat{\pt}^{*}_{c}\right).
    \end{equation*}
    Um jedoch die Isomorphie zu der gewünschten
    Hochschild-Kohomologie herzustellen, also beispielsweise
    \begin{equation*}
        H^{k}\Big(\Hom_{\mathcal{A}^{e}}(\C_{c},\mathcal{M}),d^{*}_{c}\Big)\cong
        H^{k}\Big(\Hom_{\mathcal{A}^{e}}\big(\hat{\C}_{c},\mathcal{M}\big),\hat{d}^{*}_{c}\Big)
    \end{equation*}nachzuweisen,
    benötigt man wieder Fortsetzungsargumente und hierzu die Vollständigkeit von
    $\mathcal{M}$. Dies liegt im wesentlichen daran, dass der Raum
    $\Hom_{\mathcal{A}^{e}}(\C_{c},\mathcal{M})$ für nicht vollständige
    Bimoduln $\mathcal{M}$ im Allgemeinen gehaltvoller als
    $\Hom_{\mathcal{A}^{e}}\big(\hat{\C}_{c},\mathcal{M}\big)$
    ist. Dies sieht man sofort daran, dass jedes $\psi\in
    \Hom_{\mathcal{A}^{e}}\big(\hat{\C}_{c},\mathcal{M}\big)$, vermöge
    Einschränkung, ein Element in
    $\Hom_{\mathcal{A}^{e}}(\C_{c},\mathcal{M})$ definiert, dessen
    stetige Fortsetzung es ist. Hierbei haben wir uns
    $\mathcal{M}\subseteq\hat{\mathcal{M}}$ kanonisch eingebettet
    gedacht. Umgekehrt können aber durchaus
    $\phi\in\Hom_{\mathcal{A}^{e}}(\C_{c},\mathcal{M})$ existieren,
    deren stetige Fortsetzung nicht ausschließlich nach $\mathcal{M}$
    abbildet, sondern auch Bilder in $\hat{\mathcal{M}}\backslash \mathcal{M}$ besitzt.
\end{Bemerkung}
\chapter{Hochschild-Kostant-Rosenberg-Theoreme}
\label{sec:HKRTheos}
In diesem Kapitel soll es darum gehen, Hochschild-Kostant-Rosenberg-Theoreme (vgl. \cite[Kapitel 6]{waldmann:2007a}) im Falle symmetrischer Bimoduln $\mathcal{M}$, f"ur die Kohomologie-Gruppen $HH^{k}(\SsV, \mathcal{M})$, $HH_{\cont}^{k}(\SsV, \mathcal{M})$ und $HH_{\cont}^{k}(\Hol, \mathcal{M})$ zu beweisen. Die Basis hierf"ur bilden die Sätze
\ref{satz:HochschkohmvonSym}, \ref{satz:stetigHochschSym} und
\ref{satz:HochschildHol}, die wir zun"achst etwas umformulieren wollen.
\section{Vorbereitung}
Wir benötigen die folgenden Kokettenkomplexe und Kettenabbildungen:
\begin{definition}
    \begin{enumerate}
    \item 
        Sei $\mathcal{M}$ ein$\Ss^{\bullet}(\V)-\Ss^{\bullet}(\V)$-Bimodul  und
        \begin{equation*}
            KC_{\Lambda}^{k}(\V,\mathcal{M}):=
            \begin{cases} \{0\} & k<0\\
                \mathcal{M} & k=0\\
                \Hom_{\mathbb{K}}(\Lambda^{k}(\V),\mathcal{M})& k\geq 1.
            \end{cases} 
        \end{equation*} 
        Vermöge der Links- und Rechtsmodulstruktur auf $\mathcal{M}$
        definieren wir $\mathbb{K}$-lineare Abbildungen $ \Delta_{\Lambda}^{k}\colon KC_{\Lambda}^{k}(\V, \mathcal{M})\longrightarrow KC_{\Lambda}^{k+1}(\V, \mathcal{M})$ durch:        
        \begin{equation*}
            (\Delta_{\Lambda}^{k}\phi)(u_{1}\wedge…\wedge u_{k+1})=
            \sum_{l=1}^{k+1}(-1)^{l-1}\:[u_{l}*_{L}-\:u_{l}\:*_{R}]\:\phi(u_{1}\wedge…\blacktriangle^{l}…\wedge
            u_{k+1}).
        \end{equation*}
        Des Weiteren definieren wir die Isomorphismen 
        \begin{align*}
            \Upsilon^{k}\colon \Hom_{\mathcal{A}^{e}}(\K_{k},\mathcal{M})&\longrightarrow KC_{\Lambda}^{k}(\V, \mathcal{M})\\
            \wt{\phi}&\longmapsto \left[\phi: \omega \mapsto \wt{\phi}\:(1_{e}\ot \omega)\right]
        \end{align*} mit Umkehrabbildungen
        \begin{align*}
            \Upsilon^{k}_{-1}\colon KC^{k}(\V, \mathcal{M})&\longrightarrow  \Hom_{\mathcal{A}^{e}}(\K_{k},\mathcal{M})\\
            \phi &\longmapsto \left[\wt{\phi}:a^{e}\ot \omega \mapsto a^{e}*_{e}\phi(\omega)\right].
        \end{align*}
        Nun ist
        {\allowdisplaybreaks\begin{align*}
              \left(\Upsilon^{k+1}\pt_{k}^{*}\right)\big(\wt{\phi}\big)(u_{1}&\:\wedge\dots \wedge u_{k+1})
              =\big(\pt_{k}^{*}\wt{\phi}\big)(1_{e}\ot u_{1}\wedge\dots \wedge u_{k+1})
              \\=&\:\sum_{l=1}^{k+1}(-1)^{l-1}\wt{\phi}\left([u_{l}\ot 1-1\ot u_{l}]\ot u_{1}\wedge\dots\blacktriangle^{l}\dots\wedge u_{k+1}\right)
              \\=&\:\sum_{l=1}^{k+1}(-1)^{l-1}[u_{l}\ot 1-1\ot u_{l}]*_{e}\wt{\phi}\left(1_{e}\ot u_{1}\wedge\dots\blacktriangle^{l}\dots\wedge u_{k+1}\right)
              \\=&\:\sum_{l=1}^{k+1}(-1)^{l-1}[u_{l}*_{L}- u_{l}\:*_{R}]\left(\Upsilon^{k}\wt{\phi}\right)\left(u_{1}\wedge\dots\blacktriangle^{l}\dots\wedge u_{k+1}\right)
              \\=&\: \left(\Delta_{\Lambda}^{k}\Upsilon^{k}\right)\big(\wt{\phi}\big)(u_{1}\wedge\dots\wedge u_{k+1}),
          \end{align*}}also insbesondere
        \begin{equation*}
            \Delta_{\Lambda}^{k+1}\Delta^{k}_{\Lambda}=\Upsilon^{k+2}\pt^{*}_{k+1}\pt^{*}_{k}\Upsilon^{k}_{-1}=0.
        \end{equation*} 
        Hiermit 
        induzieren die $\Upsilon^{k}$ einen Kettenisomorphismus zwischen den Kokettenkomplexen $\left(\Hom_{\mathcal{A}^{e}}(\K,\mathcal{M}),\pt^{*}\right)$ und $\left(KC_{\Lambda}(\V,\mathcal{M}),\Delta_{\Lambda}\right)$.
    \item
        Wir definieren
        \begin{equation*}
            KC^{k}(\V,\mathcal{M}):=
            \begin{cases} \{0\} & k<0\\
                \mathcal{M} & k=0\\
                \Hom^{a}_{\mathbb{K}}(\V^{k},\mathcal{M})& k\geq 1,
            \end{cases} 
        \end{equation*} die total antisymmetrischen, $\mathbb{K}$-multilinearen Abbildungen von
        $\V^{k}$ nach $\mathcal{M}$, sowie die $\mathbb{K}$-lineare Abbildungen $ \Delta^{k}:KC^{k}(\V, \mathcal{M})\longrightarrow KC^{k+1}(\V, \mathcal{M})$
        durch        
        \begin{equation*}
            (\Delta^{k}\phi)(v_{1},\dots, v_{k+1})=
            \sum_{l=1}^{k+1}(-1)^{l-1}\:[u_{l}*_{L}-\:u_{l}\:*_{R}]\:\phi(v_{1},\dots,\blacktriangle^{l},\dots,v_{k+1}).
        \end{equation*}
        Weiter definieren wir die Isomorphismen
        \begin{align*}
            \Theta^{k}\colon KC_{\Lambda}^{k}(\V, \mathcal{M})&\longrightarrow KC^{k}(\V, \mathcal{M})\\
            \phi'&\longmapsto \left[\phi\colon (v_{1},\dots,v_{k}) \mapsto \phi'(v_{1}\wedge\dots \wedge v_{k})\right]
        \end{align*}mit Umkehrabbildungen
        \begin{align*}
            \Theta^{k}_{-1}\colon KC^{k}(\V, \mathcal{M})&\longrightarrow  KC_{\Lambda}^{k}(\V, \mathcal{M})\\
            \phi &\longmapsto [\phi'\colon v_{1}\wedge\dots\wedge v_{k}\mapsto \phi(v_{1},\dots,v_{k})].
        \end{align*}
        Dabei entsprechen die letzteren gerade den Einschr"ankungen der durch die universelle Eigenschaft induzierten linearen Abbildungen $\phi_{\ot}$ auf die total antisymmetrischen Tensorelemente, denn mit der totalen Antisymmetrie von $\phi$ ist: 
        \begin{equation*}
            \phi_{\ot}(v_{1}\wedge\dots\wedge v_{k})=\frac{1}{k!}\sum_{\sigma\in S_{k}}\sign(\sigma)\:\phi(v_{\sigma(1)},\dots,v_{\sigma(k)})=\phi(v_{1},\dots,v_{k}).
        \end{equation*}Umgekehrt ist $\Theta^{k}(\phi')=\phi' \cp \mathrm{Alt}_{k} \cp\ot_{k}$.  
        Auch hier erhalten wir
        \begin{align*}
            \left(\Theta^{k+1}\Delta_{\Lambda}^{k}\right)\big(\phi'\big)(v_{1},\dots,v_{k+1})=&\: \left(\Delta^{k}_{\Lambda}\phi'\right)(v_{1}\wedge\dots\wedge v_{k+1})
            \\=&\:\sum_{l=1}^{k+1}(-1)^{l-1}[u_{l}*_{L}-u_{l}\:*_{R}]\:\phi'(v_{1}\wedge\dots\blacktriangle^{l}\dots\wedge v_{k+1})
            \\=&\:\sum_{l=1}^{k+1}(-1)^{l-1}[u_{l}*_{L}-u_{l}\:*_{R}]\:\big(\Theta^{k}\phi'\big)(v_{1},\dots,\blacktriangle^{l},\dots,v_{k+1})
            \\=&\:\left(\Delta^{k}\Theta^{k}\right)(\phi')(v_{1},\dots,v_{k+1}),
        \end{align*}also insbesondere wieder $\Delta^{k+1}\cp\Delta^{k}=0$, womit $\Theta$ ein Kettenisomorphismus zwischen den Kokettenkomplexen $\left(KC_{\Lambda}(\V,\mathcal{M}),\Delta_{\Lambda}\right)$ und $\left(KC(\V,\mathcal{M}),\Delta\right)$ ist.
    \end{enumerate}
\end{definition}
Folgendes Lemma kl"art weitere wichtige Eigenschaften obiger Definitionen.\\
\begin{lemma}
    \label{lemma:stethhh}
    \begin{enumerate}
    \item
        Gegeben ein symmetrischer $\Ss^{\bullet}(\V)-\Ss^{\bullet}(\V)$-Bimodul $\mathcal{M}$, dann gilt:
        \begin{equation*}
            H^{k}\big(KC(\V,\mathcal{M}),\Delta\big)=\Hom^{a}_{\mathbb{K}}\big(\V^{k},\mathcal{M}\big).
        \end{equation*}
    \item
        Seien $\V$ und $\mathcal{M}$ lokalkonvex und $\Lambda^{k}(\V)$ $\pi_{k}$
        topologisiert. Dann bilden sowohl die $\Upsilon^{k}$ als auch
        die $\Theta^{k}$, in beide Richtungen stetige Homomorphismen auf stetige Homomorphismen ab.
    \end{enumerate}    
    \begin{beweis}
        \begin{enumerate}
        \item 
            Für alle $\phi\in KC^{k}(\V,\mathcal{M})$ ist
            \begin{align*}
                \left(\Delta^{k}\phi\right)(v_{1},…,v_{k+1})=&\:\sum_{l=1}^{k+1}(-1)^{l-1}\:[u_{l}*_{L}-\:u_{l}\:*_{R}]\:\phi(v_{1},\dots,\blacktriangle^{l},\dots,v_{k+1})
                \\=&\:\sum_{l=1}^{k+1}(-1)^{l-1}\:[u_{l}*_{L}-\:u_{l}\:*_{L}]\:\phi(v_{1},\dots,\blacktriangle^{l},\dots,v_{k+1})
                \\=&\:0.
            \end{align*}
        \item
            Sei $\wt{\phi}\in
            \Hom^{\cont}_{\mathcal{A}^{e}}(\K_{k},\mathcal{M})$ und
            $\omega\in \Lambda^{k}(\V)$, dann folgt
            \begin{align*}
                q\left(\Big(\Upsilon^{k}\wt{\phi}\Big)(\omega)\right)=&\:q\left(\wt{\phi}\left(1_{e}\ot
                        \omega\right)\right)\leq \:c \p^{2}\ot p^{k}\left(1_{e}\ot\omega\right)=\:c\: p^{k}\left(\omega\right),
            \end{align*}was die Stetigkeit von $\Upsilon^{k}\big(\wt{\phi}\big)$
            zeigt. Sei umgekehrt $\phi\in
            \Hom^{\cont}_{\mathbb{K}}(\Lambda^{k}(\V),\mathcal{M})$, so
            zeigt Lemma \ref{lemma:AezuunittopRing}~\textit{ii.)}, dass
            \begin{align*}
                q\left(\left(\Upsilon^{k}_{-1}\phi\right)(a_{e}\ot
                    \omega)\right)=&\:q\left(a_{e}*_{e}\phi(\omega)\right)\leq c\:
                \p^{2}(a^{e})\:q'\left(\phi(\omega)\right)
                \\\leq&\: c'\: \p^{2} (a^{e})\:
                p'^{k}(\omega)\leq c'
                \p''_{k}(a^{e}\ot\omega)
            \end{align*}mit einer Halbnorm $p''\geq p,p'$ gilt.

            Für $\Theta^{k}$ sei $\phi'\in \Hom^{\cont}_{\mathbb{K}}(\Lambda^{k}(\V),\mathcal{M})$ wie eben, dann folgt:        
            \begin{align*}
                q\left(\left(\Theta^{k}\phi\right)(v_{1},…,v_{k})\right)=&\:q\big(\phi(v_{1}\wedge…\wedge
                v_{k})\big)\leq c\: p^{k}(v_{1}\wedge…\wedge
                v_{k})
                \\\leq&\:c\:p^{k}(v_{1}\ot…\ot v_{k})=c\:p(v_{1})…p(v_{k}).
            \end{align*}
            Sei umgekehrt $\phi
            \in\Hom^{a,\cont}_{\mathbb{K}}(\V^{k},\mathcal{M})$, dann
            sind
            $\Theta^{k}_{-1}\phi=\phi_{\ot}\Big|_{\Lambda^{k}(\V)}$
            und $\phi_{\ot}$ stetig mit der Charakterisierung von
            $\pi_{k}$ und es gilt:
            \begin{align*}
                q\left(\left(\Theta^{k}_{-1}\phi\right)(v_{1}\wedge…\wedge
                    v_{k})\right)=&\:
                q\big(\phi_{\ot}(v_{1}\wedge…\wedge v_{k})\big)\leq
                c \:p^{k}(v_{1}\wedge…\wedge v_{k}).
            \end{align*}
        \end{enumerate}    
    \end{beweis}
\end{lemma}
Hiermit erhalten wir folgende Umformulierungen der S"atze
\ref{satz:HochschkohmvonSym}, \ref{satz:stetigHochschSym} und \ref{satz:HochschildHol}:\\
\begin{korollar}
    \label{kor:UmformSatz}
    \begin{enumerate}
    \item
        Gegeben ein $\SsV-\SsV$-Bimodul $\mathcal{M}$, dann gilt:
        \begin{equation*}
            HH^{k}(\Ss^{\bullet}(\mathbb{V}),\mathcal{M})\cong H^{k}\left(KC(\V,\mathcal{M}),\Delta\right).
        \end{equation*}
        Ist $\mathcal{M}$ zudem symmetrisch, so ist:
        \begin{equation*}
            HH^{k}(\Ss^{\bullet}(\mathbb{V}),\mathcal{M})\cong \Hom^{a}_{\mathbb{K}}\big(\V^{k},\mathcal{M}\big).
        \end{equation*}
    \item
        Gegeben ein lokalkonvexer, $\SsV-\SsV$-Bimodul $\mathcal{M}$, dann gilt:
        \begin{equation*}
            HH^{k}_{\cont}\left(\Ss^{\bullet}(\mathbb{V}),\mathcal{M}\right)\cong
            H^{k}\big(KC^{\cont}(\V,\mathcal{M}),\Delta^{c}\big)
        \end{equation*}
        Ist $\mathcal{M}$ zudem symmetrisch, so ist:
        \begin{equation*}
            HH^{k}_{\cont}(\Ss^{\bullet}(\mathbb{V}),\mathcal{M})\cong\Hom_{\mathbb{K}}^{a,\cont}(\V^{k},\mathcal{M}).
        \end{equation*}
    \item
        Gegeben ein vollständiger, hausdorffscher, lokalkonvexer
        $\Hol-\Hol$-Bimodul $\mathcal{M}$, dann gilt:
        \begin{equation*}
            HH^{k}_{\cont}\big(\Hol,\mathcal{M}\big)\cong HH^{k}_{\cont}\big(\Ss^{\bullet}(\V),\mathcal{M}\big). 
        \end{equation*} Ist $\mathcal{M}$ zudem symmetrisch, so ist:
        \begin{equation*}        
            HH^{k}_{\cont}(\Hol,\mathcal{M})\cong \Hom_{\mathbb{K}}^{a,\cont}(\V^{k},\mathcal{M}).
        \end{equation*}
    \end{enumerate}
    \begin{beweis}
        Wegen 
        $\Delta_{\Lambda}^{k}=\Upsilon^{k+1}\pt_{k}^{*}\Upsilon^{k}_{-1}$
        sowie
        $\Delta^{k}=\Theta^{k+1}\Delta_{\Lambda}^{k}\Theta^{k}_{-1}$
        und Lemma \ref{lemma:stethhh}~\textit{ii.)} bilden sowohl
        $\Delta_{\Lambda}^{k}$ als auch $\Delta^{k}$ stetige Elemente
        auf stetige Elemente ab. Dies zeigt die Wohldefiniertheit der
        Kokettenkomplexe $(KC^{\cont}_{\Lambda},\Delta^{c}_{\Lambda})$
        und  $(KC^{\cont},\Delta^{c})$ sowie die Isomorphie ihrer
        Kohomologien. Die jeweils letzten Aussagen folgen mit Lemma \ref{lemma:stethhh}~\textit{i.)}.
    \end{beweis}
\end{korollar}
\section{Hochschild-Kostant-Rosenberg-Theoreme}
In diesem Abschnitt werden wir die jeweils zweite Isomorphie in Korollar \ref{kor:UmformSatz}
explizit ausformulieren und erhalten Analoga zu dem bekannten
Hochschild-Kostant-Rosenberg-Theorem, siehe \cite[Prop~6.2.48 ]{waldmann:2007a},\cite{cahen.gutt.dewilde:1980a}.
\begin{proposition}
    \label{prop:wichEizuHKR}
    \begin{enumerate}
    \item
        Seien $\V$ und $\mathcal{M}$ $\mathbb{K}$-Vektorräume, wobei
        $(\mathcal{M},*)$ zusätzlich ein $\Ss^{\bullet}(\V)$-Modul ist. Dann besitzt jede
        $\mathbb{K}$-multilineare Abbildung $\phi:\V^{k}\longrightarrow
        \mathcal{M}$
        eine eindeutig bestimmte, in jedem Argument derivative, $\mathbb{K}$-multilineare
        Fortsetzung
        $\phi_{D}\colon\SsV^{k}\longrightarrow \mathcal{M}$.
        Diese ist gegeben durch multilineare Fortsetzung von
        \begin{equation*}
            \label{eq:derFortsetz}
            \phi_{D}(\omega_{1},…,\omega_{k})=\sum_{m_{1}=1}^{n_{1}}…\sum_{m_{k}=1}^{n_{k}}\omega_{1}^{m_{1}}\vee…\vee\omega_{k}^{m_{k}}
            * \phi\big((\omega_{1})_{m_{1}},…,(\omega_{k})_{m_{k}}\big)
        \end{equation*}
        auf ganz $\SsV^{k}$ mit $\omega_{i}\in \Ss^{n_{i}}(\V)$ und
        $\phi_{D}\left(\omega_{1},…,1_{i},…,\omega_{k}\right)=0$ für
        $1\leq i\leq k$.
        Des Weiteren ist $\phi_{D}$ genau dann total
        antisymmetrisch, wenn $\phi$ total
        antisymmetrisch ist.
    \item
        Gegeben eine kommutative Algebra $\mathcal{A}$ und ein
        symmetrischer $\mathcal{A}-\mathcal{A}$-Bimodul $\mathcal{M}$.
        Sei weiter $\mathrm{Alt}_{k}\colon HC^{k}(\mathcal{A},\mathcal{M})\longrightarrow
        HC^{k}(\mathcal{A},\mathcal{M})$ definiert durch
        \begin{align*}
            \mathrm{Alt}_{k}(\phi)(a_{1},…,a_{k})=\frac{1}{k!}\sum_{\sigma\in S_{k}}\sign(\sigma)\:\phi(a_{\sigma(1)},…a_{\sigma(k)}).
        \end{align*}Dann gilt $\mathrm{Alt}_{k}\cp\delta^{k-1}=0$. Wegen
        $\mathrm{Alt}_{k}\cp \mathrm{Alt}_{k}=\mathrm{Alt}_{k}$ bedeutet dies insbesondere,
        dass ein total antisymmetrischer Hochschild-Kozyklus $\phi$ nur dann
        auch ein Hochschild-Korand sein kann, wenn bereits $\phi=0$ gilt.
    \item
        Sei $\phi\in \Hom^{a}_{\mathbb{K}}(\V^{k},\mathcal{M})$ und
        $\xi^{k}= \left(\ot_{k}^{*}\cp\Xi^{k}\cp G_{k}^{*} \cp
            \Upsilon^{k}_{-1}\cp \Theta^{k}_{-1}\right)$. Sei weiter
        $\mathcal{M}$ ein symmetrischer $\SsV-\SsV$-Bimodul, dann gilt:
        \begin{equation*}
            \xi^{k} (\phi)= \frac{1}{k!}\:\phi_{D}.
        \end{equation*}
    \end{enumerate}
    \begin{beweis}  
        \begin{enumerate}
        \item
            Mit der Symmetrie von
            $\omega_{1}^{m_{1}}\vee…\vee\omega_{k}^{m_{k}}$
            ist $\phi_{D}$ total antisymmetrisch, falls $\phi$ total antisymmetrisch
            ist. Dies zeigt die letzte Aussage, da die umgekehrte
            Implikation trivial ist.

            Für die Wohldefiniertheit sei $\V^{n_{i}}\ni
            \omega^{\times}_{i}=\big((\omega_{i})_{1},…,(\omega_{i})_{n_{i}}\big)$
            mit $1\leq i\leq k$ und $\omega_{i}^{m_{i}}\in
            \Ss^{n_{i-1}}(\V)$ das Element
            $(\omega_{i})_{1}\vee…\blacktriangle^{m_{i}}…\vee(\omega_{i})_{n_{i}}$.
            Dann ist
            \begin{align*}
                \phi^{n_{1},…,n_{k}}\left(\omega^{\times}_{1},…,\omega^{\times}_{k}\right)=\sum_{m_{1}=1}^{n_{1}}…\sum_{m_{k}=1}^{n_{k}}
                \omega_{1}^{m_{1}}\vee…\vee\omega_{k}^{m_{k}}
                *
                \phi\big((\omega_{1})_{m_{1}},…,(\omega_{k})_{m_{k}}\big)
            \end{align*} 
            eine wohldefinierte $\mathbb{K}$-multilineare Abbildung von
            $\V^{[n_{1}+…+n_{k}]}$ nach $\mathcal{M}$. Mit
            Lemma \ref{lemma:assTenprod}~\textit{ii.)} sowie der universellen
            Eigenschaft des Tensorproduktes erhalten wir eine lineare Fortsetzung
            \begin{equation*}
                \phi_{\ot}^{n_{1},…,n_{k}}\colon\Tt^{n_{1}}(\V)\ot…\ot\Tt^{n_{k}}(\V)\longmapsto
                \mathcal{M},
            \end{equation*}
            die verkettet mit $\ot_{n_{1},…,n_{k}}\colon\Tt^{n_{1}}(\V)\times…\times\Tt^{n_{k}}(\V)\longrightarrow
            \Tt^{n_{1}}\ot…\ot\Tt^{n_{k}}$ die Eigenschaft
            \begin{equation*}
                \phi_{\ot}^{n_{1},…n_{k}}\cp \ot_{n_{1},…,n_{k}}\big|_{\Ss^{n_{1}}(\V)\times…\times\Ss^{n_{k}}(\V)}=\phi_{D}\big|_{\Ss^{n_{1}}(\V)\times…\times\Ss^{n_{k}}(\V)}
            \end{equation*}besitzt. 
            Insgesamt folgt
            \begin{equation*}
                \phi_{D}=\sum_{n_{1},…,n_{k}}\big[\phi_{\ot}^{n_{1},…,n_{k}}\cp\ot_{n_{1},…,n_{k}}\big]\big|_{\Ss^{n_{1}}(\V)\times…\times\Ss^{n_{k}}(\V)},
            \end{equation*}also die Wohldefiniertheit von $\phi_{D}$.
            Für die Derivativität rechnen wir
            {\allowdisplaybreaks{\small\begin{align*}
                    \phi&_{D}(\omega_{1},…,\omega_{l}\vee \omega'_{l}
                    ,…,\omega_{k})
                    \\ &=\sum_{\substack{m_{i}=1\\i\neq
                        l}}^{n_{i}}\sum_{\wt{m}_{l}=1}^{\deg(\omega_{l}\vee\omega'_{l})}\omega_{1}^{m_{1}}\vee…\vee\left(\omega_{l}\vee\omega'_{l}\right)^{\wt{m}_{l}}\vee…\vee\omega_{k}^{m_{k}}*
                    \\ &\qquad\qquad\qquad\qquad\qquad\qquad\qquad\qquad\qquad\qquad\:\:\:\phi\big((\omega_{1})_{m_{1}},…,\left(\omega_{l}\vee\omega'_{l}\right)_{\wt{m}_{l}},…,(\omega_{k})_{m_{k}}\big)
                    \\ &=\omega_{l}\vee\left[\sum_{\substack{m_{i}=1\\i\neq
                            l}}^{n_{i}}\sum_{m'_{l}=1}^{\deg(\hat{\omega}{l})}\omega_{1}^{m_{1}}\vee…\vee\omega_{l}'^{m'_{l}}\vee…\vee\omega_{k}^{m_{k}}*\phi\big((\omega_{1})_{m_{1}},…,(\omega'_{l})_{m'_{l}},…,(\omega_{k})_{m_{k}}\big)\right]
                    \\ &\:\:\:+\omega'_{l}\vee\left[\sum_{\substack{m_{i}=1\\i\neq
                            l}}^{n_{i}}\sum_{m_{l}=1}^{\deg(\omega_{l})}\omega_{1}^{m_{1}}\vee…\vee\omega_{l}^{m_{l}}\vee…\vee\omega_{k}^{m_{k}}*\phi\big((\omega_{1})_{m_{1}},…,(\omega_{l})_{m_{l}},…,(\omega_{k})_{m_{k}}\big)\right]
                    \\ &=\omega_{l}\vee \phi_{D}(\omega_{1},…,\omega'_{l},…,\omega_{k})+ \omega'_{l}\vee \phi_{D}(\omega_{1},…,\omega_{l},…,\omega_{k}),
                \end{align*}}}wobei wir im zweiten Schritt die
            Moduleigenschaft von $\mathcal{M}$ benutzt haben.

            Für die Eindeutigkeit sei $\wt{\phi}$ eine weitere
            derivative Fortsetzung von $\phi$. Im Falle
            $k=1$ und mit $\omega=\omega_{1}\vee…\vee\omega_{l}$ erhalten
            wir sukzessive:
            \begin{align*}
                \wt{\phi}\:(\omega)=&\:\omega^{1}*\wt{\phi}\:(\omega_{1})+\omega_{1}*\wt{\phi}\:(\omega^{1})
                \\=&\:\omega^{1}*\wt{\phi}\:(\omega_{1})+
                \omega^{2}*\wt{\phi}\:(\omega_{2})+\omega_{1,2}*\wt{\phi}\:(\omega^{1,2})
                \\=&\:\omega^{1}*\wt{\phi}\:(\omega_{1})+…+\omega^{l-1}*\wt{\phi}\:(\omega_{l-1})+\omega_{1,…,l-1}*\wt{\phi}\:(\omega^{1,…,l-1})
                \\=&\:\omega^{1}*\wt{\phi}\:(\omega_{1})+…+\omega^{l}*\wt{\phi}\:(\omega_{l}).
            \end{align*}
            Für $k>1$ zeigt induktives Anwenden obiger Relation, dass
            {\allowdisplaybreaks\begin{align*}
                  \wt{\phi}(\omega)=&\:\sum_{m_{1}=1}^{n_{1}}…\sum_{m_{k}=1}^{n_{k}}\omega_{1}^{m_{1}}\vee…\vee\omega_{k}^{m_{k}}
                  *
                  \wt{\phi}\big((\omega_{1})_{m_{1}},…,(\omega_{k})_{m_{k}}\big)
                  \\=&\:\sum_{m_{1}=1}^{n_{1}}…\sum_{m_{k}=1}^{n_{k}}\omega_{1}^{m_{1}}\vee…\vee\omega_{k}^{m_{k}}
                  *
                  \phi\big((\omega_{1})_{m_{1}},…,(\omega_{k})_{m_{k}}\big)
                  \\=&\:\phi_{D}(\omega),
              \end{align*}}und die Derivativität erzwingt 
            \begin{align*}
                \wt{\phi}(\omega_{1},…,1_{i},…,\omega_{k})=\wt{\phi}(\omega_{1},…,1\vee
                1_{i},…,\omega_{k})=2\wt{\phi}(\omega_{1},…,1_{i},…,\omega_{k}),
            \end{align*}
            also $\wt{\phi}(\omega_{1},…,1_{i},…,\omega_{k})=0$. Dies zeigt die Eindeutigkeit besagter
            derivativer Fortsetzung.
        \item        
            Wir erhalten:
            \begin{align*}
                \left(\mathrm{Alt}_{k}\cp
                    \delta^{k-1}\right)&\big(\phi\big)(a_{1},…,a_{k})
                \\=&\: \mathrm{Alt}_{k}\left(a_{1}*_{L}\phi(a_{2},…,a_{k})\right)
                +\sum_{i=1}^{k-1}(-1)^{i}\mathrm{Alt}_{k}\left(\phi(a_{1},…,a_{i}a_{i+1},…,a_{k+1})\right)
                \\ &
                \qquad\qquad\qquad\qquad\qquad\qquad+(-1)^{k}\mathrm{Alt}_{k}(\phi(a_{1},…,a_{k-1})*_{R}a_{k})
                \\=&\: \mathrm{Alt}_{k}\left(a_{1}*_{L}\phi(a_{2},…,a_{k})\right) +(-1)^{k}(-1)^{k-1}\mathrm{Alt}_{k}(\phi(a_{2},…,a_{k})*_{R}a_{1})
                \\
                &\qquad\qquad\qquad\qquad\qquad\qquad+\sum_{i=1}^{k-1}(-1)^{i}\mathrm{Alt}_{k}\left(\phi(a_{1},…,a_{i}a_{i+1},…,a_{k})\right)
                \\=&\:0.
            \end{align*}
            Dabei verschwindet die letzte Summe wegen
            der Kommutativität von $\mathcal{A}$. 
        \item
            Sei $\phi\in \Hom^{a}_{\mathbb{K}}(\V^{k},\mathcal{M})$. Dann ist
            \begin{equation*}
                \begin{split}
                    \left(\xi^{k}\phi\right)(u_{1},\dots,u_{k})=&\:\bigg(\ot_{k}^{*}\cp\Xi^{k}\cp G_{k}^{*} \cp \Upsilon^{k}_{-1}\cp
                    \Theta^{k}_{-1}\bigg) \big(\phi\big)(u_{1},\dots,u_{k})
                    \\=&\:  \left(\Xi^{k}\cp G_{k}^{*} \cp \Upsilon^{k}_{-1}\cp
                        \Theta^{k}_{-1}\right)\big(\phi\big)(u_{1}\ot\dots\ot u_{k})
                    \\=&\: \left(G_{k}^{*} \cp \Upsilon^{k}_{-1}\cp
                        \Theta^{k}_{-1}\right)\big(\phi\big)(1\ot u_{1}\ot\dots\ot u_{k}\ot 1)
                    \\=&\: \left(\Upsilon^{k}_{-1}\cp
                        \Theta^{k}_{-1}\right)\big(\phi\big)\big(G_{k}(1\ot u_{1}\ot\dots\ot u_{k}\ot 1)\big).
                \end{split}
            \end{equation*}Für $v_{1},…,v_{k}\in \V$ folgt
            {\allowdisplaybreaks\begin{align*}
                  G_{k}(1\ot v_{1}\ot\dots\ot v_{k}\ot
                  1)=&\:\int_{0}^{1}dt_{1}…\int_{0}^{t_{k-1}}dt_{k}i\big(1\ot1\bbot
                  \overbrace{1\ot…\ot 1}^{k-mal} \bbot
                  v_{1}\wedge…\wedge v_{k}\big)
                  \\=&\:\int_{0}^{1}dt_{1}…\int_{0}^{t_{k-1}}dt_{k} 1\ot1\ot
                  v_{1}\wedge…\wedge v_{k}
                  \\=&\:\frac{1}{k!} 1\ot1\ot
                  v_{1}\wedge…\wedge v_{k},
              \end{align*}}also
            {\allowdisplaybreaks\begin{align*}
                  \left(\xi^{k}\phi\right)(v_{1},…,v_{k})=&\:\frac{1}{k!}\left(\Upsilon^{k}_{-1}\cp
                      \Theta^{k}_{-1}\right)\big(\phi\big)(1\ot
                  1\ot v_{1}\wedge\dots\wedge v_{k})
                  \\=&\:\frac{1}{k!} \Theta^{k}_{-1}\big(1\ot 1 *_{e}\phi\big)(1\ot
                  1\ot v_{1}\wedge\dots\wedge v_{k})
                  \\=&\:\frac{1}{k!}\: \phi(v_{1},…,v_{k}).
              \end{align*}}Des Weiteren ist
            $\left(\xi^{k}\phi\right)(u_{1},…,u_{k})=0$,
            falls $u_{i}=1$ für ein $1\leq i\leq k$, da dann
            $\delta(1\ot u_{1}\ot…\ot u_{k}\ot 1)=0$ gilt.

            Sei abkürzend $\Hom_{\mathcal{A}^{e}}(\K_{k},\mathcal{M})
            \ni\wt{\phi}=\left(\Upsilon^{k}_{-1}\cp\Theta^{k}_{-1}\right)\big(\phi\big)$, dann
            folgt mit \eqref{eq:DeltaDerivaufuElem} und sukzessiver
            Anwendung von 
            \begin{equation}
                \label{eq:aeTrickt}
                \begin{split}
                    \hat{i}(1\ot \mathrm{v}\ot 1) *_{e} m =& \:\big[t (\mathrm{v}\ot 1) + (1-t) (1\ot \mathrm{v}) \big] *_{e} m
                    \\=&\: t \:\mathrm{v}*_{L} m + \mathrm{v}*_{R}m - t\: \mathrm{v}*_{R} m
                    \\=&\: \mathrm{v} *_{L} m,
                \end{split}
            \end{equation}
            dass
            {\allowdisplaybreaks\begin{align*}
                  \left(\xi^{k}\phi\right)(u_{1},…,&\:u_{j}\vee
                  u'_{j},…,u_{k})
                  \\ =&\:\Big(\wt{\phi} \cp G_{k}\Big)(1\ot u_{1}\ot\dots u_{j}\vee u'_{j}\ot\dots\ot u_{k}\ot 1)
                  \\ =&\int_{0}^{1}dt_{1}\dots
                  \int_{0}^{t_{k-1}}dt_{k}\:\hat{i}_{j}(1\ot u_{j}\ot
                  1)*_{e}
                  \\ &\qquad\qquad\qquad\qquad \wt{\phi}\left(\prod_{s\neq j}(i_{s}\cp\delta)(1\ot u_{s}\ot 1) \cdot (i_{j}\cp\delta)(1\ot u'_{j}\ot 1)\right)
                  \\ & +\int_{0}^{1}dt_{1}\dots
                  \int_{0}^{t_{k-1}}dt_{k}\:\hat{i}_{j}(1\ot u'_{j}\ot
                  1)*_{e} 
                  \\ &\qquad\qquad\qquad\qquad\: \wt{\phi}\left(\prod_{s\neq
                        j}(i_{s}\cp\delta)(1\ot u_{s}\ot 1) \cdot
                      (i_{j}\cp\delta)(1\ot u_{j}\ot 1)\right)
                  \\ =&\:u_{j}*_{L} \left(\wt{\phi}\cp
                      G_{k}\right)(1\ot u_{1}\ot\dots
                  u'_{j}\ot\dots\ot u_{k}\ot 1)\:+
                  \\ &\qquad\qquad\qquad\qquad\quad u'_{j}*_{L} \left(\wt{\phi}\cp G_{k}\right)(1\ot u_{1}\ot\dots
                  u_{j}\ot\dots\ot u_{k}\ot 1)
                  \\ =&\:u_{j}*_{L} \left(\xi^{k}\phi\right)(u_{1},…,u'_{j},…,u_{k})
                  +u'_{j}*_{L} \left(\xi^{k}\phi\right)(u_{1},…,u_{j},…,u_{k}).
              \end{align*}}Mit \textit{i.)} zeigt dies die
            Behauptung. Hierfür beachte man, dass wir $\wt{\phi}$ mit den
            Integralen vertauschen dürfen, da die Intergrationen lediglich
            den verschiedenen $t$-Faktoren reelle Zahlen zuordnen und
            $\wt{\phi}$ nach Voraussetzung $\mathbb{K}$-linear ist. 
        \end{enumerate}
    \end{beweis}
\end{proposition}
Bevor wir zu dem Hauptresultat dieses Kapitels kommen, erinnern wir
an folgendes kommutatives Diagramm:
\[\begin{xy} 
    \xymatrix{  
      ...\ar[r]^{\delta^{k-2}}
      &HC^{k-1} \ar[d]_{\ot_{k-1*}}\ar[r]^{\delta^{k-1}}\ar@{->}@/^ 0.6cm/[dd]^/.6em/{\zeta^{k-1}}
      &HC^{k} \ar[d]_{\ot_{k*}}\ar[r]^{\delta^{k}}\ar@{->}@/^ 0.6cm/[dd]^/.6em/{\zeta^{k}}
      &HC^{k+1}
      \ar[d]_{\ot_{k+1*}}\ar[r]^{\delta^{k+1}}\ar@{->}@/^ 0.6cm/[dd]^/.6em/{\zeta^{k+1}} &\dots \\
      ... \ar[r]^{\delta^{k-2}_{\ot}}
      &HC^{k-1}_{\ot}\ar[d]_{\Xi^{k-1}_{-1}} \ar[r]^{\delta^{k-1}_{\ot}}
      &HC^{k}_{\ot}\ar[d]_{\Xi^{k}_{-1}} \ar[r]^{\delta^{k}_{\ot}} &HC^{k+1}_{\ot}\ar[d]_{\Xi^{k+1}_{-1}} \ar[r]^{\delta^{k+1}_{\ot}}&\dots\\
      ... \ar[r]^{d_{k-1}^{*}}
      &\C_{k-1}^{*} \ar[r]^{d_{k}^{*}}
      &\C_{k}^{*} \ar[r]^{d_{k+1}^{*}}\ar@{.>}@/^ 0.5cm/[l]^{s^{*}_{k-1}} &\C_{k+1}^{*} \ar[r]^{d_{k+2}^{*}}\ar@{.>}@/^ 0.5cm/[l]^{s^{*}_{k}}&\dots\:.
    }
\end{xy}\]
Hierbei ist $HC^{k}=HC^{k}(\SsV,\mathcal{M})$, sowie
$HC^{k}_{\ot}=HC^{k}_{\ot}(\SsV,\mathcal{M})$ die
Tensorvariante des Hochschild-Komplexes.
Der untere Komplex
ist der durch Anwendung des
$\hom_{\mathcal{A}^{e}}(\cdot,\mathcal{M})$-Funktors
erhaltenen Kokettenkomplex $(\C^{*},d^{*})$ mit
$\C_{k}^{*}=\Hom_{\mathcal{A}^{e}}(\C_{k},\mathcal{M})$ und
$d_{k+1}^{*}\phi_{k}=\phi_{k}\cp d_{k+1}$.
Die $s_{k}$
bezeichnen die in Lemma \ref{lemma:Homotopiejdfgjkf} definierten
Homotopieabbildungen und $\zeta^{k}\colon
HC^{k}(\SsV,\mathcal{M})\longrightarrow
\Hom_{\mathcal{A}^{e}}(\C_{k},\mathcal{M})$ den
Kettenisomorphismus $\zeta^{k}=\Xi^{k}_{-1}\cp \ot_{k*}$.
\begin{satz}[Hochschild-Kostant-Rosenberg]
    \label{satz:HKR}
    \begin{enumerate}
    \item
        Gegeben ein symmetrischer $\Ss^{\bullet}(\V)-\Ss^{\bullet}(\V)$-Bimodul $\mathcal{M}$.
        Dann besitzt jede Kohomologieklasse
        $\left[\eta\right]\in HH^{k}(\Ss^{\bullet}(\V),\mathcal{M})$ genau einen total
        antisymmetrischen Repräsentanten $\phi^{a,\eta}_{D}$. 
        Dieser ist derivativ in jedem Argument und gegeben durch
        $\phi^{a,\eta}_{D}=\mathrm{Alt}_{k}(\phi)$ f"ur beliebiges $\phi\in
        [\eta]$ mit $\phi^{a,0}_{D}=0$ für
        die $0$-Klasse $[0]$.  

        Insgesamt gilt für alle $\phi\in [\eta]$:
        \begin{equation}
            \label{eq:RepHKR}
            \phi=\underbrace{\phi^{a,\eta}_{D}}_{\mathrm{Alt}_{k}(\phi)}+\underbrace{\delta^{k-1}\big(\zeta^{k-1}_{-1}s^{*}_{k-1}\zeta^{k}\phi\big)}_{\phi-\mathrm{Alt}_{k}(\phi)}.
        \end{equation}
    \item
        Gegeben ein symmetrischer, lokalkonvexer
        $\Ss^{\bullet}(\V)-\Ss^{\bullet}(\V)$-Bimodul $\mathcal{M}$.
        Dann besitzt jedes
        $\left[\eta_{c}\right]\in HH^{k}_{\cont}(\Ss^{\bullet}(\V),\mathcal{M})$ genau einen total
        antisymmetrischen, stetigen Repräsentanten
        $\phi^{a,\eta}_{c,D}$. Dieser ist derivativ in jedem Argument und gegeben durch
        $\phi^{a,\eta}_{c,D}=\mathrm{Alt}_{k}(\phi_{c})$ f"ur beliebiges $\phi_{c}\in
        [\eta_{c}]$ mit $\phi^{a,0}_{c,D}=0$ für
        die $0$-Klasse $[0_{c}]$.
        Insgesamt gilt für alle $\phi_{c}\in [\eta_{c}]$:
        \begin{equation}
            \label{eq:RepHKRstet}
            \phi_{c}=\underbrace{\phi^{a,\eta}_{c,D}}_{\mathrm{Alt}_{k}(\phi_{c})}+\underbrace{\delta_{c}^{k-1}\big(\zeta^{k-1}_{-1}s^{*}_{k-1}\zeta^{k}\phi_{c}\big)}_{\phi_{c}-\mathrm{Alt}_{k}(\phi_{c})}.
        \end{equation}
    \item
        Gegeben ein vollständiger, symmetrischer, hausdorffscher, lokalkonvexer
        $\mathrm{Hol}(\V)-\Hol$-Bimodul $\mathcal{M}$. Dann besitzt jedes
        $\left[\hat{\eta}_{c}\right]\in HH^{k}_{\cont}(\Hol,\mathcal{M})$ genau einen total
        antisymmetrischen, stetigen Repräsentanten
        $\hat{\phi}^{a,\eta}_{c,D}$. Dieser ist derivativ in
        jedem Argument und gegeben durch $\hat{\phi}^{a,\eta}_{c,D}=\mathrm{Alt}_{k}\big(\hat{\phi}_{c}\big)$ f"ur beliebiges $\hat{\phi}_{c}\in
        [\hat{\eta}_{c}]$ mit $\hat{\phi}^{a,0}_{c,D}=0$ für 
        die $0$-Klasse $[\hat{0}_{c}]$.

        Insgesamt gilt für alle $\hat{\phi}_{c}\in [\hat{\eta}_{c}]$:
        \begin{equation}
            \label{eq:RepHKRstethol}
            \hat{\phi}_{c}=\underbrace{\hat{\phi}^{a,\eta}_{c,D}}_{\mathrm{Alt}_{k}(\hat{\phi}_{c})}+\underbrace{\hat{\delta}_{c}^{k-1}\widehat{\Big(\zeta^{k-1}_{-1}s^{*}_{k-1}\zeta^{k}\phi_{c}\Big)}}_{\hat{\phi}_{c}-\mathrm{Alt}_{k}(\hat{\phi}_{c})}\quad\text{
              mit }\quad \phi_{c}=\hat{\phi}_{c}\big|_{\SsV^{k}}.
        \end{equation}
    \end{enumerate}
    \begin{beweis}
        \begin{enumerate}
        \item
            Zunächst ist die Existenz eines total antisymmetrischen
            Repräsentanten $\phi_{D}^{a,\eta}$ gesichert, da die
            Abbildung $\wt{\xi^{k}}:\Hom_{\mathbb{K}}^{a}(\V^{k},\mathcal{M})\longrightarrow
            HH^{k}(\Ss^{\bullet}(\V),\mathcal{M})$
            ein Isomorphismus war und 
            $\xi^{k}(\phi)$ nach
            Proposition \ref{prop:wichEizuHKR}~\textit{iii.)} total
            antisymmetrisch und derivativ ist. Mit
            Proposition \ref{prop:wichEizuHKR}~\textit{ii.)} folgt
            \begin{equation}
                \label{eq:Altdelta}
                \mathrm{Alt}_{k}(\phi)=\mathrm{Alt}_{k}\left(\phi^{a,\eta}_{D}+\delta^{k-1}(\psi)\right)
                =\phi^{a,\eta}_{D}\qquad\quad\forall\:\phi\in[\eta],
            \end{equation}und ebenso die Eindeutigkeit. Denn für total
            antisymmetrische $\phi^{a},\hat{\phi}^{a}\in
            \left[\eta\right]$ ist
            $\phi^{a}-\hat{\phi}^{a}$ zudem exakt und somit
            $(\phi^{a}-\hat{\phi}^{a})=\mathrm{Alt}_{k}(\phi^{a}-\hat{\phi}^{a})=0$.
            Die Aussage $\phi_{D}^{a,0}=0$ ist dann wegen
            der Linearität von $ \wt{\xi^{k}}$ trivial.

            Für \eqref{eq:RepHKR}
            sei $\hat{\xi}^{k}=\Theta^{k}\cp
            \Upsilon^{k}\cp F_{k}^{*}\cp \Xi^{k}_{-1}\cp \ot_{k*}$ , womit 
            \begin{equation*}
                \wt{\hat{\xi}^{k}}\colon HH^{k}(\Ss^{\bullet}(\V),\mathcal{M})\longrightarrow
                \Hom_{\mathbb{K}}^{a}(\V^{k},\mathcal{M})
            \end{equation*}
            der zu $\wt{\xi^{k}}$ inverse Isomorphismus ist. Dann folgt
            \begin{equation*}
                \left(\xi^{k}\cp\hat{\xi}^{k}\right)(\phi)=\left(\zeta^{k}_{-1}\Omega^{*}_{k}\zeta^{k}\right)(\phi)=\phi_{D}^{a,\eta}=\mathrm{Alt}_{k}(\phi),
            \end{equation*} 
            was man unmittelbar daran sieht, dass jedes $\phi\in [\eta]$ unter $\hat{\xi}^{k}$
            das gleiche Bildelement haben muss. Alternativ rechnet man
            $\xi^{k}\cp\hat{\xi}^{k}=\mathrm{Alt}_{k}$ auch explizit
            nach (vgl. Bemerkung \ref{bem:HRKBem}~\textit{ii.)}).
            Nach
            Lemma \ref{lemma:Homotopiejdfgjkf} haben wir nun
            \begin{equation}
                \id_{\C^{*}_{k}}-\:\Omega^{*}_{k}=s^{*}_{k}d^{*}_{k+1}+d^{*}_{k}s^{*}_{k-1},
            \end{equation}
            also
            \begin{equation}
                \label{eq:1}
                \id_{HC^{k}}-\:\zeta_{-1}^{k}\Omega^{*}_{k}\zeta^{k}=\zeta_{-1}^{k}s^{*}_{k}d^{*}_{k+1}\zeta^{k}+\zeta_{-1}^{k}d^{*}_{k}s^{*}_{k-1}\zeta^{k}
            \end{equation}   
            und somit:
            \begin{equation}
                \label{eq:dgh}
                \id_{HC^{k}}-\mathrm{Alt}_{k}=\zeta_{-1}^{k}s^{*}_{k}\zeta^{k+1}\delta^{k}+\delta^{k-1}\big(\zeta_{-1}^{k-1}s^{*}_{k-1}\zeta^{k}\big).
            \end{equation}Anwenden von \eqref{eq:dgh} auf einen Korand
            $\phi\in [\eta]$ liefert
            \begin{equation*}
                \phi-\mathrm{Alt}_{k}(\phi)=\delta^{k-1}\big(\zeta_{-1}^{k-1}s^{*}_{k-1}\zeta^{k}\phi\big)
            \end{equation*}
            und zeigt somit die Behauptung.
        \item
            Alle Isomorphismen aus obigem Diagramm sind ebenfalls
            Isomorphismen auf den stetigen Unterkomplexen. Nach
            Lemma \ref{lemma:Homotopiejdfgjkf} sind die $s_{k}$
            \emph{stetig} und somit besagtes Diagramm auch auf die stetige Situation anwendbar.
            Des Weiteren sind
            \begin{align*}
                \wt{\xi^{k}_{c}}\colon\Hom^{a,\cont}_{\mathbb{K}}(\V^{k},\mathcal{M})\longrightarrow
                HH^{k}_{\cont}(\Ss^{\bullet}(\V),\mathcal{M})
            \end{align*}
            und 
            \begin{equation*}
                \wt{\hat{\xi}_{c}^{k}}\colon HH_{\cont}^{k}(\Ss^{\bullet}(\V),\mathcal{M})\longrightarrow
                \Hom_{\mathbb{K}}^{a,\cont}(\V^{k},\mathcal{M})
            \end{equation*}
            mit
            $\xi^{k}_{c}=\xi^{k}\big|_{\Hom^{a,\cont}_{\mathbb{K}}(\V^{k},\mathcal{M})}$ und
            $\hat{\xi}_{c}^{k}=\hat{\xi}^{k}\big|_{HC^{k}_{\cont}(\SsV,\mathcal{M})}$
            zueinander inverse Isomorphismen. Hiermit folgen alle
            Behauptungen analog zu \textit{i.)}. 
        \item 
            Die Eindeutigkeit folgt unmittelbar aus
            Proposition \ref{prop:wichEizuHKR}~\textit{ii.)}. Des Weiteren haben
            wir nach Satz \ref{satz:HochschildHol}, vermöge Einschr"ankung und stetiger
            Fortsetzung, eine Isomorphie
            $HC_{\cont}^{k}(\Hol,\mathcal{M})\cong
            HC_{\cont}^{k}(\SsV,\mathcal{M})$, welche die Isomorphie  $HH^{k}_{\cont}(\Hol,
            \mathcal{M})\cong
            HH^{k}_{\cont}(\Ss^{\bullet}(\V),\mathcal{M})$
            induziert. Mit der Linearität besagter Isomorphismen folgt
            dann unmittelbar:
            \begin{equation*}
                \widehat{\mathrm{Alt}_{k}(\phi_{c})}=\mathrm{Alt}_{k}\big(\hat{\phi}_{c}\big).
            \end{equation*}
            Nach \textit{ii.)} ist für jedes $\hat{\phi}_{c}\in
            [\hat{\eta}_{c}]\in HH^{k}(\Hol,\mathcal{M})$ die Einschr"ankung,\\
            $\phi_{c}=\hat{\phi}_{c}\big|_{\Ss^{\bullet}(\V)^{k}}$, darstellbar in der Form:
            \begin{equation*}
                \phi_{c}=\phi^{a,\eta}_{c,D}+\delta_{c}^{k-1}\left(\zeta^{k-1}_{-1}s^{*}_{k-1}\zeta^{k}\phi_{c}\right).
            \end{equation*}
            Hieraus folgt durch stetige Fortsetzung beider Seiten von
            \eqref{eq:muhkuhmilch}, dass
            \begin{equation*} 
                \hat{\phi}_{c}= \hat{\phi}^{a,\eta}_{c,D}+\hat{\delta}^{k-1}_{c}\left(\widehat{\zeta^{k-1}_{-1}s^{*}_{k-1}\zeta^{k}\phi_{c}}\right),
            \end{equation*}
            wobei der erste Summand wegen
            $\mathrm{Alt}_{k}\left(\hat{\phi}^{a,\eta}_{c,D}\right)=\widehat{\mathrm{Alt}_{k}\left(\phi^{a,\eta}_{c,D}\right)}=\hat{\phi}^{a,\eta}_{c,D}$
            total antisymmetrisch ist. Mit \eqref{eq:Altdelta}
            zeigt dies die Zuweisungen unter den geschweiften Klammern, da die Zerlegung
            $\hat{\phi_{c}}=\mathrm{Alt_{k}}(\hat{\phi_{c}})+\hat{\phi_{c}}-\mathrm{Alt_{k}}(\hat{\phi}_{c})$
            offenbar trivial ist.

            Es bleibt nun lediglich die Derivationseigenschaft von
            $\hat{\phi}^{a,\eta}_{c,D}$ nachzuweisen. Hierfür rechnen
            wir mit den Stetigkeiten von $*$ und $*_{L}$ , der
            Definition von $\hat{\phi}^{a,\eta}_{c,D}$ sowie der
            Derivativit"at von $\phi^{a,\eta}_{c,D}$:
            \begin{align*}
                \hat{\phi}^{a,\eta}_{c,D}(\omega_{1},\dots, \omega_{l}*\omega'_{l},&\dots,\omega_{k})
                \\=&\:\lim_{\Lambda}\: \phi^{a,\eta}_{c,D}\left((\omega_{1})_{\alpha_{1}},\dots,(\omega_{l})_{\alpha_{l}}\vee(\omega'_{l})_{\alpha'_{l}} ,\dots,(\omega_{k})_{\alpha_{k}}\right)
                \\=&\: \lim_{\Lambda}\: (\omega_{l})_{\alpha_{l}}*_{L}\phi^{a,\eta}_{c,D}\left((\omega_{1})_{\alpha_{1}},\dots,(\omega'_{l})_{\alpha'_{l}} ,\dots,(\omega_{k})_{\alpha_{k}}\right)
                \\ &+\lim_{\Lambda}\:(\omega'_{l})_{\alpha'_{l}} *_{L}\phi^{a,\eta}_{c,D}\Big((\omega_{1})_{\alpha_{1}},\dots,(\omega_{l})_{\alpha_{l}} ,\dots,(\omega_{k})_{\alpha_{k}}\Big)
                \\=&\:\omega_{l}*_{L}\hat{\phi}^{a,\eta}_{c,D}(\omega_{1},\dots,
                \omega'_{l},\dots,\omega_{k})
                +\omega'_{l}*_{L}\hat{\phi}^{a,\eta}_{c,D}(\omega_{1},\dots, \omega_{l},\dots,\omega_{k})
            \end{align*}
            mit Netzen $\SsV\supseteq\{\omega_{i}\}_{\alpha_{i}\in
              J_{i}}\rightarrow \omega_{i}\in \Hol$ $\forall$
            $1\leq i\leq k$ und $\{\omega'_{l}\}_{\alpha'_{l}\in
              J'_{l}}\rightarrow \omega'_{l}$ sowie $\Lambda=\alpha_{1}\times\dots\times
            (\alpha_{l}\times\alpha'_{l})\times\dots\times \alpha_{k}$.
        \end{enumerate}
    \end{beweis}
\end{satz}
\begin{bemerkung}
    \label{bem:HRKBem}
    \begin{enumerate}
    \item
        Obiger Satz besagt nun nicht nur, dass jedes
        $\phi\in[\eta]\in HH^{k}(\SsV,\mathcal{M})$ in der Form
        $\phi=\phi_{D}^{a,\eta}+ \delta^{k-1}(\psi)$ geschrieben werden kann, sondern legt
        uns sogar eine explizite Formel für die Berechnung eines
        derartigen $\psi\in HC^{k-1}(\SsV,\mathcal{M})$ in die
        Hand\footnote{Dieses ist wegen $\ker(\delta^{k-1})\neq
          \{0\}$ nicht eindeutig bestimmt.}. Nun ist die
        Berechnung wegen der rekursiven Definition von $s_{k}$ im
        Allgemeinen recht kompliziert, jedoch im Rahmen der
        Deformationsquantisierung, bei der man zunächst sowieso nur
        an den ersten drei Hochschild-Kohomologien interessiert ist,
        durchaus ausführbar:
        \begin{itemize}
        \item[$k=1$:]
            Hier ist $s^{*}_{0}=0$, also $[\eta]=\phi_{D}^{a,\eta}$ für alle $[\eta]\in
            HH^{1}(\SsV,\mathcal{M})$. Dies ist auch konsistent damit, dass
            wegen
            \begin{equation*}
                (\delta^{0}m)(a)=a*_{L}m-m*_{R}a=0 \quad\text{für alle}\quad m\in
                \mathcal{M}= HC^{0}(\SsV,\mathcal{M})   
            \end{equation*}$\im(\delta^{0})=0$ und somit            
            \begin{equation*}
                HH^{1}(\SsV,\mathcal{M})=\left\{\phi \in
                    \Hom_{\mathbb{K}}(\SsV,\mathcal{M})\:\big|\:\phi \text{ ist derivativ}\right\}
            \end{equation*}gilt
        \item[$k=2$:]
            In diesem Fall gilt
            $s_{1}=\ovl{h}_{1}-\ovl{h_{1}\Omega}_{1}$ mit            
            \begin{equation*}
                \ovl{h}_{1}(x_{0}\ot x_{1} \ot x_{2})=x_{0}\ot 1\ot
                x_{1}\ot x_{2} 
            \end{equation*}für $x_{0},x_{1},x_{2}\in \SsV$ und            
            \begin{equation*}
                \ovl{h_{1}\Omega}_{1}(x_{0}\ot x_{1} \ot x_{2})=\sum_{p=1}^{l}x_{0}\ot\left(\left[\int_{0}^{1}dt_{1}\hat{i}_{1}\left(1\ot
                            x_{1}^{p}\ot 1\right)\right]*_{e}1\ot (x_{1})_{p}\ot x_{2}\right).
            \end{equation*}für $\deg(x_{1})=p$, da
            \begin{align*}
                \Omega_{1}(x_{0}\ot x_{1} \ot
                x_{2})=&\:F_{1}\left(\sum_{p=1}^{l}\int_{0}^{1}dt_{1}i_{1}\left(x_{0}\ot
                        x_{1}^{p}\ot x_{2}\bbot\: (x_{1})_{p}\right)\right)
                \\=&\:\sum_{p=1}^{l}\left[\int_{0}^{1}dt_{1}\hat{i}_{1}\left(1\ot
                        x_{1}^{p}\ot 1\right)\right]*_{e}x_{0}\ot (x_{1})_{p}\ot
                x_{2}.
            \end{align*}
            Sei nun $\phi\in HC^{2}(\SsV,\mathcal{M})$ und
            $x\in \SsV$ mit $\deg(x)=p$, so folgt:            
            {\allowdisplaybreaks\begin{align*}
                  \left(\zeta^{1}_{-1}s^{*}_{1}\zeta^{2}\phi\right)(x)=&\:\left(s^{*}_{1}\zeta^{k}\phi\right)(1\ot
                  x \ot 1)
                  \\=&\:\overbrace{\big(\zeta^{2}\phi\big)(1\ot 1\ot x\ot 1)}^{\phi(1,x)}
                  \\ &+\sum_{p=1}^{l}\big(\zeta^{2}\phi\big)\left(1\ot\left(\left[\int_{0}^{1}dt_{1}\hat{i}_{1}\left(1\ot
                                  x^{p}\ot 1\right)\right]*_{e}1\ot x_{p}\ot 1\right)\right).
              \end{align*}}Für den letzten Summanden beachte man, dass
            \begin{align*}
                \hat{i}_{1}\left(1\ot x^{p}\ot 1\right)=&\: t_{1}^{p-1}x^{p}\ot 1 +…
                \\ &+ t_{1}^{(p-1)-l}(1-t_{1})^{l}\sum_{j_{1},…j_{l}}^{p-1}(x^{p})^{j_{1},…,j_{l}}\ot
                \:(x^{p})_{j_{1},…,j_{l}}+…
                \\ &+ (1-t_{1})^{p-1}1\ot x^{p}
            \end{align*}
            $\displaystyle\int_{0}^{1}dt_{1}t_{1}^{(p-1)-l}(1-t_{1})^{l}=\binom{p}{l}^{-1}$
            sowie $\big(\zeta^{2}\phi\big)(1\ot x
            \ot x_{p} \ot y)=\phi(x, x_{p})*_{R} y$
            gilt. Hiermit folgt:
            \begin{align}
                \label{eq:vbv}
                \left(\zeta^{1}_{-1}s^{*}_{1}\zeta^{2}\phi\right)(x)=&\:\phi(1,x)+\sum_{p=1}^{l}\Bigg[\frac{1}{p}\phi\left(x^{p},x_{p}\right)+…\nonumber
                \\ &+\binom{p}{l}^{-1}\sum_{j_{1},…j_{l}}^{p-1}\phi\left((x^{p})^{j_{1},…,j_{l}},x_{p}\right)*_{R}(x^{p})_{j_{1},…,j_{l}}+…\nonumber
                \\ &+\frac{1}{p}\phi(1,x_{p})*_{R}x^{p}\Bigg].
            \end{align}
        \item[$k=n$:]Wir haben
            \begin{equation*}
                s_{n}=\ovl{h}_{n}-\ovl{h_{n}\Omega}_{n}-\ovl{h_{n}s_{n-1}d}_{n}
            \end{equation*}
            und
            \begin{equation*}
                \ovl{h}_{n}(x_{0}\ot x_{1}\ot…\ot x_{n+1})= x_{0}\ot
                1 \ot x_{1}\ot…\ot x_{n+1}.
            \end{equation*}Den zweiten Summanden berechnet man wie im
            Falle $k=2$, wobei hier sehr viel mehr Kombinatorik zu
            berücksichtigen ist. 
            Im Falle $k=3$ ist der letzte Summand gleich $\ovl{h_{2}\ovl{h}_{1}d_{2}}-\ovl{h_{2}\ovl{h_{1}\Omega}_{1}d}_{2}$,
            was ebenfalls noch berechenbar ist.
        \end{itemize}
        Analoge Aussagen gelten nun natürlich auch für die anderen
        beiden Fälle, wobei f"ur $\Hol$ natürlich die
        Vervollständigung von \eqref{eq:vbv} zu nehmen ist. Des
        Weiteren ist $s^{*}$ auch für nicht
        symmetrische Bimoduln $\mathcal{M}$ gewinnbringend einsetzbar. Hier erhalten wir mit \eqref{eq:1} f"ur $\phi\in [\eta]\in HH^{k}(\SsV,\mathcal{M})$, dass
        \begin{equation*}
            \phi=\zeta_{-1}^{k}\Omega^{*}_{k}\zeta^{k}\phi + \left(\phi-\zeta_{-1}^{k}\Omega^{*}_{k}\zeta^{k}\phi\right)=\tilde{\phi}+ \delta^{k-1}\big(\zeta_{-1}^{k-1}s^{*}_{k-1}\zeta^{k}\phi\big)
        \end{equation*}mit $\tilde{\phi}=\zeta_{-1}^{k}\Omega^{*}_{k}\zeta^{k}\phi\in [\eta]$.
        Hierfür beachte man, dass wir in obiger Formel f"ur $\zeta^{1}_{-1}s^{*}_{1}\zeta^{2}\phi$
        explizit zwischen $*_{R}$ und $*_{L}$ unterschieden haben.
    \item
        Für endlich-dimensionales $\V$ ist
        Satz \ref{satz:HKR}~\textit{i.)}, von der expliziten Formel für
        den Korand, ein bereits wohl bekanntes
        Resultat. Ebenso für den Fall, dass $\mathcal{A}$ die Algebra
        der $C^{\infty}(M)$ der glatten Funktionen auf einer endlich-dimensionalen Mannigfaltigkeit $M$ ist,
        vgl. \cite{waldmann:2007a},\cite{cahen.gutt.dewilde:1980a}. Die
        Multivektorfelder nehmen hierbei den Platz der total
        antisymmetrischen, in jedem Argument derivativen
        Repräsentanten ein und in der Tat liefert dies eine zutreffende
        Analogie, da jeder derartige Repräsentant $\phi_{D}^{a,\eta}$
        ein total antisymmetrisches Element in $\DiffOpS{k}{1}$, den
        Differentialoperator der Ordnung $1$, ist\footnote{vgl. Proposition \ref{prop:MultidiffOps}}.
    \item
        Betrachtet man (\ref{eq:RepHKRstethol}), so k"onnte man den Wunsch versp"uren, $\widehat{\zeta^{k-1}_{-1}s^{*}_{k-1}\zeta^{k}}$ durch $\hat{\zeta}^{k-1}_{-1}\hat{s}^{*}_{k-1}\hat{\zeta}^{k}$  zu ersetzen.  Hierbei bezeichnen $\hat{\zeta}^{k}$ und $\hat{\zeta}^{k}_{-1}$ die f"ur $\Hol$ analog zu $\zeta^{k}$ und $\zeta^{k}_{-1}$ definierten Isomorphismen.
        Dies ist jedoch ohne weiteres nicht m"oglich, da wir f"ur die Definition von $s$ explizit die Kettenabbildung $G$ benutzt haben und somit die Einschr"ankung $\hat{s}_{k-1}|_{\C_{c}^{k-1}}$ im Allgemeinen nach $\hat{\C}_{c}^{k}$ und nicht ausschlie"slich nach $\C_{c}^{k}$  abbildet. Hierbei bezeichnet $(\C_{c},d_{c})$ den zu $\Hol$ geh"origen Bar-Komplex und $(\hat{\C}_{c},\hat{d}_{c})$ dessen Vervollst"andigung.
    \item
        Wir m"ochten f"ur Satz \ref{satz:HKR}~\textit{i.)} noch einmal auf anderen Weise argumentieren. Hierf"ur seien Eindeutigkeit und Existenz bereits gezeigt. Dann ist mit $\delta^{k}\phi=0$ ebenfalls $\left(\delta^{k}\cp \mathrm{Alt}_{k}\right)(\phi)=0$, also für einen Kozyklus
        $\phi\in [\eta]$ auch $\mathrm{Alt}_{k}({\phi})$ ein Kozyklus.
        Dies bedeutet $\mathrm{Alt}_{k}(\phi)=\phi^{a,\eta'}_{D}\in \left[\eta'\right]$ mit der
        Eindeutigkeit des total antisymmetrischen Repräsentanten in
        $\left[\eta'\right]$. Insbesondere ist dann $\mathrm{Alt}_{k}(\phi)$ derivativ, und die Aufgabe besteht nun darin, $\left[\eta'\right]=\left[\eta\right]$ nachzuweisen.
        Hierf"ur beachten wir, dass
        {\allowdisplaybreaks\begin{align*}
              \Big(\hat{\xi}^{k}\phi\Big)(v_{1},\dots,v_{k})=&\: \left(\Upsilon^{k}\cp F_{k}^{*}\cp 
                  \Xi^{k}_{-1}\cp \ot_{k*}\right)\big(\phi\big)(v_{1}\wedge\dots\wedge v_{k})
              \\=&\:\left(F_{k}^{*}\cp 
                  \Xi^{k}_{-1}\cp \ot_{k*}\right)\big(\phi\big)(1\ot 1\ot v_{1}\wedge\dots\wedge v_{k})
              \\=&\:\left(\Xi^{k}_{-1}\cp \ot_{k*}\right)\big(\phi\big)\big(F_{k}(1\ot 1\ot v_{1}\wedge\dots\wedge v_{k})\big)
              \\=&\:\sum_{\sigma\in S_{k}}\sign(\sigma)\left(\Xi^{k}_{-1}\cp \ot_{k*}\right)\big(\phi\big)(1\ot v_{\sigma(1)}\ot\dots\ot v_{\sigma(k)}\ot 1)
              \\=&\:\sum_{\sigma\in S_{k}}\sign(\sigma)\big(\ot_{k*}\phi\big)(v_{\sigma(1)}\ot\dots\ot v_{\sigma(k)})
              \\=&\:\sum_{\sigma\in S_{k}}\sign(\sigma)\:\phi\:(v_{\sigma(1)},\dots, v_{\sigma(k)})
              \\=&\: k!\: \mathrm{Alt}_{k}(\phi)\:(v_{\sigma(1)},\dots, v_{\sigma(k)}),
          \end{align*}}also $\Big(\hat{\xi}^{k}\phi\Big)=k!\:
        \mathrm{Alt}_{k}\left(\phi\big|_{\V^{k}}\right)=k!\:
        \mathrm{Alt}_{k}(\phi)\big|_{\V^{k}}$. Proposition \ref{prop:wichEizuHKR}~\textit{iii.)} zeigt dann $\left(\xi^{k}\cp\hat{\xi}^{k}\right)(\phi)=\left(\mathrm{Alt}_{k}(\phi)\big|_{\V^{k}}\right)_{D}$,
        und mit der Derivativität von $\mathrm{Alt}_{k}(\phi)$ zeigt Proposition \ref{prop:wichEizuHKR}~\textit{i.)}, dass
        $\left(\xi^{k}\cp\hat{\xi}^{k}\right)(\phi)=\mathrm{Alt}_{k}(\phi)$. Nun gilt 
        $\xi^{k}\cp \hat{\xi}^{k}\colon [\eta]\longrightarrow[\eta]$,
        also $\mathrm{Alt}_{k}(\phi)\in [\eta]$.
    \end{enumerate}
\end{bemerkung}

\newpage

\chapter{Differentielle Hochschild-Kohomologien}
\label{cha:DiffHochK}
In diesem Kapitel wollen wir uns dem Begriff der differentiellen
Hochschild-Koketten zuwenden und die Rolle der Kettenabbildungen $F^{*}$
und $G^{*}$ in diesem Rahmen diskutieren. Im Großen und Ganzen soll es hier
darum gehen, nützliche Relationen und Ideen zusammenzutragen, die als
Basis für weitergehende Analysen benutzt werden können. Wir werden
dabei besonderen Wert auf die Diskussion etwaiger
Fallstricke legen, die aus der zu naiven Betrachtung des
Unterkomplex-Begriffes resultieren. 
\section{Multidifferentialoperatoren und symmetrische Bimoduln}
In diesem Abschnitt soll es zunächst darum gehen, den Begriff des
Multidifferentialoperators in voller algebraischer Allgemeinheit
kennenzulernen (vgl.\cite[Anhang A]{waldmann:2007a}), um hiermit die
differentiellen Hochschildkomplexe sowie die stetigen, differentiellen
Hochschildkomplexe für symmetrische Bimoduln zu definieren.

Hierf"ur sei daran erinnert, dass wir unter einer $\mathbb{K}$-Algebra immer
einen $\mathbb{K}$-Vektor-raum mit assoziativer,
$\mathbb{K}$-bilinearer Algebramultiplikation verstehen. Ist von einem $\mathcal{A}$-Modul $\mathcal{M}$ die Rede,
so meinen wir einen $\mathbb{K}$-Vektorraum mit $\mathbb{K}$-bilinearer
$\mathcal{A}$-Modul-Multiplikation, wobei die Linearität im
Modul-Element für die Wohldefiniertheit des
Multi-differentialoperator-Begriffes unabdingbar ist.
Befinden wir uns im Folgenden in der Situation einer kommutativen
Algebra, so behandeln wir $\mathcal{M}$ als $\mathcal{A}$-Linksmodul
und bezeichnen die Modul-Multiplikation mit $*_{L}$, wohlwissend, dass
dies f"ur derartige Algebren keine Einschr"ankung bedeutet, siehe
Anhang \ref{sec:ringe-moduln-und}. Ist $\mathcal{A}$ kommutativ, so
setzen wir $1_{\mathcal{A}}*_{L}m=m$ für alle
$m\in \mathcal{M}$ voraus.

Als Teilresultat dieses Abschnittes erhalten wir dann die Isomorphie der stetigen,
differentiellen Kohomologien der Algebren $\Hol$ und $\SsV$ für derartige vollständige,
lokalkonvexe, symmetrischen $\Hol-\Hol$-Bimoduln.
\begin{definition}[Multidifferentialoperator] 
    \label{def:MultidiffOps}
    Gegeben eine assoziative, kommutative $\mathbb{K}$-Algebra $(\mathcal{A},*)$ und ein
    $\mathcal{A}$-Modul $(\mathcal{M},*_{L})$. Dann sind die
    Multidifferentialoperatoren
    $\mathrm{DiffOp}_{k}^{L}(\mathcal{A},\mathcal{M})$ mit Argumenten
    im $k$-fachen kartesischen Produkt $\mathcal{A}^{k}$ von
    $\mathcal{A}$ und Werten in $\mathcal{M}$, der Multiordnung
    $L=(l_{1},\dots,l_{k})\in \mathbb{Z}^{k}$, induktiv definiert durch
    \begin{equation*}
        \mathrm{DiffOp}_{k}^{L}(\mathcal{A},\mathcal{M})=\{0\}\qquad\quad \forall\: L\in \mathbb{Z}^{k} \text{ mit }l_{i}< 0 \text{ f"ur ein }1\leq i\leq k  
    \end{equation*}
    sowie
    \begin{align*}
        \mathrm{DiffOp}_{k}^{L}(\mathcal{A},\mathcal{M})=&\Big\{D\in \Hom_{\mathbb{K}}\big(\mathcal{A}^{k},\mathcal{M}\big)\:\Big|\:\forall\: a\in \mathcal{A},\:\forall\: 1\leq i\leq k \text{ gilt}:
        \\ &\qquad\qquad\qquad L_{a}\cp D - D\cp L_{a}^{i} \in \mathrm{DiffOp}_{k}^{L-e_{i}}(\mathcal{A},\mathcal{M})\Big\}.
    \end{align*}
    Dabei bedeutet $L-m\cdot e_{i}=(l_{1},\dots,l_{i}-m,\dots, l_{k})$,
    $L_{a}\cp D (a_{1},\dots,a_{k})=a*_{L}D (a_{1},\dots,a_{k})$ und
    $\left(D\cp
        L^{i}_{a}\right)(a_{1},\dots,a_{k})=D(a_{1},\dots,a_{i}*a,\dots,a_{k})$.
    Weiterhin setzen wir $|L|=\sum_{i}l_{i}$ mit $|L|=-1$ falls $l_{i}< 0$ f"ur ein $1\leq i\leq k$ und schreiben
    $L\leq L'$ falls $\:l_{i}\leq l'_{i}\:\:\forall\:1\leq i\leq k$ sowie $L=n$ falls $\:l_{i}=n \:\:\forall\:1\leq i\leq k$.
\end{definition} 
\begin{proposition}
    \label{prop:MultidiffOps}
    Unter den Voraussetzungen obiger Definition gilt:
    \begin{enumerate}
    \item
        Sei
        \begin{align*}
            []_{i}^{a}\colon\Hom_{\mathbb{K}}(\mathcal{A}^{k},\mathcal{M})&\longrightarrow \Hom_{\mathbb{K}}(\mathcal{A}^{k},\mathcal{M})\\
            D &\longmapsto L_{a}\cp D - D\cp L_{a}^{i}.
        \end{align*}
        Dann gilt $[]^{a}_{i}\cp\: []^{b}_{j}=[]_{j}^{b}\cp\: []^{a}_{i}\:$
        f"ur alle $a,b\in \mathcal{A}$ und alle $i,j\in\{1,\dots,k\}$.
    \item
        Sei $D\in
        \Hom_{\mathbb{K}}(\mathcal{A}^{k},\mathcal{M})$. Dann ist
        genau dann $D\in \DiffOp{k}{L}$, wenn:
        \begin{equation}
            \label{eq:DiffopCharack}
            [][]_{i}^{a_{1},\dots,a_{l_{i}+1}}D=0\qquad\forall\:1\leq i\leq k,\:\forall\:a_{1},\dots,a_{l_{i}+1}\in \mathcal{A}.
        \end{equation}
        Hierbei steht $[]_{i}^{a_{1},\dots,a_{l_{i}+1}}\colon\DiffOp{k}{L}\longrightarrow 0\:$
        abk"urzend f"ur $[]_{i}^{a_{1}}\cp\dots\cp
        \:[]_{i}^{a_{l_{i}+1}}$. 
    \item
        F"ur $K\leq L$ ist
        \begin{equation*}
            \DiffOp{k}{K}\subseteq \DiffOp{k}{L}
        \end{equation*}
        und 
        \begin{equation*}
            \DiffOp{k}{\bullet}=\bigcup_{L\geq 0} \DiffOp{k}{L}
        \end{equation*}
        ein filtrierter Untervektorraum von $\Hom_{\mathbb{K}}(\mathcal{A}^{k},\mathcal{M})$.
    \item
        Jedes $\DiffOp{k}{L}$ und somit $\DiffOp{k}{\bullet}$ wird verm"oge $a*_{L}D=L_{a}\cp D$ zu einem $\mathcal{A}$-Linksmodul. Des Weiteren ist $D*^{i}_{R}a=D\cp L_{a}^{i}\:$
        f"ur alle $1\leq i\leq k$
        eine Rechtsmodul-Multiplikation auf jedem $\DiffOp{k}{L}$ und somit auf $\DiffOp{k}{\bullet}$.
    \item
        Sei $D\in \DiffOp{k}{L}$ und
        $L\prec_{i}
        P:=(l_{1},…,l_{i-1},l_{i}+p_{1},…,l_{i}+p_{m},l_{i+1},…,l_{k})$.
        Dann ist:
        \begin{enumerate}
        \item
            $D_{\mathcal{A}}*_{L}D\in\DiffOp{k+m}{(P,L)}$ für $D_{\mathcal{A}}\in
            \mathrm{DiffOp}_{m}^{P}(\mathcal{A},\mathcal{A})$,
        \item
            $D\cp_{i}D_{\mathcal{A}}\in \DiffOp{k+m\:-1}{L\prec_{i}P}$
            für $D_{\mathcal{A}}\in
            \mathrm{DiffOp}_{m}^{P}(\mathcal{A},\mathcal{A})$.
        \end{enumerate}
        Insbesondere ist $\id_{\mathcal{A}}*_{L}D\in\DiffOp{k+1}{(0,L)}$
        und $D\cp_{i}*\in \DiffOp{k+1}{L\prec_{i}(0,0)}$. Hierf"ur beachte man, dass jede assoziative Algebra $\mathcal{A}$ insbesondere ein $\mathcal{A}$-Linksmodul ist. 
    \item
        Sei $D\in \Hom_{\mathbb{K}}(\mathcal{A}^{k},\mathcal{M})$ derivativ in jedem Argument. Dann ist $D\in \DiffOp{k}{1}$.
    \end{enumerate}
    \begin{beweis}
        \begin{enumerate}
        \item
            Zun"achst ist das Bild unter $[]_{i}^{a}$ in der Tat $\mathbb{K}$-multilinear, da $*_{L}$ linear im $\mathcal{M}$-Argument und $*$ bilinear ist. Die behauptete Vertauschungsrelation folgt unmittelbar aus der Kommutativit"at von  $\mathcal{A}$.
        \item
            Sei $D\in \DiffOp{k}{L}$. Dann gilt per Definition:
            \begin{equation*} 
                []_{i}^{a_{1},\dots,a_{l_{i}+1}}D\in \DiffOp{k}{(l_{1},\dots,-1_{i},\dots,l_{k})}=0.
            \end{equation*}
            F"ur die umgekehrte Implikation sei $[]_{j}^{a}D=0$ f"ur alle $1\leq j\leq k$. Dann folgt unter Ber"ucksichtigung von $a*_{L} 0=0_{\mathbb{K}}\cdot [a*_{L} 0]=0$, dass
            \begin{equation*}
                D'=[]^{a_{1}}_{1}\cp \dots \cp []^{a_{k}}_{k}D=0\in \DiffOp{k}{-1},
            \end{equation*}
            also $[]^{a_{2}}_{2}\cp \dots\cp []^{a_{k}}_{k}D\in \DiffOp{k}{(0,-1,\dots,-1)}$ und induktiv $D\in \DiffOp{k}{0}$.
            Sei nun die Aussage f"ur $L$ korrekt und \eqref{eq:DiffopCharack} f"ur $L'=L+e_{i}$ erf"ullt. Sei weiter $D'= []_{i}^{a}D$, so ist nach Voraussetzung $[]_{i}^{a_{1},\dots,a_{l_{i}+1}}D'=0$ und f"ur $j\neq i$ folgt:
            \begin{align*}
                []_{j}^{a_{1},\dots,a_{l_{j}+1}}D' \glna{\textit{i.)}}\:[]_{i}^{a}\cp\: []_{j}^{a_{1},\dots,a_{l_{j}+1}}D'=[]_{i}^{a}0=0.
            \end{align*}
            Dies zeigt $D'\in \DiffOp{k}{L}$ und somit $D\in \DiffOp{k}{L'}$.
        \item
            Die erste Inklusion folgt unmittelbar aus \textit{ii.)} und $[]_{i}^{a}0=0$. F"ur die zweite Behauptung beachte man, dass die $[]_{i}^{a}$ lineare Abbildungen sind und somit \eqref{eq:DiffopCharack} ein lineares Kriterium ist. Hiermit ist $\DiffOp{k}{\bullet}$ ein Untervektorraum von $\Hom_{\mathbb{K}}(\mathcal{A}^{k},\mathcal{M})$ und besagte Inklusion liefert die Filtrationseigenschaft. 
        \item
            Dies folgt sofort aus \textit{ii.)}, $[]_{i}^{a}0=0$ sowie den Vertauschungsrelationen $ []_{i}^{b}\left[L_{a} D\right] =L_{a} \left[[]_{i}^{b}D\right]$ und $[]_{i}^{b}\left[D\cp L^{i}_{a}\right]  =\left[[]_{i}^{b}D\right]\cp L^{i}_{a}$.
        \item
            Zunächst sind alle Kompositionen $\mathbb{K}$-multilineare Abbildungen
            nach $\mathcal{M}$.
            \begin{enumerate}
            \item
                Mit der Kommutativität von $\mathcal{A}$ und den Modul-Multiplikationsregeln gilt:
                \begin{equation*}
                    []_{j}^{a}\big[D_{\mathcal{A}}*_{L}D\big]=\big[a*D_{\mathcal{A}}\big]*_{L}D-\big[D_{\mathcal{A}}\cp L_{a}^{j}\big]*_{L}D =\big[[]_{j}^{a}D_{\mathcal{A}}\big]*_{L}D
                \end{equation*}
                für $1\leq j\leq m\:$ und 
                \begin{equation*}
                    []_{j}^{a}\big[D_{\mathcal{A}}*_{L}D\big]=D_{\mathcal{A}}*_{L}\big[a*_{L}D\big]-D_{\mathcal{A}}*_{L}\big[D\cp
                    L_{a}^{j}\big]=D_{\mathcal{A}}*_{L}\big[[]_{j}^{a}D\big]
                \end{equation*}
                f"ur $m+1\leq j\leq k+m$.
                Die Behauptung folgt dann unmittelbar aus \textit{ii.)}.\\
            \item
                Seien $|L|,|P|\neq -1$, andernfalls ist die Aussage trivial. Wir zeigen diese per Induktion über $|L|+|P|$. 
                Sei hierf"ur $L=m=0$. Dann ist $[]_{j}^{a}D=0$ für
                $1\leq j\leq k$ und
                $[]_{j}^{a}D_{\mathcal{A}}=0$ für $1\leq
                j\leq m$. In den Fällen $1\leq j< i$ und $i+m-1< j\leq
                k+m-1$ folgt
                {\small\begin{align*}
                      []_{j}^{a}\big[D\cp_{i} D_{\mathcal{A}}\big]=\big[[]_{j}^{a}D\big]\cp_{i} D_{\mathcal{A}}=0
                  \end{align*}}und für $i\leq j\leq i+m-1$ gilt
                {\small\begin{align*}
                      \big[D\cp_{i}D_{\mathcal{A}}\big]\cp L_{a}^{j}=&\:D\cp_{i}\big[D_{\mathcal{A}}\cp
                      L^{j-i+1}_{a}\big]=D\cp_{i}\big[a*_{L}D_{\mathcal{A}}\big]
                      =\big[D\cp L_{a}^{i}\big]\cp
                      D_{\mathcal{A}}\\=&\:a*_{L}\big[D\cp_{i}D_{\mathcal{A}}\big],
                  \end{align*}}also $[]_{j}^{a}\big[D\cp_{i}D_{\mathcal{A}}\big]=0$.
                Sei nun die Aussage für $|L|+|P|-1$ korrekt. Dann \- gilt in den Fällen
                $1\leq j< i$ und $i+m-1< j\leq k+m-1$:
                \begin{equation*}
                    []_{j}^{a}\big[D\cp_{i}D_{\mathcal{A}}\big]=\big[[]_{j}^{a}D\big]\cp_{i}D_{\mathcal{A}}\in
                    \DiffOp{k+m\:-1}{[L-e_{j}]\prec_{i}P}
                \end{equation*}
                nach Induktionsvorraussetzung und für $i\leq j\leq i+m-1$ erhalten wir:
                {\small\begin{align*}
                      []_{j}^{a}\big[D\cp_{i}D_{\mathcal{A}}&\:\big]=a*_{L}\big[D\cp_{i}D_{\mathcal{A}}\big]
                      \overbrace{-\big[D\cp L_{a}^{i}\big]\cp_{i} D_{\mathcal{A}} + \big[D\cp
                        L_{a}^{i}\big]\cp_{i} D_{\mathcal{A}}}^{0}-
                      \big[D\cp_{i}D_{\mathcal{A}}\big]\cp L^{j}_{a}
                      \\=&\:\big[a*_{L}D\big]\cp_{i}D_{\mathcal{A}}
                      -\big[D\cp L_{a}^{i}\big]\cp_{i} D_{\mathcal{A}} + D\cp_{i} \big[a*D_{\mathcal{A}}\big]- D\cp_{i}\big[D_{\mathcal{A}}\cp L^{j-i+1}_{a}\big]
                      \\=&\:\underbrace{\big[[]_{i}^{a}D\big]\cp_{i}D_{\mathcal{A}}}_{\DiffOp{k+m\:-1}{[L-
                          e_{i}]\prec_{i} P}}+\underbrace{D\cp_{i}\big[[]_{j-i+1}^{a}D_{\mathcal{A}}\big]}_{\DiffOp{k+m\:-1}{L\prec_{i}[P-e_{(j-i+1)}]}}.
                  \end{align*}}Hierbei gelten die Zugeh"origkeiten unter den geschweiften Klammern nach Induktionsvorraussetzung.
                
                Im ersten Fall: $j\neq \{i,\dots,i+m-1\}$ ist
                $[L-e_{j}]\prec_{i}P=[L\prec_{i}P]-e_{j}$ und für $i\leq
                j\leq i+m-1$ gilt $L\prec_{i}[P-e_{(j-i+1)}]=
                [L\prec_{i}P]-e_{j}$ sowie $[L-e_{i}]\prec_{i}P=[L\prec_{i}
                P]-e_{i}-…-e_{i+m-1}\leq [L\prec_{i}P]-e_{j}$.
                Mit \textit{iii.)} zeigt dies $[]_{j}^{a}\big[D\cp_{i}D_{\mathcal{A}}\big]\in
                \DiffOp{k+m-1}{[L\prec_{i}P]-e_{j}}$ und Definition \ref{def:MultidiffOps} liefert schlie"slich $D\cp_{i} D_{\mathcal{A}}\in \DiffOp{k+m-1}{L\prec_{i} P}$. 
            \end{enumerate}
            Die letzte Behauptung folgt mit dem bereits Gezeigten und mit $\id_{\mathcal{A}}\in
            \mathrm{DiffOp}_{1}^{0}(\mathcal{A},\mathcal{A})$ sowie 
            $*\in\mathrm{DiffOp}_{2}^{0}(\mathcal{A},\mathcal{A})$.
        \item
            F"ur jedes $1\leq j\leq k$ ist
            $\big[[]_{j}^{a}D\big](a_{1},\dots,a_{k})=-\:a_{i}*_{L}D(a_{1},\dots,a,\dots,a_{k})$,
            also $[]_{j}^{a,b}D=0$.
        \end{enumerate}
    \end{beweis}
\end{proposition}
\begin{bemerkung}
    \label{bem:DiffOpCRn}
    Die obige rein algebraische Definition der
    Multidifferentialoperatoren scheint zun"achst etwas befremdlich. Es
    l"asst sich jedoch zeigen (vgl.\cite[Anhang A]{waldmann:2007a}), dass diese f"ur die assoziative, kommutative Algebra $\mathcal{A}=C^{\infty}(M)$ auf einer glatten Mannigfaltigkeit $M$ gerade mit der "ublichen analytischen Definition "ubereinstimmt. Beispielsweise ist genau dann $D\in \mathrm{DiffOp}_{1}^{n}(\mathcal{A},\mathcal{A})$, wenn ein mit der auf $M$ gegebenen differenzierbaren Struktur vertr"aglicher $C^{\infty}$-Atlas $\mathbf{A}$ von $M$ derart existiert, dass f"ur
    jede Karte $(U,x)\in \mathbf{A}$, in den Indizes $i_{1},\dots,i_{k}$ symmetrische Funktionen $D_{U}^{i_{1},\dots,i_{k}}\in C^{\infty}(U)$ existieren, so dass:
    \begin{equation*}
        D\left(f\big|_{U}\right)=\sum_{r=0}^{n}\sum_{i_{1},\dots,i_{r}}\frac{1}{r!}D_{U}^{i_{1},\dots,i_{k}}\frac{\pt^{r}f\big|_{U}}{\pt x^{i_{1}}\dots\pt x^{i_{r}}}.
    \end{equation*}
\end{bemerkung}
\begin{definition}[Stetige Multidifferentialoperatoren] 
    \label{def:MultidiffOpsstet}
    Gegeben eine assoziative, kommutative, lokalkonvexe $\mathbb{K}$-Algebra $(\mathcal{A},*)$ und ein
    lokal-konvexer $\mathcal{A}$-Modul $(\mathcal{M},*_{L})$, so sind die
    stetigen Multidifferentialoperatoren induktiv definiert durch
    \begin{equation*}
        \mathrm{DiffOp}_{k}^{L,\cont}(\mathcal{A},\mathcal{M})=\{0\}\qquad\quad \forall\: L\in \mathbb{Z}^{k} \text{ mit }l_{i}< 0 \text{ f"ur ein }1< i\leq k  
    \end{equation*}
    sowie
    \begin{align*}
        \mathrm{DiffOp}_{k}^{L,\cont}(\mathcal{A},\mathcal{M})=&\Big\{D\in \Hom^{\cont}_{\mathbb{K}}\big(\mathcal{A}^{k},\mathcal{M}\big)\:\Big|\:\forall\: a\in \mathcal{A},\:\forall\: 1\leq i\leq k \text{ gilt}:
        \\ &\qquad\qquad\qquad\qquad\quad L_{a}\cp D - D\cp L_{a}^{i} \in \mathrm{DiffOp}_{k}^{L-e_{i},\cont}(\mathcal{A},\mathcal{M})\Big\}.
    \end{align*}
\end{definition}
\begin{bemerkung}
    \label{bem:DeriBeiStet}
    \begin{enumerate}
    \item
        Proposition \ref{prop:MultidiffOps} überträgt sich sinngemäß auf $\DiffOpc{k}{\bullet}$, 
        da in der Situation von Definition \ref{def:MultidiffOpsstet} sowohl $*$ als auch $*_{L}$ stetige Abbildungen sind und somit $[]_{i}^{a}\colon\DiffOpc{k}{L}\longrightarrow
        \DiffOpc{k}{L-e_{i}}$ gilt.
        In der Tat gewährleistet dies im
        Induktionsschritt zu Proposition \ref{prop:MultidiffOps}~\textit{ii.)}, dass $D' \in
        \Hom_{\mathbb{K}}^{\cont}(\mathcal{A}^{k},\mathcal{M})$ stetig
        ist. Proposition \ref{prop:MultidiffOps}~\textit{ii.)} zeigt dann
        insbesondere, dass genau dann $\phi\in
        \DiffOpc{k}{L}$ gilt, wenn $\phi\in
        \Hom_{\mathbb{K}}^{\cont}(\mathcal{A}^{k},\mathcal{M})$ und
        $\phi\in\DiffOp{k}{L}$ ist. 
    \item
        Man beachte, dass die Forderung der Stetigkeit der Differentialoperatoren in der Tat eine echte Zusatzbedingung liefert. F"ur den Fall $\mathcal{M}=\mathcal{A}=C^{\infty}(\mathbb{R}^{n},\mathbb{R})$, in welchem $\mathcal{A}$ durch die "ublichen Halbnormen 
        \begin{equation*}
            p_{K,l}:\phi\mapsto \sup_{\substack{x\in K\\ |\alpha|\leq l}}\left|\frac{\pt^{\alpha}\phi}{\pt x^{\alpha}}\right|
        \end{equation*}
        mit $K\subseteq \mathbb{R}^{n}$ kompakt und $l\in \mathbb{N}$ topologisiert ist, sind Multidifferentialoperatoren eben nur wegen ihrer analytischen Form, also ihrer unmittelbaren "Ahnlichkeit zu obigen Halbnormen stetig. W"ahlt man hier ein anderes Halbnormensystem, so ist deren Stetigkeit auch f"ur diese Algebra im Allgemeinen nicht gew"ahrleistet. Um dies noch deutlicher zu machen, sei dem Leser nahegelegt zu versuchen, die Stetigkeit der Derivation $i_{u}\in \mathrm{DiffOp}_{1}^{1}(\Ss^{\bullet}(\V),\Ss^{\bullet}(\V))$ mit
        \begin{align*}
            i_{u}:v_{1}\vee\dots\vee v_{k}\longmapsto \sum_{i=1}^{k} |u(v_{i})|\cdot v_{1}\vee\dots\blacktriangle^{i}\dots\vee v_{k}
        \end{align*}und $i_{u}(1)=0$ f"ur beliebiges $u\in \V^{*}$
        nachzuweisen. Im Falle $u\in V'$ ist dies allerdings kein Problem.
    \end{enumerate}
\end{bemerkung}
\begin{lemma}
    \label{lemma:stetfortDiffOp}
    Gegeben ein vollständiger, hausdorffscher, lokalkonvexer $\Hol$-Modul
    $\mathcal{M}$. Dann gilt
    \begin{equation*}
        \mathrm{DiffOp}_{k}^{\bullet,\cont}(\Hol,\mathcal{M})\cong\DiffOpS{k}{\bullet,\cont}    
    \end{equation*}
    vermöge Einschränkung und
    stetiger Fortsetzung.
    \begin{beweis}
        Mit Proposition \ref{prop:MultidiffOps}~\textit{ii.)} folgt unmittelbar $\phi\big|_{\SsV^{k}}\in \DiffOpS{k}{\bullet,\cont}$
        für alle $\phi\in\mathrm{DiffOp}_{k}^{\bullet,\cont}(\Hol,\mathcal{M})$.
        Für den Rest der Behauptung reicht es zu zeigen, dass $\hat{\phi}\in\mathrm{DiffOp}_{k}^{\bullet,
          \cont}(\Hol,\mathcal{M})$, falls $\phi \in
        \DiffOpS{k}{\bullet, \cont}$. Hierzu beachten wir, dass
        \begin{align*}
            []_{i_{1},…,i_{l}}^{\bullet_{1},…,\bullet_{l}}\phi:(a_{1},…,a_{l+k})&\longmapsto
            []_{i_{1},…,i_{l}}^{a_{1},…,a_{l}}\phi(a_{l+1},\dots,a_{k+l})\quad\text{
              sowie}\\
            []_{i_{1},…,i_{l}}^{\bullet_{1},…,\bullet_{l}}\hat{\phi}:(a_{1},…,a_{l+k})&\longmapsto
            []_{i_{1},…,i_{l}}^{a_{1},…,a_{l}}\hat{\phi}(a_{l+1},\dots,a_{k+l})
        \end{align*}
        beide stetig sind und $\left([]_{i_{1},…,i_{l}}^{\bullet_{1},…,\bullet_{l}}\hat{\phi}\right)\Big|_{\SsV^{k}}=[]_{i_{1},…,i_{l}}^{\bullet_{1},…,\bullet_{l}}\phi$
        gilt. Mit der Eindeutigkeit der stetigen Fortsetzung zeigt
        dies
        $[]_{i_{1},…,i_{l}}^{\bullet_{1},…,\bullet_{l}}\hat{\phi}=\widehat{\left([]_{i_{1},…,i_{l}}^{\bullet_{1},…,\bullet_{l}}\phi\right)}$
        und für $\phi\in \DiffOpS{k}{L,\cont}$ folgt
        \begin{equation*}
            []_{i_{1},…,i_{l+1}}^{\bullet_{1},…,\bullet_{l+1}}\hat{\phi}=\widehat{\left([]_{i_{1},…,i_{l+1}}^{\bullet_{1},…,\bullet_{l+1}}\phi\right)}=0,
        \end{equation*} 
        also $\hat{\phi}\in \mathrm{DiffOp}_{k}^{L,
          \cont}(\Hol,\mathcal{M})$. 
    \end{beweis}
\end{lemma}

\begin{definition}[Differentieller Hochschild-Komplex]
    \label{def:DiffKomplexe}
    Gegeben eine kommutative Algebra $\mathcal{A}$ und ein symmetrischer $\mathcal{A}-\mathcal{A}$-Bimodul $\mathcal{M}$. Wir betrachten die $\mathbb{K}$-Vektorräume
    \begin{equation*}
        HC_{\diff}^{k}(\mathcal{A},\mathcal{M}):=
        \begin{cases} \{0\} & k<0\\
            \mathcal{M} & k=0\\
            \DiffOp{k}{\bullet}& k\geq 1
        \end{cases} 
    \end{equation*} 
    sowie die durch \eqref{eq:Hochschilddelta} definierten $\mathbb{K}$-lineare Abbildungen: 
    \begin{equation*}
        \delta^{k}_{\diff}\colon HC_{\diff}^{k}(\mathcal{A},\mathcal{M})\longrightarrow HC_{\diff}^{k+1}(\mathcal{A},\mathcal{M}).
    \end{equation*}
    Hierf"ur beachte man, dass im symmetrischen Falle $*_{L}=*_{R}$ gilt und Proposition \ref{prop:MultidiffOps}~\textit{v.)} dann zeigt, dass die $\delta_{\diff}^{k}$ in der Tat in die behauptete Menge abbilden.
    Sind $\mathcal{A}$ und $\mathcal{M}$ lokalkonvex, so definieren wir
    \begin{equation*}
        HC_{\mathrm{c,d}}^{k}(\mathcal{A},\mathcal{M}):=
        \begin{cases} \{0\} & k<0\\
            \mathcal{M} & k=0\\
            \DiffOp{k}{\bullet, \cont}& k\geq 1
        \end{cases} 
    \end{equation*} 
    sowie die zugeh"origen Kettendifferentiale:
    \begin{equation*}
        \delta^{k}_{\mathrm{c,d}}: HC_{\mathrm{c,d}}^{k}(\mathcal{A},\mathcal{M})\longrightarrow HC_{\mathrm{c,d}}^{k+1}(\mathcal{A},\mathcal{M}).
    \end{equation*}
\end{definition}
Abschließend erhalten wir folgendes Korollar:
\begin{korollar}
    \label{kor:DiffKohom}
    \begin{enumerate}
    \item
        Sei $\mathcal{M}$ ein symmetrischer $\Ss^{\bullet}(\V)-\Ss^{\bullet}(\V)$-Bimodul. Dann induzieren $\xi$ und $\hat{\xi}$ wohldefinierte Kettenabbildungen  $\wt{\xi^{k}}$ und $\wt{\hat{\xi}^{k}}$ zwischen $(HC_{\diff}(\SsV,\mathcal{M}),\delta_{\diff})$ und\\  $(KC(\V,\mathcal{M}),\Delta)$.
        Des Weiteren ist $\wt{\xi^{k}}$ injektiv und $\wt{\hat{\xi}^{k}}$ surjektiv.
    \item
        Sei $\mathcal{M}$ ein symmetrischer, lokalkonvexer
        $\Ss^{\bullet}(\V)-\Ss^{\bullet}(\V)$-Bimodul. Dann induzieren $\xi$
        und $\hat{\xi}$ wohldefinierte Kettenabbildungen zwischen
        $(HC_{\mathrm{c,d}}(\SsV,\mathcal{M}),\delta_{\mathrm{c,d}})$
        und\\  $(KC^{\cont}(\V,\mathcal{M}),\Delta)$. Des Weiteren ist
        $\wt{\xi^{k}}$ injektiv und $\wt{\hat{\xi}^{k}}$ surjektiv.
    \item
        Sei $\mathcal{M}$ ein vollständiger, symmetrischer, hausdorffscher, lokalkonvexer 
        $\Hol-\Hol$-Bimodul, so sind
        $\big(HC_{\mathrm{c,d}}(\Hol,\mathcal{M}),\hat{\delta}_{\mathrm{c,d}}\big)$
        und $\big(HC_{\mathrm{c,d}}(\SsV,\mathcal{M}),\delta_{\mathrm{c,d}}\big)$ kettenisomorph
        vermöge Einschränkung und stetiger
        Fortsetzung. Des Weiteren gilt:
        \begin{equation*}
            HH^{k}_{\mathrm{c,d}}(\Hol,\mathcal{M})\cong HH^{k}_{\mathrm{c,d}}(\SsV,M).
        \end{equation*}
    \end{enumerate}
    \begin{beweis}
        \begin{enumerate}
        \item[\textit{i.),ii.)}]
            Zunächst ist klar, dass sowohl
            $\hat{\xi}^{k}\colon HC_{\diff}(\SsV,\mathcal{M})\longrightarrow
            KC(\V,\mathcal{M})$ als auch $\hat{\xi}^{k}\colon HC_{\mathrm{c,d}}(\SsV,\mathcal{M})\longrightarrow KC^{\cont}(\V,\mathcal{M})$ gilt. Des Weiteren ist das Bild
            unter $\xi^{k}$ nach Proposition \ref{prop:wichEizuHKR}~\textit{iii.)}
            derivativ in jedem Argument, also
            nach Proposition \ref{prop:MultidiffOps}~\textit{vi.)} ein Element in
            $\DiffOpS{k}{1}$. Dies zeigt $\xi^{k}\colon KC(\V,\mathcal{M})\longrightarrow
            HC_{\diff}(\SsV,\mathcal{M})$, und da $\xi^{k}$ im lokal
            konvexen Fall stetige Elemente auf stetige Elemente abbildet,
            gilt gleichermaßen $\xi^{k}\colon KC^{\cont}(\V,\mathcal{M})\longrightarrow
            HC_{\mathrm{c,d}}(\SsV,\mathcal{M})$. Die Injektivität von
            $\wt{\xi^{k}}$ sowie die Surjektivität von $\wt{\hat{\xi}^{k}}$ folgen
            in beiden Fällen wieder unmittelbar aus Lemma \ref{lemma:Fkettenabb}~\textit{i.)}.
        \item[\textit{iii.)}]
            Dies folgt mit Lemma \ref{lemma:stetfortDiffOp} analog zu
            Satz \ref{satz:HochschildHol}, da auch hier für
            \begin{align*}
                &\hat{\delta}^{k}_{\mathrm{c,d}}\colon HC^{k}_{\mathrm{c,d}}\big(\Hol,\mathcal{M}\big)\longrightarrow
                HC^{k+1}_{\mathrm{c,d}}\big(\Hol,\mathcal{M}\big)\\
                &\delta^{k}_{\mathrm{c,d}}\colon HC^{k}_{\mathrm{c,d}}\big(\Ss^{\bullet}(\V),\mathcal{M}\big)\longrightarrow
                HC^{k+1}_{\mathrm{c,d}}\big(\Ss^{\bullet}(\V),\mathcal{M}\big)
            \end{align*}gilt, dass:
            \begin{equation*}
                \hat{\delta}_{\mathrm{c,d}}^{k}\left(\hat{\phi}\right)\Big|_{\SsV^{k+1}}= \delta_{\mathrm{c,d}}^{k}\left(\hat{\phi}\big|_{\SsV^{k}}\right).
            \end{equation*}
        \end{enumerate}
    \end{beweis}
\end{korollar} 
\begin{bemerkung}
    Es ist im Allgemeinen nicht klar, dass die $\wt{\xi^{k}}$
    Isomorphismen sind, dass also
    \begin{align*}
        &HH^{k}_{\diff}(\SsV,\mathcal{M})\cong
        \Hom_{\mathbb{K}}^{a}\big(\V^{k},\mathcal{M}\big)\cong
        HH^{k}(\SsV,\mathcal{M})\\
        &HH^{k}_{\mathrm{c,d}}(\SsV,\mathcal{M})\cong
        \Hom_{\mathbb{K}}^{a,\cont}\big(\V^{k},\mathcal{M}\big)\cong
        HH_{\cont}^{k}(\SsV,\mathcal{M})
    \end{align*}gilt. Dies ist ein Phänomen, dass 
    bei Unterkomplexen immer auftreten kann.
    Hierfür beachte man, dass es wegen $\ker\left(\delta^{k}_{\diff}\right)\subseteq\ker\left(\delta^{k}\right)$
    Elemente  $[\nu]\in HH^{k}(\SsV,\mathcal{M})$
    geben kann, in denen kein $[\eta_{\diff}]\in HH^{k}_{\diff}(\SsV,\mathcal{M})$ 
    enthalten ist. Wegen 
    \begin{equation*}
        HH^{k}_{\diff}(\SsV,\mathcal{M})\stackrel{\wt{\hat{\xi}^{k}}}{\longrightarrow}\Hom_{\mathbb{K}}^{a}(\V^{k},\mathcal{M})\cong HH^{k}(\SsV,\mathcal{M})
    \end{equation*}
    und der Surjektivität von
    $\wt{\hat{\xi}^{k}}$ ist die bei uns aber nicht der Fall. Des
    Weiteren kann es wegen
    $\im\left(\delta^{k-1}_{\diff}\right)\subseteq\im\left(\delta^{k-1}\right)$
    passieren, dass $[\eta_{\diff}],[\mu_{\diff}]\in
    HH^{k}_{\diff}(\SsV,\mathcal{M})$ existieren, für die sowohl
    $[\eta_{\diff}]\neq [\mu_{\diff}]$ als auch
    $[\eta_{\diff}],[\mu_{\diff}] \subseteq [\nu] \in
    HH^{k}(\SsV,\mathcal{M})$ gilt. In unserem Fall ist eben dies das Problem, da $\wt{\hat{\xi}^{k}}$ nicht notwendigerweise
    injektiv ist.
    Abhilfe würde hier die Homotopie $s$ schaffen, wenn
    gewährleistet wäre, dass
    \begin{equation*}
        \zeta_{-1}^{k}s_{k}^{*}\zeta^{k+1}\colon HC^{k+1}_{\diff}(\SsV,\mathcal{M})\longrightarrow
        HC^{k}_{\diff}(\SsV,\mathcal{M})  
    \end{equation*}
    gilt. In der Tat
    erhielten wir dann analog zu \eqref{eq:dgh}, dass
    \begin{equation}
        \label{eq:ggg}
        \id_{HC_{\diff}^{k}}-\:\xi^{k}\cp\:\hat{\xi}^{k}=\big(\zeta_{-1}^{k}s^{*}_{k}\zeta^{k+1}\big)\delta^{k}+\delta^{k-1}\big(\zeta_{-1}^{k-1}s^{*}_{k-1}\zeta^{k}\big),
    \end{equation}
    also $\id_{HC_{\diff}}\sim \xi^{k}\cp\hat{\xi}^{k}\:$ und
    $\id_{HH^{k}_{\diff}}=\wt{\xi^{k}\cp\hat{\xi}^{k}}=\wt{\xi^{k}}\cp\wt{\hat{\xi}^{k}}$,
    mithin die Surjektivität von $\wt{\xi^{k}}$ und die
    Injektivität von $\wt{\hat{\xi}^{k}}$. In der Tat könnte dann
    obiger Fall nicht mehr eintreten, denn für
    $[\eta_{\diff}],[\mu_{\diff}]\subseteq [\nu]$ wäre jede
    Differenz $\phi=\phi_{\eta}-\phi_{\mu}$ von Repräsentanten
    $\phi_{\eta}\in[\eta_{\diff}]$ und
    $\phi_{\mu}\in[\mu_{\diff}]$ ein differentieller
    Korand. Unter Berücksichtigung von \eqref{eq:ggg} erhielten
    wir 
    $\phi=\delta^{k-1}\big(\zeta_{-1}^{k-1}s^{*}_{k-1}\zeta^{k}\phi\big)$
    mit $\zeta_{-1}^{k-1}s^{*}_{k-1}\zeta^{k}\phi \in
    HC^{k-1}_{\diff}(\SsV,\mathcal{M})$, also
    $[\eta_{\diff}]=[\mu_{\diff}]$. 

    Nun gilt, dass $\zeta_{-1}^{k}h^{*}_{k}\zeta^{k+1}$, $\zeta_{-1}^{k}\Omega^{*}_{k}\zeta^{k}$
    und $\zeta_{-1}^{k}d^{*}_{k}\zeta^{k}$ die Eigenschaft
    besitzen, differentielle Elemente auf differentielle Elemente
    abzubilden und dass die $\mathcal{A}^{e}$-Linearisierung
    einer derartigen Abbildung weiterhin diese Eigenschaft
    besitzt. Jedoch dürfen wir in
    $\zeta_{-1}^{k}\Omega_{k}^{*}h^{*}_{k}\zeta^{k+1}$ nicht
    einfach die Eins $\zeta^{k}\cp\zeta_{-1}^{k}$ einfügen, da
    $h_{k}$ nicht $\mathcal{A}^{e}$-linear ist. Nun könnte
    trotzdem 
    $\zeta_{-1}^{k}\Omega_{k}^{*}h^{*}_{k}\zeta^{k+1}\colon HC^{k+1}_{\diff}(\SsV,\mathcal{M})\longrightarrow
    HC^{k}_{\diff}(\SsV,\mathcal{M})$ richtig sein. Jedoch ist
    \begin{align*}
        \left(\zeta_{-1}^{k}\Omega_{k}^{*}h^{*}_{k}\zeta^{k+1}\right)\big(\phi\big)(u_{1},…,u_{k})=&\:\left(\Omega_{k}^{*}h^{*}_{k}\zeta^{k+1}\right)\big(\phi\big)(1\ot
        u_{1}\ot…\ot u_{k}\ot 1)
        \\=&\:\left(h^{*}_{k}\zeta^{k+1}\right)\big(\phi\big)(\Omega_{k}(1\ot
        u_{1}\ot…\ot u_{k}\ot 1))
        \\=&\:\left(\zeta^{k+1}\phi\right)(1\ot\Omega_{k}(1\ot
        u_{1}\ot…\ot u_{k}\ot 1)),
    \end{align*}
    also die $\mathcal{A}^{e}$-Linearität von
    $\left(\zeta^{k+1}\phi\right)$ in
    Kombination mit \eqref{eq:aeTrickt} nur
    noch für Elemente $\phi\in \DiffOpS{k}{(0,l_{2},…,l_{k+1})}$
    nutzbringend einsetzbar. In diesen Fällen ist dann
    $\left(\zeta_{-1}^{k}\Omega_{k}^{*}h^{*}_{k}\zeta^{k+1}\right)\big(\phi\big)\in
    \DiffOpS{k}{1}$,
    was man mit der $\mathcal{A}^{e}$-Linearität von $F_{k}$,
    durch eine ähnliche Rechnung wie in
    Proposition \ref{prop:wichEizuHKR}, sieht.\\
    Im Falle $\mathcal{A}=C^{\infty}(V)$ mit einer konvexen
    Teilmenge $V\subseteq \mathbb{R}^{n}$ kann gezeigt werden
    (vgl. \cite[Kapitel 5]{Weissarbeit}), dass
    $\zeta_{-1}^{k}s^{*}_{k}\zeta^{k+1}$ tatsächlich die
    gewünschte Eigenschaft besitzt, differentielle Elemente
    auf differentielle Elemente abzubilden. In diesem Fall ist
    dies aber der speziellen Beschaffenheit des
    Differentialoperator-Begriffes geschuldet, der wegen der
    endlichen Dimension von $\mathbb{R}^{n}$, konsistent zur algebraischen
    Definition, durch Verkettung
    von partiellen Ableitungen definiert werden kann, siehe
    \cite[Def~5.3.2]{Weissarbeit}. Insbesondere gelten dann
    Kettenregeln der Form $\pt_{y} f(tx +(1-t)y)=f'(tx
    +(1-t)y)(1-t)$, die im Beweis zu
    \cite[Prop~5.6.6]{Weissarbeit} von essentieller Bedeutung
    sind. Um also die Vorgehensweise aus \cite[Kapitel
    5]{Weissarbeit} auf unsere Situation zu übertragen, könnte
    man sich im differentiellen Hochschild-Komplex von Anfang
    an auf Differentialoperatoren beschränken, die als
    endliche Summe in der Form 
    \begin{equation*}
        \phi=\sum_{l=0}^{s}\sum_{|\alpha|=l}\delta_{\alpha_{1}}^{|\alpha_{1}|}…\delta_{\alpha_{k}}^{|\alpha_{k}|}*_{L}m^{\alpha_{1},…,\alpha_{k}}
    \end{equation*}mit Derivationen
    $\delta_{\alpha_{i}}\in\mathrm{DiffOp}_{1}{1}(\SsV,\SsV)$
    geschrieben werden können. Hierbei darf $k$ für jeden Summanden
    variieren, und mit $\delta^{|\alpha_{i}|}_{\alpha_{i}}$ ist
    die $|\alpha_{i}|$-fache Anwendung von
    $\delta_{\alpha_{i}}$ gemeint. Dabei ist die Reihenfolge der Verkettungen wegen
    Derivationseigenschaft der $\delta_{\alpha_{i}}$ unwichtig.
    Für eine Derivation $\delta$ gilt dann mit
    $\delta_{2}(x\ot y):=x\ot \delta(y)$ ebenfalls
    \begin{equation*}
        \delta_{2}(\hat{i}_{1}(1\ot x\ot
        1))=(1-t_{1})\hat{i}_{1}(1\ot \delta(x)\ot 1),
    \end{equation*}also die Kettenregel. Um nun jedoch
    sicherzustellen, dass $G_{k}^{*}$ und somit $\xi^{k}$ in
    diesen Unterkomplex abbildet, wird man sich im allgemeinen
    auch auf einen Unterkomplex von $(\K^{*}, \pt^{*})$
    beschränken müssen.\\\\
    Abseits dieser ganzen Diskussion besteht natürlich
    durchaus auch die Möglichkeit, dass $\id_{HC_{\diff}}\sim
    \xi^{k}\cp\hat{\xi}^{k}$ vermöge anderer Homotopieabbildung
    $\mathrm{s}\colon  HC^{k+1}_{\diff}(\SsV,\mathcal{M})\longrightarrow
    HC^{k}_{\diff}(\SsV,\mathcal{M})$ gilt. Dies würde dann
    die Surjektivität von $\wt{\xi^{k}}$ und die Injektivität
    von $\wt{\hat{\xi}^{k}}$ zeigen.
\end{bemerkung}
\section{Differentielle Bimoduln}
Motiviert durch Korollar \ref{kor:DiffKohom} wollen wir in diesem Abschnitt
der Frage nachgehen, inwiefern obige Aussagen auch f"ur
nicht-symmetrische Bimoduln zu erwarten sind. Seien hierf"ur
$(\mathcal{A},*)$ eine kommutative Algebra und $(\mathcal{M},*_{L})$
ein $\mathcal{A}$-Modul wie im letzten Abschnitt. F"ur $1\leq l\leq s$
seien $\mathbb{K}$-bilineare Abbildungen $D_{l}\colon \mathcal{A}\times \mathcal{M}\longrightarrow \mathcal{M}$ derart gegeben, dass folgende Konsistenzbedingungen erf"ullt sind:
\begin{itemize}
\item[\textbf{a.)}]
    $D_{l}(a,b*_{L}m)=b*_{L}D_{l}(a,m)\qquad\qquad\forall\:a,b\in \mathcal{A},\:\forall\: m\in \mathcal{M},\forall\:1\leq l\leq s$
\item[\textbf{b.)}]
    F"ur festes $a\in \mathcal{A}$ und $1\leq l\leq s$ bezeichne
    $D_{l}^{a}\colon\mathcal{M}\longrightarrow \mathcal{M}$ die
    $*_{L}$-lineare Abbildung $D_{l}^{a}\colon m\longmapsto D_{l}(a,m)$. Sei
    des Weiteren $D_{l_{1},\dots,l_{p}}^{a_{1},\dots,a_{p}}=D_{l_{1}}^{a_{1}}\cp
    \dots \cp D_{l_{p}}^{a_{p}}$, dann soll f"ur alle $a_{i}\in \mathcal{A}$ gelten, dass:
    \begin{equation*}
        D_{l_{1},\dots,l_{p}}^{a_{1},\dots,a_{p}}= 0,\qquad \text{ falls } \displaystyle\sum_{i=1}^{p}l_{i}> s.
    \end{equation*}
\item[\textbf{c.)}]
    F"ur $1\leq l\leq s$ gilt:
    \begin{align*}
        D_{l}(a*b,m)=&\:b*_{L}D_{l}(a,m)+ D_{1}(b,D_{l-1}(a,m))+D_{2}(b,D_{l-2}(a,m))+\dots\\ &\qquad+D_{l-2}(b,D_{2}(a,m))+ D_{l-1}(b,D_{1}(a,m))+a*_{L}D_{l}(b,m). 
    \end{align*}
\item[\textbf{d.)}]
    F"ur fixiertes $m\in \mathcal{M}$ ist $D_{l}(\cdot,m)\in \DiffOp{1}{l}\qquad \forall\:m\in \mathcal{M},\:\forall\: 1\leq l\leq s$.
\end{itemize}
Hiermit erhalten wir folgendes Lemma:
\begin{lemma}
    \label{lemma:defBim}
    Gegeben eine kommutative Algebra $(\mathcal{A},*)$ und ein $\mathcal{A}$-Modul $(*_{L},\mathcal{M})$. Seien weiter $D_{1},\dots,D_{s}$ Abbildungen, die \textit{i.)} - \textit{iv.)} erf"ullen. Dann wird $\mathcal{M}$ verm"oge 
    \begin{equation*}
        *_{R}=*_{L}+ D_{1}+\dots+D_{s} 
    \end{equation*}
    zu einem $\mathcal{A}-\mathcal{A}$-Bimodul. 
    \begin{beweis}
        Mit  \textbf{a.)} folgt unmittelbar, dass $a*_{L}(m *_{R} b)=
        (a*_{L}m) *_{R}b$ gilt, und f"ur die Bedingung $m*_{R}(a*b)=(m*_{R}a)*_{R}b$ rechnen wir:
        \begin{align*}
            (m*_{R}a)*_{R}b=\left[a*_{L}m+D_{1}(a,m)+D_{2}(a,m)+\dots+D_{s}(a,m)\right]*_{R}b.
        \end{align*}
        Durch Ausmultiplizieren und Anwenden von \textbf{a.)} und \textbf{b.)} ergibt dies:\\\\
        {\footnotesize
          \begin{array}[t]{cccccccccccc} 
              &(a*b)*_{L} m &+& a*_{L}D_{1}^{b}(m) &+& a*_{L} D_{2}^{b}(m)  &+&\dots &+&a*_{L}D_{s-1}^{b}(m)& +&a*_{L}D_{s}^{b}(m)\\\\
              +&b*_{L}D_{1}^{a}(m)&+&D_{1,1}^{b,a}(m)&+&D_{2,1}^{b,a}(m) &+& \dots&+&D_{s-1,1}^{b,a}(m)&+& \cancel{D_{s,1}^{b,a}(a,m)}\\\\
              +&b*_{L}D_{2}^{a}(m)&+&D_{1,2}^{b,a}(m)&+&D_{2,2}^{b,a}(m) &+& \dots&+&\cancel{D_{s-1,2}^{b,a}(m)}&+& \cancel{D_{s,2}^{b,a}(m)}\\\\
              +&b*_{L}D_{3}^{a}(m)&+&D_{1,3}^{b,a}(m)&+&D_{2,3}^{b,a}(m) &+& \dots&+&\cancel{D_{s-1,3}^{b,a}(m)}&+& \cancel{D_{s,3}^{b,a}(m)}\\\\
              &&&&&&\vdots&&&&&\\\\
              +&b*_{L}D_{s-1}^{a}(m)&+& D_{1,s-1}^{b,a}(m)&+&\cancel{D_{2,s-1}^{b,a}(m)} &+& \dots&+&\cancel{D_{s-1,s-1}^{b,a}(m)}&+& \cancel{D_{s,s-1}^{b,a}(m)}\\\\
              +&b*_{L}D_{s}^{a}(m)&+& \cancel{D_{1,s}^{b,a}(m)}&+&\cancel{D_{2,s}^{b,a}(m)} &+& \dots&+&\cancel{D_{s-1,s}^{b,a}(m)}&+& \cancel{D_{s,s}^{b,a}(m)}.
          \end{array}}\\\\\\Durch Zusammenfassen der Diagonalen von links unten nach recht oben folgt mit \textbf{c.)}:
        \begin{align*}
            (m*_{R}a)*_{R}b=&\:(a*b)*_{L}m+\Big[b*_{L}D_{1}(a,m)+
            a*_{L}D_{1}(b,m)\Big]
            \\ &+\Big[b*_{L}D_{2}(a,m)+D_{1}(b,D_{1}(a,m))+a*_{L}D_{2}(b,m)\Big]+\dots
            \\ &+ \Big[b*_{L}D_{s}(a,m)+ D_{1}(b,D_{s-1}(a,m))+D_{2}(b,D_{s-2}(a,m))+\dots\\ &\qquad+D_{s-2}(b,D_{2}(a,m))+ D_{s-1}(b,D_{1}(a,m))+a*_{L}D_{s}(b,m)\Big]
            \\=&\:(a*b)*_{L}m+D_{1}(a*b,m)+\dots+D_{s}(a*b,m)
            \\=&\: m*_{R}(a*b).
        \end{align*}
        Schließlich ist $*_{R}$ bilinear, da $*_{L}$ und alle $D_{l}$
        bilinear sind. Des Weiteren folgt f"ur alle $l$ mit $a=b=1$ aus \textbf{c.)},
        dass $D_{l}(1,m)=0$, also
        $m*_{R}1_{\mathcal{A}}=1_{\mathcal{A}}*_{L}m$ gilt, womit
        $\mathcal{M},*_{L},*_{R}$ alle unsere Anforderungen an einen
        $\mathcal{A}-\mathcal{A}$-Bimodul erfüllt.
    \end{beweis}
\end{lemma}
Wir geben nun die zentrale Definition dieses Kapitels:
\begin{definition}
    Gegeben die kommutative Algebra $\Ss^{\bullet}(\V)$ und ein $\Ss^{\bullet}(\V)-\Ss^{\bullet}(\V)$-Bimodul wie in Lemma \ref{lemma:defBim}.
    Wir nennen $\mathcal{M}$ einen differentiellen Bimodul
    "uber $\Ss^{\bullet}(\V)$, falls folgende Zusatzbedingung erfüllt ist:
    \begin{itemize}
    \item[\textbf{e.)}]
        F"ur alle $m\in \mathcal{M}$ und alle $\omega\in \SsV$ mit $\deg(\omega)< l$ ist $D_{l}(\omega,m)=0$.
    \end{itemize}
    Mit \textbf{c.)} ist dies gleichbedeutend mit der Forderung, dass
    f"ur alle $m\in \mathcal{M}$ und alle $l\geq 2$
    $D_{l}(\mathcal{\vv},m)=0$, falls $\deg(\vv)=1$.
\end{definition}
\begin{bemerkung}
    Sei $\mathcal{M}$ ein differentieller $\SsV-\SsV$-Bimodul, so ist
    mit $\DiffOpS{k}{\bullet}$ im Folgenden immer der
    Differentialoperator-Begriff bezüglich der $*_{L}$-Multiplikation gemeint.
\end{bemerkung}
\begin{beispiel}
    \begin{enumerate}
    \item
        Sei $(\mathcal{M},*_{L})$ ein $\Ss^{\bullet}(\V)$-Modul, $s=2$ und $\wt{\mathcal{M}}=\mathcal{M}\times\mathcal{M} \times\mathcal{M}$. Sei weiter $i_{u}$ wie in Bemerkung \ref{bem:DeriBeiStet} und:
        \begin{align*}
            \mathrm{sh}\colon (m_{1},m_{2},m_{3})&\longmapsto (0,m_{1},m_{2}) \\
            *_{L}\colon\big(\omega, (m_{1},m_{2},m_{3})\big)&\longmapsto (\omega *_{L}m_{1},\omega *_{L}m_{2},\omega *_{L}m_{3}).
        \end{align*}
        Wir setzen $D_{1}(\omega,\wt{m})=\sqrt{2}\cdot i_{u}(\omega)*_{L}
        \mathrm{sh}^{1}(\wt{m})$ und
        $D_{2}(\omega,\wt{m})=i_{u}^{2}*_{L}\mathrm{sh}^{2}(\wt{m})$. Dann
        sind \textbf{a.)} und \textbf{b.)} per Definition erf"ullt, und f"ur \textbf{c.)} rechnet man:
        \begin{align*}
            D_{2}(v\vee w,\wt{m})=&\:i_{u}(v\vee i_{u}(w)+w\vee i_{u}(v))*_{L}\mathrm{sh}^{2}(\wt{m})\\
            =&\:w\vee i_{u}^{2}(v)*_{L}\mathrm{sh}^{2}(\wt{m})+2\cdot i_{u}(v)\vee i_{u}(w)*_{L}\mathrm{sh}^{2}(\wt{m})\\
            &+v\vee i_{u}^{2}(w)*_{L}\mathrm{sh}^{2}(\wt{m})\\
            =&\: w*_{L} D_{2}(v,\wt{m})+D_{1}(w,D_{1}(v,\wt{m}))+v*_{L} D_{2}(w,\wt{m}).
        \end{align*}
        Des Weiteren ist
        $D_{2}(\mathrm{v},\wt{m})=i_{u}^{2}(\mathrm{v})*_{L}\mathrm{sh}^{2}(\wt{m})=(0,0,0)$
        falls $\deg{\mathrm{v}}=1$, und für \textbf{d.)} beachte man, dass $i_{u}\in \mathrm{DiffOp}_{1}^{1}(\Ss^{\bullet}(V),\Ss^{\bullet}(V))$ sowie $i_{u}^{2}\in \mathrm{DiffOp}_{1}^{2}(\Ss^{\bullet}(V),\Ss^{\bullet}(V))$, womit 
        \begin{align*}
            &[]_{2}^{a,b} D_{1}(\cdot, m)=\Big([]_{2}^{a,b}i_{u}\Big) *_{L}\mathrm{sh}^{1}(\wt{m})=0\\
            &[]_{3}^{a,b,c} D_{2}(\cdot, m)=\Big([]_{3}^{a,b,c}i_{u}^{2}\Big) *_{L}\mathrm{sh}^{2}(\wt{m})=0.
        \end{align*}
    \item 
        Sei $(\mathcal{A},*)=(C^{\infty}(\mathbb{R}^{n},\mathbb{R}),\cdot)$ und
        $\mathcal{M}=\mathrm{DiffOp}_{1}^{s}(\mathcal{A},\mathcal{A})$,
        versehen mit den Modul-Multiplikationen aus
        Proposition \ref{prop:MultidiffOps}~\textit{iii.)}. Dann ist
        $\mathcal{M}$ in der Tat ein
        $\mathcal{A}-\mathcal{A}$-Bimodul, und unter Verwendung der
        Multiindex-Konventionen
        \begin{equation*}
            |\alpha|=\sum_{i=1}^{n}\alpha_{i}\quad\qquad
            \pt^{\alpha}=\frac{\pt^{\alpha_{1}}}{\pt
              x_{1}}…\frac{\pt^{\alpha_{n}}}{\pt x_{n}}\quad\qquad\alpha!=\prod_{i=1}^{n}\alpha_{i}!
        \end{equation*}
        \begin{equation*}
            \alpha+\beta=(\alpha_{1}+\beta_{1},…\alpha_{n}+\beta_{n})\quad\qquad \binom{\alpha}{\beta}=\frac{\alpha!}{(\alpha-\beta)!\:\beta!}=\prod_{i=1}^{n}\binom{\alpha_{i}}{\beta_{i}}
        \end{equation*} für $\alpha,\beta \in \mathbb{N}^{n}$
        erhalten wir aus Bemerkung \ref{bem:DiffOpCRn}, dass wir jedes $m
        \in\mathrm{DiffOp}_{1}^{s}(\mathcal{A},\mathcal{A})$ in der
        Form $\displaystyle
        m=\sum_{l=0}^{s}\sum_{|\alpha|=l}\phi_{\alpha}\pt^{\alpha}$
        mit Elementen $\phi_{\alpha}\in
        C^{\infty}(\mathbb{R}^{n},\mathbb{R})$
        schreiben können. Mit der Derivationseigenschaft der
        partiellen Ableitungen folgt
        \begin{equation*}
            \pt^{\alpha}(f\cdot g)=\sum_{\beta\leq
              \alpha}\binom{\alpha}{\beta}\pt^{\beta}f \cdot
            \pt^{\alpha-\beta}g\qquad\forall\:f,g\in C^{\infty}(\mathbb{R}^{n},\mathbb{R})
        \end{equation*}und wir erhalten für
        $m=\phi_{\alpha}\pt^{\alpha}$, dass
        \begin{align*}
            (m *_{R}f)(g) = \phi_{\alpha}\sum_{\beta\leq\alpha}\binom{\alpha}{\beta}\pt^{\beta}f \cdot
            \pt^{\alpha-\beta}g
            =\sum_{l=0}^{|\alpha|}\overbrace{\sum_{\substack{|\beta|=l\\\beta\leq
                  \alpha}}\phi_{\alpha}\pt^{\beta}f
              \pt^{\alpha-\beta}}^{D_{l}(f,m)}g
        \end{align*}
        mit $D_{0}(f,m)=f\cdot\phi_{\alpha}\pt^{\alpha}=f*_{L}m$ gilt. Hierbei sind die
        $D_{l}$ ganz allgemein durch lineare Fortsetzung auf
        ganz $\mathrm{DiffOp}_{1}^{s}(\mathcal{A},\mathcal{A})$ von
        \begin{align*}
            D_{l}\colon
            \left(f,\phi_{\alpha}\pt^{\alpha}\right)&\longmapsto\sum_{\substack{|\beta|=l\\\beta\leq
                \alpha}}\binom{\alpha}{\beta}\phi_{\alpha}\pt^{\beta}f
            \pt^{\alpha-\beta}\in \mathrm{DiffOp}_{1}^{|\alpha|-l}(\mathcal{A},\mathcal{A})
        \end{align*}mit $D_{l}|_{\mathcal{A}\times\mathrm{DiffOp}_{1}^{m\leq
            l}(\mathcal{A},\mathcal{A})}=0$ definiert. Insgesamt zeigt dies \textbf{a.)} und
        \textbf{b.)}. Nun ist $D_{l}(\cdot,m)$ linear und wegen
        Proposition \ref{prop:MultidiffOps}~\textit{ii.)} folgt 
        \textbf{d.)} unmittelbar aus:
        \begin{equation*}
            []_{l+1}^{f_{1},…,f_{k+1}}D_{l}(\cdot,m)=\sum_{\substack{|\beta|=l\\\beta\leq
                \alpha}}\binom{\alpha}{\beta}\phi_{\alpha}\left([]_{l+1}^{f_{1},…,f_{k+1}}\pt^{\beta}\right) \pt^{\alpha-\beta}=0.
        \end{equation*}

        Für \textbf{c.)} beachten wir, dass $\mathcal{M}$ ein Bimodul ist, also
        $m*_{R}(f\cdot g)=(m*_{R}f)*_{R} g$ gilt. Wir betrachten nun
        das Schema aus Lemma \ref{lemma:defBim}, welches wir durch
        ausmultiplizieren von $(m*_{R}f)*_{R} g$ erhielten. Es ist
        dann zu zeigen, dass die $l$-te Diagonale mit $D_{l}(f\cdot
        g,m)$ übereinstimmt. Hierfür reicht, es diese
        Aussage für Elemente der Form
        $m_{k}=\sum_{|\alpha|=k}\phi_{\alpha}\pt^{\alpha}$ mit
        $k\leq s$, welche wir
        im Folgenden als "`exakt der Ordnung $k$"' bezeichnen wollen, nachzuweisen.
        Denn jedes
        $m\in\mathrm{DiffOp}_{1}^{s}(\mathcal{A},\mathcal{A})$ kann
        offenbar als eindeutige Linearkombination solcher Elemente dargestellt
        werden. Für ein derartiges $m_{k}$ ist $D_{l}(f,m_{k})$ exakt
        der Ordnung $k-l$ und ebenfalls ist $D_{l_{2}}(g,D_{l_{1}}(f,m_{k}))$ exakt
        der Ordnung $k-l_{1}-l_{2}$. Hiermit enthält die $l$-te
        Diagonale nur exakte Elemente der Ordnung $l$, was
        \textbf{c.)} zeigt. Schränken wir uns auf die
        Unteralgebra $\Pol(\mathbb{R}^{n},\mathbb{R})\subseteq
        C^{\infty}(\mathbb{R}^{n},\mathbb{R})$ ein, so ist
        schließlich auch \textbf{e.)} erfüllt. 
    \item
        In Analogie zu \textit{ii.)} betrachten wir den Unterraum $\mathcal{M}$ aller
        Differentialoperatoren
        $m\in\mathrm{DiffOp}_{1}^{s}(\SsV,\SsV)$, die als eine endliche
        Summe der Form
        \begin{equation*}
            m=\sum_{l=0}^{s}\sum_{|\alpha|=l}\delta_{\alpha_{1}}^{|\alpha_{1}|}…\delta_{\alpha_{k}}^{|\alpha_{k}|}
        \end{equation*}mit Derivationen
        $\delta_{\alpha_{i}}\in\mathrm{DiffOp}_{1}{1}(\SsV,\SsV)$
        geschrieben werden können. Hierbei ist in der zweiten Summe
        $\alpha \in \mathbb{N}^{k}$, wobei $k$ für jeden Summanden
        variieren darf. Mit $\delta^{|\alpha_{i}|}_{\alpha_{i}}$ ist
        die $|\alpha_{i}|$-fache Anwendung von
        $\delta_{\alpha_{i}}$ gemeint, und wegen der
        Derivationseigenschaft ist die Reihenfolge Verkettungen
        unwichtig. Der Summand für $l=0$ soll dann lediglich aus einem
        Element $m_{0}\in \SsV$ bestehen. 
        Vermöge Proposition \ref{prop:MultidiffOps} wird
        $\mathcal{M}$ zu einem $\SsV-\SsV$-Bimodul, und wir erhalten
        für
        $m=\delta_{\alpha_{1}}^{|\alpha_{1}|}…\delta_{\alpha_{k}}^{|\alpha_{k}|}=:\delta^{\alpha}$
        mit $|\alpha|\leq s$ sowie $v,w\in\SsV$, dass
        \begin{align*}
            (m *_{R}v)(w) = \sum_{\alpha\geq\beta\in \mathbb{N}^{k}}\binom{\alpha}{\beta}\delta^{\beta}(v) \vee
            \delta^{\alpha-\beta}(w)
            =\sum_{l=0}^{|\alpha|}\overbrace{\sum_{\substack{|\beta|=l\\\alpha\geq\beta\in \mathbb{N}^{k}}}\binom{\alpha}{\beta}\delta^{\beta}(v)\vee
              \delta^{\alpha-\beta}}^{D_{l}(v,m)}(w),
        \end{align*}also $*_{R}=\displaystyle\sum_{l=1}^{s}D_{l}$ mit Abbildungen
        \begin{equation*}
            D_{l}\colon\left(v,\delta^{\alpha}\right)\longmapsto \begin{cases}
                0\qquad\qquad\qquad\qquad\qquad \text{ falls }l>|\alpha|,\\
                \sum_{\substack{|\beta|=l\\\alpha\geq\beta\in \mathbb{N}^{k}}}\binom{\alpha}{\beta}\delta^{\beta}(v)\vee
                \delta^{\alpha-\beta}\:\quad\text{ sonst}.
            \end{cases}
        \end{equation*}gilt. Nun folgt \textbf{a.)} unmittelbar aus
        $\delta^{0}(v)=v$ und \textbf{b.)} mit
        $\deg(D_{l}(v,m))=|\alpha|-l$. Die Bedingungen \textbf{d.)} und \textbf{c.)}
        folgen analog zu \textit{ii.)} und \textbf{e.)} ist
        wegen $\delta^{\alpha}(v)=0$ für $\deg(v)<|\alpha|$ ebenfalls klar.
    \end{enumerate}
\end{beispiel}

\begin{lemma}
    \label{lemma:DiffBimodWichEi}
    Für differentielle
    $\Ss^{\bullet}(\V)-\Ss^{\bullet}(\V)$-Bimoduln $\mathcal{M}$ erhalten
    wir nun folgende Aussagen:
    \begin{enumerate}
    \item  
        Sei $m\in \mathcal{M}$ und $u\in \Ss^{\bullet}(\V)$, dann gilt:
        \begin{equation*}
            \hat{i}(1\ot u\ot 1)*_{e} m= u*_{L}m+ (1-t) D_{1}(u,m)+\dots+ (1-t)^{s} D_{s}(u,m).
        \end{equation*}
    \item
        Sei $\wt{\phi} \in \Hom_{\mathcal{A}^{e}}(\Ss^{\bullet}(\V),\mathcal{M})$ und $\mathcal{D}=D_{l_{1},\dots,l_{k}}^{a_{1},\dots,a_{k}}$ mit $q=s-\displaystyle\sum_{i=1}^{p}l_{i}\geq 0$. Dann ist die Abbildung
        \begin{align*}
            \tau_{\mathcal{D}}^{\wt{\phi}}\colon(u_{1},\dots,u_{k})\longmapsto \mathcal{D}\int_{0}^{1}dt_{1}\dots\int_{0}^{t_{k-1}}dt_{k}\:\mathrm{p}(t)\:\wt{\phi}\left(\prod_{s=1}^{k}(i_{s}\cp\delta)(1\ot u_{s}\ot 1)\right)
        \end{align*}
        f"ur  jedes $\mathrm{p}(t)\in \mathrm{Pol}(t_{1},\dots,t_{k})$
        ein Element in $\DiffOpS{k}{q+1}$. Des Weiteren ist $\xi^{k}(\phi) \in \DiffOpS{k}{s+1}$ f"ur alle $\phi\in \Hom_{\mathbb{K}}^{a}(\V^{k},\mathcal{M})$.
    \item
        Unter den gegebenen Voraussetzungen sind die Kokettenkomplexe aus Definition \ref{def:DiffKomplexe} ebenfalls wohldefiniert. 
    \end{enumerate}
    \begin{beweis}
        \begin{enumerate}
        \item
            Wir zeigen dies per Induktion "uber $\deg(u)$. Sei hierf"ur $\deg(\mathrm{v})=1$, so erhalten wir mit \textbf{e.)}:
            \begin{align*}
                \hat{i}(1 \ot \mathrm{v} \ot 1)*_{e}m=&\:\left[t \vv \ot 1 +(1-t)1\ot \vv\right]*_{e} m
                \\=&\:t \vv*_{L}m+ (1-t)\vv*_{L} m +(1-t) D_{1}(\vv,m)
                \\=&\: \vv*_{L}m +(1-t) D_{1}(\vv,m)+\dots+(1-t)^{s}D_{s}(\vv,m).
            \end{align*}
            Sei nun die Aussage f"ur $\deg(u)=l$ korrekt, dann folgt gleicherma"sen:
            {\allowdisplaybreaks\small\begin{align*}
                  \hat{i}(1\ot &\vv\vee\: u \ot 1)*_{e}m\\
                  =&\:\hat{i}(1\ot \vv \ot 1)*_{e}\Big(\hat{i}(1\ot u \ot 1)*_{e} m\Big)\\
                  =&\:\left[t\vv\ot 1+(1-t)1\ot\vv\right]*_{e}\left[u*_{L}m+ (1-t) D_{1}(u,m)+\dots+(1-t)^{l}D_{l}(u,m)\right]
                  \\=&\:t \vv\vee\: u *_{L} m +(1-t)\: \vv\vee\: u *_{L} m + (1-t)\: u*_{L}D_{1}(v,m) + \cancel{\dots}
                  \\ &+t(1-t) \vv*_{L}D_{1}(u,m)+(1-t)^{2}\vv*_{L}D_{1}(u,m)+(1-t)^{2}D_{1}(\vv,D_{1}(u,m))+ \cancel{\dots}
                  \\ &+t(1-t)^{2} \vv*_{L}D_{2}(u,m)+(1-t)^{3}\vv*_{L}D_{2}(u,m)+(1-t)^{3}D_{1}(\vv,D_{2}(u,m))+ \cancel{\dots}
                  \\ &+t(1-t)^{3} \vv*_{L}D_{3}(u,m)+(1-t)^{4}\vv*_{L}D_{3}(u,m)+(1-t)^{4}D_{1}(\vv,D_{3}(u,m))+ \cancel{\dots} 
                  \\ &+\dots
                  \\ &+t(1-t)^{l} \vv*_{L}D_{l}(u,m)+(1-t)^{l+1}\vv*_{L}D_{l}(u,m)+(1-t)^{l+1}D_{1}(\vv,D_{l}(u,m))\:. 
              \end{align*}}Fasst man den jeweils letzten Term mit den ersten beiden Termen der n"achsten Reihe zusammen, so folgt mit $(1-t)^{m}=t(1-t)^{m}+ (1-t)^{m+1}$:
            {\begin{equation}
                  \label{eq:klupop}
                  \begin{split}
                      \hat{i}(1\ot \vv\vee\: u \ot 1)*_{e}m=&\:\vv\vee\: u *_{L} m 
                      \\ &+(1-t)\:u*_{L}D_{1}(\vv,m)+(1-t)\vv*_{L}D_{1}(u,m)
                      \\ &+(1-t)^{2}D_{1}(\vv,D_{1}(u,m))+(1-t)^{2}\vv*_{L}D_{2}(u,m)
                      \\ &+(1-t)^{3}D_{1}(\vv,D_{2}(u,m))+(1-t)^{3}\vv*_{L}D_{3}(u,m)
                      \\ &+\dots
                      \\ &+(1-t)^{l}D_{1}(\vv,D_{l-1}(u,m))+(1-t)^{l}\vv*_{L}D_{l}(u,m)
                      \\ &+(1-t)^{l+1}D_{1}(\vv,D_{l}(u,m)).
                  \end{split}
              \end{equation}}
            Unter Ber"ucksichtigung von \textbf{c.)} und \textbf{e.)} erhalten wir hieraus im Falle $l<s$\::
            \begin{align*}
                \hat{i}(1\ot \vv\vee \:u \ot 1&)*_{e}m
                \\=&\:\vv\vee\: u *_{L} m +(1-t) D_{1}(\vv\vee\: u,m) +\dots + (1-t)^{l+1}D_{l+1}(\vv\vee \:u,m)
                \\ &+ \cancel{(1-t)^{l+2}D_{l+2}(\vv\vee \:u,m)}+\dots+\cancel{(1-t)^{s}D_{s}(\vv\vee\: u,m)}
            \end{align*}
            Im Falle $\deg(u) \geq s$ gilt \eqref{eq:klupop} mit s anstelle von l und wegen \textbf{b.)} verschwindet der letzte Summand. Dies zeigt die Behauptung.
        \item
            F"ur den Induktionsanfang sei $q=0$. Dann folgt mit
            \begin{align*}
                \wt{\phi}_{\omega_{j}}:=&\:\wt{\phi}\left(\prod_{s\neq
                      j}(i_{s}\cp\delta)(1\ot u_{s}\ot 1) \cdot
                    (i_{j}\cp\delta)(1\ot \omega_{j}\ot 1)\right)
            \end{align*} für $u_{1},…,u_{k},\omega\in \SsV$, dass
            {\allowdisplaybreaks
              \footnotesize
              \begin{align*}
                  \tau_{\mathcal{D}}^{\wt{\phi}}(u_{1}&,\dots, u_{j}\vee u'_{j},\dots,u_{k})
                  \\ =&\:\mathcal{D}\int_{0}^{1}dt_{1}\dots \int_{0}^{t_{k-1}}dt_{k}\:\mathrm{p}(t)\:\hat{i}_{j}(1\ot u_{j}\ot 1)*_{e} \wt{\phi}\left(\prod_{s\neq j}(i_{s}\cp\delta)(1\ot u_{s}\ot 1) \cdot (i_{j}\cp\delta)(1\ot u'_{j}\ot 1)\right)
                  \\ & +\mathcal{D}\int_{0}^{1}dt_{1}\dots \int_{0}^{t_{k-1}}dt_{k}\:\mathrm{p}(t)\:\hat{i}_{j}(1\ot u'_{j}\ot 1)*_{e} \wt{\phi}\left(\prod_{s\neq j}(i_{s}\cp\delta)(1\ot u_{s}\ot 1) \cdot (i_{j}\cp\delta)(1\ot u_{j}\ot 1)\right)
                  \\ =&\:\mathcal{D}\int_{0}^{1}dt_{1}\dots \int_{0}^{t_{k-1}}dt_{k}\:\mathrm{p}(t)\left[u_{j}*_{L}\wt{\phi}_{u'_{j}}+\:(1-t_{j})D_{1}^{u_{j}}\left(\wt{\phi}_{u'_{j}}\right)+\dots+(1-t_{j})^{s}D_{s}^{u_{j}}\left(\wt{\phi}_{u'_{j}}\right) \right] 
                  \\ & +\mathcal{D}\int_{0}^{1}dt_{1}\dots \int_{0}^{t_{k-1}}dt_{k}\:\mathrm{p}(t)\left[u'_{j}*_{L}\wt{\phi}_{u_{j}}+\:(1-t_{j})D_{1}^{u'_{j}}\left(\wt{\phi}_{u_{j}}\right)+\dots+(1-t_{j})^{s}D_{s}^{u'_{j}}\left(\wt{\phi}_{u_{j}}\right) \right]
                  \\ =&\int_{0}^{1}dt_{1}\dots \int_{0}^{t_{k-1}}dt_{k}\:\mathrm{p}(t)\:\mathcal{D}\left[u_{j}*_{L}\wt{\phi}_{u'_{j}}+\:(1-t_{j})D_{1}^{u_{j}}\left(\wt{\phi}_{u'_{j}}\right)+\dots+(1-t_{j})^{s}D_{s}^{u_{j}}\left(\wt{\phi}_{u'_{j}}\right) \right] 
                  \\ & +\int_{0}^{1}dt_{1}\dots \int_{0}^{t_{k-1}}dt_{k}\:\mathrm{p}(t)\:\mathcal{D}\left[u'_{j}*_{L}\wt{\phi}_{u_{j}}+\:(1-t_{j})D_{1}^{u'_{j}}\left(\wt{\phi}_{u_{j}}\right)+\dots+(1-t_{j})^{s}D_{s}^{u'_{j}}\left(\wt{\phi}_{u_{j}}\right) \right]
                  \\ =&\int_{0}^{1}dt_{1}\dots \int_{0}^{t_{k-1}}dt_{k}\:\mathrm{p}(t)\:u_{j}*_{L}\mathcal{D}\left(\wt{\phi}_{u'_{j}}\right)
                  +\int_{0}^{1}dt_{1}\dots \int_{0}^{t_{k-1}}dt_{k}\:\mathrm{p}(t)\:u'_{j}*_{L}\mathcal{D}\left(\wt{\phi}_{u_{j}}\right)
                  \\ =&\:u_{j}*_{L}\mathcal{D}\int_{0}^{1}dt_{1}\dots \int_{0}^{t_{k-1}}dt_{k}\:\mathrm{p}(t)\:\wt{\phi}_{u'_{j}}
                  +\:u'_{j}*_{L}\mathcal{D}\int_{0}^{1}dt_{1}\dots \int_{0}^{t_{k-1}}dt_{k}\:\mathrm{p}(t)\:\wt{\phi}_{u_{j}}
                  \\=&\: u_{j}*_{L} \tau_{\mathcal{D}}^{\wt{\phi}}(u_{1},\dots,u'_{j},\dots,u_{k})+u'_{j}*_{L} \tau_{\mathcal{D}}^{\wt{\phi}}(u_{1},\dots,u_{j},\dots,u_{k}), 
              \end{align*}}also $\tau_{\mathcal{D}}^{\wt{\phi}}\in
            \DiffOpS{k}{1}$ nach
            Proposition \ref{prop:MultidiffOps}~\textit{v.)} gilt. Hierbei durften wir $\mathcal{D}$ wegen seiner $\mathbb{K}$-Linearität mit den Integralen vertauschen.
            Sei nun $s-\displaystyle\sum_{i=1}^{p}l_{i}=q$ und die Aussage f"ur $q-1$ korrekt. Dann ist:
            \begin{align*}
                \Big([]_{j}^{a_{q+2}}\tau_{\mathcal{D}}^{\wt{\phi}}\Big)(u_{1},\dots,u_{k})
                =&\:a_{q+2}*_{L}\tau_{\mathcal{D}}^{\wt{\phi}}(u_{1},\dots,u_{k}) - \tau_{\mathcal{D}}^{\wt{\phi}}(u_{1},\dots, a_{q+2}\vee u_{j},\dots,u_{k}).
            \end{align*}
            Der zweite Summand ergibt ausgeschrieben:
            {\footnotesize
              \begin{align*}
                  \tau_{\mathcal{D}}^{\wt{\phi}}&(u_{1},\dots, a_{q+2}\vee u_{j},\dots,u_{k})
                  \\ =&\int_{0}^{1}dt_{1}\dots \int_{0}^{t_{k-1}}dt_{k}\:\mathrm{p}(t)\:\mathcal{D}\left[a_{q+2}*_{L}\wt{\phi}_{u_{j}}+\:(1-t_{j})D_{1}^{a_{q+2}}\left(\wt{\phi}_{u_{j}}\right)+\dots+(1-t_{j})^{s}D_{s}^{a_{q+2}}\left(\wt{\phi}_{u_{j}}\right) \right] 
                  \\   &+\int_{0}^{1}dt_{1}\dots \int_{0}^{t_{k-1}}dt_{k}\:\mathrm{p}(t)\:\mathcal{D}\left[u_{j}*_{L}\wt{\phi}_{a_{q+2}}+\:(1-t_{j})D_{1}^{u_{j}}\left(\wt{\phi}_{a_{q+2}}\right)+\dots+(1-t_{j})^{s}D_{s}^{u_{j}}\left(\wt{\phi}_{a_{q+2}}\right) \right]
                  \\ =&\:a_{q+2}*_{L}\tau_{\mathcal{D}}^{\wt{\phi}}+ \mathcal{D} D_{1}^{a_{q+2}}\mathbf{\int}\:\mathrm{p}(t)(1-t_{j})\wt{\phi}_{u_{j}}+\dots+\mathcal{D} D_{q}^{a_{q+2}}\mathbf{\int}\:\mathrm{p}(t)(1-t_{j})^{q}\wt{\phi}_{u_{j}}
                  \\ & + u_{j}*_{L}\tau_{\mathcal{D}}^{\wt{\phi}}+ \mathcal{D} D_{1}^{u_{j}}\mathbf{\int}\:\mathrm{p}(t)(1-t_{j})\wt{\phi}_{a_{q+2}}+\dots+\mathcal{D} D_{q}^{u_{j}}\mathbf{\int}\:\mathrm{p}(t)(1-t_{j})^{q}\wt{\phi}_{a_{q+2}},
              \end{align*}}so dass insgesamt:
            {\allowdisplaybreaks\small\begin{align*}
                  \Big([]_{j}^{a_{q+2}}\tau_{\mathcal{D}}^{\wt{\phi}}\Big)&(u_{1},\dots,u_{k})
                  \\ =&\: -\underbrace{\mathcal{D} D_{1}^{a_{q+2}}\mathbf{\int}\:\overbrace{\mathrm{p}(t)(1-t_{j})}^{\mathrm{p}'(t)}\:\wt{\phi}_{u_{j}}}_{\DiffOpS{k}{q}}-\dots-\underbrace{\mathcal{D} D_{q}^{a_{q+2}}\mathbf{\int}\:\mathrm{p}(t)(1-t_{j})^{q}\wt{\phi}_{u_{j}}}_{\DiffOpS{k}{1}}
                  \\ & - u_{j}*_{L}\tau_{\mathcal{D}}^{\wt{\phi}}- \mathcal{D} D_{1}^{u_{j}}\mathbf{\int}\:\mathrm{p}(t)(1-t_{j})\:\wt{\phi}_{a_{q+2}}-\dots-\mathcal{D} D_{q}^{u_{j}}\mathbf{\int}\:\mathrm{p}(t)(1-t_{j})^{q}\wt{\phi}_{a_{q+2}}.
              \end{align*}}Die Summanden in der ersten Reihe sind nach Induktionsannahme Differentialoperatoren der Ordnung $L=q,\dots,1$, verschwinden also nach Anwendung von $[]_{j}^{a_{1},\dots,a_{q+1}}$. 
            Das gleiche gilt f"ur die Terme in der zweiten Reihe, denn mit der $*_{L}$-Linearit"at der $D_{l}$ in $\mathcal{M}$ folgt:
            \begin{equation*} 
                a*_{L}\mathcal{D}D_{l}(u,m)-\mathcal{D}D_{l}(a*u,m)=\mathcal{D}\big[a*_{L}D_{l}(u,m)-D_{l}(a*u,m)\big].
            \end{equation*}
            Insgesamt ist somit
            $[]_{j}^{a_{1},\dots,a_{q+2}}\tau_{\mathcal{D}}^{\wt{\phi}}=0$
            f"ur alle $1\leq j\leq s$, und
            Proposition \ref{prop:MultidiffOps}~\textit{ii.)} zeigt dann, dass
            $\tau_{\mathcal{D}}^{\wt{\phi}}\in \DiffOpS{k}{q+1}$ gilt.

            Um die letzte Behauptung einzusehen erinnern wir daran, dass
            \begin{align*}
                \left(\xi^{k}\phi\right)(u_{1},\dots,u_{k})=\big(\wt{\phi}\cp G_{k}\big)(1\ot u_{1}\ot \dots \ot u_{k}\ot 1)
            \end{align*}
            mit $\wt{\phi}=\left(\left(\Upsilon^{k}\right)^{-1}\cp
                \left(\Theta^{k}\right)^{-1}\right)(\phi)$ gilt.

            Hieraus wird durch Anwendung von $[]_{j}^{a_{s+2}}$:
            \begin{align*}
                \left([]_{j}^{a_{s+2}}\xi^{k}\phi\right)&(u_{1},\dots,u_{k})= -\underbrace{D_{1}^{a_{s+2}}\mathbf{\int}\:(1-t_{j})\:\wt{\phi}_{u_{j}}}_{\DiffOpS{k}{s}}-\dots-\underbrace{D_{s}^{a_{s+2}}\mathbf{\int}\:(1-t_{j})^{s}\wt{\phi}_{u_{j}}}_{\DiffOpS{k}{1}}
                \\ & \qquad- u_{j}*_{L}\tau_{\mathcal{D}}^{\wt{\phi}}- D_{1}^{u_{j}}\mathbf{\int}\:(1-t_{j})\:\wt{\phi}_{a_{s+2}}-\dots- D_{s}^{u_{j}}\mathbf{\int}\:(1-t_{j})^{s}\wt{\phi}_{a_{s+2}}.
            \end{align*}
            Aus dem bisher Gezeigten folgt nun unmittelbar $[]^{a_{1},\dots,a_{s+2}}_{j}\left(\xi^{k}\phi\right)=0$ f"ur alle $1\leq j\leq k$, also $\xi^{k}(\phi)\in \DiffOpS{k}{s+1}$, wie behauptet.
        \item
            Sei $\phi\in \DiffOpS{k}{\bullet}$. Dann ist:
            \begin{align*}
                \big(\delta^{k}\phi\big)(a_{1},…,a_{k+1})=a_{1}*_{L}\phi(a_{2},…,a_{k+1})&+\sum_{j=1}^{k}(-1)^{j}\phi(a_{1},…,a_{i}a_{j+1},…,a_{k+1})\\
                &+(-1)^{k+1}\phi(a_{1},…,a_{k})*_{R}a_{k+1}.
            \end{align*}
            Die differentielle Natur der ersten beiden Summanden hatten wir bereits eingesehen und der letzte ergibt ausgeschrieben:
            \begin{align*}
                \phi(a_{1},…,a_{k})*_{R}a_{k+1}=a_{k+1}*_{L}\phi(a_{1},…,a_{k})&+D_{1}(a_{k+1},\phi(a_{1},…,a_{k}))+\dots\\ &+D_{s}(a_{k+1},\phi(a_{1},…,a_{k})). 
            \end{align*}
            Die Behauptung folgt nun unmittelbar aus Proposition \ref{prop:MultidiffOps}~\textit{ii.)}, \textbf{d.)} und mit:
            \begin{align*}
                []_{j}^{a}D_{l}(a_{k+1},\phi(a_{1},…,a_{k}))&=D_{l}(a_{k+1},[]_{j}^{a}\phi(a_{1},…,a_{k}))\qquad\quad\forall\:j\neq k+1,
            \end{align*}da hiermit $\delta^{k}\colon\DiffOpS{k}{\bullet}\longrightarrow \DiffOpS{k}{\bullet}$ gilt. Der lokalkonvexe Fall folgt analog.
        \end{enumerate}
    \end{beweis}
\end{lemma}
Mit Lemma \ref{lemma:DiffBimodWichEi} erhalten wir abschlie"send folgendes Resultat:
\begin{satz}
    \begin{enumerate}
    \item
        Sei $\mathcal{M}$ ein differentieller $\Ss^{\bullet}(\V)-\Ss^{\bullet}(\V)$-Bimodul. Dann besitzt jede Kohomologieklasse $[\eta]\in HH^{k}(\SsV,\mathcal{M})$ mindestens einen Repr"asentanten $\phi\in \DiffOpS{k}{s+1}$. Des Weiteren induzieren $\xi$ und $\hat{\xi}$ Kettenabbildungen zwischen $\left(HC_{\diff}(\SsV,\mathcal{M}),\delta_{\diff}\right)$ und  $(KC(\V,\mathcal{M}),\Delta)$.
        Hierbei ist $\wt{\xi^{k}}$ injektiv und $\wt{\hat{\xi}^{k}}$ surjektiv.
    \item
        Sei $\mathcal{M}$ ein differentieller, lokalkonvexer
        $\Ss^{\bullet}(\V)-\Ss^{\bullet}(\V)$-Bimodul. Dann besitzt
        jede Kohomologieklasse $[\eta]\in
        HH_{\cont}^{k}(\SsV,\mathcal{M})$ mindestens einen
        differentiellen Repr"asentanten $\phi\in \DiffOpS{k}{s+1,\cont}$. Des Weiteren induzieren $\xi$
        und $\hat{\xi}$ wohldefinierte Kettenabbildungen zwischen
        $(HC_{\mathrm{c,d}}(\SsV,\mathcal{M}),\delta_{\mathrm{c,d}})$
        und  $(KC^{\cont}(\V,\mathcal{M}),\Delta)$. Hierbei ist
        $\wt{\xi^{k}}$ injektiv und $\wt{\hat{\xi}^{k}}$ surjektiv.
    \end{enumerate}
\end{satz} 
\clearpage
\thispagestyle{empty}
\appendix 
\chapter{Algebraische Grundlagen}
\label{sec:AlgebraischeDefinitionen}
\section{Ringe, Moduln und Kategorien}
\label{sec:ringe-moduln-und}
Dieser Abschnitt soll die algebraischen Grundbegriffe und Konventionen bereitstellen,
die wir im Haupttext benötigen werden.
\begin{Definition}[Gruppe]
    Eine Gruppe ist eine Menge $G$, versehen mit einer Abbildung
    $\circ\colon G\times G\longrightarrow G$ derart, dass: 
    \begin{enumerate}
    \item 
        $a\circ(b\circ c)=(a\circ b)\circ c\qquad\quad \forall\: a,b,c\in G$.
    \item
        Es existiert ein eindeutig bestimmtes neutrales Element
        $e\in G$, so dass\\ $a\circ e=e\circ a=a$ für alle $a\in G$.
    \item
        Zu jedem $a\in G$ existiert ein eindeutig bestimmtes $a^{-1}\in G$
        derart, dass\\ $a\circ a^{-1}=a^{-1}\circ a=e$.
    \end{enumerate}
    Eine Gruppe heißt abelsch oder kommutativ, falls $a\circ b=b\circ
    a$ für alle $a,b\in G$. Man schreibt dann oft auch $+$ anstelle
    $\circ$.
    \begin{bemerkung}
        Um obige Definition zu erhalten, reicht es in der Tat bereits
        aus, zusätzlich zu \textit{i.)} entweder zu fordern, dass:
        \begin{enumerate}
        \item[$ii'.)$]
            $\exists\: e\in G\text{ mit }a\cp e=a\quad\forall\: a\in G$,
        \item[$iii'.)$]
            $\text{zu }a\in G,\:\exists\: a^{-1}\in G\text{ mit
            }a^{-1}\cp a=e$
        \end{enumerate} oder
        \begin{enumerate}
        \item[$ii''.)$]
            $\exists\: e\in G\text{ mit }e\cp a=a\quad\forall\: a\in G$,
        \item[$iii''.)$]
            $\text{zu }a\in G,\:\exists\: a^{-1}\in G\text{ mit }a\cp a^{-1}=e$.
        \end{enumerate} 
    \end{bemerkung}
\end{Definition}
\begin{definition}
    Eine Abbildung $\phi\colon G\longrightarrow H$, zwischen Gruppen $(G,\cp_{G})$
    und $(H,\cp_{H})$, heißt Gruppen-Homomorphismus oder einfach
    Homomorphismus, falls
    \begin{equation*}
        \phi(f\cp_{G} g)=\phi(f)\:\cp_{H}\:\phi(g)\qquad\qquad \forall\:f,g\in G.
    \end{equation*}
\end{definition}
\begin{Definition}[Ring]
    Ein Ring ist eine Menge $R$, versehen mit zwei Abbildungen 
    $+\colon R\times R\longrightarrow R$ und\\ $*\colon R\times R\longrightarrow R$
    derart, dass $(R, + )$ eine abelsche Gruppe ist und $*$ das
    Assoziativgesetz erfüllt. Zudem gelten folgende Distributivgesetze:
    \begin{align*}
        a * (b+c)& = a * b + a * c \\
        (a+b) * c &= a * c + b * c\quad\quad  \forall\:a,b,c\in R.
    \end{align*}
    Das neutrale Element $0_{R}$ von $(R, + )$ heißt Nullelement von
    $R$.
    Ein Ring heißt unitär, falls er ein Einselement $1_{R}$ bezüglich
    $*$ mit
    \begin{equation*}
        1_{R}*a=a*1_{R}=a\quad\quad \forall\: a\in R
    \end{equation*}besitzt. Dieses ist, vermöge
    $1_{R}=1_{R}*\widehat{1}_{R}=\widehat{1}_{R}$, bereits
    eindeutig bestimmt.
\end{Definition}
\begin{Definition}[Modul]
    Sei $R$ ein Ring.
    \begin{enumerate}
    \item
        Ein $R$-Linksmodul ist eine abelsche Gruppe $(\mathcal{M},+)$,
        versehen mit einer Abbildung $\cdot\colon R\times \mathcal{M}\longrightarrow \mathcal{M}$
        derart, dass für alle $r,s\in R$ und alle $x,y\in \mathcal{M}$:
        \begin{enumerate}
        \item
            $r\cdot(x+ y)=r\cdot x + r\cdot y$
        \item
            $(r+s)\cdot x=r\cdot x+ s\cdot x$
        \item
            $(r*s)\cdot x=r\cdot(s\cdot x)$ $\quad$ also $\quad$ $\cdot(r*s,x)=\cdot(r,\cdot(s,x))$
        \end{enumerate}gilt.
    \item
        Ein $R$-Rechtsmodul ist eine abelsche Gruppe $(\mathcal{M},+)$,
        versehen mit einer Abbildung $\cdot\colon\mathcal{M}\times R\longrightarrow \mathcal{M}$
        derart, dass für alle $r,s\in R$ und alle $x,y\in \mathcal{M}$:
        \begin{enumerate}
        \item
            $(x+ y)\cdot r=x\cdot r + y\cdot r$
        \item
            $x\cdot(r+s)= x\cdot r+ x \cdot s$
        \item
            $x\cdot (r*s)=(x\cdot r)\cdot s$ $\quad$ also $\quad$
            $\cdot(x,r*s)=\cdot(\cdot(x,r),s)$
        \end{enumerate}gilt.
    \item 
        Ist $R$ kommutativ, so ist jeder $R$-Linksmodul, vermöge $m \cdot_{R}r=r\cdot_{L}m$, ebenfalls ein
        $R$-Rechtsmodul, denn es gilt:
        {\small\begin{equation*}
              (m\cdot_{R}r)\cdot_{R}s=(r\cdot_{L}m)\cdot_{R}s=s\cdot_{L}(r\cdot_{L}m)=(s*r)\cdot_{L}m=(r*s)\cdot_{L}m=m\cdot_{R}(r*s).
          \end{equation*}}Ebenso ist in dieser Situation jeder Rechtsmodul, vermöge $r\cdot_{L}m=m\cdot_{R}r$,
        ein Linksmodul. Daher ist es für derartige Ringe legitim, jeden
        $R$-Modul als Linksmodul zu behandeln.
    \item
        Gegeben Ringe $R_{1}$ und $R_{2}$, so ist ein $R_{1}-R_{2}$-Bimodul $\mathcal{M}$ ein $R_{1}$-Linksmodul, der gleichzeitig ein
        $R_{2}$-Rechtmodul ist und der folgende zusätzliche Bedingung erfüllt:
        \begin{equation*}
            (r_{1}\cdot_{L}m)\cdot_{R}r_{2}=r_{1}\cdot_{L}(m\cdot_{R}r_{2})\quad\quad\forall\:
            r_{1}\in R_{1},\:r_{2}\in R_{2}.
        \end{equation*}Hierbei bezeichnen $\cdot_{L/R}$ die Links-
        bzw. Rechtsmodul-Multiplikation. 

        Ein $R-R$-Bimodul
        heißt symmetrisch, falls $r*_{L} m= m*_{R} r$ für alle $r\in
        R$ und alle $m\in \mathcal{M}$.
    \item
        Sei $R$ unitär, so wollen wir im Folgenden immer die $R$-Vertr"aglichkeit von $\mathcal{M}$ voraussetzen. Dies bedeutet, dass $1_{R}\cdot_{L} m= m$ bzw.  $m\cdot_{R} 1_{R}= m$ 
        für alle $m\in \mathcal{M}$ gilt . F"ur einen $R_{1}-R_{2}$-Bimodul soll dies dann f"ur beide Modulstrukturen erf"ullt sein.
    \end{enumerate}
\end{Definition} 
\begin{bemerkung}
    \begin{enumerate}
    \item
        Sei $\mathbb{K}$
        ein Körper, also insbesondere ein kommutativer Ring, so ist
        ein $\mathbb{K}$-Modul $\mathbb{V}$ gerade ein
        $\mathbb{K}$-Vektorraum. Dabei kommt es wegen \textit{iii.)} nicht
        darauf an, ob $\mathbb{V}$ Links- oder Rechtmodul ist. 
    \item
        Ist klar, um welchen Ring und welchen
        Modul es sich handelt, so schreibt man oft auch einfach $rs$ für
        $r*s$ und $r m$ anstelle $r\cdot m$. 
        Für Bimoduln schreibt man oft auch
        einfach nur $r_{1} m r_{2}$ und unterdrücken die Klammerung. Will man nicht weiter spezifizieren, ob
        von einem $R$-Links-/ oder einem $R$-Rechtsmodul die Rede ist,
        spricht man schlicht von einem $R$-Modul.
    \item
        Für einen $R$-Linksmodul $\mathcal{M}$ gilt:
        \begin{equation*}
            ((r*s)*t)*_{L}m=(r*s)*_{L}(t*_{L}m)=r*_{L}(s*_{L}(t*_{L}m))=((r*s)*t)*_{L}m     
        \end{equation*} und für einen $R$-Rechtmodul:
        $m*_{R}((r*s)*t)=m*_{R}(r*(s*t))$. In beiden Fällen
        liefert dies eine kosmetische Begründung
        dafür, dass man Moduln gerade über Ringen definiert, in welchen
        die Assoziativität von $*$ Definitionsgemäß gegeben ist.
    \end{enumerate}
\end{bemerkung}
\begin{Definition}
    Eine Abbildung $\phi\colon\mathcal{M}\longrightarrow \mathcal{N}$ zwischen $R$-Moduln heißt
    $R$-Homomorphismus oder einfach nur Homomorphismus, falls
    $\phi(rx+sy)=r\phi(x)+s\phi(y)$ für alle $r,s\in R$ und $x,y\in
    \mathcal{M}$. Einen surjektiven $R$-Homomorphismus bezeichnet man als
    Epimorphismus. Ein injektiver $R$-Homomorphismus heißt Monomorphismus. 
    Die Menge aller $R$-Homomorphismen von $\mathcal{M}$ nach $\mathcal{N}$
    bezeichnen wir mit $\Hom_{R}(\mathcal{M},\mathcal{N})$. 
\end{Definition}
\begin{definition}
    \label{def:Algebra}
    \begin{enumerate}
    \item 
        Eine Algebra über einem Ring $R$ ist ein $R$-Modul $\mathcal{A}$, versehen mit einer $R$-bilinearen
        Verknüpfung:
        $*\colon\mathcal{A}\times \mathcal{A}\longrightarrow \mathcal{A}$.
        Es gilt also für alle $r\in R$ und alle $a,b,c\in \mathcal{A}$: 
        \begin{align*}
            &r(a*b)=(ra)*b=a*(rb),\\
            &c*(a+b)=c*a +c*b,\\
            &(a+b)*c=a*c+b*c.
        \end{align*}
        Motiviert durch
        $(rs)(a*b)=(ra)*(sb)=(sr)(a*b)$,
        betrachtet man oft auch nur Algebren über
        kommutativen Ringen. In diesem Rahmen sind
        $\mathbb{K}$-Algebren gerade 
        $\mathbb{K}$-Vektorräume mit Algebramultiplikation.
    \item 
        $\mathcal{A}$ heißt unitär, falls ein Element $e\in \mathcal{A}$ derart
        existiert, dass $e*a=a*e=a$ für alle $a\in
        \mathcal{A}$. Dieses ist dann wieder eindeutig bestimmt.
    \end{enumerate}
\end{definition}
\begin{korollar}
    Jede assoziative $R$-Algebra $\mathcal{A}$ definiert einen Ring.
\end{korollar}
\begin{definition}
    \label{def:vertrABimod}
    Gegeben eine assoziative $R$-Algebra $\mathcal{A}$ und ein
    $\mathcal{A}$-Modul $(\mathcal{M},*)$, der selbst ein $R$-Modul ist. Dann setzen wir im Folgenden immer die $R$-Bilinearit"at von  $*$ vorraus. Ist $(\mathcal{M},*_{L},*_{R})$ ein $\mathcal{A}-\mathcal{A}$-Bimodul, fordern wir sowohl die $R$-Bilinearit"at von $*_{L}$ als auch die von $*_{R}$.

\end{definition}
\begin{Definition}[projektiver Modul]
    Ein $R$-Modul $\mathrm{P}$ heißt projektiv, falls für jeden
    Homomorphismus $f\in \Hom_{R}(\mathcal{M},\mathcal{N})$ und
    jeden Epimorphismus $h\in \Hom_{R}(\mathcal{M},\mathcal{N})$ mit $R$-Moduln $\mathcal{M}$ und
    $\mathcal{N}$, ein $f'\in \Hom_{R}(\mathrm{P},\mathcal{M})$ derart existiert, dass
    folgendes Diagramm kommutiert:
    $$ 
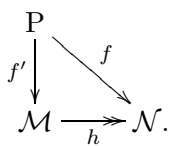
\diagram \mathrm{P} \dto_{f'} \drto^{f}  &  \\
    \mathcal{M} \rto_{h}|>>\tip      & \mathcal{N}. 
    \enddiagram $$ 
\end{Definition}
\begin{Definition}
    Gegeben ein $R$-Modul $\mathcal{M}$. Dann ist eine Basis von $\mathcal{M}$ eine Teilmenge $\{e_{\alpha}\}_{\alpha\in I}\subseteq \mathcal{M}$
    derart, dass jedes $m\in \mathcal{M}$ eine eindeutige, endliche Darstellung
    $m=\displaystyle\sum_{i=1}^{n}r_{i}e_{\alpha_{i}}$ mit
    Ko-effizienten $r_{i}\in R$ besitzt.
\end{Definition}
\begin{lemma}
    \label{lemma:freieModulnProjektiv}
    Jeder $R$-Modul $\mathrm{P}$ mit Basis ist projektiv.
    \begin{beweis}
        Gegeben $R$-Moduln $\mathcal{M},\mathcal{N}$, $f\in
        \Hom_{R}(\mathrm{P},\mathcal{N})$ und ein Epimorphismus $h\in \Hom_{R}(\mathcal{M},\mathcal{N})$.
        Wir wählen dann eine Basis $\{e_{\alpha}\}_{\alpha\in I}$ von
        $\mathrm{P}$ und setzen $f_{\alpha}=f(e_{\alpha})$. Weiterhin wählen wir $m_{\alpha}\in h^{-1}(f_{\alpha})$, was wegen der
        Surjektivität von $h$ ohne weiteres möglich ist. Für
        $p\in \mathrm{P}$ mit
        $p=\displaystyle\sum_{i=1}^{n}r_{i}\:e_{\alpha_{i}}$
        definieren wir $f'(p)=\displaystyle\sum_{i=1}^{n}r_{i}\:m_{\alpha_{i}}$.
        Aus der Eindeutigkeit
        der Basisdarstellung jedes $p\in \mathrm{P}$ folgen
        sowohl Wohldefiniertheit, als auch $R$-Linearität von
        $f'$, und per Konstruktion ist klar, dass $h\circ f'=f$. 
    \end{beweis}
\end{lemma}
\begin{definition}
    \label{def:freiMod}
    Ein $R$-Modul heißt frei, falls er isomorph zu einer direkten
    Summe von Kopien von $R$ ist.
\end{definition}
\begin{bemerkung}[Direkte Summe, Kartesisches Produkt]
    \begin{enumerate}
    \item
        Gegeben eine durch eine Indexmenge $I$ indizierte
        Sammlung von Mengen $\{T_{\alpha}\}_{\alpha\in I}$, so ist deren
        kartesisches Produkt $\prod_{\alpha\in I}T_{\alpha}$
        definiert, als die Menge aller $|I|$-Tupel mit Einträgen in den
        $T_{\alpha}$. Dies ist so zu verstehen, dass jedes Element aus
        $\prod_{\alpha\in I}T_{\alpha}$ einer Abbildungsvorschrift
        entspricht, die jedem $\alpha\in I$ genau ein $t_{\alpha}\in
        T_{\alpha}$ zuordnet. Im Falle $I=\mathbb{N}$ entsprechen die
        Elemente in $\prod_{k=1}^{\infty}T_{n}$ gerade den
        $|\mathbb{N}|$-Tupeln $(t_{0},t_{1},t_{2},…)$ mit $t_{k}\in
        T_{k}$ für alle $k\in \mathbb{N}$.
    \item
        Sind alle $T_{\alpha}$ Mengen mit Null-Elementen
        $0_{\alpha}\in T_{\alpha}$, wie dies beispielsweise für Gruppen, Moduln oder
        Vektorräume der Fall ist, so definieren wir die direkte Summe
        $\bigoplus_{\alpha\in I}T_{\alpha}$ als die Teilmenge aller $|I|$-Tupel
        aus $\prod_{\alpha\in I}$ , bei denen nur für endlich viele
        $t_{\alpha}\neq 0_{\alpha}$ gilt. Sind alle $T_{\alpha}$ Gruppen,
        so wird die direkte Summe, wie auch deren kartesisches Produkt, vermöge komponentenweiser Addition
        \begin{equation*}
            [\alpha\rightarrow t_{\alpha}]+[\alpha\rightarrow t'_{\alpha}]=[\alpha\rightarrow t_{\alpha}+t'_{\alpha}],
        \end{equation*}
        ebenfalls zu einer Gruppe. Sind alle $T_{\alpha}$ $R$-Linksmoduln, so
        wird $\bigoplus_{\alpha\in I}T_{\alpha}$ durch
        \begin{equation*}
            r*_{L}
            [\alpha\rightarrow t_{\alpha}]=[\alpha\rightarrow
            r*_{L}t_{\alpha}]
        \end{equation*}
        zu einem
        $R$-Linksmodul. Dies folgt analog für Rechtsmoduln. Für obige Definition
        beachte man, dass jeder Ring sowohl Links-, als auch
        Rechtsmodul über sich selbst ist. Sind alle $T_{\alpha}=T$ so schreiben wir auch $T^{|I|}$, anstelle
        $\bigoplus_{\alpha\in I}T_{\alpha}$. Für eine endlichen Menge
        $\{t_{\alpha_{1}},…,t_{\alpha_{n}}\}$ mit $t_{\alpha_{i}}\in
        T_{\alpha_{i}}$ sei dann:
        {\small\begin{equation*}
              \bigoplus_{\alpha\in
                I}T_{\alpha}\ni\bigoplus_{i=1}^{n}t_{\alpha_{i}}=
              \begin{cases}
                  \alpha_{i}\rightarrow t_{\alpha_{i}}\\
                  \beta\rightarrow 0_{\beta}\: \text{ falls}\: \beta\neq
                  \alpha_{i}\:\forall\:1\leq i\leq n .
              \end{cases}
          \end{equation*}}Im Falle, dass
        $T_{\alpha}=T$ für alle $\alpha\in I$, definieren wir für $t\in
        T$:
        {\small\begin{equation*}
              \oplus_{\alpha} t=\begin{cases}
                  \alpha \rightarrow t_{\alpha}\\
                  \beta\rightarrow 0_{\beta}\: \text{ falls}\: \beta\neq
                  \alpha_{i}.
              \end{cases}
          \end{equation*}}Oft benutzen wir f"ur die Darstellung von Elementen auch das Symbol $\sum$ anstelle von $\bigoplus$.
    \end{enumerate}
\end{bemerkung}

\begin{lemma}
    \label{lemma:freunibasis}
    Gegeben ein $R$-Modul $\mathcal{M}$, dann gilt:
    \begin{enumerate}
    \item
        Besitzt $\mathcal{M}$ eine Basis, dann ist $\mathcal{M}$ frei.
    \item
        Ist $\mathcal{M}$ frei und $R$ unitär, dann besitzt $\mathcal{M}$ eine Basis.
    \end{enumerate}
    \begin{beweis}  
        \begin{enumerate}
        \item
            Wir wählen eine Basis $\net{e}{\alpha}$ und definieren {\small$\phi\colon\displaystyle\sum_{i=1}^{n}r_{\alpha_{i}}e_{\alpha_{i}}\longmapsto \bigoplus_{i=1}^{n}r_{\alpha_{i}}$}. 

            Wegen der Eindeutigkeit der Basiszerlegung ist $\phi$
            wohldefiniert und injektiv. Die Surjektivität und
            Linearität ist klar.
        \item
            $\mathcal{M}$ ist frei und somit isomorph zu einer direkten Summe
            von Kopien von $R$. Sei $\phi$ dieser Isomorphismus. Dann
            ist
            {\small$\phi(m)=\displaystyle\bigoplus_{i=1}^{n}r_{\alpha_{i}}$},
            und es folgt: 
            {\small\begin{align*}
                  m=\phi^{-1}(\phi(m))=\phi^{-1}\left(\bigoplus_{i=1}^{n}r_{\alpha_{i}}\right)&=\sum_{i=1}^{n}r_{\alpha_{i}}\underbrace{\phi^{-1}\left(\bigoplus
                        1_{\alpha_{i}}\right)}_{e_{\alpha_{i}}}=\sum_{i=1}^{n} r_{\alpha_{i}}e_{\alpha_{i}}.
              \end{align*}}Dabei wurde im vorletzten Schritt die Unitarität
            von $R$ benutzt. Folglich hat jedes $m\in \mathcal{M}$ eine eindeutige
            Darstellung als endliche Linearkombination der
            $e_{\alpha}\in \mathcal{M}$, was die Behauptung zeigt.
        \end{enumerate}
    \end{beweis}
\end{lemma}
\begin{korollar}
    \label{kor:freiemoduberunitRingensindProjektiv}
    Freie Moduln über unitären Ringen sind projektiv.
\end{korollar}
\begin{definition}[Klasse]
    \label{def:Klasse}
    Unter einer Klasse wollen wir im Folgenden eine Sammlung von
    Objekten verstehen, die eine bestimmte Eigenschaft gemein
    haben. Insbesondere ist dann jede Menge eine Klasse, da
    ihre Elemente die Eigenschaft gemein haben, Elemente dieser
    Menge zu sein. 
    Die umgekehrte Inklusion gilt nicht. Betrachten wir
    nämlich die durchaus vernünftige Klasse aller Mengen, die sich
    nicht selbst enthalten, so folgt
    unmittelbar, dass diese per Definition keine Menge
    sein kann. Wäre sie nämlich eine Menge, so enthielte sie sich entweder, oder eben nicht. Beide Annahmen widersprechen
    sich selbst. Für eine ausführliche Diskussion des Mengenbegriffes
    siehe beispielsweise \cite{1234567}.  
\end{definition}

\begin{definition}[Kategorie]
    Eine Kategorie $\mathcal{C}$ ist ein Konstrukt, bestehend aus
    folgenden Daten:
    \begin{enumerate}
    \item
        Einer Klasse $\Ob_{\mathcal{C}}$ von Objekten.
    \item
        Je einer Menge $\Mor_{C}(X,Y)$ von Morphismen zu jedem Paar $(X,Y)$ von
        Objekten. 
        Es gilt dabei die Disjunktheit von $\Mor_{\mathcal{C}}(X,Y)$ und
        $\Mor_{\mathcal{C}}(X',Y')$, falls $X\neq X'$ oder $Y\neq Y'$. Für
        $\phi\in \Mor_{\mathcal{C}}(X,Y)$ bezeichnen $\mathrm{\dom}(\phi)=X$ die
        Quelle und $\cod(\phi)=Y$ das Ziel des Morphismus.
    \item
        Einer assoziativen Abbildung $\circ\colon\Mor_{\mathcal{C}}(X,Y)\times
        \Mor_{\mathcal{C}}(Y,Z)\longrightarrow \Mor_{\mathcal{C}}(X,Z)$ für
        alle $X,Y,Z\in \Ob_{\mathcal{C}}$.
    \end{enumerate}
    Des Weiteren existiert zu jedem $X\in \Ob_{\mathcal{C}}$ ein Element
    $\id_{X}\in \Mor_{\mathcal{C}}(X,X)$ mit $f\circ \id_{X}=f$, falls
    $\cod(f)=X$ und $\id_{X}\cp f=f$ für $\dom(f)=X$. 
\end{definition}
\begin{beispiel}
    \begin{enumerate}
    \item
        Jede Klasse von Mengen als Objekte, zusammen mit der Menge der Abbildungen
        zwischen diesen als Morphismen.
    \item 
        Gegeben ein Ring $R$, dann bilden die $R$-Moduln zusammen
        mit den $R$-Ho\-mo\-mor\-phis\-men die Kategorie $\Rm$. 
    \item
        Die Kategorie $\Ab$ mit den abelschen Gruppen als Objekten und
        den Gruppenhomomorphismen als Morphismen.
    \end{enumerate} 
\end{beispiel}
\begin{definition}[Funktor ko-/kontravariant]
    Ein Funktor $\mathcal{F}$ von einer Kategorie
    $\mathcal{C}$ in eine Kategorie $\mathcal{C}'$ ist eine
    strukturerhaltende Operation, bestehend aus folgenden Daten:
    \begin{enumerate}
    \item
        Einer Abbildung $\mathcal{F}_{\Ob}\colon\Ob_{\mathcal{C}}\longrightarrow
        \Ob_{\mathcal{C}'}$.
    \item
        Einer Abbildung
        $\mathcal{F}_{\Mor}\colon\Mor_{\mathcal{C}}(X,Y)\longrightarrow \Mor_{\mathcal{C}'}(\mathcal{F}_{Ob}(X),\mathcal{F}_{Ob}(Y))$.
    \end{enumerate}
    Diese erfüllen folgende Kompatibilitätsbedingungen:
    \begin{enumerate}
    \item
        $\mathcal{F}_{\Mor}(f\circ g)=\mathcal{F}_{\Mor}(f)\circ
        \mathcal{F}_{\Mor}(g)\quad$ kovarianter Funktor\\
        $\mathcal{F}_{\Mor}(f\circ g)=\mathcal{F}_{\Mor}(g)\circ
        \mathcal{F}_{\Mor}(f)\quad$ kontravarianter Funktor
    \item
        $\mathcal{F}(\id_{X})=\id_{\mathcal{F}(X)}$
    \end{enumerate}
\end{definition}
\begin{Definition}[Additivität]
    \begin{enumerate}
    \item
        Wir wollen im Folgenden eine Kategorie $\mathcal{C}$ als
        additiv bezeichnen, falls für jedes Paar $(X,Y)$ von Objekten
        von $\mathcal{C}$ eine abelsche Gruppenstruktur
        $(\Mor_{\mathcal{C}}(X,Y),+,0_{X,Y})$ mit einer kommutativen
        Abbildung
        \begin{equation*}
            +\colon\Mor_{\mathcal{C}}(X,Y)\times
            \Mor_{\mathcal{C}}(X,Y)\longrightarrow \Mor_{\mathcal{C}}(X,Y)
        \end{equation*} und einem 0-Element $0_{X,Y}$ derart existiert, dass für $f,f_{1},f_{2}\in \Mor_{\mathcal{C}}(X,Y)$ und
        $g,g_{1},g_{2}\in \Mor_{\mathcal{C}}(Y,Z)$ mit $Z\in
        \Ob_{\mathcal{C}}$ folgende algebraische Identitäten erfüllt sind:
        \begin{align}
            \label{eq:addkatDistlaws1}
            (g_{1}+g_{2})\circ f&=g_{1}\circ f+g_{2}\circ f,\\
            \label{eq:addkatDistlaws2}
            g\circ(f_{1}+f_{2})&=g\circ f_{1}+g\circ f_{2}.
        \end{align}
    \item
        Gegeben ein Funktor $\mathcal{F}$ zwischen additiven Katgorien
        $\mathcal{C}$ und $\mathcal{C}'$, so heißt $\mathcal{F}$
        additiv falls:
        \begin{equation*}
            \mathcal{F}_{\Mor}(f+g)=\mathcal{F}_{\Mor}(f)+\mathcal{F}_{\Mor}(g)\quad\quad
            \forall\:f,g\in \Mor_{\mathcal{C}}(X,Y),
        \end{equation*} wenn $\mathcal{F}$ also verträglich ist mit
        den abelschen Gruppenstrukturen auf $\mathcal{C}$ und $\mathcal{C}'$.
    \end{enumerate}
\end{Definition}
\begin{beispiel}
    Die Kategorie $\Rm$ ist additiv vermöge 
    \begin{equation*}
        (\phi+\psi)(m)=\phi(m)+\psi(m)\qquad\qquad\forall\:m\in \mathcal{M},
    \end{equation*} für $\phi,\psi\in \Mor_{\Rm}(\mathcal{M},\mathcal{N})=\Hom_{R}(\mathcal{M},\mathcal{N})$
    mit $R$-Moduln $\mathcal{M},\mathcal{N}$. 
\end{beispiel}
%
%
%
%
%
%
\section{Homologische Algebra}
\label{sec:HomologAlgebr}
Ziel dieses Abschnittes ist die Bereitstellung der
homologisch-algebraischen Begrifflichkeiten und Zusammenhänge, die
dieser Arbeit als Ausgangspunkt dienen sollen. Er enthält haupsächlich
Resultate aus \cite[Kapitel 6]{jacobson:1989a}.
\subsection{Komplexe und Homologien}
\label{subsec:kompundhomolog}
Aufbauend auf \ref{sec:AlgebraischeDefinitionen}.1 beginnen wir mit folgenden elementaren Definitionen:
\begin{definition}[Komplex]
    \label{def:Komplex}
    Gegeben ein Ring $R$.
    \begin{enumerate}
    \item
        Ein $R$-Komplex ist eine Menge $\{C_{i},d_{i}\}_{i\in
          \mathbb{Z}}$ von Paaren $(C_{i},d_{i})$, von $R$-Moduln
        $C_{i}$ und $R$-Homomorphismen $d_{i}\colon C_{i}\longrightarrow
        C_{i-1}$ derart, dass $d_{i-1}\circ d_{i}=0$ für alle $i\in
        \mathbb{Z}$.
    \item
        Ein $R$-Kettenkomplex ist ein $R$-Komplex mit
        $C_{i}=\{0\}\:\forall\: i<0$ und
        $d_{i}=0 \:\forall\: i\leq 0$.
    \item
        Ein $R$-Kokettenkomplex ist ein $R$-Komplex, für den
        $C_{i}=\{0\},\: d_{i}=0$ und $\forall\: i>0$. Man setzt dann
        $(C^{i},d^{i}):=(C_{-i},d_{-i})$, womit $d^{i}\colon C^{i}\longrightarrow C^{i+1}$.
    \item
        Gegeben zwei $R$-Komplexe $(C,d)$ und $(C',d')$, so heißt eine
        Menge $\alpha=\{\alpha_{i}\}_{i\in \mathbb{Z}}$ von
        $R$-Homomorphismen $\alpha_{i}\colon C_{i}\longrightarrow C_{i}'$
        Kettenabbildung von $(C,d)$ nach $C',d')$, falls folgendes Diagramm für alle $i\in
        \mathbb{Z}$ kommutiert:
        $$ \diagram C_{i} \rto^{d_{i}} \dto_{\alpha_{i}} &C_{i-1} \dto^{\alpha_{i-1}} \\
        C'_{i} \rto_{d'_{i}}               &C'_{i-1}.
        \enddiagram $$ 
    \end{enumerate}
\end{definition}
\begin{bemerkung}
    \begin{enumerate}
    \item
        Gegeben ein Ring $R$, so bilden die $R$-Komplexe zusammen mit
        den Kettenabbildungen die Kategorie $\Rcc$.
    \item
        Aus der Additivität von $\Rm$ erhalten wir die von $\Rcc$ vermöge
        \begin{equation*}
            (\alpha+\beta)_{i}:=\alpha_{i}+\beta_{i}
        \end{equation*} für Kettenabbildungen $\alpha,\beta\colon
        (C,d)\longrightarrow (C',d')$. In der Tat liefert diese Definition mit 
        \begin{equation*}
            (\alpha_{i-1}+\beta_{i-1})d_{i}=\alpha_{i-1}d_{i}+\beta_{i-1}d_{i}=d'_{i}\alpha_{i}+d'_{i}\beta_{i}=d'_{i}(\alpha_{i}+\beta_{i})
        \end{equation*} wieder eine Kettenabbildung zwischen besagten Komplexen. Diese Addition ist zudem
        kommutativ, und es ist klar, dass auch die
        Distributivgesetze \eqref{eq:addkatDistlaws1} und
        \eqref{eq:addkatDistlaws2} erfüllt sind. 
    \end{enumerate}
\end{bemerkung}
\begin{definition}[Ko-/Homologiemodul]
    \label{def:kohomol}
    \begin{enumerate}
    \item 
        Gegeben ein $R$-Kettenkomplex $(C,d)$. Sei $Z_{i}=\ker(d_{i})\subseteq C_{i}$, $B_{i}=\im(d_{i+1})\subseteq
        C_{i}$,\\ dann sind sowohl $Z_{i}$, als auch $B_{i}$ Untermoduln von
        $C_{i}$. Mit $d_{i}\cp d_{i+1}=0$ gilt zudem $B_{i}\subseteq Z_{i}$,
        womit $B_{i}$ 
        sogar ein Untermodul von $Z_{i}$ ist.
        Wir betrachten
        den Quotienten $H_{i}=Z_{i}/B_{i}$, womit genau dann $\alpha\in
        [z_{i}]\subseteq Z_{i}$ ist, wenn
        $\alpha=z_{i}+b_{i}$ für ein $b_{i}\in B_{i}$ gilt. Die
        Elemente in $Z_{i}$ nennt man $i$-Zykeln oder
        geschlossen, die Elemente aus $B_{i}$ $i$-Ränder oder exakt. Die
        Elemente $[\eta_{i}]\in H_{i}$ hei"sen $i$-te Homologieklassen und
        $H_{i}$ selbst $i$-ter Homologiemodul. Dabei sind
        Modul-Addition und Modul-Multiplikation durch
        $[\eta_{i}]+[\mu_{i}]=[\eta_{i}+\mu_{i}]$ sowie
        $r[\eta_{i}]:=[r\eta_{i}]$ auf Repräsentanten-Niveau
        definiert. Die Wohldefiniertheit dieser Operationen folgt
        dabei unmittelbar aus der Untermoduleigenschaft von
        $B_{i}\subseteq Z_{i}$. Man setzt dann
        $H_{0}=C_{0}/\im(d_{1})$.  
    \item
        Für einen $R$-Kokettenkomplex definieren wir analog
        $Z^{i}=\ker(d^{i})$, $B^{i}=\im(d^{i-1})$ und
        $H^{i}=Z^{i}/B^{i}$. Die Elemente $[\eta_{i}]\in H^{i}$ heißen 
        $i$-te Kohomologieklassen und $H^{i}$ selbst $i$-ter
        Kohomologiemodul mit der Konvention $H^{0}=\ker(d^{0})$.
    \item
        Sprechen wir ganz allgemein von einem $R$-Komplex, so benutzen
        wir die Nomenklatur aus \textit{i.)}.
    \end{enumerate}
\end{definition}
\begin{definition}
    \label{def:exakt}
    Ein $R$-Komplex $(C,d)$ heißt exakt, falls
    $\ker(d_{i})=\im(d_{i+1})$, und nach Definition \ref{def:kohomol} ist dies gleichbedeutend mit $H_{i}=\{0\}\:\forall\:i\in \mathbb{Z}$.  
\end{definition}
Wir wollen nun aufzeigen, inwiefern uns eine Kettenabbildung $\alpha$ zwischen zwei
$R$-Komplexen $(C,d)$ und $(C',d')$, Abbildungen zwischen deren
Homologiemoduln $H_{i}$ und $H'_{i}$, und sogar Funktoren von $\Rcc$ nach
$\Rm$ definiert.
\begin{lemma}
    \label{lemma:kettenabzu}
    Gegeben zwei $R$-Komplexe $(C,d)$ und $(C',d')$ und eine
    Kettenabbildung $\alpha$ von $(C,d)$ nach $(C',d')$, dann gilt:
    \begin{enumerate}
    \item
        Es ist $\alpha_{i}(Z_{i})\subseteq Z'_{i}$ und $\alpha_{i}(B_{i})\subseteq B'_{i}$.
    \item
        Die Abbildungen 
        \begin{align*}
            \widetilde{\alpha}_{i}\colon H_{i}&\longrightarrow H'_{i}\\
            [z_{i}]&\longmapsto [\alpha_{i}(z_{i})]
        \end{align*} sind $R$-Homomorphismen. Ist $\alpha_{i}$ ein
        Isomorphismus, so auch $\widetilde{\alpha}_{i}$.
    \item
        Die Abbildungen
        \begin{align*}
            \sim^{Ob}_{i}\colon R\text{-Kompl.}& \longrightarrow R\text{-Moduln}\\
            (C,d)&\longmapsto H_{i}(C)
        \end{align*} und
        \begin{align*}
            \sim^{Mor}_{i}:\Hom_{R}(C,C')&\longrightarrow \Hom_{R}(H_{i}(C),H_{i}(C'))\\
            \alpha& \longmapsto \widetilde{\alpha}_{i}
        \end{align*} definieren einen additiven, kovarianten Funktor
        $\sim_{i}$ von $\Rcc$
        nach $\Rm$. Den $i$-ten Homologie-Funktor. 
    \end{enumerate}
    \begin{beweis}  
        \begin{enumerate}
        \item
            Mit Definition \ref{def:Komplex}~\textit{iv.)} erhalten wir:
            \begin{equation*}
                (d'_{i}\cp\alpha_{i})(z_{i})=(\alpha_{i-1}\cp d_{i})(z_{i})=\alpha_{i-1}(0)=0
                \qquad\quad \forall\:z_{i}\in Z_{i}, 
            \end{equation*}
            \begin{equation*}
                \alpha_{i}(b_{i})=(\alpha_{i}\cp
                d_{i+1})(b_{i+1})=d'_{i+1} (\alpha_{i+1}\cp
                b_{i+1})\in B'_{i}\quad\quad \text{für }b_{i+1}\in
                C_{i+1},\:b_{i}\in B_{i}.
            \end{equation*}
        \item
            Die $R$-Linearität ist klar mit der von $\alpha_{i}$ und der
            Definition der Modul-Mul\-ti\-pli\-ka\-tion und Addition in
            $H'_{i}$. Für die Wohldefiniertheit sei $z'_{i}\in [z_{i}]$ mit
            $z_{i}\in Z$. Dann ist zu zeigen, dass
            $[\alpha_{i}(z'_{i})]=[\alpha_{i}(z_{i})]$. Dies folgt mit
            \begin{align*}
                [\alpha_{i}(z'_{i})]=[\alpha_{i}(z_{i}+b_{i})]\glna{\textit{i.)}}[\alpha_{i}(z_{i})+b'_{i}]=[\alpha_{i}(z_{i})]+[b'_{i}]=[\alpha_{i}(z_{i})].
            \end{align*} Sind die $\alpha_{i}$ Isomorphismen, so gilt $\alpha_{i-1}^{-1}\cp d_{i}'=d_{i}\cp\alpha_{i}^{-1}$, verm"oge Definition \ref{def:Komplex}~\textit{v.)}, womit auch die $\alpha_{i}^{-1}$
            Kettenabbildungen sind. Mit \textit{i.)} folgt unmittelbar $Z_{i}\stackrel{\alpha_{i}}{\cong}Z'_{i}$ und $B_{i}\stackrel{\alpha_{i}}{\cong}B'_{i}$, und für $[z'_{i}]\in H'_{i}$ finden wir somit
            $z_{i}\in Z_{i}$ mit
            $\widetilde{\alpha_{i}}([z_{i}])=[\alpha_{i}(z_{i})]=[z'_{i}]$,
            was die Surjektivität von $\wt{\alpha}_{i}$ zeigt. 
            Für die Injektivität nehmen wir
            an, es wäre
            $\widetilde{\alpha_{i}}([z_{i}])=\widetilde{\alpha_{i}}([z'_{i}])$
            mit $[z_{i}]\neq[z'_{i}]$. Dann folgt
            $\alpha_{i}(z_{i})=\alpha_{i}(z'_{i})+b'_{i}$ mit
            $b'_{i}=\alpha_{i}(b_{i})$ für ein $b_{i}\in B_{i}$ und somit
            $z_{i}-z'_{i}-b_{i}=0$. Im Widerspruch
            zu $[z_{i}]\neq[z'_{i}]$.
        \item
            Zunächst ist klar, dass
            $\wt{\id_{C_{i}}}=\id_{H_{i}}$. Weiter haben wir zu
            zeigen, dass
            \begin{equation*}
                \sim^{Mor}_{i}(\beta\circ\alpha):=\widetilde{(\beta\circ\alpha)}_{i}\:\glna{!}\:
                \widetilde{\beta}_{i}\circ\widetilde{\alpha}_{i}
                =:\:\sim^{Mor}_{i}(\beta)\:\circ\sim^{Mor}_{i}(\alpha)
            \end{equation*}
            für Kettenabbildungen $\alpha\colon (C,d)\longrightarrow (C',d')$ und $\beta:(C',d')\longrightarrow (C'',d'')$.
            Sei hierf"ur $[z_{i}]\in H_{i}$, dann folgt 
            \begin{align*}
                \widetilde{(\beta\circ\alpha)}_{i}\big([z_{i}]\big)&=
                \big[(\beta_{i}\circ
                \alpha_{i})(z_{i})\big]=[\beta_{i}(\alpha_{i}(z_{i}))]\\
                &=\widetilde{\beta}_{i}([\alpha_{i}(z_{i})])=\widetilde{\beta}_{i}\big(\widetilde{\alpha}_{i}([z_{i}])\big)=(\widetilde{\beta}_{i}\circ\widetilde{\alpha}_{i})([z_{i}]).
            \end{align*} Für die Additivität beachte man, dass sowohl
            $\Rcc$, als auch $\Rm$ additive Kategorien sind, und es
            ist zu zeigen, dass:
            \begin{equation*}
                \sim^{Mor}_{i}(\alpha+\beta)=\:\sim^{Mor}_{i}(\alpha)\:+\sim^{Mor}_{i}(\beta)\qquad\quad\forall\: \alpha,\beta \in \Hom_{R}(C,C').
            \end{equation*} Dies folgt sofort mit
            \begin{align*}
                \widetilde{(\alpha+\beta)_{i}}([\eta_{i}])=[(\alpha_{i}+\beta_{i})(\eta_{i})]=[\alpha_{i}(\eta_{i})]+[\beta_{i}(\eta_{i})]=\widetilde{\alpha}_{i}([\eta_{i}])+\widetilde{\beta}_{i}([\eta_{i}]).
            \end{align*}
        \end{enumerate}
    \end{beweis}
\end{lemma}
\subsection{Homotopie}
\label{subsec:homotopie}
Ein Kriterium, das festlegt, ob zwei Kettenabbildungen
$\alpha,\beta\colon (C,d)\longrightarrow (C',d')$ auf Homologie-Niveau die
gleiche Abbildung induzieren, wird durch die sogenannte
Homotopieeigenschaft bereitgestellt:
\begin{definition}[Homotopie]
    \label{def:homotopie}
    Gegeben zwei Kettenabbildungen $\alpha,\beta\colon(C,d)\longrightarrow
    (C',d')$ von $R$-Komplexen $(C,d)$ und $(C',d')$, so heißen $\alpha$
    und $\beta$ homotop ($\alpha\sim \beta$), falls $R$-Homomorphismen $\{s_{i}\}_{i\in
      \mathbb{Z}}$ mit $s_{i}\colon C_{i}\longrightarrow C'_{i+1}$ derart existieren, dass 
    \begin{equation}
        \label{eq:homotopiebed}
        \alpha_{i}-\beta_{i}=d'_{i+1}s_{i}+s_{i-1}d_{i}\qquad\qquad
        \forall i\in \mathbb{Z}.
    \end{equation}
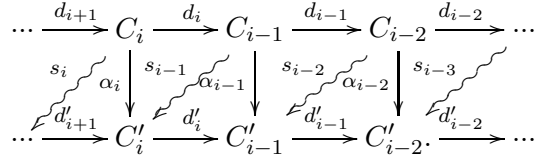
    \[\begin{xy} 
        \xymatrix{  
          ...\ar[r]^{d_{i+1}} &C_{i}\ar[d]_{\alpha_{i}}
          \ar@{~>}[dl]_{s_{i}}\ar[r]^{d_{i}}  &C_{i-1}
          \ar[d]_{\alpha_{i-1}}\ar@{~>}[dl]_{s_{i-1}}\ar[r]^{d_{i-1}} &C_{i-2} \ar[d]_{\alpha_{i-2}}\ar@{~>}[dl]_{s_{i-2}}\ar[r]^{d_{i-2}}
          &...\ar@{~>}[dl]_{s_{i-3}} &\\
          ...\ar[r]^{d'_{i+1}} &C'_{i} \ar[r]^{d'_{i}}  &C'_{i-1} \ar[r]^{d'_{i-1}} &C'_{i-2} \ar[r]^{d'_{i-2}}.
          &... &
        }
    \end{xy}\]
    $\bold{Warnung}$: Die gewellten Pfeile sind hier und im
    Folgenden keine
    kommutativen Elemente des Diagramms.
\end{definition}
\begin{bemerkung}
    \label{bem:kokettenhomotopiekram}
    \begin{enumerate}
    \item
        Für $R$-Kettenkomplexe gilt \eqref{eq:homotopiebed} mit der
        Zusatzbedingung $s_{i}=0$ f"ur $i< 0$.
    \item
        Für $R$-Kokettenkomplexe $(C^{i},d^{i})$ übersetzt sich
        Definition \ref{def:homotopie} zu    
        \begin{equation}
            \label{eq:homotopiebedkokett}
            \alpha^{i}-\beta^{i}=d'^{i-1}s^{i}+s^{i+1}d^{i}\quad\quad
            \forall  i\in \mathbb{Z}
        \end{equation} mit $R$-Homomorphismen $s^{i}\colon C^{i}\longrightarrow
        C^{i-1}$ und $s^{i}=0$ f"ur $i\leq0$.
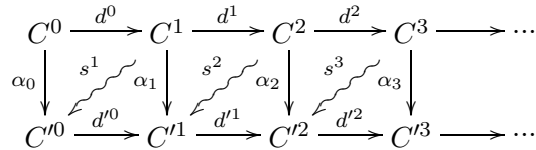
        \[\begin{xy} 
            \xymatrix{ 
              C^{0}\ar[d]_{\alpha_{0}} \ar[r]^{d^{0}}  &C^{1} \ar[d]_{\alpha_{1}}\ar[r]^{d^{1}}\ar@{~>}[dl]_{s^{1}} &C^{2} \ar[d]_{\alpha_{2}}\ar[r]^{d^{2}}\ar@{~>}[dl]_{s^{2}}
              &C^{3} \ar[d]_{\alpha_{3}}\ar@{~>}[dl]_{s^{3}}\ar[r]&...\\
              C'^{0} \ar[r]^{d'^{0}}  &C'^{1} \ar[r]^{d'^{1}} &C'^{2} \ar[r]^{d'^{2}}
              &C'^{3}  \ar[r] &... 
            }
        \end{xy}\]
    \end{enumerate}
\end{bemerkung}

\begin{korollar}
    Homotopie induziert eine Äquivalenzrelation.
    \begin{beweis}
        Die Symmetrie ist klar, ebenso die Reflexivität mit der
        Wahl $s_{i}=0$. Für die Transitivität sei $\alpha\sim \beta$ via $s$ und
        $\beta\sim \gamma$ via $t$, dann folgt $\alpha_{i}-\beta_{i}=d'_{i+1}s_{i}+s_{i-1}d_{i}$ sowie $\beta_{i}-\gamma_{i}=d'_{i+1}t_{i}+t_{i-1}d_{i}$ und somit $\alpha_{i}-\gamma_{i}=d'_{i+1}(s_{i}+t_{i})+(s_{i-1}+t_{i-1})\:d_{i}$. Dies zeigt $\alpha\sim \gamma$ via
        $u=s+t= \{s_{i}+t_{i}\}_{i \in \mathbb{Z}}$.
    \end{beweis}
\end{korollar}
\begin{lemma}
    \label{lemma:tildeabbeind}
    Ist in der Situation von Definition \ref{def:homotopie} $\alpha\sim \beta$,
    so gilt
    $\widetilde{\alpha}_{i}=\widetilde{\beta}_{i}$.
    \begin{beweis}        
        \begin{align*}
            \widetilde{\alpha}_{i}([z_{i}])=[\alpha_{i}(z_{i})]
            =[\beta_{i}(z_{i})]+[(d'_{i+1}s_{i})(z_{i})]+\underbrace{[(s_{i-1}d_{i})(z_{i})]}_{[0]}
            =[\beta_{i}(z_{i})]+[b'_{i}]=\widetilde{\beta}_{i}([z_{i}])
        \end{align*} für $z_{i}\in Z_{i},\:B'_{i}\ni b_{i}'= (d'_{i+1}s_{i})(z_{i})$.
    \end{beweis}
\end{lemma}
\subsection{Auflösungen}
\label{subsec:aufloesung}
\begin{definition}
    \label{def:komplexubermodul}
    Gegeben ein $R$-Modul $\mathcal{M}$, so bezeichnen wir mit einem Komplex $(C,d,\epsilon)$ über $\mathcal{M}$ einen
    $R$-Kettenkomplex $(C,d)$ zusammen mit einem $R$-Epimorphismus
    $\epsilon:C_{0}\longrightarrow \mathcal{M}$ derart, dass
    $\epsilon\circ d_{1}=0$. Solch ein $\epsilon$ bezeichnet man auch als
    Augmentierung von $(C,d)$. Grafisch bedeutet dies:
    $$ \diagram  
    ...\rto^{d_{i+1}} &C_{i} \rto^{d_{i}}  &C_{i-1} \rto^{d_{i-1}}
    &... \rto^{d_{2}} &C_{1} \rto^{d_{1}} &C_{0} \rto^{\epsilon}|>>\tip   &\mathcal{M},  
    \enddiagram $$ wobei der letzte Pfeil die
    Surjektivität von $\epsilon$ beutet. In obiger Kette von
    Homomorphismen ergibt somit die Hintereinanderausführung je
    zweier aufeinander folgender die $0$-Abbildung.
\end{definition}
\begin{definition}
    Gegeben ein Komplex $(C,d,\epsilon)$ über einem $R$-Modul $\mathcal{M}$. 
    \begin{enumerate}
    \item
        $(C,d,\epsilon)$ heißt Auflösung, falls die Kette von
        Homomorphismen aus obiger Definition 
exakt ist,
        also zusätzlich zu der Exaktheit von $(C,d)$ auch
        $\im(d_{1})=\ker(\epsilon)$ gilt.
        Das bedeutet $H_{i}(C)=0$ für $i>0$ und
        $H_{0}:=C_{0}/\im(d_{1})=C_{0}/\ker(\epsilon)\cong \mathcal{M}$.
    \item
        $(C,d,\epsilon)$ heißt projektiv, falls alle $C_{i}$
        projektive $R$-Moduln sind.
    \end{enumerate}
\end{definition}
\begin{satz}
    \label{satz:AufluProjKompKettab}
    Gegeben ein projektiver Komplex $(C,d,\epsilon)$ über einem $R$-Modul
    $\mathcal{M}$ und eine Auflösung $(C',d',\epsilon')$ eines $R$-Moduls
    $\mathcal{M}'$. Sei weiter $\mu: \mathcal{M}\rightarrow \mathcal{M}'$ ein $R$-Homomorphismus.
    Dann existiert eine Kettenabbildung $\alpha:C\longrightarrow C'$
    derart, dass folgendes Diagramm kommutiert:
    $$ \diagram  
    ...\rto^{d_{i+1}} &C_{i}\dto^{\alpha_{i}} \rto^{d_{i}}  &C_{i-1} \dto^{\alpha_{i-1}}\rto^{d_{i-1}} &C_{i-2} \dto^{\alpha_{i-2}}\rto^{d_{i-2}}
    &... \rto^{d_{1}} &C_{0} \dto^{\alpha_{0}}\rto^{\epsilon}|>> \tip
    &\mathcal{M}
    \dto^{\mu}\\
    ...\rto^{d'_{i+1}} &C'_{i} \rto^{d'_{i}}  &C'_{i-1} \rto^{d'_{i-1}} &C'_{i-2} \rto^{d'_{i-2}}
    &... \rto^{d'_{1}} &C'_{0} \rto^{\epsilon'}|>> \tip &\mathcal{M}'.
    \enddiagram $$
    Weiterhin sind alle derartigen Kettenabbildungen zueinander homotop.
    \begin{beweis}
        Nach Voraussetzung ist $\epsilon'$ surjektiv und $C_{0}$ projektiv. Hermit finden wir ein
        $\alpha_{0}\in \Hom_{R}(C_{0},C'_{0})$ derart, dass 
        $$ \diagram C_{0} \dto_{\alpha_{0}} \drto^{\mu\circ\epsilon}  &  \\
        C'_{0} \rto_{\epsilon'}|>>\tip      & \mathcal{M}' 
        \enddiagram $$ kommutiert. Dies bildet den Induktionsanfang,
        und wir müssen dann lediglich, für $\alpha_{i-1}$ vorgegeben,
        die  Existenz eines $\alpha_{i}$ nachweisen, welches die
        zugehörige Zelle zum kommutieren bringt. Hierzu wollen wir
        zunächst $C'_{i-1}$ durch $\im(d'_{i})$ ersetzen dürfen, da dann $d'_{i}$ surjektiv w"are, und mit der
        Projektivität von $C_{i}$ folgte abermals die Existenz eines solchen
        $\alpha_{i}$. Dafür reicht es zu zeigen, dass
        $\im(\alpha_{i-1}\:d_{i})\subseteq \im(d'_{i})$. Nun
        folgt
        \begin{equation*}
            d'_{i-1}\alpha_{i-1}\:d_{i}=\alpha_{i-2}\:d_{i-1}d_{i}=0,
        \end{equation*} womit $\im(\alpha_{i-1}\cp d_{i})\subseteq
        \ker(d'_{i-1})=\im(d'_{i})$ mit der Exaktheit von
        $(C',d',\epsilon')$. Das zeigt die erste Aussage des Satzes.

        Für die zweite seien
        $\alpha,\:\beta$ so, dass sie obiges Diagramm zum
        kommutieren bringen. Wir wollen dann eine Homotopie $s$ derart finden,
        dass:
        \begin{equation}
            \label{eq:homotopbed}
            \gamma_{i}=\alpha_{i}-\beta_{i}=d'_{i+1}s_{i}+s_{i-1}d_{i}.
        \end{equation} Zunächst folgt unmittelbar:
        \begin{align}
            \label{eq:kerepsilonstrich}
            \epsilon'\gamma_{0}&=\epsilon'\alpha_{0}-\epsilon'\beta_{0}=\mu\:\epsilon-\mu\:\epsilon=0,\\
            \label{eq:vertasuschiGamma}
            d'_{i}\gamma_{i}&=\gamma_{i-1}d_{i}\quad i\geq 1.
        \end{align}
        Wir betrachten den Diagrammausschnitt: 
        $$ \diagram &C_{0} \dto_{\gamma_{0}}   &\\
        C'_{1}\rto^{d'_{1}} &C'_{0} \rto^{\epsilon'}|>>\tip  & \mathcal{M}' 
        \enddiagram $$ und beachten, dass mit
        \eqref{eq:kerepsilonstrich} und der Exaktheit von $(C',d',\epsilon')$: $\im(\gamma_{0})\subseteq \ker(\epsilon')=\im(d'_{1})$ gilt, 
        womit mittels der Projektivität von $C_{0}$:
        $$ \diagram &C_{0} \dlto_{s_{0}} \dto_{\gamma_{0}}   &\\
        C'_{1}\rto^{d'_{1}}|>>\tip   &\im(d'_{1}).  
        \enddiagram $$ Durch $\gamma_{0}=d'_{1}s_{0}$ mit $i=0$ und $s_{-1}=0$ ist die
        Homotopiebedingung \eqref{eq:homotopbed} erfüllt.
        Sei nun wieder \eqref{eq:homotopbed} für $0\leq
        i\leq n-1$ korrekt. Wir definieren dann die Hilfsabbildung
        $\widetilde{\gamma}_{n}\colon C_{n}\longrightarrow C'_{n}$ induktiv durch
        $\widetilde{\gamma}_{n}:=\gamma_{n}-s_{n-1}d_{n}$ und erhalten:
        \begin{align*}
            d'_{n}\widetilde{\gamma}_{n}&=d'_{n}(\gamma_{n}-s_{n-1}d_{n})=d'_{n}\gamma_{n}-d'_{n}s_{n-1}d_{n}\glna{\eqref{eq:vertasuschiGamma}}\gamma_{n-1}d_{n}-d'_{n}s_{n-1}d_{n}\\
            &=(\gamma_{n-1}-d'_{n}s_{n-1})d_{n}\glna{\eqref{eq:homotopbed}}(\gamma_{n-1}-\gamma_{n-1}+s_{n-2}d_{n-1})d_{n}=0.
        \end{align*} Es ist abermals
        $\im(\widetilde{\gamma}_{n})\subseteq \ker(d'_{n})=\im(d'_{n+1})$,
        und mit der Projektivität von $C_{n}$ erhalten wir ein
        $s_{n}\colon C_{n}\longrightarrow C'_{n+1}$ derart, dass
        $d'_{n+1}s_{n}=\widetilde{\gamma}_{n}=\gamma_{n}-s_{n-1}d_{n}$:
        $$ \diagram &C_{n} \dlto_{s_{n}} \dto_{\wt{\gamma}_{n}}   &\\
        C'_{n+1}\rto^{d'_{n+1}}|>>\tip   &\im(d'_{n+1}).  
        \enddiagram $$ 
        Es folgt unmittelbar
        $\gamma_{n}=d'_{n+1}s_{n}+s_{n-1}d_{n}$ und somit \eqref{eq:homotopbed}.
    \end{beweis}
\end{satz}
\subsection{Rechtsinduzierte Funktoren}
Zunächst wollen wir die in Abschnitt \ref{subsec:kompundhomolog} und
\ref{subsec:homotopie} behandelten Begrifflichkeiten und Zusammenhänge
von $R$-Moduln auf abelsche Gruppen übertragen ($R$-Moduln waren ja
lediglich derartige Gruppen mit Zusatzstruktur). Aus Definition \ref{def:Komplex}
wird:
\begin{definition}
    \label{def:KomplexG}
    \begin{enumerate}
    \item
        Ein Gruppenkomplex ist eine Menge $\{G_{i},d_{i}\}_{i\in
          \mathbb{Z}}$ von Paaren $(G_{i},d_{i})$ von abelschen Gruppen
        $G_{i}$ und Homomorphismen $d_{i}\colon G_{i}\longrightarrow
        G_{i-1}$ mit $d_{i-1} \cp d_{i}=0$ für alle $i\in
        \mathbb{Z}$.
    \item
        Ein Gruppen-Kettenkomplex ist ein Gruppenkomplex, für den
        $G_{i}=\{0\}\:\forall\:i<0$ und $d_{i}=0 \:\forall\: i\leq 0$.
    \item
        Ein Gruppen-Kokettenkomplex ist ein Gruppenkomplex, f"ur den
        $G_{i}=\{0\}$ und $d_{i}=0$ f"ur alle $i>0$. Man setzt
        $(G^{i},d^{i}):=(G_{-i},d_{-i})$, womit $d^{i}:G^{i}\longrightarrow G^{i+1}$.
    \item
        Gegeben zwei Gruppenkomplexe $(G,d)$ und $(G',d')$, so heißt eine
        Menge $\alpha=\{\alpha_{i}\}_{i\in \mathbb{Z}}$ von
        Gruppenhomomorphismen $\alpha_{i}\colon G_{i}\longrightarrow G_{i}'$
        Kettenabbildung von $(G,d)$ nach $(G',d')$, falls folgendes
        Diagramm für alle $i\in \mathbb{Z}$ kommutiert:
        $$ \diagram G_{i} \rto^{d_{i}} \dto_{\alpha_{i}} &G_{i-1} \dto^{\alpha_{i-1}} \\
        G'_{i} \rto_{d'_{i}}               &G'_{i-1}\:.
        \enddiagram $$ 
    \end{enumerate}
\end{definition}
Definition \ref{def:kohomol} wird zu:
\begin{definition}[Ko-/Homologiegruppe]
    \label{def:kohomolG}
    \begin{enumerate}
    \item 
        Gegeben ein Gruppen-Kettenkomplex $(G,d)$. Sei $Z_{i}=\ker(d_{i})\subseteq G_{i}$ sowie $B_{i}=\im(d_{i+1})\subseteq
        G_{i}$. Dann ist $B_{i}\subseteq Z_{i}$ eine Untergruppe von $Z_{i}$. Wir betrachten wieder den Quotienten $H_{i}=Z_{i}/B_{i}$
        und nennen die $[g_{i}]\in H_{i}$ $i$-te Homologieklassen sowie
        $H_{i}$ selbst $i$-te Homologiegruppe. Man setzt dann
        $H_{0}=G_{0}/\im(d_{1})$.
    \item
        Für einen Guppen-Kokettenkomplex definieren wir 
        $Z^{i}=\ker(d_{i})$, $B^{i}=\im(d^{i-1})$ und
        $H^{i}=Z^{i}/B^{i}$. Die Elemente $[g_{i}]\in H^{i}$ heißen
        $i$-te Kohomologieklassen und $H^{i}$ selbst $i$-te
        Kohomologiegruppe mit der Konvention $H^{0}=\ker(d^{0})$.
    \item
        Sprechen wir von einem Gruppen-Komplex, so benutzen wir die
        Nomenklatur aus \textit{i.)}.
    \end{enumerate}
\end{definition}
Definition \ref{def:exakt} und Lemma \ref{lemma:kettenabzu} übertragen sich
sinngemäß, ebenso der gesamte Abschnitt \ref{subsec:homotopie}.
In Lemma \ref{lemma:kettenabzu}~\textit{iv.)} haben wir dann additive,
kovariante Funktoren
$\sim_{i}$ 
\begin{align*}
    \sim^{Ob}_{i}\colon\text{Gruppenkompl.}&\longrightarrow \text{ab. Gr.}\\
    (G,d)&\longmapsto H_{i}(G)
\end{align*} und
\begin{align*}
    \sim^{Mor}_{i}\colon\Hom(G,G')&\longrightarrow \Hom(H_{i}(G),H_{i}(G'))\\
    \alpha& \longmapsto\widetilde{\alpha}_{i}.\nonumber
\end{align*}
Für einen R-Modul $\mathcal{M}$, einen Komplex $(C,d,\epsilon)$ über $\mathcal{M}$ und einen kontravarianten,
additiven Funktor $\mathcal{F}\colon\Rm\longrightarrow \Ab$ erhalten wir
durch Anwendung dieses Funktors auf $(C,d,\epsilon)$ einen
Gruppen-Kokettenkomplex
$(\mathcal{F}C,\mathcal{F}d,\mathcal{F}\epsilon)$:
$$ \diagram  
\mathcal{F}M \rto^{\mathcal{F}\epsilon} &\mathcal{F}C_{0}
\rto^{\mathcal{F}d_{1}} &\mathcal{F}C_{1} \rto^{\mathcal{F}d_{2}}
&\mathcal{F}C_{2} \rto^{\mathcal{F}d_{3}} &...\:.
\enddiagram $$
In der Tat sind mit der Additivität von $\mathcal{F}$ alle Pfeile
Gruppenhomomorphismen. Des Weiteren gilt $\mathcal{F}d_{i+1}\circ
\mathcal{F}d_{i}=\mathcal{F}(d_{i}\circ
d_{i+1})=\mathcal{F}(0)=0$. Ist $(C,d,\epsilon)$ exakt, so überträgt sich dies
nicht notwendigerweise auf
$(\mathcal{F}C,\mathcal{F}d\mathcal{F}\epsilon)$ und wir erhalten im Allgemeinen
nicht-triviale Kohomologiegruppen $H^{i}(\mathcal{F}C)$ sowie
$H^{0}(\mathcal{F}C):=\ker(\mathcal{F}d_{1})$.
\begin{lemma} 
    \label{lemma:GruppenKOhomsausprojaufloesundFunktoren}
    Gegeben $R$-Moduln $\mathcal{M},\mathcal{M}'$, ein projektiver Komplex
    $(C,d,\epsilon)$ über $\mathcal{M}$,
    eine Auflösung $(C',d',\epsilon')$ von $\mathcal{M}'$ und ein
    additiver, kontravarianter Funktor $\mathcal{F}\colon\Rm \longrightarrow \Ab$. Sei des Weiteren $\mu$
    ein Homomorphismus $\mu\colon\mathcal{M}\longrightarrow \mathcal{M}'$. Dann gilt:
    \begin{enumerate}
    \item Es existieren
        nur von diesen Daten abhängige Gruppenhomomorphismen\\
        $\widetilde{\mathcal{F}\alpha_{i}}\colon H^{i}(\mathcal{F}C')\longrightarrow
        H^{i}(\mathcal{F}C)$.
    \item
        Sind
        $(C,d,\epsilon)$ und
        $(\ovl{C},\ovl{d},\ovl{\epsilon})$ 
        projektive
        Auflösungen des selben $R$-Moduls $\mathcal{M}$, so gilt $H^{i}(\mathcal{F}C)\simeq H^{i}(\mathcal{F}\ovl{C})$.
    \end{enumerate} 
    \begin{beweis}
        \begin{enumerate}
        \item 
            Zunächst haben wir mit Satz \ref{satz:AufluProjKompKettab} eine bis auf Homotopie eindeutig
            bestimmte Kettenabbildung
            $\alpha\colon (C,d,\epsilon)\longrightarrow (C',d',\epsilon')$
            derart, dass
            $$ \diagram  
            ...\rto^{d_{3}}
            &C_{2} \dto_{\alpha_{2}}\rto^{d_{2}}
            &C_{1} \dto_{\alpha_{1}}\rto^{d_{1}}
            &C_{0}
            \dto_{\alpha_{0}}\rto^{\epsilon}|>>\tip &\mathcal{M}
            \dto_{\mu}\\
            ... \rto^{d'_{3}}
            &C'_{2} \rto^{d'_{1}}\rto^{d'_{2}}
            &C'_{1} \rto^{d'_{1}} &C'_{0} \rto^{\epsilon'}|>>\tip &\mathcal{M}'
            \enddiagram $$ kommutiert. Sei weiter $\alpha\sim \beta$
            vermöge $s_{i}\colon C_{i}\longrightarrow C'_{i+1}$, dann erhalten wir durch Anwendung
            von $\mathcal{F}$:
            \[\begin{xy}
                \xymatrix{
                  \mathcal{F}M'
                  \ar[d]_{\mathcal{F}\mu}\ar[r]^{\mathcal{F}\epsilon'}
                  &\mathcal{F}C'_{0}
                  \ar[d]_{\mathcal{F}\alpha_{0}}\ar[r]^{\mathcal{F}d'_{1}}
                  &\mathcal{F}C'_{1}\ar@{~>}[dl]_{\mathcal{F}s_{0}}
                  \ar[d]_{\mathcal{F}\alpha_{1}}\ar[r]^{\mathcal{F}d'_{2}}
                  &\mathcal{F}C'_{2}\ar@{~>}[dl]_{\mathcal{F}s_{1}}
                  \ar[d]_{\mathcal{F}\alpha_{2}}\ar[r]^{\mathcal{F}d'_{3}} &…\ar@{~>}[dl]_{\mathcal{F}s_{2}}\\
                  \mathcal{F}M \ar[r]^{\mathcal{F}\epsilon} &\mathcal{F}C_{0}
                  \ar[r]^{\mathcal{F}d_{1}} &\mathcal{F}C_{1}
                  \ar[r]^{\mathcal{F}d_{2}} &\mathcal{F}C_{2} \ar[r]^{\mathcal{F}d_{3}} &…\:.
                }
            \end{xy}\]
            Wegen der Additivität und Kontravarianz von $\mathcal{F}$ folgt:
            \begin{equation*}
                \mathcal{F}\alpha_{i}-\mathcal{F}\beta_{i}=\mathcal{F}(\alpha_{i}-\beta_{i})=\mathcal{F}(d'_{i+1}s_{i}+s_{i-1}d_{i})=\mathcal{F}
                d_{i}\cp \mathcal{F}s_{i-1}+\mathcal{F}s_{i}\cp \mathcal{F}d'_{i+1}.
            \end{equation*} Mit den Definitionen
            $d^{i}=\mathcal{F}d'_{i+1}$; 
            $d'^{i}=\mathcal{F}d_{i+1}$; $s^{i}=\mathcal{F}s_{i-1}$
            zeigt dies
            \begin{equation*}
                \mathcal{F}\alpha_{i}-\mathcal{F}\beta_{i}=d'^{i-1}s^{i}+s^{i+1}d^{i},
            \end{equation*}
            also $\mathcal{F}\alpha\sim
            \mathcal{F}\beta$ nach \eqref{eq:homotopiebedkokett}. Die Gruppen-Version von
            Lemma \ref{lemma:tildeabbeind} liefert schließlich
            $\widetilde{\mathcal{F}\alpha_{i}}=\widetilde{\mathcal{F}\beta_{i}}$,
            und somit die erste Aussage.
        \item
            Wir haben für $\mu=\id_{\mathcal{M}}$   
            $$ \diagram  
            ...
            &\lto_{\mathcal{F}d_{3}}\mathcal{F}C_{2}
            &\lto_{\mathcal{F}d_{2}}\mathcal{F}C_{1}
            &\lto_{\mathcal{F}d_{1}}\mathcal{F}C_{0}
            &\lto_{\mathcal{F}\epsilon}\mathcal{F}\mathcal{M}\\
            ...
            &\lto_{\mathcal{F}\ovl{d}_{3}}\mathcal{F}\ovl{C}_{2}\uto_{\mathcal{F}\beta_{2}}
            &\lto_{\mathcal{F}\ovl{d}_{2}}\mathcal{F}\ovl{C}_{1}\uto_{\mathcal{F}\beta_{1}} 
            &\lto_{\mathcal{F}\ovl{d}_{1}}\mathcal{F}\ovl{C}_{0}\uto_{\mathcal{F}\beta_{0}} 
            &\lto_{\mathcal{F}\ovl{\epsilon}}\mathcal{F}\mathcal{M}\uto_{\id_{\mathcal{F}\mathcal{M}}}\\
            ... 
            &\lto_{\mathcal{F}d_{3}}\mathcal{F}C_{2} \uto_{\mathcal{F}\widehat{\beta}_{2}}
            &\lto_{\mathcal{F}d_{2}}\mathcal{F}C_{1} \uto_{\mathcal{F}\widehat{\beta}_{1}}
            &\lto_{\mathcal{F}d_{1}}\mathcal{F}C_{0} \uto_{\mathcal{F}\widehat{\beta}_{0}} 
            &\lto_{\mathcal{F}\epsilon}\mathcal{F}\mathcal{M}\uto_{\id_{\mathcal{F}\mathcal{M}}},
            \enddiagram $$ womit
            \begin{equation*} 
                \id_{\mathcal{F}C_{i}}=\mathcal{F}(\id_{C_{i}})\sim
                \mathcal{F}\big(\widehat{\beta}_{i}\circ\beta_{i}\big).
            \end{equation*}
            Mit Lemma \ref{lemma:tildeabbeind} und
            \ref{lemma:kettenabzu}~\textit{iv.)} erhalten wir
            \begin{equation*}
                \id_{H^{i}(\mathcal{F}C)}=\widetilde{\id_{\mathcal{F}C_{i}}}=\widetilde{\mathcal{F}\big(\widehat{\beta}_{i}\circ\beta_{i}\big)}=\widetilde{\mathcal{F}\beta_{i}\circ\mathcal{F}\widehat{\beta}_{i}}
                =\widetilde{\mathcal{F}\beta_{i}}\circ\widetilde{\mathcal{F}\widehat{\beta}_{i}}.
            \end{equation*} 
            Analog folgt $\id_{H_{i}(\mathcal{F}\ovl{C})}=\widetilde{\mathcal{F}\widehat{\beta}_{i}}\circ\widetilde{\mathcal{F}\beta}_{i}$ womit $\widetilde{\mathcal{F}\beta}_{i}=\widetilde{\mathcal{F}\widehat{\beta}_{i}}^{-1}$, also $H^{i}(C)\simeq H^{i}(\ovl{C})$
            vermöge
            $\eta^{i}=\widetilde{\mathcal{F}\beta_{i}}\colon H^{i}(\ovl{C})\longrightarrow
            H^{i}(C)$.
        \end{enumerate} 
    \end{beweis}
\end{lemma} 
\begin{korollar}
    \label{kor:Extkor}
    Sind in der Situation von
    Lemma \ref{lemma:GruppenKOhomsausprojaufloesundFunktoren} beide Komplexe
    $(C,d,\epsilon)$, $(C',d',\epsilon')$ projektive Auflösungen, so
    sind die zugehörigen Kohomologiegruppen $H^{i}(\mathcal{F}C)$,
    $H^{i}(\mathcal{F}C')$ sowie die Homomorphismen aus
    Lemma \ref{lemma:GruppenKOhomsausprojaufloesundFunktoren}~\textit{i.)} bereits
    bis auf Verkettung mit Isomorphismen eindeutig bestimmt durch $\mathcal{M},\mathcal{M}', \mu$ und
    $\mathcal{F}$, hängen also nicht von der Wahl der Auflösungen ab.
    \begin{beweis}
        Die Isomorphie der Kohomologiegruppen folgt sofort aus Teil
        \textit{ii.)} obigen Lemmas, da wir zwischen verschiedenen
        Auflösungen des selben Moduls via Isomorphismen wechseln können. Mit der Funktoreigenschaft von
        $\mathcal{F}$ und $\sim_{i}$ kann dann für Kettenabbildungen
        \begin{align*}
            \beta\colon(C,d,\epsilon)&\longrightarrow
            (\ovl{C},\ovl{d},\ovl{\epsilon})\\
            \beta'\colon(C',d',\epsilon')&\longrightarrow
            (\ovl{C}',\ovl{d}',\ovl{\epsilon}')
            \\\alpha\colon(C,d,\epsilon)&\longrightarrow (C',d',\epsilon')
        \end{align*} der Kohomologiegruppen-Homomorphismus $\tau^{i}=(\eta^{i})^{-1}\widetilde{\mathcal{F}\alpha_{i}}\eta'^{i}$ geschrieben werden, als $\widetilde{\mathcal{F}(\beta'_{i}\alpha_{i}\widehat{\beta}_{i})}$. Da nun aber
        $\beta'\alpha\widehat{\beta}$: $(\ovl{C},\ovl{d},\ovl{\epsilon})\longrightarrow
        (\ovl{C}',\ovl{d}',\ovl{\epsilon}')$ selbst eine
        Kettenabbildung ist, stimmt $\tau$ mit allen anderen durch
        $\mathcal{F}$ gewonnenen Kettenabbildungen überein. Dies zeigt die
        zweite Aussage des Korollars.
    \end{beweis}
\end{korollar}
Das nächste Lemma liefert ein Kriterium welches garantiert,
dass überhaupt eine projektive Auflösung eines $R$-Moduls existiert.
\begin{lemma}
    \label{lemma:unitarereRingModulnhabenfreieAufloesung}
    Gegeben ein unitärer Ring $R$ und ein $R$-Modul $\mathcal{M}$, so existiert eine freie
    Auflösung $(C,d,\epsilon)$ von $\mathcal{M}$. Dabei bedeutet frei, dass die
    $C_{i}$ freie $R$-Moduln sind.
    \begin{beweis}
        Wir wählen eine Indizierung $J$ der Elemente von $\mathcal{M}$ und
        betrachten die direkte Summe
        $C_{0}=\bigoplus^{|J|}R$. Bezeichne
        $\alpha_{m}\in J$ den zu $m\in \mathcal{M}$ gehörigen Index, so definieren wir eine surjektive Abbildung
        $\epsilon:\bigoplus^{|J|}R \longrightarrow \mathcal{M}$ durch
        \begin{align*}
            \bigoplus 1_{\alpha_{m}}\longrightarrow m, 
        \end{align*} und setzen diese
        $R$-linear auf ganz $C_{0}$ fort. $C_{0}$ ist dann per
        Definition ein freier $R$-Modul (je nachdem ob $\mathcal{M}$ ein Links-
        oder Rechtsmodul ist,
        definiert man die $R$-Multiplikation in $C_{0}$ in
        entsprechender Weise), und wir erhalten
        $$ \diagram  
        &  \ker(\epsilon) \rto|<\ahook^{i} &C_{0} \rto|>>\tip
        ^{\epsilon} &\mathcal{M},
        \enddiagram $$ wobei $i$ die Injektion von $\ker(\epsilon)$ in
        $C_{0}$ bezeichnet. Da $\ker(\epsilon)$ ein Untermodul und somit
        selbst ein Modul ist, dürfen wir $C_{1}=\ker(\epsilon)$ und
        $d_{0}=i$ setzen, und
        erhalten $\epsilon\cp d_{0}=0$ sowie $\im(i)=\ker(\epsilon)$. Die
        Iteration von Schritt 1 mit $C_{1}$ an Stelle von $\mathcal{M}$ liefert eine
        freie Auflösung über $\mathcal{M}$.
    \end{beweis}
\end{lemma}
\begin{korollar}
    \label{kor:exprojAufl}
    Gegeben ein unitärer Ring $R$ und ein $R$-Modul $\mathcal{M}$, so existiert eine
    projektive Auflösung $(C,d,\epsilon)$ von $\mathcal{M}$.
    \begin{beweis}
        Das folgt sofort mit
        Lemma \ref{lemma:unitarereRingModulnhabenfreieAufloesung} und Korollar \ref{kor:freiemoduberunitRingensindProjektiv}.
    \end{beweis}
\end{korollar}
\begin{definition}[Rechtsinduzierte Funktoren]
    Gegeben ein unitärer Ring $R$ und ein kontravarianter Funktor
    $\mathcal{F}\colon \Rm\longrightarrow \Ab$, so ist der $i$-te rechtsinduzierte
    Funktor $R^{i}\mathcal{F}\colon \Rm\longrightarrow \Ab$ definiert durch
    \begin{align*}
        R^{i}\mathcal{F}^{Ob}\colon R\text{-Moduln}&\longrightarrow
        \text{abelsche Gruppen}\\
        \mathcal{M}&\longmapsto H^{i}(\mathcal{F}C)
    \end{align*}
    \begin{align*}
        R^{i}\mathcal{F}^{Mor}\colon\Hom_{R}(\mathcal{M},\mathcal{M}') &\longrightarrow \Hom_{G}(H^{i}(\mathcal{F}C'),H^{i}(\mathcal{F}C))\\
        \mu&\longmapsto \widetilde{\mathcal{F}\alpha_{i}}.
    \end{align*} Dessen Wohldefiniertheit folgt aus
    Korollar \ref{kor:Extkor}, wobei wir beliebige Auflösungen
    $(C,d,\epsilon)$ und $(C',d',\epsilon')$ von $\mathcal{M}$ und $\mathcal{M}'$ wählen dürfen.
\end{definition}
Wir wollen nun ein Beispiel vorstellen, welches eine essentielle Rolle im Haupttext spielen wird.
\begin{beispiel}
    \label{bsp:ExtBeisp}
    Gegeben ein $R$-Modul $\mathcal{N}$, dann ist der additive, kontravariante
    Funktor $\mathrm{hom}_{R}(\cdot,\mathcal{N})$ definiert durch
    \begin{equation}
        \label{eq:homfktOb}
        \begin{split}
            \mathrm{hom}_{R}(\cdot,\mathcal{N})^{Obj}\colon R\text{-Moduln}&\longrightarrow
            \text{abelsche Gruppen}\\
            \mathcal{M}&\longmapsto \Hom_{R}(\mathcal{M},\mathcal{N})
        \end{split}
    \end{equation}
    \begin{equation}
        \label{eq:homfktrMor}
        \begin{split}
            \mathrm{hom}_{R}(\cdot,\mathcal{N})^{Mor}\colon\Hom_{R}(\mathcal{M},\mathcal{M}')&\longrightarrow \Hom_{R}\left(\Hom_{R}(\mathcal{M}',\mathcal{N}),\Hom_{R}(\mathcal{M},\mathcal{N})\right)\\
            \mu &\longmapsto
            [\alpha\mapsto\alpha\circ \mu].
        \end{split}
    \end{equation}
    Die Kontravarianz ist klar und die Additivität
    sieht man mit der Bilinearität von $\circ$ bezüglich der Addition
    in $\Hom_{R}(\mathcal{M}',\mathcal{N})$ und $\Hom_{R}(\mathcal{M},\mathcal{N})$.\\\\
    Für einen unitären Ring $R$ und einen $R$-Modul $\mathcal{N}$ definieren wir den Funktor
    \begin{align*}
        Ext^{i}_{R}(\cdot,\mathcal{N})=R^{i} \mathrm{hom}_{R}(\cdot,\mathcal{N}).
    \end{align*}
    Es ist dann
    \begin{equation*}
        Ext^{i}_{R}(\cdot,\mathcal{N})(\mathcal{M})=H^{i}(\mathrm{hom}_{R}(\cdot,\mathcal{N})C)=H^{i}(\Hom_{R}(C,\mathcal{N}),d^{*}),
    \end{equation*} für eine projektive Auflösung $(C,d,\epsilon)$ von
    $\mathcal{M}$. Hierbei bezeichnet $(\Hom_{R}(C,\mathcal{N}),d^{*})$ den durch Anwendung des $\hom_{R}$-Funktors auf
    $(C,d)$ entstehenden Gruppen-Koketten-komplex mit Kettengliedern
    $\mathcal{F}C_{i}=\mathrm{hom}_{R}(\cdot,\mathcal{N})C_{i}=\Hom_{R}(C_{i},\mathcal{N})$
    und Kettendifferentialen $d^{i}=d_{i+1}^{*}=\hom_{R}(\cdot,\mathcal{M})d_{i+1}$.
    Als Anwendung berechnen wir $Ext^{0}_{R}(\cdot,\mathcal{N})(\mathcal{M})$:\\\\ Wir haben 
    $$ \diagram  
    ...\rto^{d_{3}} &C_{2} \rto^{d_{2}} &C_{1} \rto^{d_{1}}
    &C_{0}\rto^{\epsilon}|>>\tip 
    &\mathcal{M} \\
    \enddiagram $$ 
    $$ \diagram  
    \mathrm{Hom}_{R}(\mathcal{M},\mathcal{N})\rto^{\epsilon^{*}}
    &\mathrm{Hom}_{R}(C_{0},\mathcal{N})\rto^{d_{1}^{*}} &\mathrm{Hom}_{R}(C_{1},\mathcal{N})\rto &…\\
    \enddiagram $$ und es folgt
    \begin{align*}
        Ext^{0}_{R}(\cdot,\mathcal{N})(\mathcal{M})&
        =H^{0}(\Hom_{R}(C,\mathcal{N}))=\ker(\overbrace{\mathrm{hom}_{R}^{Mor}(\cdot,\mathcal{N})d_{1}}^{d_{1}^{*}}) \\
        &=\Menge{\alpha\in \Hom_{R}(C_{0},\mathcal{N})}{\alpha\circ d_{1}=0}\\ &=
        \Menge{\alpha\in \Hom_{R}(C_{0},\mathcal{N})}{\ker(\alpha)\supseteq
          \im(d_{1})=\ker(\epsilon)},
    \end{align*} womit
    $Ext^{0}_{R}(\cdot,\mathcal{N})(\mathcal{M})\cong\Hom_{R}(C_{0}/\ker(\epsilon),\mathcal{N})$. Da nun
    $C_{0}/\ker(\epsilon)\cong \im(\epsilon)=\mathcal{M}$, folgt
    $Ext^{0}_{R}(\cdot,\mathcal{N})(\mathcal{M})\cong \Hom_{R}(\mathcal{M},\mathcal{N})$.
\end{beispiel} 
\chapter{Lokalkonvexe Analysis}
\label{sec:lokalkonvAnal}
Dieser Abschnitt soll die analytischen Mittel bereitstellen, die im
Haupttext benötigt werden. Dabei werden wir den Begriff des
topologischen Raumes als bekannt voraussetzen. Beginnen werden wir mit grundlegenden
Definitionen und wichtigen Sätzen (ohne Beweis). Der zweite Teil
behandelt die Konzepte der Vollständigkeit und der stetigen
Fortsetzungen. Enden wird dieser
mit zwei wichtigen Sätzen, die garantieren, dass man in lokal
konvexen Vektorräumen stetige multilineare Abbildungen
stetig multilinear auf eine immer existierende Vervollständigung
besagten lokalkonvexen Vektorraumes fortsetzen kann. Der letzte
Abschnitt behandelt topologische Tensorprodukte lokalkonvexer
Vektorräume. Sei hierfür im Folgenden $\mathbb{K}$ immer $\mathbb{R}$ oder $\mathbb{C}$.
\section{Grundlagen} 
Die hier dargestellten Zusammenhänge sind auch in \cite{rudin:1991a} und
\cite{blanchard.bruening:1993a} zu finden.\\\\
\begin{definition}[Halbnorm]
    \label{def:Halbnormdef}
    Gegeben ein $\mathbb{K}$-Vektorraum $\mathbb{V}$ und eine Abbildung
    $p\colon\mathbb{V}\longrightarrow \mathbb{R}$, dann heißt $p$
    Halbnorm auf $\V$, wenn folgende Eigenschaften erfüllt sind:
    \begin{enumerate}
    \item
        $p(\alpha x)=|\alpha|p(x)\qquad \forall\:\alpha\in
        \mathbb{K}, x\in \V$;
    \item
        $p(x+y)\leq p(x)+p(y)\qquad \forall\:x,y\in \V$.
    \end{enumerate}
    Es folgt unter anderem $p(x)\geq 0$, $p(0)=0$, $p(x)=p(-x)$,
    $|p(x)-p(y)|\leq p(x-y)$ für alle $x,y \in \V$, zudem sind 
    die Halbnormenbälle $B_{p,\epsilon}(0):=\{x\in \V|\: p(x-p)<\epsilon\}$ konvex,
    ausgeglichen und absorbierend. Dabei heißt eine Teilmenge
    $U\subseteq \mathbb{V}$
    \begin{enumerate}
    \item
        $\emph{konvex}$, falls: 
        \begin{equation*}
            \sum_{i=1}^{n}t_{i}U\subseteq U\quad\quad\forall\:n\in
            \mathbb{N} \text{ und alle }
            t_{1},…,t_{n}>0\:\mathrm{mit}\:
            \sum_{i=1}^{n}t_{i}=1;
        \end{equation*}
    \item
        $\emph{ausgeglichen}$, falls:
        \begin{equation*}
            \lambda U\subseteq U\quad\quad \forall\:\lambda\in
            \mathbb{K},\:|\lambda|\leq 1;
        \end{equation*}
    \item
        $\emph{absorbierend}$, falls für jedes $x\in \mathbb{V}$ ein
        $\lambda>0$ derart existiert, dass für alle $\alpha\geq \lambda$
        $x\in \alpha U:=\{y\in\V|\: \exists\: u\in U \text{ mit } y=\alpha u\}$ gilt. 
    \end{enumerate}
    Hierfür beachte man, dass hier und im Folgenden mit $\alpha\geq 0$
    insbesondere immer $\alpha \in
    \mathbb{R}\subseteq \mathbb{C}$ gemeint ist.
\end{definition}
\begin{definition}
    \label{def:Halbnormensysteme}
    Gegeben ein $\mathbb{K}$-Vektorraum $\mathbb{V}$ und
    ein Halbnormensystem  (d.h eine Menge von Halbnormen) $P$ auf $\mathbb{V}$.
    \begin{enumerate}
    \item
        \label{item:separierend} 
        $P$ heißt separierend, falls zu jedem $x\in \mathbb{V}$
        ein $p \in P$ derart existiert, dass $p(x)\neq 0$.
    \item
        \label{item:filtrierend} 
        $P$ heißt filtrierend, falls zu jeder endlichen
        Teilmenge $\{p_{\alpha}\}_{\alpha\in I} \subseteq P$ von
        Halbnormen ein $p_{I} \in P$ existiert, so dass
        $p_{I}(x)\geq p_{\alpha}(x)\:\forall\alpha\in I,\:\forall\:
        x\in \mathbb{V}$.
    \end{enumerate}
\end{definition}
\begin{bemerkung}
    \label{bem:fdgds}    
    \begin{enumerate}
    \item
        Folgender Satz zeigt, dass jedes Halbnormensystem $P$ auf
        einem $\mathbb{K}$-Vektorraum
        $\V$ diesen zu einem topologischen Vektorraum macht
        (Vektoraddition und Skalar-multiplikation sind stetig in
        den jeweiligen Produkttopologien). Die
        Separationseigenschaft von $P$ garantiert dann, dass
        besagte Topologie hausdorffsch ist. Abweichend zu \cite{rudin:1991a} wollen wir also die Abgeschlossenheit von Punkten $x\in \V$, aus der bereits die Hausdorff-Eigenschaft eines topologischen Vektorraumes folgt, nicht voraussetzen
    \item
        Es ist klarerweise immer möglich zu jeder endlichen Teilmenge $\{p_{\alpha}\}_{\alpha\in
          I} \subseteq P$ durch
        $p_{I}(x):=\max_{\alpha\in I}(p_{\alpha}(x))$ eine
        Halbnorm zu definieren, für die $p_{I}(x)\geq p_{\alpha}(x)$
        für alle $\alpha\in I,x\in \V$ gilt. Mit
        $\ovl{P}$ als die Menge aller dieser Halbnormen erhalten
        wir so ein filtrierendes Halbnormensystem auf
        $\mathbb{V}$. Folgender Satz besagt dann ebenfalls, dass
        sowohl $P$ als auch $\ovl{P}$ dieselbe Topologie
        auf $\mathbb{V}$ definieren. 

        Man beachte, dass zudem folgende Gleichheit erfüllt ist:
        \begin{equation*}
            B_{p_{I},\epsilon}(x)=\bigcap_{\alpha \in I}B_{p_{\alpha},\epsilon}(x).
        \end{equation*}
    \item
        Sei $\mathcal{M}$ eine Menge.
        Eine Subbasis einer Topologie $\T$ auf
        $\mathcal{M}$ ist ein Teilsystem
        $O \subseteq \T$ derart, dass jedes $U\in \T$ als beliebige
        Vereinigung von endlichen Schnitten von Elementen aus $O$ geschrieben
        werden kann. Des Weiteren erhalten wir aus jeder beliebigen Sammlung
        $O\subseteq \mathcal{P}(X)$ von Teilmengen aus $\V$
        eine Topologie $\T_{O}$ auf $\V$ indem wir $\T_{O}$ als das
        System aller Mengen definieren, die als beliebige
        Vereinigung von endlichen Schnitten von Elementen aus $O$ geschrieben
        werden können. Diese wollen wir im Folgenden als die durch $O$
        auf $\V$ induzierte Topologie bezeichnen. Per Definition
        ist dann insbesondere klar, dass $O$ eine Subbasis von $\T_{O}$ ist.
    \item
        Eine Teilmenge $\mathcal{B}\subseteq \T$ heißt Umgebungsbasis
        von $x\in \mathcal{M}$, falls für jede offene Umgebung $V$ von $x$
        ein $U\in \mathcal{B}$ derart existiert, dass $U\subseteq V$.
    \end{enumerate}
\end{bemerkung}
Der folgende Satz fasst die für uns relevanten topologischen Aspekte aus
\cite{blanchard.bruening:1993a} zusammen. Der Beweis
ist nicht wirklich schwer, aber sehr technisch und wir wollen an dieser
Stelle darauf verzichten. 
\begin{satz}
    \label{satz:wichtigerSatzueberHalbnormentopologien}
    Sei $\mathbb{V}$ ein $\mathbb{K}$-Vektorraum und $P$
    ein System von Halbnormen auf $\mathbb{V}$, dann gilt: 
    \begin{enumerate}
    \item
        \label{item:topologVr}
        Die durch die Sammlung aller Halbnormenbälle auf $\V$
        induzierte Topologie $\T_{P}$ macht
        $\mathbb{V}$ zu einem topologischen
        Vektorraum $(\mathbb{V},\T_{P})$.
    \item 
        \label{item:Topologiengleich} 
        Es ist $\T_{P}=\T_{\ovl{P}}$.
    \item
        \label{item:offenheitvontauPMengen} 
        Sei $U\subseteq \mathbb{V}$, so ist genau dann $U\in
        \T_{P}$, wenn $U$ mit jedem $x\in \V$ auch einen ganzen
        Halbnormball $B_{\ovl{p},\epsilon}(x)$ mit
        $\epsilon>0$ und $\ovl{p}\in \ovl{P}$ enthält.
    \item
        \label{item:separierendgdwhausdorffsch} 
        Es ist $(\mathbb{V},\T_{P})$ genau dann hausdorffsch,
        wenn $P$ separierend ist.
    \item
        \label{item:lokaleBasis}
        $(\mathbb{V}, \T_{P})$ besitzt eine konvexe, ausgeglichene Umgebungsbasis der 
        0.
    \item
        \label{item:HNstet}
        Gegeben eine Halbnorm $q$ auf $\mathbb{V}$, so ist diese
        genau dann bezüglich $\T_{P}$ stetig, wenn ein $\ovl{p}\in
        \ovl{P}$ und ein $c>0$ derart existiert, dass
        \begin{equation*}
            q(x)\leq c\: \ovl{p}(x) \quad\quad \forall\:x\in \mathbb{V}.
        \end{equation*}
    \end{enumerate}
\end{satz}

\begin{korollar} 
    \label{kor:HNTop}
    \begin{enumerate}
    \item 
        Halbnormen sind stetig in der von ihnen induzierten
        Topologie.
    \item
        Bezeichne $\tilde{P}$ das filtrierende System aller bez"uglich $\T_{P}$ stetigen Halbnormen. Dann gilt $\T_{\tilde{P}}=\T_{P}$.
    \item
        Gegeben zwei Halbnormensysteme $P$ und $Q$ auf einem
        $\mathbb{K}$-Vektorraum $\mathbb{V}$, so gilt genau dann $\T_{Q}\subseteq\T_{P}$,
        wenn für jedes $q\in Q$ ein $\ovl{p}\in \ovl{P}$ derart existiert, dass:
        \begin{equation*}
            q(x)\leq c\:\ovl{p}(x)\quad\quad \forall\: x\in \mathbb{V}.
        \end{equation*}
    \item
        Ist $P$ breits filtrierend, so gilt \textit{iii.)} auch mit $p\in P$ anstelle von $\ovl{p}\in \ovl{P}$. 
    \end{enumerate}
    \begin{beweis}
        \begin{enumerate}
        \item
            Dies folgt sofort aus
            Satz \ref{satz:wichtigerSatzueberHalbnormentopologien}~\textit{vi.)}, da $P\subseteq \ovl{P}$.
        \item
            Die Aussage $\T_{P}\subseteq \T_{\tilde{P}}$ folgt unmittelbar aus \textit{i.)}, da hiermit $P\subseteq \tilde{P}$ gilt und somit eine Subbasis von $\T_{P}$ in $\T_{\tilde{P}}$ enthalten ist. F"ur die umgekehrte Inklusion sei $\tilde{p}\in \tilde{P}$ stetig bez"uglich $\T_{P}$. Dann ist $B_{\tilde{p},\epsilon}(0)= \tilde{p}^{-1}([0,\epsilon))$ offen in
            $\T_{P}$, da $[0,\epsilon)$ offen in
            $\mathbb{R}_{\geq 0}$ ist. Mit der Stetigkeit der Addition folgt die separate Stetigkeit der Addition und hiermit die Stetigkeit der Translationsoperation auf
            Teilmengen von $\V$. Damit sind translierte offene Mengen offen und mit
            \begin{equation*}
                B_{\tilde{p},\epsilon}(x)=x+B_{\tilde{p},\epsilon}(0)
            \end{equation*}folgt, dass eine Subbasis von $\T_{\tilde{P}}$ in
            $\T_{P}$ enthalten ist. Dies zeigt $\T_{\tilde{P}}\subseteq \T_{P}$.
        \item
            Gilt besagte Absch"atzbarkeit, so ist nach Satz \ref{satz:wichtigerSatzueberHalbnormentopologien}~\textit{vi.)} jedes $q\in Q$ stetig bez"uglich $\T_{P}$, also $Q\subseteq \tilde{P}$ uns somit $\T_{Q}\subseteq \T_{\tilde{P}}=\T_{P}$. F"ur die umkehrte Implikation sei $q\in Q$. Dann ist $q$ stetig bez"uglich $\T_{Q}$ und wegen $\T_{Q}\subseteq \T_{P}$ auch stetig bez"uglich $\T_{P}$. Damit folgt die Ungleichung unmittelbar aus Satz \ref{satz:wichtigerSatzueberHalbnormentopologien}~\textit{vi.)}. 
        \item
            Gilt die Absch"atzbarkeit f"ur $P$, so offenbar auch f"ur $\ovl{P}$.  Ist $P$ filtrierend, so l"asst sich jedes $\ovl{p}\in \ovl{P}$ durch ein $p\in P$ absch"atzen. Dies zeigt die andere Richtung.
        \end{enumerate}
    \end{beweis}
\end{korollar}
\begin{bemerkung}
    \label{bem:Minkowski}
    Wir wollen einen $\mathbb{K}$-Vektorraum $\V$ lokalkonvex nennen, wenn er ein topologische Vektorraum $(\V,\T_{\V})$ ist, dessen Topologie durch ein Halbnormensystem $P$ erzeugt wird, also
    $\T_{\V}=\T_{P}$ für ein Halbnormensystem $P$ auf $\V$
    gilt.
    Dies ist
    äquivalent dazu, dass die Topologie $\T_{\mathbb{V}}$ eine
    konvexe Umgebungsbasis $\mathcal{B}$ der $0$ besitzt (jede konvexe
    0-Umgebung enthält eine ausgeglichene konvexe 0-Umgebung, vgl. \cite[Thm~1.14]{rudin:1991a}) 
    Die eine Richtung folgt dabei sofort aus
    Satz \ref{satz:wichtigerSatzueberHalbnormentopologien}~\textit{v.)}. 

    Für
    die Andere müssen wir uns aus einer derartigen lokalkonvexen
    Basis ein $\T_{\V}$ erzeugendes Halbnormensystem verschaffen. Dies
    geschieht mit Hilfe des sogenannten
    Minkowski-Funktionals, dass f"ur eine absorbierende Teilmenge $U\subseteq \V$ definiert ist durch:
    \begin{align}
        \label{eq:MinkFunkt}
        \mu_{U}(x)= \inf_{t\in \mathbb{R}^{+}}\left\{x\in tU\right\}\quad\quad
        x\in \mathbb{V}
    \end{align}
    Dieses ist im Falle eines konvexen, ausgeglichenen
    $U\in \T_{\mathbb{V}}$ eine Halbnorm auf $\mathbb{V}$.

    In
    der Tat sind vermöge der Stetigkeit der Skalarmultiplikation alle
    offene Teilmengen von topologischen Vektorräumen absorbierend und
    somit \eqref{eq:MinkFunkt} wohldefiniert. Die Konvexität von $U$
    induziert hierbei die Dreiecksungleichung, und die Ausgeglichenheit
    sichert Eigenschaft \textit{i.)}. Für
    ausgeglichenes, konvexes $U\in \T_{\V}$ zeigt man:
    \begin{enumerate}
    \item        
        $U=B_{\mu_{U},1}(0)$;    
    \item
        $\mu_{U}$ ist stetig in $\T_{\mathbb{V}}$.
    \end{enumerate} Bezeichne $\T_{\mu}$ die aus allen derartigen
    Minkowski-Halbnormen erhaltene Topologie auf $\mathbb{V}$, so zeigt
    \textit{i.)}, dass $\mathcal{B}\subseteq \T_{\mu}$ und mit der Stetigkeit
    der Translation auch $\T_{\mathbb{V}}\subseteq \T_{\mu}$, da wir hiermit
    aus einer
    Umgebungsbasis der $0$ vermöge Translation eine Umgebungsbasis
    jedes Punktes erhalten.

    Umgekehrt folgt mit
    \textit{ii.)}, dass $B_{\mu_{U},\epsilon}(0)=\mu_{U}^{-1}([0,\epsilon))$
    offen in $\T_{\mathbb{V}}$ und somit $\T_{\mu}\subseteq
    \T_{\mathbb{V}}$ wieder mit der Stetigkeit der Translation und demselben Subbasis-Argument wie in Korollar \ref{kor:HNTop}~\textit{ii)}. Insgesamt
    folgt $\T_{\mathbb{V}}=\T_{\mu}$ und somit die behauptete
    Äquivalenz.
\end{bemerkung}
Folgender Satz offeriert ein handhabbares Stetigkeitskriterium für
multilineare Abbildungen zwischen lokalkonvexen Vektorräumen, welche wir
im Folgenden auch abkürzend als lkVR's bezeichnen werden.
\begin{satz}
    \label{satz:stetmultabb}
    Gegeben lokalkonvexe Vektorräume $(E_{1},
    \T_{E_{1}}),…,(E_{k},\T_{E_{k}})$ und $(F,\T_{F})$ und filtrierend
    gewählte Halbnormensysteme $P_{1},…,P_{k},\:Q$, die obige
    Topologien erzeugen. Sei 
    \begin{equation*}
        \Phi \colon E_{1}\times…\times E_{k}\longmapsto
        F
    \end{equation*}
    eine multilineare Abbildung,
    dann sind folgende Aussagen äquivalent:
    \begin{enumerate}
    \item
        \label{item:multistetig} 
        $\Phi$ ist stetig;
    \item
        \label{item:multistetigbei0} 
        $\Phi$ ist stetig bei 0;
    \item
        \label{item:multiabsch}
        Für jedes $q\in Q$ existiert ein Satz $\{p_{1},…,p_{k}\}$ mit
        $p_{i}\in P_{i},\:1\leq i\leq k$ und eine Konstante $c>0$, so
        dass
        \begin{equation*}
            q\left(\Phi(x_{1},…,x_{k})\right)\leq
            c\prod_{i=1}^{k}{p_{i}(x_{i})}\quad\quad \forall\:
            x_{i}\in E_{i},\: 1\leq i\leq k.
        \end{equation*}
    \end{enumerate}
    \begin{beweis}
        Einen Beweis f"ur den lineare Fall findet man in \cite[Satz
        1.12]{blanchard.bruening:1993a}. Die Verallgemeinerung auf den
        multilinearen Fall ist rein technischer Natur. 
    \end{beweis}  
\end{satz}
\section{Vervollständigungen}
Eine konsistente Darstellung der hier behandelten Zusammenh"ange unter Verwendung des Filter-Begriffes findet man beispielsweise in  \cite[Kapitel
5]{treves:1967a}. Wir geben hier eine eigene Darstellung mit
Hilfe von Netzen.
\label{subsec:Vervollstanhang}
\begin{definition}[Netze, Vollständigkeit]
    \label{def:NetzNetzkonvergenz}
    \begin{enumerate}
    \item
        \label{item:gerichteteMenge} 
        Eine gerichtete Menge $(I, \geq)$ ist eine Menge $I$ mit einer Relation $\geq$ derart, dass:        
        \begin{itemize}
        \item[a.)] 
            $\alpha \geq \alpha\:\forall\:\alpha \in I$;
        \item[b.)] 
            $\alpha\geq\beta,\:\beta\geq\gamma\Rightarrow\alpha\geq\gamma
            \qquad\forall\:\alpha,\beta,\gamma \in I$;
        \item[c.)]
            $\alpha,\:\beta \in I \Rightarrow \exists\: \gamma\in I$ mit
            $\gamma\geq\alpha,\:\gamma\geq\beta$.
        \end{itemize}
    \item
        \label{item:Netz}
        Ein Netz $\{x_{\alpha}\}_{\alpha \in I}$ in einer Menge $X$ ist
        eine Abbildung 
        \begin{align*}
            \{x_{\alpha}\}_{\alpha \in I}\colon I &\longrightarrow X\\
            \alpha &\longmapsto x_{\alpha},
        \end{align*}
        von einer gerichteten Menge $I$
        nach $X$.
    \item
        \label{item:Netzkonvergenz}
        Gegeben ein topologischer Raum $(X,\T_{X})$ und ein Netz
        $\net{x}{I} \subseteq X$, so heißt $\net{x}{I}$ konvergent gegen
        $x \in X$, wenn für jedes $U\in \T_{X}$ mit $x\in U$ ein $\alpha_{U} \in
        I$ existiert, so dass $x_{\alpha}\in U$ für alle $\alpha\geq
        \alpha_{U}$. Wir schreiben dann $\{x_{\alpha}\}_{\alpha\in I}\rightarrow x$ und im hausdorffschen Fall $\displaystyle\lim_{\alpha}{x_{\alpha}}=x$, da dann besagter Konvergenzpunkt eindeutig ist. Ganz allgemein bezeichnen wir mit $\displaystyle\lim_{\alpha} x_{\alpha}\subseteq X$ die Menge aller Konvergenzpunkte von  $\net{x}{I}$.
    \item
        Gegeben ein topologischer Vektorraum
        $(\mathbb{V},\T_{\V})$, so heißt ein Netz
        $\net{x}{I}\subseteq \mathbb{V}$ ein Cauchynetz, wenn
        für jede $0$-Umgebung $U\in
        \T_{P}$ ein $\gamma_{U}\in I$ derart existiert, dass:
        \begin{equation*}
            (x_{\alpha}-x_{\beta})\in U\quad\quad
            \forall\:\alpha,\beta \geq \gamma_{U}\in I. 
        \end{equation*}Handelt es sich hierbei um einen lokalkonvexen Vektorraum $(\mathbb{V},\T_{P})$, so ist dies mit Satz \ref{satz:wichtigerSatzueberHalbnormentopologien}~\textit{iii.)} gleichbedeutend damit, dass f"ur jedes $\epsilon> 0$ und jedes $\ovl{p}\in \ovl{P}$ ein
        $\gamma_{\ovl{p},\epsilon}\in I$ existiert, so dass
        \begin{equation*}
            p(x_{\alpha}-x_{\beta})< \epsilon\quad\quad
            \forall\:\alpha,\beta \geq \gamma_{\ovl{p},\epsilon}\in I 
        \end{equation*}gilt. Wegen $\ovl{p}=\max_{\beta\in J}p_{\beta}$ mit $p_{\beta}\in P$ und
        $B_{\ovl{p},\epsilon}(0)=\displaystyle\bigcap_{\beta \in J}B_{p_{\beta},\epsilon}(0)$\footnote{vgl. Bemerkung \ref{bem:fdgds}~\textit{ii.)}}, reicht es mit der Endlichkeit des Schnittes sogar aus, obige Absch"atzbarkeit nur f"ur alle $p\in P$ zu fordern.
    \item
        Ein topologischer Vektorraum heißt vollständig, falls für
        jedes Cauchynetz\\ $\net{x}{I}\subseteq \mathbb{V}$ gilt, dass
        $\{x_{\alpha}\}_{\alpha\in I}\rightarrow x$ für ein $x\in X$.
    \end{enumerate}
\end{definition}
\begin{bemerkung}[Netze]
    \label{bem:Netzbem}
    Die Bedeutung von Netzen sieht man an den folgenden drei
    essentiellen Punkten:
    \begin{enumerate}
    \item 
        Punkt eins ist der, dass es
        durchaus auch lokalkonvexe Vektorräume mit überabzählbaren
        Halbnormensystemen geben kann. Für filtrierendes $P$ bilden dann nämlich die Halbnormbälle $B_{p,\frac{1}{n}}(x)$ mit Radius
        $r=\frac{1}{n}$ eine Umgebungsbasis von $x\in \mathbb{V}$. Mit einer Folge alleine
        ist nun dieser Punkt a priori  nicht approximierbar, denn wir brauchen ja in
        der Tat für jedes $p$ eine ganze Folge, die bezüglich dieser
        Ballsorte $B_{p,\cdot}$ gegen $x$ konvergiert. Für abzählbare Halbnormensysteme
        $P=\{p_{n}\}_{n\in \mathbb{N}}$ liegt der Fall hingegen anders, denn
        sei
        \begin{equation*}
            \widetilde{p}_{n}(x)=\max_{l\leq n}(p_{l}(x)),
        \end{equation*}
        so bilden die abzählbar vielen Bälle
        $B_{\widetilde{p}_{n},\frac{1}{n}}(x)$ eine Umgebungsbasis von $x$ und
        wir kommen mit nur einer Folge aus, wenn wir $x$ approximieren
        wollen. Diese vage Vorstellung ist dabei Ausdruck des allgemeinen Faktes, dass ein topologischer Vektorraum $(\V,\T_{\V})$ mit einer abz"ahlbaren $0$-Umgebungsbasis $\mathcal{B}$ genau dann vollst"andig ist, wenn er folgenvollst"andig ist, also jede Cauchyfolge $\{x_{n}\}_{n\in \mathbb{N}}\subseteq \V$ gegen ein $x\in \V$ konvergiert. Dies ist dann insbesondere f"ur normierte Vektorr"aume der Fall. 
    \item
        Eine Abbildung $f$ zwischen
        zwei topologischen Räumen $X$ und $Y$ ist genau dann stetig,
        wenn für jeden Punkt $x\in X$ und jedes Netz
        $\{x_{\alpha}\}_{\alpha\in I}\rightarrow x$ gilt, dass
        $\{f(x_{\alpha})\}_{\alpha\in I}\rightarrow f(x)\in
        Y$.

        Diese Aussage ist diffizil, denn in der Tat muss man sich darüber im
        klaren sein, dass im Nicht-Hausdorff-Fall sowohl
        $\displaystyle\lim_{\alpha}x_{\alpha}$, als auch
        $\displaystyle\lim_{\alpha}f(x_{\alpha})$ 
        mehrelementige Mengen sein können.
    \item
        \label{item:Netzbemdrei}
        Sei $(X,\T_{X})$ ein topologischer Raum und $Y\subseteq X$. Dann ist
        der topologische Abschluss $\ovl{Y}$ von $Y$ bezüglich $(X,\T_{X})$ definiert, als die kleinste abgeschlossene
        Teimenge von $X$, in der $Y$ enthalten ist. Dann gilt        
        \begin{equation*}
            \ovl{Y}=\left\{x\in X\:\big|\:\forall\:U\in\T_{X}\text{
                  mit }x\in U\text{ ist }U\cap Y\neq \emptyset\right\}
        \end{equation*}und   
        hiermit zeigt man, dass $\ovl{Y}$ gerade
        die Menge aller $x\in X$ ist, für
        die es Netze $\{y^{x}_{\alpha}\}_{\alpha\in I}\subseteq Y$
        derart gibt, dass $\displaystyle
        \lim_{\alpha}y^{x}_{\alpha}=x$ gilt. Ist $\ovl{Y}=X$, so heißt
        $Y$ dicht in $X$. 
    \end{enumerate}
\end{bemerkung}
\begin{definition}[$\sim$, Kanonischer Repräsentant, Isometrie]
    \label{def:kanNetzIsoetc}
    Gegeben ein lokalkonvexer Vektorraum
    $(\mathbb{V},\T_{P})$ und bezeichne $\mathcal{N_{C}}(\mathbb{V})$ die
    Klasse\footnote{Für jede Menge $M$ wird deren Potenzmenge
      $\mathcal{P}(M)$, ausgestattet mit der Inkusionsrelation, zu einer 
      gerichteten Menge. Somit steht diese Sammlung von Mengen in
      Bijektion zu der Klasse der Mengen, von der wir in Definition \ref{def:Klasse} gesehen hatten,
      dass sie keine Menge ist.} aller Cauchynetze in $\mathbb{V}$.
    \begin{enumerate}
    \item
        \label{item:auqivCN}
        Wir definieren eine Äquivalenzrelation $\sim$ auf
        $\mathcal{N_{C}}(\mathbb{V})$ wie folgt. Es sei genau dann
        \begin{equation*}
            \net{x}{I}\sim \{y_{\beta}\}_{\beta\in J},
        \end{equation*}
        wenn für jedes
        $p\in P$ das Netz $\left\{p_{\alpha,\beta}\right\}_{I\times J}:=
        \left\{p(x_{\alpha}-y_{\beta})\right\}_{I\times J}$
        gegen $0$ konvergiert. Hierfür sei $I\times J$ mit der
        Relation
        \begin{equation*}
            (\alpha,\beta)\geq
            (\alpha',\beta')\Longleftrightarrow \alpha\geq
            \alpha'\:\wedge\:\beta\geq \beta'
        \end{equation*}
        bestückt. Es ist dann unmittelbar klar,
        dass $\sim$ reflexiv und symmetrisch ist. Die Transitivität
        folgt mit der Dreiecksungleichung für Halbnormen. Ist
        $(\mathbb{V},\T_{\mathbb{V}})$ hausdorffsch und existiert
        der eindeutige Limes $\displaystyle\lim_{\alpha}x_{\alpha}=x$, so gilt: 
        \begin{equation} 
            \label{eq:AequddwLimsgleich}
            \net{x}{I}\sim \{y_{\beta}\}_{\beta\in
              J}\Longleftrightarrow \lim_{\alpha}x_{\alpha}=\lim_{\beta}y_{\beta}.
        \end{equation}
        Für "`$\Longrightarrow$"' beachten wir, dass
        \begin{align*} 
            p(y_{\beta}-x)\leq
            \overbrace{p(y_{\beta}-x_{\alpha})}^{<\frac{\epsilon}{2}\:\text{für}\:(\alpha,\beta)\geq (\alpha',\beta')}+\quad\overbrace{p(x_{\alpha}-x)}^{<\frac{\epsilon}{2}\:\text{für}\:\alpha\geq \alpha''}<\epsilon,
        \end{align*} 
        falls $(\alpha,\beta)\geq (\wt{\alpha},\beta')\in
        I\times J$ mit $\wt{\alpha}\geq \alpha',\alpha''$.

        "`$\Longleftarrow$"' erhalten wir mit
        \begin{equation*}
            p(y_{\beta}-x_{\alpha})\leq p(y_{\beta}-z)+p(z-x_{\alpha})<\epsilon
        \end{equation*}
        für alle $(\alpha, \beta)\geq (\alpha',\beta')\in I\times J$.
    \item
        Gegeben ein Cauchynetz $\net{x}{I}\subseteq \mathbb{V}$. Wir
        statten die Menge $P\times \mathbb{N}$ mit der
        Relation
        \begin{equation*}
            (p,n)\geq(p',n')\Longleftrightarrow p\geq p'\:\wedge\:n\geq n',
        \end{equation*}aus, 
        wobei $p\geq p'$ bedeuten soll, dass $p(x)\geq
        p'(x)$ für alle $x\in \mathbb{V}$.\\\\
        Für festes $p$ finden
        wir per Definition ein $\gamma_{p,n}\in I$ derart,
        dass $(x_{\alpha}-x_{\beta})\in B_{p,\frac{1}{2n}}(0)$ für alle
        $\alpha,\beta \geq \gamma_{p,n}$, und mit
        Definition \ref{def:NetzNetzkonvergenz}~\textit{i.)}~c.) erreichen wir zudem die
        Vergleichbarkeit: 
        \begin{equation*}
            …\geq \gamma_{p,n+1} \geq \gamma_{p,n} \geq \gamma_{p,n-1}\geq…\:. 
        \end{equation*}
        Vermöge
        \begin{equation*}
            \left\{x_{p,n}\right\}_{P\times \mathbb{N}}:= \left\{x_{\gamma_{p,n}}\right\}_{P\times\mathbb{N}}
        \end{equation*} erhalten wir so ein zu
        $\net{x}{I}$ äquivalentes Cauchynetz. Dieses bezeichnen wir im Folgenden als das zu
        $\net{x}{I}$ gehörige kanonische Netz bzw. den zu 
        $\left[ \net{x}{I}\right]\\\in \mathcal{N_{C}}(\mathbb{V})/\sim$
        gehörigen kanonischen Repräsentanten.

        In der Tat folgt für $(p',n'),(p'',n'')\geq (p,n)$ mit
        $\alpha\geq_{I} \gamma_{n',p'},\:\gamma_{n'',p''}$:
        \begin{align*}
            p\left(x_{p',n'}-x_{p'',n''}\right)\leq&\:
            p\left(x_{p',n'}-x_{\alpha}\right)+p\left(x_{\alpha}-x_{p'',n''}\right)\\
            \leq&\: p'\left(x_{p',n'}-x_{\alpha}\right)+p''\left(x_{\alpha}-x_{p'',n''}\right)\\
            <&\: \frac{1}{2n'}+\frac{1}{2n''}
            \leq \frac{1}{n},
        \end{align*} was die Cauchyeigenschaft zeigt. Dabei bezeichnet $\geq_{I}$ die ursprüngliche
        Ordnungsrelation auf $I$. Die Äquvivalenz folgt mit
        \begin{align*}
            p\left(x_{p',n'}-x_{\alpha}\right)\leq&\:
            p\left(x_{p',n'}-x_{p,n}\right)+p\left(x_{p,n}-x_{\alpha}\right)\\
            <&\:\frac{1}{n}+\frac{1}{2n}=\:\frac{3}{2}\frac{1}{n}
        \end{align*} für alle $(p',n')\geq (p,n)$,
        $\alpha\geq_{I}\gamma_{p,n}$ und somit
        $\displaystyle\lim_{([p,n],\alpha)}p\left(x_{p,n}-x_{\alpha}\right)=0$.

        Diese Festlegung
        auf die feste gerichtete Menge $P\times \mathbb{N}$ zeigt
        insbesondere, dass besagter Quotient eine echte Menge ist. Denn in der Tat
        ist somit $\mathcal{N_{C}}(\mathbb{V})/\sim$ auffassbar als
        Teilmenge von $\left(P\times \mathbb{N}\right)\times \V$.
    \item
        Gegeben zwei lkVR's $(X,P)$, $(X',P')$ und eine
        Zuordnung $i$, bestehend aus Abbildungen:
        \begin{align*}
            i_{X}\colon X&\longrightarrow X'\\
            i_{P}\colon P&\longrightarrow P'.
        \end{align*} Dann heißt $i$ Isometrie, falls $i_{X}$ linear,
        $i_{P}$ bijektiv und für alle $x\in X$ und alle $p\in
        P$ folgende Gleichheit erfüllt ist:
        \begin{equation*}
            i_{P}(p)(i_{X}(x))=p(x).
        \end{equation*}
        Da keine Verwechslungen zu befürchten sind, werden wir im Folgenden werden die Indizes unterdrücken
        und einfach nur noch $i$ anstelle von $i_{P}$ und $i_{X}$
        schreiben.
    \end{enumerate} 
\end{definition}
Wir benötigen folgende vorbereitende Proposition:
\begin{proposition}
    \label{prop:wichEiVervollst}
    Gegeben ein lkVR $(X,P)$.
    \begin{enumerate}
    \item
        \label{item:HNormNetzLimeserhalt}
        Gegeben ein Netz $\net{x}{I}\subseteq X$. Existiert ein
        $\epsilon>0$ und für jedes $p\in P$ ein $\alpha_{p,\epsilon}$
        derart, dass $p(x_{\alpha})< \epsilon$ für alle $\alpha\geq \alpha_{\epsilon}\in
        I$ gilt. Dann existiert der Limes und es ist
        $\displaystyle\lim_{\alpha}p(x_{\alpha})\leq \epsilon$.
    \item
        \label{item:ProduktlimeslkVRundR} 
        Sei $(X,P)$ hausdorffsch und $\nett{x}{\alpha,\beta}{I\times J}\subseteq
        X$ ein Netz derart, dass $\displaystyle\lim_{\beta}x_{\alpha,\beta}$
        für alle $\alpha\in I$ existiert. Dann gilt:
        \begin{itemize}
        \item
            Existiert der Limes 
            $\displaystyle\lim_{\alpha\times\beta}x_{\alpha,\beta}$,
            so gilt:
            \begin{equation*}
                \lim_{\alpha}\Big[\lim_{\beta}x_{\alpha,\beta}\Big]=\lim_{\alpha\times\beta}x_{\alpha,\beta}.
            \end{equation*}
        \item
            Existiert der Limes
            $\displaystyle\lim_{\alpha}\Big[\displaystyle\lim_{\beta}x_{\alpha,\beta}\Big]$
            und ist $\{x_{\alpha,\beta}\}_{I\times
              J}=\{x_{\alpha}\}_{\alpha\in
              I}+\{y_{\beta}\}_{\beta\in J}$ mit Netzen $\{x_{\alpha}\}_{\alpha\in
              I},\{y_{\beta}\}_{\beta\in J}\subseteq X$, so gilt:
            \begin{equation*}
                \lim_{\alpha\times\beta}x_{\alpha,\beta}=\lim_{\alpha}\Big[\lim_{\beta}x_{\alpha,\beta}\Big].
            \end{equation*}
        \end{itemize}
    \item
        Sei $(X,P)$ hausdorffsch, dann ist jede Isometrie injektiv.
    \item
        Isometrien sind stetig und sogar gleichmäßig stetig.
    \item
        \label{item:eindverv}
        Sei $(X,P)$ hausdorfsch und $(X',P')$, $(X'',P'')$
        vollständige hlkVR's mit Isometrien
        \begin{align*}
            i'\colon X&\longrightarrow X'\\ 
            i''\colon X&\longrightarrow X''
        \end{align*}derart, dass $\overline{i'(X)}=X'$ und  
        $\ovl{i''(X)}=X''$ gilt. Dann sind $(X',P')$ und $(X'',P'')$
        zueinander linear
        homöomorph.
    \end{enumerate}
    \begin{beweis}
        \begin{enumerate}
        \item
            Zunächst ist $\left\{p(x_{\alpha})\right\}_{\alpha\in
              I}\subseteq \mathbb{R}$ wegen
            \begin{equation*}
                |p(x_{\alpha})-p(x_{\beta})|\leq p(x_{\alpha}-x_{\beta})\leq
                p(x_{\alpha})+p(x_{\beta})< 2\epsilon \qquad\quad \forall\:\alpha,\beta\geq \alpha_{\epsilon},
            \end{equation*}
            ein Cauchynetz und mit der Vollständigkeit von
            $\mathbb{R}$ zeigt dies, dass der Limes existiert.
            Für die Abschätzung argumentieren wir durch Widerspruch. Sei hierfür
            $\displaystyle\lim_{\alpha}p(x_{\alpha})=\delta>\epsilon$.
            Dann existiert $\Delta>0$, so dass $B_{\Delta}(\delta)\cap
            \big[0,\epsilon\big]=\emptyset$ sowie ein $\gamma_{\Delta}\in
            I$ derart, dass $p(x_{\alpha})\in B_{\Delta}(\delta)$ für alle
            $\alpha\geq \gamma_{\Delta}$ gilt.

            Dies bedeutet insbesondere $p(x_{\alpha})>
            \epsilon\:\:\forall\: \alpha \geq \gamma_{\Delta}$. Nun
            finden wir aber $\alpha'\in I$ mit $\alpha'\geq \alpha_{\epsilon}$
            und $\alpha'\geq \gamma_{\Delta}$, womit
            $p(x_{\alpha'})<\epsilon$ und gleichzeitig
            $p(x_{\alpha'})>\epsilon$ gelten müsste, was ein
            Widerspruch ist.
        \item
            Für die erste Aussage sei
            $\displaystyle\lim_{\alpha\times\beta}x_{\alpha,\beta}=x$
            und $x_{\alpha}=\displaystyle\lim_{\beta}
            x_{\alpha,\beta}$. Dann gilt:
            \begin{align*}
                p(x_{\alpha}-x)=p\left(\lim_{\beta}
                    x_{\alpha,\beta}-x\right)\glna{\mathit{p}\text{ stetig}}\lim_{\beta}p(x_{\alpha,\beta}-x).
            \end{align*} Nun existiert
            $(\alpha_{\epsilon},\beta_{\epsilon})\in I\times J$, so dass
            $p(x_{\alpha,\beta}-x)<\frac{\epsilon}{2}$ für alle
            $(\alpha,\beta)\geq
            (\alpha_{\epsilon},\beta_{\epsilon})$ und
            $\textit{i.)}$ zeigt dass:
            \begin{equation*}
                \lim_{\beta}p(x_{\alpha,\beta}-x)\leq \frac{\epsilon}{2}<\epsilon\qquad\quad\forall\:\alpha\geq\alpha_{\epsilon}.
            \end{equation*}
            Für die zweite Behauptung beachten wir, dass
            $\displaystyle\lim_{\beta}x_{\alpha,\beta}$ existiert und
            mit der Stetigkeit der Addition 
            $\displaystyle\lim_{\beta}\left[x_{\alpha}+y_{\beta}\right]=x_{\alpha}+\lim_{\beta}y_{\beta}=x_{\alpha}+
            y$ für ein $y\in X$ gilt. Durch Wiederholung dieser
            Argumentation folgt:           
            \begin{equation*}
                \lim_{\alpha}\Big[\lim_{\beta}x_{\alpha,\beta}\Big]=\lim_{\alpha}\Big[x_{\alpha}+y\Big]=\lim_{\alpha}x_{\alpha}+y=x+y.
            \end{equation*}
            Dann existieren $\alpha_{\epsilon}\in I$ und
            $\beta_{\epsilon}\in J$, so dass $p(x_{\alpha}-x)<
            \frac{\epsilon}{2}$ für alle $\alpha\geq
            \alpha_{\epsilon}$ und $p(y_{\beta}-y)<
            \frac{\epsilon}{2}$ für alle
            $\beta\geq\beta_{\epsilon}$. Dies zeigt           
            \begin{align*}
                p(x_{\alpha,\beta}-x+y)\leq
                p(x_{\alpha}-x)+p(y_{\beta}-y)< \frac{\epsilon}{2}+\frac{\epsilon}{2}=\epsilon
            \end{align*}für alle $(\alpha,\beta)\geq (\alpha_{\epsilon},\beta_{\epsilon})$.
        \item
            Sei $i\colon (X,P)\longrightarrow (X',P')$ eine Isometrie und
            $(X',P')$ ein lkVR. Nun ist $X$ hausdorffsch und somit $P$ separierend. Hiermit
            existiert für alle $x,y\in X$ mit
            $i(x)=i(y)$ und $x\neq y$ ein $p\in P$, so dass $p(x-y)\geq
            0$ gilt. Dies liefert einen Widerspruch, denn wir erhalten:
            \begin{align*}
                0=i(x)-i(y)\Longrightarrow
                0=i(p)(i(x)-i(y))=i(p)(i(x-y))=p(x-y)\geq 0. 
            \end{align*}
        \item
            Sei $p'\in P'$. Dann ist
            $p'(i(x)-i(y))=i(p)(i(x)-i(y))=p(x-y)$ für ein $p\in P$, womit: 
            \begin{align*}
                p(x-y)<\delta \Longrightarrow
                p'(i(x)-i(y))<\epsilon=\delta\quad\quad\forall\:x,y\in X.
            \end{align*} Das zeigt die gleichmäßige Stetigkeit und
            insbesondere die Stetigkeit. Alternativ folgt diese so:

            Sei $x\in X$ mit $\net{x}{I}\rightarrow x$, dann finden
            wir für $p'\in P'$ ein $p\in P$ mit $i(p)=p'$ und erhalten
            \begin{equation*}
                p'(i(x_{\alpha})-i(x))=p(x_{\alpha}-x)<
                \epsilon\qquad\qquad \forall\:\alpha \geq \gamma_{\epsilon}.
            \end{equation*}
            Dies zeigt $\left\{i(x_{\alpha})\right\}_{\alpha\in I}\rightarrow
            i(x)$ und somit die Stetigkeit.
        \item
            Zunächst sind Abbildungen $i'_{X}\colon X\longrightarrow
            i'(X)\subseteq X'$,
            $i''_{X}\colon X\longrightarrow i''(X)\subseteq X''$
            bijektiv und somit $h=i''\cp i'^{-1}$ eine bijektive
            Isometrie $h\colon i'(X)\longrightarrow i''(X)$ mit
            $h(p')=\left(i''_{P}\cp i'^{-1}_{P}\right)(p')$.
            Wir definieren 
            \begin{equation}
                \label{eq:habschldef}
                \begin{split}
                    \ovl{h}:X'&\longrightarrow X''\\
                    x'&\longmapsto\lim_{\alpha} h(x'_{\alpha})
                \end{split}
            \end{equation}
            für ein Netz $\net{x'}{I}\subseteq i'(X)$
            mit $\displaystyle\lim_{\alpha} x'_{\alpha}=x'$. Dabei ist
            die Existenz eines
            solchen Netzes gesichert, da nach Voraussetzung
            $\ovl{i'(X)}=X'$ gilt. Für die Existenz von
            $\displaystyle\lim_{\alpha} h(x'_{\alpha})$ beachtet man,
            dass für jedes $p''\in P''$ ein $p'\in P'$ existiert, so dass
            \begin{align*}
                p''(h(x'_{\alpha})-h(x'_{\beta}))=\:h(p')(h(x'_{\alpha}-x'_{\beta}))
                =\:p'(x'_{\alpha}-x'_{\beta})<\:\epsilon
            \end{align*}für alle $\alpha,\beta \geq \gamma_{\epsilon}\in
            I$ gilt. Hiermit ist $\{h(x)_{\alpha}\}_{\alpha\in I}$ ein
            Cauchynetz in $X''$ und konvergiert nach
            Voraussetzung. Mit der Hausdorff-Eigenschaft von $X''$
            ist besagter Limes zudem eindeutig bestimmt.\\\\
            Um die Wohldefiniertheit von \eqref{eq:habschldef} 
            nachzuweisen, betrachten wir Netze $\net{x'}{I}$,
            $\nett{y'}{\beta}{J}\subseteq i'(X)$ mit $\net{x'}{I}\rightarrow
            x'$ und $\nett{y'}{\beta}{J}\rightarrow x'$. Sei
            $x''=\displaystyle\lim_{\alpha} h(x'_{\alpha})$, so folgt
            mit \eqref{eq:AequddwLimsgleich}:
            \begin{align*}
                p''(h(y'_{\beta})-x'')\leq&\:p''\left(h(y'_{\beta})-h(x'_{\alpha})\right)+p''\left(h(x'_{\alpha})-x''\right)
                \\=&\: h(p')\left(h(y'_{\beta})-h(x'_{\alpha})\right)+
                p''\left(h(x'_{\alpha})-x''\right)
                \\=&\:p'\left(y'_{\beta}-x'_{\alpha}\right)+p''\left(h(x'_{\alpha})-x''\right).
            \end{align*}
            Wegen $x''=\displaystyle\lim_{\alpha} h(x'_{\alpha})$ existiert ein $\alpha_{\epsilon}\in I$, so dass
            $p''\left(h(x'_{\alpha})-x''\right)<\frac{\epsilon}{2}$ für alle
            $\alpha\geq \alpha_{\epsilon}$ und nach
            Definition \ref{def:kanNetzIsoetc}~\textit{i.)} ein
            $(\alpha',\beta')\in I\times J$, so dass
            $p'(y'_{\beta}-x'_{\alpha})< \frac{\epsilon}{2}$ für alle
            $(\alpha,\beta)\geq
            (\alpha',\beta')$. Mit Definition \ref{def:NetzNetzkonvergenz}~\textit{i.)~c.})
            erhalten wir ein $\alpha''\geq \alpha',\alpha_{\epsilon}$,
            womit $p''(h(y'_{\beta})-x'')< \epsilon$ für alle $(\alpha,\beta)\geq
            (\alpha'',\beta')$. Dies zeigt
            $\{h(y'_{\beta})\}_{\beta\in J}\rightarrow x''$ und
            somit die Wohldefiniertheit von \eqref{eq:habschldef}.

            Wir wollen nun zeigen, dass $\ovl{h}$ eine Isometrie ist. Sei hierfür $\net{x'}{I}\subseteq i'(X)$ mit
            $\net{x'}{I}\rightarrow x'$, dann folgt
            \begin{align*}
                \ovl{h}(p')\left(\ovl{h}(x')\right)=\lim_{\alpha}h(p')\big(h(x'_{\alpha})\big)=\lim_{\alpha}p'(x_{\alpha})=p'(x'),
            \end{align*} wobei wir im letzten Schritt
            Korollar \ref{kor:HNTop}~\textit{i.)} benutzt haben. 

            Die Linearität von $\ovl{h}$
            folgt mit der Linearität von $h$, 
            Proposition \ref{prop:wichEiVervollst}~\textit{ii.)},
            der Unabhängigkeit von \eqref{eq:habschldef} von der Wahl
            des Netzes und der Stetigkeit der Addition in $X'$:
            \begin{align*}
                \ovl{h}(x')+\ovl{h}(y')=&\lim_{\alpha}h(x'_{\alpha})+
                \lim_{\beta} h(y'_{\beta})
                =\lim_{\alpha}\left[\lim_{\beta} \left(h(x'_{\alpha}+y'_{\beta})\right)\right]
                \\=&\lim_{\alpha\times\beta}h(x'_{\alpha}+y'_{\beta})
                =\ovl{h}(x'+y').
            \end{align*}
            Mit der Hausdorff-Eigenschaft von $(X',P')$ folgen
            Stetigkeit und Injektivität von $\ovl{h}$ nun unmittelbar
            aus Teil \textit{iii.)} und \textit{iv.)} und es bleibt zu zeigen, dass $\ovl{h}$ surjektiv und $\ovl{h}^{-1}$ ebenfalls eine Isometrie
            darstellt. Dann wäre nämlich $\ovl{h}$ ein linearer
            Homöomorphismus und die Behauptung gezeigt.\\\\
            Für die Surjektivität sei $x''\in X''$ vorgegeben. Da
            $\ovl{i''(X)}=X''$ finden wir ein Netz
            $\net{x''}{I}\subseteq i''(X'')$ mit $\displaystyle\lim_{\alpha}
            x''_{\alpha}=x''$ und definieren:
            \begin{equation}
                \label{eq:netzt}
                \net{x'}{I}=\{h^{-1}(x''_{\alpha})\}_{\alpha\in I}\subseteq i'(X').
            \end{equation}Nun ist jedes konvergente Netz insbesondere
            ein Cauchynetz, womit für jedes $\epsilon>0$ ein $\gamma_{\epsilon}\in I$ derart
            existiert, dass
            \begin{equation*}
                p'(x'_{\alpha}-x'_{\beta})=h(p')\big(h(x'_{\alpha})-h(x'_{\beta})\big)=h(p')\big(x''_{\alpha}-x''_{\beta}\big)<
                \epsilon\quad\quad \forall\:\alpha,\beta\leq,
            \end{equation*}gilt. Hiermit ist \eqref{eq:netzt}
            ebenfalls ein Cauchynetz und konvergiert folglich gegen
            ein $x'\in X'$. Wir erhalten dann
            \begin{align*}
                \ovl{h}(x')=\lim_{\alpha}h\big(h^{-1}(x''_{\alpha})\big)=\lim_{\alpha}
                x_{\alpha}=x'',
            \end{align*} 
            also die Surjektivität von $\ovl{h}$.
            Die Isometrieeigenschaft von $\ovl{h}^{-1}$
            folgt mit
            \begin{align*}
                \ovl{h}^{-1}\big(p''\big)\Big(\ovl{h}^{-1}\big(x''\big)\Big)=\Big(\ovl{h}\cp\ovl{h}^{-1}\Big)\big(p''\big)\left(\ovl{h}\cp\ovl{h}^{-1}\big(x''\big)\right)=p''\big(x''\big).
            \end{align*}
        \end{enumerate}
    \end{beweis}
\end{proposition}
\begin{satz}
    \label{satz:vervollsthlkVR}
    Gegeben ein hlkVR $(X,P)$, dann existiert ein
    vollständiger lkVR $(\hat{X},\hat{P})$
    zusammen mit einer Isometrie $i:X\rightarrow \hat{X}$ derart, dass
    $\ovl{i(X)}=\hat{X}$. Besagte Vervollständigung ist ebenfalls
    hausdorffsch und bis auf lineare Homöomorphie eindeutig bestimmt. 
    \begin{beweis}
        Die Eindeutigkeit ist klar mit
        Proposition \ref{prop:wichEiVervollst}~\textit{v.)}. Für die
        Existenz betrachten wir die Menge
        $\hat{X}=\mathcal{N_{C}}(X)/\sim$ und statten diese mit
        den Vektorraumoperationen $\cdot\colon \big(\lambda, [\hat{x}]\big)\longmapsto
        [\lambda\hat{x}]$ und $+\colon \big([\hat{x}],
        [\hat{y}]\big)\longmapsto [\hat{x}+\hat{y}]$ mit
        \begin{align*}
            \lambda \net{x}{I}&:= \{\lambda x_{\alpha}\}_{\alpha\in I,}\\
            \net{x}{I}+\{y_{\beta}\}_{\beta\in
              I}&:=\left\{x_{\alpha}+y_{\beta}\right\}_{I\times J}.
        \end{align*}aus. Die Wohldefiniertheit dieser Abbildungen folgt
        unmittelbar aus den Halbnormeneigenschaften und
        der Definition von $\sim$.
        Des Weiteren topologisieren wir $\hat{X}$ mit dem System $\hat{P}$, bestehend aus den Halbnormen:
        \begin{equation}
            \label{eq:VollHNDef}
            \hat{p}\left(\left[\hat{x}\right]\right)=\lim_{\alpha}p(x_{\alpha})\quad
            \text{mit }\:\net{x}{I}\in [\hat{x}]\in \hat{X}.
        \end{equation} Obige Isometrie erhalten wir dann durch die
        Zuordnungsvorschrift: $i:x\mapsto \big[\{x,x,…\}\big]$ ($x\in X$ wird
        die Äquivalenzklasse der konstanten Folge $\{x_{n}\}_{n\in \mathbb{N}}=(x,x,…)$
        zugeordnet). Man sieht nebenbei bemerkt, dass $i$ in der Tat nicht injektiv ist,
        falls $P$ nicht hausdorffsch. Denn gilt für ein $x\in X$ 
        $p(x)= 0$ für alle $p\in P$, so folgt bereits $i(x)=\hat{0}=[0]$.\\\\
        Wir müssen nun zunächst die Existenz des Limes in
        \eqref{eq:VollHNDef} und die Wohldefiniertheit dieser
        Zuordnungsvorschrift nachweisen.
        Erstere folgt dabei unmittelbar aus
        \begin{align}
            \label{eq:ungl}
            |p(x_{\alpha})-p(x_{\beta})|\leq p(x_{\alpha}-x_{\beta}),
        \end{align} womit $\{p(x_{\alpha})\}_{\alpha\in I}$ ein
        Cauchynetz in $\mathbb{R}$ ist und mit dessen Vollständigkeit
        konvergiert.\\\\
        Für die Wohldefiniertheit sei $\net{x}{I}\sim
        \{y_{\beta}\}_{\beta\in J}$, dann folgt mit der Stetigkeit der
        Addition und $||$:
        \begin{align*}
            \left|\lim_{\alpha}p(x_{\alpha})-\lim_{\beta}p(y_{\beta})\right|=&\:
            \left|\lim_{\alpha}\Big[\lim_{\beta}\left[p(x_{\alpha})-p(y_{\beta})\right]\Big]\right|=\left|\lim_{\alpha\times
                  \beta}\left[p(x_{\alpha})-p(y_{\beta})\right]\right|
            \\=&\:\lim_{\alpha\times
              \beta}\left|p(x_{\alpha})-p(y_{\beta})\right|
            \stackrel{*,\eqref{eq:ungl}}{\leq}\lim_{\alpha\times\beta}p(x_{\alpha}-y_{\beta})
            \\ =&\:0  
        \end{align*}
        Der zweite Schritt folgt dabei mit
        Proposition \ref{prop:wichEiVervollst}~\textit{ii.)} und  $*$ bedeutet den Erhalt von $\leq$ unter Limesbildung.\\\\
        Um die Halbnormeigenschaften von $\hat{p}$ nachzuweisen
        rechnen wir
        \begin{align*}
            \hat{p}(\lambda \hat{x})=\lim_{\alpha}p(\lambda
            x_{\alpha})=\lambda
            \lim_{\alpha}p(x_{\alpha})=\lambda \hat{p}(\hat{x})
        \end{align*} und
        \begin{align*}
            \hat{p}(\hat{x}+\hat{y})&=\lim_{\alpha\times\beta}p\left(x_{\alpha}+
                y_{\beta}\right)
            \stackrel{*}{\leq}\lim_{\alpha\times \beta}\Big(p(x_{\alpha})+p(y_{\beta})\Big)\\
            &= \lim_{\alpha}\left[\lim_{\beta}
                \Big(p(x_{\alpha})+p\left(y_{\beta}\right)\Big)\right]
            =\hat{p}(\hat{x})+\hat{p}(\hat{y}),
        \end{align*} wobei wir bei $*$ wieder den Erhalt von $\leq$
        unter Limesbildung benutzt haben, im dritten Schritt
        Proposition \ref{prop:wichEiVervollst}~\textit{ii.)} und im letzten Schritt die
        Stetigkeit der Addition.\\\\
        Für $\ovl{i(X)}=\hat{X}$
        sei $[\hat{x}]\in \mathcal{N_{C}}(X)/\sim$ und $\net{x}{I}\in
        [\hat{x}]$, dann existiert ein $\gamma_{p,n}\in I$ derart, dass
        \begin{align*}
            p(x_{\alpha}-x_{\beta})<\frac{1}{n}\quad\quad \forall\:
            \alpha,\beta\geq \gamma_{p,n}\in I
        \end{align*}
        und es folgt
        \begin{equation*}
            \hat{p}\Big(\left[\{x_{\gamma_{p,n+1}}\}\big]-\big[\net{x}{I}\right]\Big)
            =\lim_{\alpha}p(x_{\gamma_{n+1,p}}-x_{\alpha})
            \stackrel{**}{\leq} \frac{1}{n+1}<\frac{1}{n}
        \end{equation*} mit $**=$
        Proposition \ref{prop:wichEiVervollst}~\textit{i.)}. Dabei
        bezeichnet
        $\big[\{x_{\gamma_{n+1,p}}\}\big]$ die Äquivalenzklasse der konstanten Folge
        $\{x_{\gamma_{n+1,p}},x_{\gamma_{n+1,p}},…\}\subseteq X$.\\ 
        Dies
        bedeutet $\big[\{x_{\gamma_{n+1,p}}\}\big]\in
        B_{\hat{p},\frac{1}{n}}(\hat{x})\cap i(X)$. Also $B_{\hat{p},\frac{1}{n}}(\hat{x})\cap i(X)\neq \emptyset$,
        womit $[\hat{x}]\in \ovl{i(X)}$ und folglich 
        $\hat{X}\subseteq \ovl{i(X)}$. Die umgekehrte Inklusion ist
        klar mit Bemerkung \ref{bem:Netzbem}~\textit{iii.)}.\\\\
        Für die Hausdorff-Eigenschaft sei $\hat{p}([\hat{x}])=0$ für
        alle $\hat{p}\in \hat{P}$. Sei $\net{x}{I}\in [\hat{x}]$, so
        folgt
        $\displaystyle\lim_{\alpha}p(x_{\alpha})=0\:\:\forall\:p\in P$,
        mithin $\net{x}{I}\sim \{0\}$ für jedes $0$-Netz $\{0\}$. Das
        zeigt $[\hat{x}]=\hat{0}$.
        Es bleibt die Vollständigkeit nachzuweisen. Sei hierfür
        $\net{[\hat{x}]}{I}\subseteq \hat{X}$ ein Cauchynetz. Wir müssen
        dann zeigen, dass dieses Cauchynetz von Äquivalenzklassen von Cauchynetzen in $X$
        bezüglich $\hat{P}$ gegen eine Äquivalenzklasse
        eines Cauchynetzes in $X$ konvergiert.

        Mit den Erläuterungen in
        Definition \ref{def:kanNetzIsoetc}~\textit{i.)} reicht es dazu aus, dies für das zu $\net{[\hat{x}]}{I}$
        äquivalente kanonische Cauchynetz
        $\{[\hat{x}]_{\hat{p},n}\}_{\hat{P}\times \mathbb{N}}$ zu
        zeigen.\\ Jedes $[\hat{x}]_{\hat{p},n}$ ist nun eine
        Äquivalenzklasse von Cauchynetzen in $X$, und auch hier wählen
        wir den kanonischen Repräsentanten $\left\{\big([x]_{\hat{p},n}\big)_{p',n'}\right\}_{P\times
          \mathbb{N}}$ bezüglich
        $P$. Es gilt also:
        \begin{equation*}
            \left[\hat{x}\right]_{\hat{p},n}=\left[\left\{\big([x]_{\hat{p},n}\big)_{p,n}\right\}_{P\times
                  \mathbb{N}}\right].
        \end{equation*}
        Da $P\cong \hat{P}$ können wir ohne
        Schwierigkeiten
        \begin{equation}
            \label{eq:Grenznetz}
            y_{p,n}=\big([\hat{x}]_{\hat{p},n}\big)_{p,n}
        \end{equation} definieren und behaupten, dass dann
        $\{[\hat{x}]_{\hat{p},n}\}_{\hat{P}
          \times\mathbb{N}}\longrightarrow \big[\{y_{p,n}\}_{P\times
          \mathbb{N}}\big]$ gilt.
        Zunächst ist jedoch nachzuweisen, dass $\{y_{p,n}\}_{\mathbb{N}\times
          P}\subseteq X$ selbst ein Cauchynetz ist. Hierfür beachten wir
        \begin{align*}
            p\big(y_{p',n'}-y_{p'',n''}\big)\leq&\:\:
            p\Big(y_{p',n'}-\left([\hat{x}]_{\hat{p}',n'}\right)_{\delta'}\Big)\\
            &+p\Big(\left([\hat{x}]_{\hat{p}',n'}\right)_{\delta'}-\left([\hat{x}]_{\hat{p}'',n''}\right)_{\delta''}\Big)+p\Big(\left([\hat{x}]_{\hat{p}'',n''}\right)_{\delta''}-y_{\hat{p}'',n''}\Big)
        \end{align*} mit $\delta',\delta''\in P \times\mathbb{N}$.\\\\
        Sei zunächst
        $\frac{1}{n}<\frac{\epsilon}{3}$, dann folgt für die
        äußeren Summanden mit der Definition des kanonischen Repräsentanten und \eqref{eq:Grenznetz}, dass
        \begin{equation}
            \begin{split}
                \label{eq:Huckepack}
                &p\Big(y_{p',n'}-\left([\hat{x}]_{\hat{p}',n'}\right)_{\delta'}\Big)<\frac{\epsilon}{3}
                \qquad\quad\text{ und}\\
                &p\Big(\left([\hat{x}]_{\hat{p}'',n''}\right)_{\delta''}-y_{\hat{p}'',n''}\Big)
                <\frac{\epsilon}{3},
            \end{split} 
        \end{equation} falls $\delta',(p',n'),\delta'',(p'',n'') \geq
        (p,n)$.
        Für den mittleren Summanden beachte man, dass per Definition
        \begin{align*}
            \lim_{\delta'\times \delta''}p\Big(\left(\hat{x}_{\hat{p}',n'}\right)_{\delta'}-\left(\hat{x}_{\hat{p}'',n''}\right)_{\delta''}\Big)
            =\:\hat{p}\Big([\hat{x}]_{\hat{p}',n'}-[\hat{x}]_{\hat{p}'',n''}\Big)
            <\:\frac{1}{n}<\frac{\epsilon}{3}
        \end{align*} für $(\hat{p}',n'),(\hat{p}'',n'')\geq
        (\hat{p},n)$ gilt. Wir finden dann 
        $\tilde{\delta}'$ und $\tilde{\delta}''$ abhängig von
        $(\hat{p}',n')$ und $(\hat{p}'',n'')$ derart, dass
        \begin{align*}
            p\Big(\left(\hat{x}_{\hat{p}',n'}\right)_{\delta'}-\left(\hat{x}_{\hat{p}'',n''}\right)_{\delta''}\Big)<\frac{\epsilon}{3}
        \end{align*} für alle $\delta'\geq \tilde{\delta}'$,
        $\delta''\geq \tilde{\delta}''$. Mit
        Definition \ref{def:NetzNetzkonvergenz}~\textit{i.)~c.)}
        finden wir $\delta'_{\epsilon}\geq
        (n,p),\tilde{\delta}'$ sowie $\delta''_{\epsilon}\geq
        (n,p),\tilde{\delta}''$, so dass insgesamt
        \begin{equation}
            \label{eq:cauchybew}
            p\left(y_{p',n'}-y_{p'',n''}\right)<\epsilon
        \end{equation}gilt. Da wir gezeigt haben, dass dies für alle
        $(p',n'),(p'',n'')\geq (n,p)$ erreichbar ist, folgt
        die Cauchyeigenschaft.\\\\
        Für die Konvergenz sei  $\big[\{y_{p',n'}\}\big]$ die Äquivalenzklasse der
        konstanten Folge $\{y_{p',n'}\}$, dann gilt:
        \begin{align*}
            \hat{p}\Big(\big[\{y_{p,n}\}_{P\times\mathbb{N}}\big]-[\hat{x}]_{\hat{p}',n'}\Big)\leq&\:
            \hat{p}\Big(\big[\{y_{p,n}\}_{P\times\mathbb{N}}\big]-\big[\{y_{p',n'}\}\big]\Big)+\hat{p}\Big(\big[\{y_{p',n'}\}\big]-[\hat{x}]_{\hat{p}',n'}\Big).
        \end{align*}
        Für den ersten Summanden erhalten wir mit
        \eqref{eq:cauchybew} und
        Proposition \ref{prop:wichEiVervollst}~\textit{i.)}
        \begin{equation*}
            \hat{p}\Big(\big[\{y_{p,n}\}_{P\times\mathbb{N}}\big]-\big[\{y_{p',n'}\}\big]\Big)<\frac{\epsilon}{2},\qquad\text{falls}\qquad
            (p',n')\geq (p,\wt{n})  
        \end{equation*}
        für $\wt{n}$ groß genug. Für den zweiten Summanden folgt mit
        der gleichen Argumentation angewandt auf \eqref{eq:Huckepack}
        \begin{equation*}
            \hat{p}\Big(\big[\{y_{p',n'}\}\big]-[\hat{x}]_{\hat{p}',n'}\Big)<\frac{\epsilon}{2}\qquad\text{für}\qquad
            (p',n')\geq (p,\wt{n}'),
        \end{equation*}
        also insgesamt:
        \begin{equation*}
            \hat{p}\Big(\big[\{y_{p,n}\}_{P\times\mathbb{N}}\big]-[\hat{x}]_{\hat{p}',n'}\Big)<\epsilon\quad\text{
              für alle }\quad(p',n')\geq \left(p,\max\left(\wt{n},\wt{n}'\right)\right). 
        \end{equation*}
        Das beweist die Vollständigkeit, und wir sind fertig.
    \end{beweis}
\end{satz}
\begin{satz}
    \label{satz:stetfortsetz}
    Gegeben lkVR's $(X,P)$ und $(Y,Q)$ und sei $(Y,Q)$ zudem
    vollständig und hausdorffsch. Sei weiter $f\colon\eta\longrightarrow Y$ eine
    gleichmäßig stetige Abbildung von einer dichten Teilmenge
    $\eta\subseteq X$ nach $Y$. Dann existiert eine eindeutig bestimmte, gleichmäßig
    stetige Abbildung $\hat{f}\colon X\longrightarrow Y$ mit der
    Eigenschaft $\hat{f}\big|_{\eta}=f$.
    \begin{beweis}
        Wir definieren:
        \begin{equation*}
            \hat{f}(x)=\lim_{\alpha}f(x_{\alpha})\quad\text{für ein
              Netz}\quad\eta\supseteq\net{x}{I}\rightarrow x\in X.
        \end{equation*} 
        Dann ist Existenz und Eindeutigkeit des Limes sowie
        die Unabhängigkeit der Definition von der Wahl des Netzes
        nachzuweisen.\\\\
        \textbf{Existenz}: Zunächst ist $\net{x}{I}$ als
        konvergentes Netz insbesondere ein Cauchynetz. Mit der
        Vollständigkeit von $Y$ reicht es dann zu zeigen, dass dies für
        $\{f(x_{\alpha})\}_{\alpha\in I}$ ebenso der Fall ist. Sei
        hierfür $q\in Q$ und $\epsilon>0$ vorgegeben. Dann finden wir
        mit der gleichmäßigen Stetigkeit von $f$ ein $p\in P$ derart,
        dass für alle $x_{\alpha},x_{\beta}$ :
        \begin{align}
            \label{eq:glstetungl}
            q\left(f(x_{\alpha})-f(x_{\beta})\right)<\epsilon\quad\text{ falls
            }\quad p\left(x_{\alpha}-x_{\beta}\right)<\delta.
        \end{align}
        Da $p(x_{\alpha}-x_{\beta})<\delta$ für
        $\alpha,\beta\geq \gamma_{\delta}$ folgt daraus die Behauptung. Die
        Eindeutigkeit des Limes ist mit der Hausdorff-Eigenschaft
        von $Y$ klar.\\\\
        $\textbf{Wohldefiniertheit}$: Gegeben
        $\net{x}{I},\:\{y\}_{\beta\in J}\subseteq \eta$ beide
        konvergent gegen $x\in X$, so erhalten wir mit der Stetigkeit
        von $q\in Q$:
        \begin{align*}
            q\Big(\lim_{\alpha}f(x_{\alpha})-\lim_{\beta}f(y_{\beta})\Big)=&\: q\left(\lim_{\alpha}\left[\lim_{\beta} \left[f(x_{\alpha})-f(y_{\beta})\right]\right]\right)
            =q\left(
                \lim_{\alpha\times\beta}\left[f(x_{\alpha})-f(y_{\beta})\right]\right)
            \\=&\:\lim_{\alpha\times\beta}q\left(f(x_{\alpha})-f(y_{\beta})\right)
            =0.
        \end{align*} Die letzte Gleichheit folgt dabei aus
        \eqref{eq:glstetungl} und \eqref{eq:AequddwLimsgleich}. Die
        zweite Gleichheit folgt mit Proposition \ref{prop:wichEiVervollst}~\textit{ii.)}.\\\\
        Für $\hat{f}\big|_{\eta}=f$ sei $\eta\supseteq
        \{x_{\alpha}\}_{\alpha\in I}$ mit $\{x_{\alpha}\}_{\alpha\in
          I}\rightarrow x\in \eta$, dann folgt
        \begin{equation*}
            \hat{f}\left(x\right)=\displaystyle\lim_{\alpha}f\left(x_{\alpha}\right)=f\left(\lim_{\alpha}x_{\alpha}\right)=f(x).
        \end{equation*}
        mit der Stetigkeit von $f$ und der
        Hausdorff-Eigenschaft von $Y$, siehe auch Bemerkung \ref{bem:Netzbem}~\textit{ii.)}.\\\\
        $\textbf{Gleichmäßige Stetigkeit}$: Es ist zu
        zeigen, dass für alle $q\in Q$ und $\epsilon>0$ vorgegeben,
        ein $p\in P$ und ein $\delta>0$ derart existieren, dass für
        alle  $x,y\in X$
        \begin{equation*}
            q\left(\hat{f}(x)-\hat{f}(y)\right)<\epsilon\quad\text{
              falls }\quad p(x-y)<\delta
        \end{equation*}gilt. Zunächst haben wir nach Voraussetzung für
        $\delta'$ vorgegeben, ein $\epsilon'$, so dass:
        \begin{equation*}
            q\left(f(x)-f(y)\right)<\epsilon'<\epsilon\quad\text{
              falls }\quad p(x-y)<\delta<\delta'.
        \end{equation*}
        Sei $p(x-y)<\delta$, so folgt mit Dreiecksungleichung
        \begin{equation*}
            p(x_{\alpha}-y_{\beta})\leq p(x-y)+p(x-x_{\alpha})+p(y-y_{\beta})<\delta+\Delta<\delta'
        \end{equation*} für $\alpha\geq \alpha_{\Delta}\in I$, $\beta \geq
        \beta_{\Delta}\in J$ mit Netzen $\eta\supseteq\net{x}{I}\rightarrow
        x$ und $\eta\supseteq \{y_{\beta}\}_{\beta\in J}\rightarrow
        y$. 
        Wir erhalten dann mit der Stetigkeit von $q$ und der Addition, dass
        \begin{align*}
            q\left(\hat{f}(x)-\hat{f}(y)\right)=&\:
            q\left(\lim_{\alpha}\left[\lim_{\beta}\left[f(x_{\alpha})-f(y_{\beta})\right]\right]\right)=q\left(\lim_{\alpha\times\beta}\left[f(x_{\alpha})-f(y_{\beta})\right]\right)
            \\=&\:\lim_{\alpha\times\beta}q\Big(f(x_{\alpha})-f(y_{\beta})\Big)
            \leq\epsilon'<\epsilon,
        \end{align*} wobei wir abermals 
        Proposition \ref{prop:wichEiVervollst}~\textit{i.),~ii.)} und den Erhalt von
        $\leq$ unter Limesbildung benutzt haben. Dies zeigt die
        gleichmäßige Stetigkeit.\\\\
        Die Eindeutigkeit besagter Abbildung folgt aus
        $\ovl{\eta}=X$, denn für jede weitere derartige Abbildung
        $\hat{f}'$ ist $\left(\hat{f}'-\hat{f}\right)\Big|_{\eta}=0$
        und somit für $\eta
        \supseteq\net{x}{I}\rightarrow x\in X$
        \begin{equation*}
            \left(\hat{f}'-\hat{f}\right)(x)=\lim_{\alpha}\left(\hat{f}'-\hat{f}\right)(x_{\alpha})=0
        \end{equation*}
        mit der Eindeutigkeit des Limes in $Y$.
    \end{beweis}
\end{satz}
\begin{bemerkung}
    \label{bem:stetfortmult}
    \begin{enumerate}
    \item
        Man beachte, dass wir als Voraussetzung nicht benötigt
        haben, dass $(X,P)$ hausdorffsch ist. Das liegt daran, dass für
        stetiges $f$ und $x\neq x'\in X$ mit $f(x)\neq f(x')$, vermöge der
        Hausdorff-Eigenschaft von $Y$, disjunkte Umgebungen von $f(x)$
        und $f(x')$ existieren, deren Urbilder disjunkte Umgebungen von $x$
        und $x'$ sind .
        
        Sprich, ein solches $f$ kann sowieso nur dann stetig sein, wenn es
        untrennbare Punkte in $X$ auf das selbe Element in $Y$ abbildet.
    \item 
        Die besondere Wichtigkeit dieses Satzes liegt unter anderem darin,
        dass jede stetige, lineare Abbildung insbesondere gleichmäßig stetig
        ist. Für diese ist dann obiger Satz auch leichter mit
        $q(\phi(x))\leq c\:p(x)$ beweisbar.
        
        Für stetige bilineare
        Abbildungen $q(\phi(x,y))\leq c\:p_{1}(x)p_{2}(y)$ und analog für
        stetige multilineare Abbildungen liefert folgender Satz
        sinngemäße Aussagen.
    \end{enumerate}
\end{bemerkung}
\begin{satz}
    \label{satz:stetfortsBillphi}
    Gegeben lkVR's $(X_{1},P_{1}),(X_{2},P_{2})$ und $(Y,Q)$ mit
    $(Y,Q)$ zudem vollständig und hausdorffsch sowie
    dichte Teilmengen $\eta_{1}\subseteq X_{1}$, $\eta_{2}\subseteq X_{2}$. 
    Sei weiter $\phi:\eta_{1}\times
    \eta_{2}\longrightarrow Y$ eine stetige, bilineare Abbildung, so
    existiert eine eindeutig bestimmte, stetige bilineare Abbildung:
    \begin{equation*}
        \hat{\phi}:X_{1}\times X_{2}\longrightarrow Y\qquad\text{mit}\qquad \hat{\phi}\big|_{\eta_{1}\times \eta_{2}}=\phi.
    \end{equation*}
    \begin{beweis}
        Wir definieren
        $\hat{\phi}(x,y)=\displaystyle\lim_{\alpha\times\beta}\overbrace{\phi(x_{\alpha},y_{\beta})}^{\phi_{\alpha,\beta}}$
        für Netze $\eta_{1}\supseteq\net{x}{I}\rightarrow x$ und\\
        $\eta_{2}\supseteq\{y_{\beta}\}_{\beta\in J}\rightarrow
        y$.
        Besagter Limes ist dann mit der
        Hausdorff-Eigenschaft von $Y$ eindeutig.

        \textbf{Existenz}: Sei $p\in P$ vorgegeben. Dann ist
        $p\left(\phi(x,y)\right)\leq c\: p_{1}(x)p_{2}(y)$ für alle
        $x\in X_{1}$ und alle $y\in X_{2}$. Weiterhin beachten wir,
        dass $\net{x}{I}$ und $\{y_{\beta}\}_{\beta\in
          J}$ insbesondere Cauchynetze sind. Wir finden dann Konstanten $c',c''$ sowie
        Elemente $\alpha_{c''}\in I$ und $\beta_{c'}\in J$ derart,
        dass $p_{1}(x_{\alpha})\leq c''$ für alle $\alpha\geq
        \alpha_{c''}$ und $p_{2}(y_{\beta})\leq c'$ für alle $\beta\geq
        \beta_{c'}$. Des Weiteren existieren
        $\alpha_{\frac{\epsilon}{2cc''}}\in I$ und
        $\beta_{\frac{\epsilon}{2cc'}}\in J$, so dass
        $p_{1}(x_{\alpha}-x_{\alpha'})\leq \frac{\epsilon}{2cc''}$ für
        alle $\alpha,\alpha'\geq \alpha_{\frac{\epsilon}{2cc''}}$ 
        und
        $p_{2}(y_{\beta}-y_{\beta'})\leq \frac{\epsilon}{2cc'}$ für
        alle $\beta,\beta'\geq \beta_{\frac{\epsilon}{2cc'}}$
        gilt. Wir wählen $\wt{\alpha}\geq
        \alpha_{c''},\alpha_{\frac{\epsilon}{2cc''}}$ sowie
        $\wt{\beta}\geq
        \beta_{c'},\beta_{\frac{\epsilon}{2cc'}}$ und erhalten: 
        \begin{align*}
            q\left(\phi_{\alpha,\beta}-\phi_{\alpha',\beta'}\right)\leq&\:
            q\left(\phi(x_{\alpha},y_{\beta}-y_{\beta'})\right)+q\left(\phi(x_{\alpha}-x_{\alpha'},y_{\beta'})\right)
            \\\leq&\:c\:p_{1}(x_{\alpha})\:p_{2}(y_{\beta}-y_{\beta'})+c\:p_{1}(x_{\alpha}-x_{\alpha'})\:p_{2}(y_{\beta'})
            \\\leq&\:cc'\:p_{2}(y_{\beta}-y_{\beta'})+cc''\:p_{1}(x_{\alpha}-x_{\alpha'})
            \\\leq&\:\epsilon
        \end{align*} für alle $\alpha,\alpha'\geq \wt{\alpha}$ und $\beta,\beta'\geq
        \wt{\beta}$. Die Wohldefiniertheit folgt mit
        \eqref{eq:AequddwLimsgleich} auf die
        gleiche Weise, und für die Stetigkeit rechnen wir mit
        Proposition \ref{prop:wichEiVervollst}~\textit{ii.)} und dem Erhalt von
        $\leq$ unter Limesbildung:
        \begin{align*}
            q\big(\hat{\phi}(x,y)\big)=&\lim_{\alpha\times\beta}q(\phi(x_{\alpha},y_{\beta}))
            \leq\lim_{\alpha\times\beta}c\: p_{1}(x_{\alpha})\:p_{2}(y_{\beta})
            \\=&\:c\lim_{\alpha}p_{1}(x_{\alpha})\lim_{\beta}p(y_{\beta})
            =c\:p_{1}(x)\:p_{2}(y).
        \end{align*}
        Die Eindeutigkeit von $\hat{\phi}$ folgt wie in
        Satz \ref{satz:stetfortsetz}, ebenso die Behauptung
        $\hat{\phi}\big|_{\eta_{1}\times \eta_{2}}=\phi$, da mit
        $\eta_{1}\supseteq\{x_{\alpha}\}_{\alpha\in I}\rightarrow x\in
        X_{1}$ und $\eta_{2}\supseteq\{y_{\beta}\}_{\beta\in J}\rightarrow y\in
        X_{2}$ das Netz
        $\{(x_{\alpha},y_{\beta})\}_{I\times J}$ in der
        Produkttopologie von $X_{1}\times X_{2}$ gegen $(x,y)$
        konvergiert. Alternativ kann man hier auch wieder
        Proposition \ref{prop:wichEiVervollst}~\textit{ii.)} benutzen.
    \end{beweis} 
\end{satz}

\section{Tensorprodukte lokalkonvexer
  Vektorräume und deren Topologien}
\label{subsec:TenprodLkvVr}
Zu Beginn dieses Abschnittes erinnern wir zunächst  an folgende
wohlbekannte Tatsachen.
\begin{satz}[Hahn-Banach, vgl.\cite{hirzebruch.scharlau:1991a}, Kapitel 2.6]
    \label{satz:HahnBanach}
    Gegeben ein $\mathbb{K}$-Vektorraum $X$ und eine Halbnorm $p\colon X\rightarrow
    \mathbb{R}$. Sei des Weiteren $f\colon L\rightarrow \mathbb{K}$ eine
    lineare Abbildung von einem Unterraum $L\subseteq X$ nach
    $\mathbb{K}$ und es gelte $|f|\leq p\big|_{L}$. Dann existiert
    eine lineare Abbildung $F\colon X\rightarrow \mathbb{K}$ mit $|F|\leq p$
    und $F\big|_{L}=f$.
\end{satz}
\begin{definition}[Universelle Eigenschaft]
    \label{def:UniverselleEig}
    Gegeben $\mathbb{K}$-Vektorräume
    $\mathbb{V}_{1},…,\mathbb{V}_{k}$, so heißt ein Tupel
    $(\ot_{k},\mathbb{V}_{1}\ot…\ot \mathbb{V}_{k})$, bestehend aus einem
    $\mathbb{K}$-Vektorraum $\mathbb{V}_{1}\ot…\ot \mathbb{V}_{k}$ und
    einer $\mathbb{K}$-multilinearen Abbildung\\
    $\ot_{k}\colon \mathbb{V}_{1}\times…\times
    \mathbb{V}_{k}\longrightarrow \mathbb{V}_{1}\ot…\ot \mathbb{V}_{k}$
    genau dann ein Tensorprodukt von
    $\V_{1},…,\V_{k}$, wenn für jedes
    weitere derartige Tupel $(\phi,\mathbb{M})$ eine eindeutig
    bestimmte lineare Abbildung $\tau\colon\mathbb{V}_{1}\ot…\ot \mathbb{V}_{k} \longrightarrow
    \mathbb{M}$ existiert, so dass folgendes Diagramm kommutiert:
    $$ \diagram \mathbb{V}_{1}\times…\times \mathbb{V}_{k} \rrto^{\ot_{k}}\drrto_{\phi}&  &\mathbb{V}_{1}\ot…\ot \mathbb{V}_{k} \dto_{\tau} \\
    &        &\mathbb{M}
    \enddiagram $$
    Mit $v_{1}\ot…\ot v_{k}$ bezeichnen wir im Folgenden das Bild von  $v_{1},…,v_{k}\in \V_{1}\times…\times
    \V_{k}$ unter der Abbildung $\ot_{k}$.
\end{definition}
\begin{lemma}[Isomorphie und Assoziativit"at des Tensorproduktes]
    \label{lemma:assTenprod}
    \begin{enumerate}
    \item
        Je zwei Tensorprodukte $(\ot_{k},\mathbb{V}_{1}\ot…\ot \mathbb{V}_{k})$ und $(\ot'_{k},\mathbb{V}_{1}\ot'…\ot' \mathbb{V}_{k})$ von $\V_{1},…,\V_{k}$ sind isomorph.
    \item
        Gegeben $\mathbb{K}$-Vektorr"aume $\V_{1},\V_{2},\V_{3}$, dann
        gilt $(\V_{1}\ot \V_{2})\ot \V_{3} \cong \V_{1}\ot \V_{2}\ot
        \V_{3}$ vermöge linearer Fortsetzung von:
        \begin{align*}
            \tau\colon  \V_{1}\ot \V_{2}\ot
            \V_{3}&\longrightarrow(\V_{1}\ot \V_{2})\ot \V_{3}\\
            v_{1}\ot v_{2}\ot v_{3}&\longmapsto(v_{1}\ot v_{2})\ot v_{3}
        \end{align*}
        Allgemein ist $\left(\V_{1}\ot\dots\ot \V_{l}\right)\ot
        \:(\V_{l+1}\ot\dots\ot \V_{k})\cong \V_{1}\ot\dots\ot \V_{k}$
        vermöge linearer Fortsetzung von $\mu\colon v_{1}\ot…\ot v_{k}\longmapsto(v_{1}\ot…\ot v_{l})\ot\: (v_{l+1}\ot…\ot v_{k})$.
    \end{enumerate}
    \begin{beweis}
        \begin{enumerate}
        \item
            Mit der universellen Eigenschaft beider Tensorprodukte
            folgt die Kommutativität von
            $$ \diagram \mathbb{V}_{1}\times…\times \mathbb{V}_{k} \rrto^{\ot_{k}}\drrto_{\ot'_{k}}&  &\mathbb{V}_{1}\ot…\ot \mathbb{V}_{k} \dto_{\tau_{\ot_{k}'}}\\
            &        &\mathbb{V}_{1}\ot'…\ot' \mathbb{V}_{k}\uto<-1ex>_{\tau_{\ot_{k}}} 
            \enddiagram $$ mit eindeutig bestimmten linearen Abbildungen $\tau_{\ot_{k}'}$ und $\tau_{\ot_{k}}$. F"ur diese gilt sowohl $\tau_{\ot_{k}'}\cp \tau_{\ot_{k}}= \id_{\mathbb{V}_{1}\ot'…\ot' \mathbb{V}_{k}}$ als auch $\tau_{\ot_{k}}\cp \tau_{\ot_{k}'} = \id_{\mathbb{V}_{1}\ot…\ot \mathbb{V}_{k}}$, da beide Identit"aten die eindeutig bestimmten Isomorphismen sind, die
            $$ \diagram \mathbb{V}_{1}\times…\times \mathbb{V}_{k} \rrto^{\ot'_{k}}\drrto_{\ot'_{k}}&  &\mathbb{V}_{1}\ot'…\ot' \mathbb{V}_{k} \dto^{\id_{\mathbb{V}_{1}\ot'…\ot' \mathbb{V}_{k}}}\\
            &        &\mathbb{V}_{1}\ot'…\ot' \mathbb{V}_{k}
            \enddiagram $$
            bzw. 
            $$ \diagram \mathbb{V}_{1}\times…\times \mathbb{V}_{k} \rrto^{\ot_{k}}\drrto_{\ot_{k}}&  &\mathbb{V}_{1}\ot…\ot \mathbb{V}_{k} \dto^{\id_{\mathbb{V}_{1}\ot…\ot \mathbb{V}_{k}}}\\
            &        &\mathbb{V}_{1}\ot…\ot\mathbb{V}_{k}
            \enddiagram $$
            zum Kommutieren bringen. Insgesamt zeigt dies, dass $\tau_{\ot_{k}}$ und $\tau_{\ot_{k}'}$ zueinander inverse Isomorphismen sind.
        \item 
            Nach \textit{i.)} reicht es zu zeigen, dass $(\V_{1}\ot
            \V_{2})\ot \V_{3}$ ein Tensorprodukt von
            $\V_{1},\V_{2}$, $\V_{3}$ ist. Denn es ist $(\tau \cp
            \ot_{3})(v_{1},v_{2},v_{3})=\ot_{1}(\ot_{2}(v_{1},v_{2}),v_{3})$
            und somit der gesuchte Isomorphismus die eindeutig
            bestimmte lineare Fortsetzung von $\tau$. Sei hierf"ur $\phi\colon
            \V_{1}\times\V_{2}\times \V_{3}\longrightarrow \M$
            trilinear, so definieren wir f"ur festes $v_{3}\in \V_{3}$ die bilineare Abbildung $\tau_{v_{3}}\colon \V_{1}\times \V_{2}\longrightarrow \M$ durch $\tau_{v_{3}}(v_{1},v_{2})=\phi(v_{1},v_{2},v_{3})$. Mit der universellen Eigenschaft existiert dann ein eindeutig bestimmtes, lineares $\tau_{v_{3}}^{\ot}\colon \V_{1}\ot \V_{2}\longrightarrow \M$ mit $\tau_{v_{3}}^{\ot}(v_{1}\ot v_{2})=\tau_{v_{3}}(v_{1},v_{2})=\phi(v_{1},v_{2},v_{3})$. Wir behaupten nun, dass
            \begin{align*}
                \wt{\tau}\colon(\V_{1}\ot \V_{2})\times \V_{3}&\longrightarrow \M
                \\(z, v_{3})&\longmapsto \tau_{v_{3}}^{\ot}(z)
            \end{align*}bilinear ist. Die Linearit"at im ersten
            Argument ist klar und die Linearit"at im zwei-ten Argument erhalten wir
            f"ur
            $z\in \V_{1}\ot \V_{2}$, $v,w\in \V_{3}$ und $\lambda \in
            \mathbb{K}$ mit: 
            \begin{align*}
                \wt{\tau}(z, \lambda v+ w)=&\:\sum_{i=1}^{n}\wt{\tau}\left(v_{1}^{i}\ot v_{2}^{i}, \lambda v+ w\right)= \sum_{i=1}^{n} \tau^{\ot}_{\lambda v + w}\left(v_{1}^{i}\ot v_{2}^{i}\right)
                \\=&\:\sum_{i=1}^{n} \phi\left(v_{1}^{i}, v_{2}^{i},\lambda v+ w\right)
                =\lambda \sum_{i=1}^{n} \phi\left(v_{1}^{i}, v_{2}^{i},v\right)+\sum_{i=1}^{n} \phi\left(v_{1}^{i}, v_{2}^{i},w\right)\\=&\:\lambda \wt{\tau}(z, v)+\wt{\tau}(z, w).
            \end{align*}Somit existiert ein eindeutig bestimmtes $\widehat{\tau}: (\V_{1}\ot \V_{2})  \ot \V_{3} \longrightarrow \M$ mit 
            \begin{equation*}
                \widehat{\tau}((v_{1}\ot v_{2})\ot v_{3})=\wt{\tau}((v_{1}\ot v_{2}), v_{3})=\tau^{\ot}_{v_{3}}(v_{1}\ot v_{2})=\phi(v_{1},v_{2},v_{3})
            \end{equation*}für alle $v_{1}\in \V_{1}, v_{2}\in \V_{2}$
            und $v_{3}\in \V_{3}$, wie gewünscht. Die allgemeine Aussage zeigt man auf analoge Weise.     
        \end{enumerate}
    \end{beweis}
\end{lemma}
\begin{bemerkung}
    \label{bem:TenprodBasis}
    Man kann zeigen, dass, $\V_{1},…,\V_{k}$
    vorgegeben, mindestens immer ein solches Tensorprodukt
    existiert. Eine
    äquivalente und im Einzelfall praktischere Definition (besonders
    wenn es darum geht, die Tensorprodukteigenschaft eines solchen
    Tupels nachzuweisen) erhält man
    wie folgt:

    Gegeben $\mathbb{K}$-Vektorräume
    $\mathbb{V}_{1},…,\mathbb{V}_{k}$, so heißt ein Tupel
    $(\ot_{k},\mathbb{V}_{1}\ot…\ot \mathbb{V}_{k})$ bestehend aus einem
    $\mathbb{K}$-Vektorraum $\mathbb{V}_{1}\ot…\ot \mathbb{V}_{k}$ und
    einer $\mathbb{K}$-multilinearen Abbildung\\
    $\ot_{k}\colon \mathbb{V}_{1}\times…\times
    \mathbb{V}_{k}\longrightarrow \mathbb{V}_{1}\ot…\ot \mathbb{V}_{k}$ genau dann  ein Tensorprodukt von
    $\mathbb{V}_{1},…,\mathbb{V}_{k}$, wenn:
    \begin{enumerate}
    \item
        $\mathbb{V}_{1}\ot…\ot \mathbb{V}_{k}=\mathrm{span}(\im(\ot_{k}))$.
    \item
        Sind $\{v^{j}_{i_{j}}\}_{1\leq i_{j}\leq n_{j}}\subseteq \mathbb{V}_{j}$ 
        linear unabhängig in den
        $\mathbb{V}_{j}$ für alle $1\leq j\leq k$, so ist die Menge
        $\left\{\ot_{k}\left(v^{1}_{i_{1}},…,v^{k}_{i_{k}}\right)\right\}_{\substack{1\leq
            i_{j}\leq n_{j}\\ 1\leq j\leq k}}\subseteq \mathbb{V}_{1}\ot…\ot
        \mathbb{V}_{k}$ ebenfalls linear unabhängig.
    \end{enumerate}
    Grob gesagt entspricht dabei \textit{i.)} der Eindeutigkeit der Abbildung $\tau$ und \textit{ii.)}
    sichert deren Existenz. Die Aussage von \textit{i.)} und \textit{ii.)} ist dabei
    im Wesentlichen die, dass die Menge aller Elemente
    $e^{1}_{i_{1}}\ot…\ot
    e^{k}_{i_{k}}=\ot_{k}\left(e^{1}_{i_{1}},…,e^{k}_{i_{k}}\right)$
    für Basen $\left\{e^{j}_{i_{j}}\right\}_{i_{j}\in I_{j}}$ der
    $\mathbb{V}_{j}$ ein Basis von $\mathbb{V}_{1}\ot…\ot \mathbb{V}_{k}$
    ist.
\end{bemerkung}
Im Gegensatz zur Charakterisierung des Tensorproduktes in
Bemerkung \ref{bem:TenprodBasis}, legt uns die Definition über die
universelle Eigenschaft eine
ausgesprochen praktische Möglich-keit in die Hand, lineare Abbildungen von
Tensorprodukten in andere $\mathbb{K}$-Vektorräume zu definieren: 
\begin{korollar}
    \label{kor:WohldefTensorprodabbildungen}
    Gegeben $\mathbb{K}$-Vektorräume
    $\mathbb{V}_{1},…,\mathbb{V}_{k}, \mathbb{M}$ und ein Tensorprodukt
    $(\ot_{k},\mathbb{V}_{1}\ot…\ot \mathbb{V}_{k})$ von\\ $\V_{1},…,\V_{k}$,
    so liefert jede Abbildungsvorschrift $\phi\colon \im(\ot_{k})\longrightarrow
    \mathbb{M}$, die 
    \begin{equation}
        \label{eq:linFortsetzTensProdAbb}
        \begin{split}
            \phi(v_{1}\ot…\ot v_{i}+\lambda v'_{i}\ot…\ot
            v_{k})=&\:\phi(v_{1}\ot…\ot v_{i}\ot…\ot
            v_{k})
            \\ &+\lambda\phi(v_{1}\ot…\ot v'_{i}\ot…\ot
            v_{k})
        \end{split}
    \end{equation} für $1\leq i\leq k$ und alle $\lambda\in \mathbb{K}$ erfüllt,
    eine eindeutig bestimmte, wohldefinierte lineare Fortsetzung $\phi_{\ot_{k}}$ auf
    ganz $\mathbb{V}_{1}\ot…\ot \mathbb{V}_{k}$.
    \begin{beweis}
        $\phi$ definiert eine eindeutige $\mathbb{K}$-multilineare Abbildung
        \begin{align*}
            \phi_{\times}\colon \V_{1}\times…\times \V_{k}&\longrightarrow
            \mathbb{M}\\
            (v_{1},…,v_{k})&\longmapsto \phi(\ot_{k}(v_{1},…,v_{k})),
        \end{align*} 
        und mit der universellen Eigenschaft existiert somit
        eine eindeutig bestimmte lineare Abbildung $\phi_{\ot_{k}}\colon
        \V_{1}\ot…\ot \V_{k}\rightarrow \mathbb{M}$ derart, dass
        $\phi_{\ot_{k}}\cp \ot_{k}=\phi_{\times}=\phi\cp \ot_{k}$, also\\ $\phi_{\ot_{k}}\big|_{\im(\ot_{k})}=\phi$ gilt.
    \end{beweis}
\end{korollar}
\begin{definition}[$\pi_{k}$-Topologie]
    \label{def:pitop}
    Gegeben lkVR's $(\mathbb{V}_{1},P_{1}),…,(\mathbb{V}_{k},P_{k})$ und ein Tensorprodukt $(\ot_{k},\mathbb{V}_{1}\ot…\ot \mathbb{V}_{k}$) von
    $\mathbb{V}_{1},…,\mathbb{V}_{k}$, so ist eine $\pi_{k}$-Topologie auf $\mathbb{V}_{1}\ot…\ot \mathbb{V}_{k}$ eine
    Toplogie derart, dass jede lineare Abbildung
    \begin{equation*}
        \tau\colon \mathbb{V}_{1} \pite…\pite \mathbb{V}_{k}\longrightarrow \mathbb{M}
    \end{equation*} in einen weiteren lokalkonvexen Vektorraum $(\mathbb{M},Q)$ genau dann stetig ist, wenn die Abbildung $\tau\cp \ot_{k}$
    bezüglich der Produkttopologie auf $\mathbb{V}_{1}\times…\times
    \mathbb{V}_{k}$ stetig ist. Hierbei bezeichnet
    $\mathbb{V}_{1}\pite…\pite\mathbb{V}_{k}$
    den topologischen Raum $(\mathbb{V}_{1}\ot…\ot \mathbb{V}_{k},\pi_{k})$.
\end{definition}
\begin{satz}
    \label{satz:PiTopsatz}
    Gegeben lkVR's $(\mathbb{V}_{1},P_{1}),…,(\mathbb{V}_{k},P_{k})$, so gilt:
    \begin{enumerate}
    \item 
        Es ist $\ot_{k}$ stetig in jeder $\pi_{k}$-Topologie auf
        $\mathbb{V}_{1}\ot…\ot\mathbb{V}_{k}$.  
    \item
        Sofern sie existiert, ist die $\pi_{k}$-Topologie auf $\mathbb{V}_{1}\ot…\ot\mathbb{V}_{k}$ eindeutig bestimmt.
    \item
        Die $\pi_{k}$-Toplologie auf $\mathbb{V}_{1}\ot…\ot\mathbb{V}_{k}$ existiert, ist lokalkonvex
        und wird induziert durch das Halbnormensystem
        $\Pi_{\mathbf{P}}$, $(\mathbf{P}=P_{1}\times…\times P_{k})$:
        \begin{equation*}  
            \pi_{\mathbf{p}}(z):=\inf\left\{\sum_{i=1}^{n}p_{1}(x^{i}_{1})…p_{k}(x^{i}_{k})\right\}\qquad\qquad  \mathbf{P}\ni\mathbf{p}=(p_{1},…,p_{k}).
        \end{equation*} 
        Dabei ist das Infimum über alle Zerlegungen
        $z=\displaystyle\sum_{i=1}^{n}x^{i}_{1}\ot…\ot x^{i}_{k}$ zu nehmen.

        Für separables $z=x_{1}\ot…\ot x_{k}$ folgt
        \begin{equation}
            \label{eq:pisepei}
            \pi_{\mathbf{p}}(z)=p_{1}(x_{1})… p_{k}(x_{k}).
        \end{equation} 
        Die Halbnorm $\pi_{\mathbf{p}}$ heißt Tensorprodukt der
        Halbnormen $p_{1},…,p_{k}$ und man schreibt oft auch einfach
        $p_{1}\ot…\ot p_{k}$ anstelle $\pi_{\mathbf{p}}$.
    \item
        $\pi_{k}$ ist die feinste lokalkonvexe Topologie auf
        $\V_{1}\ot…\ot \V_{k}$, bezüglich der die Abbildung $\ot_{k}$ stetig ist.
    \item        
        $(\mathbb{V}_{1}\ot…\ot \mathbb{V}_{k},\pi_{k})$ ist genau dann hausdorffsch, wenn alle $(\mathbb{V}_{j},P_{j})$ hausdorffsch sind.
    \item
        Sind $P_{1},\dots, P_{k}$ filtrierend, so auch $\prod_{\mathbf{P}}$.
    \item
        Sind alle $(\mathbb{V}_{j},P_{j})$ hausdorffsch, so gilt
        \begin{equation*}
            \widehat{\mathbb{V}_{1}\pite…\pite
              \mathbb{V}_{k}}\cong\widehat{\hat{\V}_{1}\pite…\pite
              \hat{\V}_{k}}. 
        \end{equation*} 
        Dabei bedeutet $\cong$ lineare Homöomorphie.
        
        Wir dürfen somit $\hat{\V}_{1}\pite…\pite
        \hat{\V}_{k}$ als dichte Teilmenge von $\widehat{\mathbb{V}_{1}\pite…\pite\mathbb{V}_{k}}$ auffassen.
    \end{enumerate}
    \begin{beweis}
        Die Beweise von \eqref{eq:pisepei} und \textit{v.)} finden sich
        auch in \cite[Kapitel
        43]{treves:1967a} bzw.\\ \cite[Kapitel 15]{jarchow:1981a}.
        \begin{enumerate}
        \item 
            Mit Definition \ref{def:pitop} und der Kommutativität von 
            $$ \diagram
            \mathbb{V}_{1}\times…\times \mathbb{V}_{k}
            \rrto^{\ot_{k}}\drrto_{\ot_{k}}&
            &\mathbb{V}_{1}\pite…\pite \mathbb{V}_{k} \dto^{\mathrm{id}} \\
            & &\mathbb{V}_{1}\pite… \pite\mathbb{V}_{k},
            \enddiagram $$
            ist $\mathrm{id}\cp \ot_{k} =\ot_{k}$ genau dann stetig, wenn
            $\mathrm{id}$ stetig ist. Das ist aber klar, da beide
            Räume dieselbe Topologie tragen.
        \item
            Wir haben
            $$ \diagram
            \mathbb{V}_{1}\times…\times \mathbb{V}_{k}
            \rrto^{\ot_{k}}\drrto_{\ot_{k}}&
            &\mathbb{V}_{1}\pite…\pite \mathbb{V}_{k} \dto^{\mathrm{id}} \\
            & &\mathbb{V}_{1}\ot_{\hat{\pi}}… \ot_{\hat{\pi}}\mathbb{V}_{k},
            \enddiagram $$
            und die Stetigkeit von $\ot_{k}$ nach \textit{i).} Vermöge $\mathrm{id}\cp \ot_{k}=\ot_{k}$ ist auch $\mathrm{id}\cp
            \ot_{k}$ stetig und mit Definition \ref{def:pitop} $\mathrm{id}$
            selbst. Das zeigt $\pi=\hat{\pi}$ und somit die
            Behauptung.
        \item
            Zunächst sind die $\pi_{\mathbf{p}}$ in der Tat Halbnormen, denn
            es ist:
            \begin{align*}
                \pi_{\mathbf{p}}(\lambda
                z)=&\:\inf\left\{\sum_{i=1}^{n}|\lambda|\:p_{1}\big(x^{i}_{1}\big)…p_{k}\big(x^{i}_{k}\big)\right\}
                \\=&\:|\lambda|\inf\left\{\sum_{i=1}^{n}\:p_{1}\big(x^{i}_{1}\big)…p_{k}\big(x^{i}_{k}\big)\right\}
                =|\lambda|\:\pi_{\mathbf{p}}(z).
            \end{align*}Weiterhin gibt mindestens so viele Zerlegungen von
            $\tilde{z}=z+z'$, wie man durch Addition von Zerlegungen von $z$ und
            $z'$ erhält. Damit folgt: 
            \begin{align*}
                \pi_{\mathbf{p}}(z+z')=&\inf\left\{\sum_{i=1}^{\tilde{n}}p_{1}\left(\tilde{x}^{i}_{1}\right)…p_{k}\big(\tilde{x}^{i}_{k}\big)\right\}\\\leq&\inf\left\{\sum_{i=1}^{n}p_{1}\left(x^{i}_{1}\right)…p_{k}\big(x^{i}_{k}\big)+\sum_{i=1}^{n'}p_{1}\big(x'^{i}_{1}\big)…p_{k}\big(x'^{i}_{k}\big)\right\}
                \\=
                &\inf\left\{\sum_{i=1}^{n}p_{1}\big(x^{i}_{1}\big)…p_{k}\big(x^{i}_{k}\big)\right\}+\inf\left\{\sum_{i=1}^{n'}p_{1}\big(x'^{i}_{1}\big)…p^{k}\big(x'^{i}_{k}\big)\right\}
                \\=&\: \pi_{\mathbf{p}}(z)+\pi_{\mathbf{p}}(z').
            \end{align*}
            Mit Satz \ref{satz:wichtigerSatzueberHalbnormentopologien}
            erzeugt dieses Halbnormensystem eine lokalkonvexe
            Topologie auf $\mathbb{V}_{1}\ot…\ot \mathbb{V}_{k}$, die diesen Raum
            zu einem topologischen Vektorraum macht.

            Für \eqref{eq:pisepei} sei $z=x_{1}\ot…\ot x_{k}$, dann ist $\pi_{\mathbf{p}}(z)\leq
            p_{1}(x_{1})…p_{k}(x_{k})$ per Definition. Für die
            umgekehrte Abschätzung definieren wir lineare Abbildungen
            $u_{j}$ auf den Unterräumen $L_{j}=\mathrm{span}(x_{j})\subseteq
            \V_{j}$ durch $u_{j}(\lambda x_{j})=\lambda\:
            p_{j}(x_{j})$. 

            Dies bedeutet $|u_{j}|\leq p_{j}\big|_{L_{j}}$ und
            mit Satz \ref{satz:HahnBanach} finden wir Fortsetzungen
            $|U_{j}|\leq p_{j}$ mit $U_{j}\big|_{L_{j}}=u_{j}$. Es
            folgt für die durch 
            \begin{align*}
                U_{1}\ot…\ot U_{k}\colon \V_{1}\ot…\ot
                \V_{k}&\longrightarrow \mathbb{K}\\
                x_{1}\ot…\ot x_{k}&\longmapsto U_{1}(x_{1})\cdot…\cdot
                U_{k}(x_{k}),
            \end{align*}
            nach Korollar \ref{kor:WohldefTensorprodabbildungen},
            wohldefinierte lineare Abbildung, dass
            \begin{align*}
                p_{1}(x_{1})…p_{k}(x_{k})=&\:|U_{1}(x_{1})\cdot…\cdot U_{k}(x_{k})|=\left|U_{1}\ot…\ot
                    U_{k}\right|(z)\\\leq& \:\sum_{i=1}^{n}\left|U_{1}\ot…\ot
                    U_{k}\right|(y^{i}_{1}\ot…\ot y^{i}_{k})
                \\\leq&\:\sum_{i=1}^{n}p_{1}\big(y^{i}_{1}\big)…p_{k}\big(y^{i}_{k}\big)
            \end{align*}
            für alle Zerlegungen
            $z=\displaystyle\sum_{i=1}^{n}y^{i}_{1}\ot…\ot y^{i}_{k}$ gilt.
            Dies zeigt
            \begin{equation*}
                p_{1}(x_{1})…p_{k}(x_{k})\leq \inf\left\{\displaystyle\sum_{i=1}^{n}p_{1}\big(y^{i}_{1}\big)…p_{k}\big(y^{i}_{k}\big)\right\}=\pi_{\mathbf{p}}(z)
            \end{equation*}
            und somit \eqref{eq:pisepei}.\\\\
            Es bleibt nachzuweisen, dass $(\mathbb{V}_{1}\ot…\ot
            \mathbb{V}_{k},\pi_{k})$ Definition \ref{def:pitop} erfüllt. Sei hier-f"ur $(\mathbb{M},Q)$ lokalkonvex  und $P_{1},\dots,P_{k},Q$ filtrierend gew"ahlt. Sei weiter 
            \begin{equation*}
                \tau\colon
                \mathbb{V}_{1}\ot…\ot\mathbb{V}_{k}\longrightarrow \mathbb{M}
            \end{equation*}
            linear und stetig. Dann folgt mit Satz \ref{satz:stetmultabb} für jede Halbnorm
            $q\in Q$, dass
            \begin{equation*}
                q(\tau(x_{1}\ot…\ot x_{k}))\leq
                c\:\pi_{\mathbf{p}}(x_{1}\ot…\ot
                x_{k})\glna{\eqref{eq:pisepei}}c\:p_{1}(x_{1})…p_{k}(x_{k})
            \end{equation*} für eine besagter Halbnormen $\pi_{\mathbf{p}}\in
            \Pi_{\mathbf{P}}$ gilt. Hierbei haben wir \textit{vi.)} benutzt. Dies ist nach selbigem
            Satz das Stetigkeitskriterium für die $\mathbb{K}$-multilineare Abbildung
            $\tau\cp \ot_{k}\colon\mathbb{V}_{1}\times…\times \mathbb{V}_{k}\longrightarrow \mathbb{M}$, was diese Richtung zeigt.
            
            Für die umgekehrte Implikation sei
            $\tau\cp\ot_{k}$ stetig, dann folgt zunächst für separable Elemente
            \begin{align*}
                q(\tau(x_{1}\ot…\ot x_{k}))=&\:q((\tau\cp
                \ot_{k})(x_{1},…,x_{k}))\leq\: c\:
                p_{1}(x_{1})…p_{k}(x_{k})
            \end{align*} und somit für $z\in
            \mathbb{V}_{1}\ot…\ot \mathbb{V}_{k}$ beliebig, dass
            \begin{align*}
                q(\tau(z))\leq&\:
                \sum_{i=1}^{n}q\left(\tau\big(x^{i}_{1}\ot…\ot
                    x^{i}_{k}\big)\right)
                \leq c\sum_{i=1}^{n}p_{1}\big(x^{i}_{1}\big)…p_{k}\big(x^{i}_{k}\big)
            \end{align*} für alle Zerlegungen von $z$. Dies zeigt 
            \begin{equation*}
                q(\tau(z))\leq c\: \inf\left\{\sum_{i=1}^{n}p_{1}\big(x^{1}_{i}\big)…p_{k}\big(x^{k}_{i}\big)\right\}=c\:\pi_{\mathbf{p}}(z)
            \end{equation*}und somit die Stetigkeit von $\tau$.
        \item
            Angenommen es gäbe eine feinere lokalkonvexe Topologie $\T_{\tilde{P}}$
            auf $\V_{1}\ot…\ot \V_{k}$, bezüglich derer $\ot_{k}$ stetig ist.
            Dann gäbe es ein $\tilde{p}\in \tilde{P}$, welches durch keine
            bezüg-lich $\pi_{k}$ stetige
            Halbnorm abschätzbar wäre, siehe Korollar \ref{kor:HNTop}~\textit{ii.)}. 

            Aus der Stetigkeit von $\ot_{k}$ folgt aber
            $\tilde{p}(x_{1}\ot…\ot x_{k})\leq
            c\:p_{1}(x_{1})…p_{k}(x_{k})$, mithin $\tilde{p}(z)\leq
            c\:\displaystyle\sum_{i=1}^{n}p_{1}(x^{i}_{1})…p_{k}(x^{i}_{k})$ für
            alle Zerlegungen von $z\in \V_{1}\ot…\ot \V_{k}$. Das zeigt $\tilde{p}\leq
            c\:\pi_{\mathbf{p}}$ im Widerspruch zur Annahme.
        \item 
            Ist $(\mathbb{V}_{1}\ot…\ot \mathbb{V}_{k},\pi_{k})$
            hausdorffsch, so ist $\Pi_{\mathbf{P}}$ separierend und
            somit auch jedes $P_{j}$. Hiermit sind alle $(\V_{j},
            P_{j})$ hausdorffsch.

            Für die umgekehrte Richtung reicht es, die
            Separationseigenschaft für $\Pi_{\mathbf{P}}$ nach\-zu\-wei\-sen. Sei
            hierfür $z=\displaystyle\sum_{i=1}^{n}x^{i}_{1}\ot…\ot x^{i}_{k}$ eine Zerlegung
            von $z\in \V_{1}\ot…\ot \V_{k}$ derart, dass die
            $\big\{x^{i}_{j}\big\}_{1\leq i\leq n}$ für jedes
            $1\leq j\leq k$ linear unabhängig sind. Wir betrachten
            die endlichdimensionalen Unterräume
            $L_{j}=\mathrm{span}\left(x^{1}_{j},…,x^{n}_{j}\right)$ und die linearen
            Abbildungen
            \begin{align*}
                u_{j}\colon L_{j}&\longrightarrow \mathbb{K}\\
                x^{i}_{j}&\longmapsto \delta_{1,i}. 
            \end{align*} Da die $\V_{j}$ hausdorffsch sind, finden wir 
            Halbnormen $p_{j}\in P_{j}$ derart, dass die
            Einschränkungen $p_{j}\big|_{L_{j}}$ echte Normen
            sind. Auf endlichdimensionalen normierten $\mathbb{K}$-Vektorräumen sind alle linearen Abbildungen in die
            komplexen Zahlen stetig und es folgt $|u_{j}|\leq
            c_{j}\:p_{j}(u_{j})$ auf dem Unterraum $L_{j}$.

            Mit dem Satz
            von Hahn-Banach erhalten wir lineare Abbildungen $U_{j}:
            \V_{j}\rightarrow \mathbb{K}$ derart, dass
            $U_{j}\big|_{L_{j}}=u_{j}$ und $|U_{j}|\leq c_{j}\:p_{j}$
            auf ganz $\V_{j}$.

            Für die lineare Abbildung $U_{1}\ot…\ot U_{k}$ folgt 
            \begin{align*}
                \big|U_{1}\ot…\ot U_{k}\big|\:(x_{1}\ot…\ot x_{k})=\:\big|U_{1}(x_{1})…U_{k}(x_{k})\big|\leq
                \overbrace{\left[\prod_{j=1}^{k}c_{j}\right]}^{c} p_{1}(x_{1})…p_{k}(x_{k})
            \end{align*} und somit $\big|U_{1}\ot…\ot U_{k}\big|\leq
            c\:\pi_{\mathbf{p}}$ mit $\mathbf{p}=(p_{1},…,p_{k})$. Nun
            gilt per
            Konstruktion, dass $|U_{1}\ot…\ot U_{k}|\:(z)=1$, also folgt insgesamt            
            \begin{equation*}
                \pi_{\mathbf{p}}(z)\geq \frac{1}{c}\: \big|U_{1}\ot…\ot U_{k}\big|\:(z)=\frac{1}{c}>0
            \end{equation*}
            und somit die Behauptung.
        \item
            Sind $P_{1},\dots,P_{k}$ filtrierend und $p_{1}\ot\dots\ot p_{k},p'_{1}\ot\dots\ot p'_{k} \in \prod_{\mathbf{P}}$. Dann finden wir $q_{j}\in P_{j}$ mit $q_{j}\geq p_{j},p'_{j}$ f"ur alle $1\leq j \leq k$. Dann zeigt \eqref{eq:pisepei}, dass
            \begin{align*}
                p_{1}\ot\dots\ot p_{k}\:(x_{1}\ot\dots\ot x_{k})&\leq q_{1}\ot\dots\ot q_{k}\:(x_{1}\ot\dots\ot x_{k})\quad\text{sowie}\\ 
                p'_{1}\ot\dots\ot p'_{k}\:(x_{1}\ot\dots\ot x_{k})&\leq q_{1}\ot\dots\ot q_{k}\:(x_{1}\ot\dots\ot x_{k})
            \end{align*}
            f"ur alle separablen $x_{1}\ot\dots\ot x_{k}\in \V_{1}\pite\dots\pite\V_{k}$.  Hiermit folgt f"ur\\ $z\in \V_{1}\pite \dots\pite\V_{k}$ beliebig, dass
            \begin{align*}
                p_{1}\ot\dots\ot p_{k}\:(z)\leq&\:\sum_{i=1}^{n} p_{1}\ot\dots\ot p_{k}\left(x^{i}_{1}\ot\dots\ot x^{i}_{k}\right)
                \\\leq&\:\sum_{i=1}^{n} q_{1}\ot\dots\ot q_{k}\left(x^{i}_{1}\ot\dots\ot x^{i}_{k}\right)
            \end{align*} f"ur alle Zerlegungen $z=\displaystyle\sum_{i=1}^{n}x^{i}_{1}\ot\dots\ot x^{i}_{k}$. Dies zeigt 
            \begin{equation*}
                p_{1}\ot\dots\ot p_{k}\:(z)\leq q_{1}\ot\dots\ot q_{k}\:(z)
            \end{equation*}
            und analog folgt $p'_{1}\ot\dots\ot p'_{k}\:(z) \leq q_{1}\ot\dots\ot q_{k}\:(z)$.
        \item 
            Sind alle $(\mathbb{V}_{j},P_{j})$ hausdorffsch, so  mit \textit{v.)} auch
            $\V_{1}\pite…\pite \V_{k}$ und dessen
            Vervollständigung existiert. Angenommen wir hätten die
            Aussage für $k-1$ bereits gezeigt, so folgte 
            \begin{align*} 
                \widehat{\V_{1}\pite…\pite
                  \V_{k}}=&\widehat{\Big[\left(\V_{1}\pite…\pite
                      \V_{k-1}\right)\pite \V_{k}\Big]}
                =\widehat{\Big[\widehat{\left(\V_{1}\pite…\pite
                        \V_{k-1}\right)}\pite \hat{\V}_{k}\Big]}
                \\=&\widehat{\left[\left(\hat{\V}_{1}\pite…\pite\hat{\V}_{k-1}\right)\pite \hat{\V}_{k}\right]}
                =\:\widehat{\hat{\V}_{1}\pite…\pite\hat{\V}_{k}},
            \end{align*} wobei die Gleichheitszeichen lineare
            Homöomorphien bedeuten. In der Tat sind 
            $\V_{1}\pite…\pite \V_{k}$ und $(\V_{1}\pite…\pite
            \V_{k-1})\pite \V_{k}$ als $\mathbb{K}$-Vektorräume
            nach
            Lemma \ref{lemma:assTenprod} isomorph. Sei $\cong$ besagter
            Isomorphismus in der ersten Gleichheit, dann folgt diese mit
            \begin{align*}
                \pi_{p_{1},…,p_{k-1}}\ot p_{k}\:(\:\cong(x_{1}\pite…\pite
                x_{k}))=&\:\pi_{p_{1},…,p_{k-1}}\ot
                p_{k}\:((x_{1}\pite…\pite x_{k-1})\pite x_{k})
                \\=&\:\pi_{p_{1},…,p_{k}}(x_{1}\pite…\pite x_{k}),
            \end{align*} 
            da dann $\cong$ eine Isometrie mit der offensichtlichen
            Zuordnung der Halbnormen ist und somit Satz \ref{satz:vervollsthlkVR}
            die Homöomorphie beider Vervollständigungen zeigt. Die
            zweite und dritte die Gleichheit liefert die
            Induktionsvoraussetzung und die letzte folgt wie die erste.
            
            Um die Behauptung für $k=2$ zu zeigen, reicht es nach
            Satz \ref{satz:vervollsthlkVR} die Existenz einer
            Isometrie 
            \begin{equation*}
                i:\V_{1}\pite \V_{2}\longrightarrow
                \widehat{\hat{\V}_{1}\pite \hat{\V}_{2}}
            \end{equation*}mit $\overline{i\left(\V_{1}\pite
                  \V_{2}\right)}=\widehat{\hat{\V}_{1}\pite \hat{\V}_{2}}$
            nachzuweisen. Hierfür definieren wir 
            \begin{align*}
                j_{X}:\V_{1}\pite \V_{2}&\longrightarrow \hat{\V}_{1}\pite
                \hat{\V}_{2}\\
                x\pite y&\longmapsto \big[\{x\}\big]\pite \big[\{y\}\big]
            \end{align*} durch Korollar \ref{kor:WohldefTensorprodabbildungen}
            mit $\big[\{x\}\big]$ die
            Äquivalenzklasse der konstanten Folge\\ $\{x,x,x,…\}$ und setzen
            \begin{align*}
                &i_{X}:z\longmapsto \big[\{ j_{X}(z)\}\big]\\
                &i_{P}:p_{1}\ot p_{2}\longmapsto \widehat{\hat{p}_{1}\ot \hat{p}_{2}}.
            \end{align*} 
            Für die Isometrieeigenschaft rechnen wir mit $z\in
            \V_{1}\pite \V_{2}$:
            {\allowdisplaybreaks
              \begin{align*}
                  \widehat{\hat{p}_{1}\ot
                    \hat{p}_{2}}\Big(\big[\{j_{X}(z)\}\big]\Big)=&\:\hat{p}_{1}\ot
                  \hat{p}_{2}\:\big(j_{X}(z)\big)
                  =\:\inf\left(\sum_{i}\hat{p}_{1}\Big(\big[\{x^{i}\}\big]\Big)\hat{p}_{2}\Big(\big[\{y^{i}\}\big]\Big)\right)
                  \\=&\:\inf\left(\sum_{i}p_{1}\big(x^{i}\big)p_{2}\big(y^{i}\big)\right)
                  =\:p_{1}\ot p_{2}\:(z).
              \end{align*}}Es bleibt zu zeigen, dass $\overline{i\left(\V_{1}\pite
                  \V_{2}\right)}=\widehat{\hat{\V}_{1}\pite
              \hat{\V}_{2}}$ gilt. Dabei ist per Definition des topologischen
            Abschlusses bereits klar, dass $\overline{i\left(\V_{1}\pite
                  \V_{2}\right)}\subseteq\widehat{\hat{\V}_{1}\pite
              \hat{\V}_{2}}$ ist.

            Für die umgekehrte Inklusion sei
            $z=\big[\{z_{\alpha}\}_{\alpha\in I}\big]\in \widehat{\hat{\V}_{1}\pite
              \hat{\V}_{2}}$ und wir müssen nachweisen, dass
            $B_{\widehat{\hat{p}_{1}\ot
                \hat{p}_{2}},\epsilon}\:(z)\cap i\left(\V_{1}\pite
                \V_{2}\right)\neq \emptyset$ für jeden solchen
            Ball um $z$
            
            erfüllt ist.
            Nun finden wir für $\epsilon-\Delta>0$ mit $\Delta>0$ ein
            $\alpha_{\frac{\epsilon}{3}}\in I$ derart, dass
            \begin{equation*}
                \hat{p}_{1}\ot
                \hat{p}_{2}\:(z_{\alpha}-z_{\beta})<\frac{\epsilon-\Delta}{3}\qquad\qquad
                \forall\:\alpha,\beta\geq \alpha_{\frac{\epsilon}{3}}\in I,
            \end{equation*} womit
            \begin{equation*}
                \widehat{\hat{p}_{1}\ot
                  \hat{p}_{2}}\left(\left[\net{z}{I}\right]-\big[\{z_{\alpha'}\}\big]\right)\leq\frac{\epsilon-\Delta}{3}<\frac{\epsilon}{3}\qquad\quad
                \forall\:\alpha'\geq\alpha_{\frac{\epsilon}{3}}\in I.
            \end{equation*} Nun ist $z_{\alpha'}\in \hat{\V}_{1}\pite
            \hat{\V}_{2}$, also
            $z_{\alpha'}=\displaystyle\sum_{i=1}^{n}\big[\{x^{i}_{\alpha_{i}}\}_{\alpha_{i}\in
              J_{i}}\big]\pite \big[\{y^{i}_{\beta_{i}}\}_{\beta_{i}\in J'_{i}}\big]$
            und folglich
            \begin{align*}
                p_{1}\left(x^{i}_{\gamma}-x^{i}_{\delta}\right)&<\frac{\epsilon-\Delta}{\hat{p}_{2}\big(\big[\{y^{i}_{\beta_{i}}\}_{\beta_{i}\in
                    J'_{i}}\big]\big)3n}\quad\quad
                \forall\:\gamma,\delta\geq \alpha_{i}'\in J_{i}\\
                p_{2}\left(y^{i}_{\gamma}-y^{i}_{\delta}\right)&<\frac{\epsilon-\Delta}{\hat{p}_{1}\left(\left[\{x^{i}_{\alpha_{i}}\}_{\alpha_{i}\in
                            J_{i}}\right]\right)3n}\quad\quad
                \forall\:\gamma,\delta\geq \beta_{i}'\in J'_{i},
            \end{align*}
            sowie
            \begin{align*}
                \hat{p}_{1}\left(\big[\{x^{i}_{\alpha_{i}}\}_{\alpha_{i}\in
                      J_{i}}\big]-\big[\{x^{i}_{\alpha_{i}'}\}\big]\right)&<
                \frac{\epsilon}{\hat{p}_{2}\big(\big[\{y^{i}_{\beta_{i}}\}_{\beta_{i}\in
                    J'_{i}}\big]\big)3n}\\
                \hat{p}_{2}\left(\big[\{y^{i}_{\beta_{i}}\}_{\beta_{i}\in
                      J'_{i}}\big]-\big[\{y^{i}_{\beta_{i}'}\}\big]\right)&< \frac{\epsilon}{\hat{p}_{1}\left(\left[\{x^{i}_{\alpha_{i}}\}_{\alpha_{i}\in
                            J_{i}}\right]\right)3n}.
            \end{align*} Dann ist
            $\tau=\displaystyle\sum_{i=1}^{n}x^{i}_{\alpha_{i}'}\pite
            y^{i}_{\beta_{i}'}\in \V_{1}\pite \V_{2}$ mit $i(\tau)=\left[\left\{\displaystyle\sum_{i=1}^{n}\Big[\big\{x^{i}_{\alpha_{i}'}\big\}\Big]\pite
                    \left[\big\{y^{i}_{\beta_{i}'}\big\}\right]\right\}\right]$ und wir
            erhalten
            {\allowdisplaybreaks\begin{align*}
                  \widehat{\hat{p}_{1}\ot 
                    \hat{p}_{2}}\:(z-i(\tau))\leq& \:\widehat{\hat{p}_{1}\ot
                    \hat{p}_{2}}\:\left(z-\big[\{z_{\alpha'}\}\big]\right)\\ &+\widehat{\hat{p}_{1}\ot
                    \hat{p}_{2}}\:\left(\big[\{z_{\alpha'}\}\big]-\left[\left\{\sum_{i=1}^{n}\big[\{x^{i}_{\alpha_{i}}\}_{\alpha_{i}\in
                                J_{i}}\big]\pite
                              \left[\big\{y^{i}_{\beta_{i}'}\big\}\right]\right\}\right]\right)
                  \\ &+\widehat{\hat{p}_{1}\ot\hat{p}_{2}}\:\left(\left[\left\{\sum_{i=1}^{n}\big[\{x^{i}_{\alpha_{i}}\}_{\alpha_{i}\in
                                J_{i}}\big]\pite
                              \left[\{y^{i}_{\beta_{i}'}\}\right]\right\}\right]-i(\tau)\right)\displaybreak
                  \\<&\:\frac{\epsilon}{3}+ \hat{p}_{1}\ot
                  \hat{p}_{2}\:\left(\sum_{i=1}^{n}\left[\{x^{i}_{\alpha_{i}}\}_{\alpha_{i}\in
                            J_{i}}\right]\pite \left[\{y^{i}_{\beta_{i}}\}_{\beta_{i}\in
                            J'_{i}}-\{y^{i}_{\beta_{i}'}\}\right]\right)
                  \\ &+\hat{p}_{1}\ot
                  \hat{p}_{2}\:\left(\sum_{i=1}^{n}\left[\{x^{i}_{\alpha_{i}}\}_{\alpha_{i}\in
                            J_{i}}-\{x^{i}_{\alpha_{i}'}\}\right]\pite
                      \left[\{y^{i}_{\beta_{i}'}\}\right]\right)
                  \\<&\:\frac{\epsilon}{3}+\sum_{i=1}^{n}\hat{p}_{1}\left(\left[\{x^{i}_{\alpha_{i}}\}_{\alpha_{i}\in
                            J_{i}}\right]\right)\frac{\epsilon}{\hat{p}_{1}\left(\left[\{x^{i}_{\alpha_{i}}\}_{\alpha_{i}\in
                              J_{i}}\right]\right)3n}
                  \\ &+ \sum_{i=1}^{n}\frac{\epsilon}{\hat{p}_{2}\left(\left[\{y^{i}_{\beta_{i}}\}_{\beta_{i}\in
                              J'_{i}}\right]\right)3n}\hat{p}_{2}\left(\left[\{y^{i}_{\beta_{i}}\}_{\beta_{i}\in
                            J'_{i}}\right]\right)
                  \\=&\:\epsilon,
              \end{align*}}wobei wir in der letzten Ungleichheit die Halbnormeigenschaft von $\hat{p}_{1}\ot
            \hat{p}_{2}$ benutzt haben.
            Dies zeigt $i(\tau)\in B_{\widehat{\hat{p}_{1}\ot
                \hat{p}_{2}},\epsilon}\:(z)\cap i\left(\V_{1}\pite
                \V_{2}\right)\neq \emptyset$
            und somit die Behauptung.
        \end{enumerate} 
    \end{beweis}
\end{satz}

%
%

 \cleardoublepage
\backmatter
\bibliographystyle{alpha} 
\bibliography{dqarticle,dqbook,dqprocentry,dqproceeding,dqKlops}
\thispagestyle{empty}

\cleardoublepage
\chapter*{Danksagung}
\thispagestyle{empty}
Danken möchte ich meiner Mutter, die mir im Laufe dieses Studiums weit mehr
als nur moralischen Beistand geleistet hat. Danken möchte ich auch
allen anderen, die während meiner Zeit in Freiburg unerschütterlich an
meiner Seite gestanden haben und hier besonders
Michael Schulze und Iosas Schindler, die mich motivierten, das Wagnis dieses
Studiums einzugehen. 

Vielen Dank an Stefan Waldmann für seine freundliche und
kompetente Betreuung und dafür, dass er mich in seiner
Abteilung aufgenommen hat. Großen Dank an Nikolai
Neumaier für seine
exzelenten Vorlesungen, durch die ich den Weg in die mathematische
Physik gefunden habe.

Herzlichen Dank an meine Mitstreiter Dominic
Maier, Patricia Calon, Jan Paki, Alexander Held, Gregor Schaumann, Torsten Kirk,
Alexander Huss und
Markus Hecht für ihre Korrekturarbeiten und die sowohl technische als auch moralische
Unterstützung und dafür, dass sie meine Späße ertragen haben. 

\end{document}